\documentclass[a4paper,11pt]{article}
\pdfoutput=1

\usepackage{jheppub}
\usepackage[T1]{fontenc}
\usepackage{slashed}
\usepackage[utf8]{inputenc}
\usepackage{array}
\usepackage{wrapfig}
\usepackage{multirow}
\usepackage{tabu}
\usepackage{amsmath,amssymb,amsthm}
\usepackage{enumitem}
\usepackage{graphicx}
\usepackage{caption}
\usepackage{placeins}
\usepackage{longtable}
\usepackage{float}
\usepackage{amsfonts}

\title{\boldmath A forest formula to subtract infrared singularities in amplitudes for wide-angle scattering}

\author{Yao Ma}

\affiliation[]{C.N. Yang Institute for Theoretical Physics and Department of Physics and Astronomy,\\Stony Brook University, Stony Brook, 11794, NY, USA}

\emailAdd{yao.ma@stonybrook.edu}

\abstract{For any hard QCD amplitude with massless partons, infrared (IR) singularities arise from pinches in the complex planes of loop momenta, called pinch surfaces. To organize and study their leading behaviors in the neighborhoods of these surfaces, we can construct approximation operators for collinear and soft singularities. A BPHZ-like forest formula can be developed to subtract them systematically.

In this paper, we utilize the position-space analysis of Erdo$\breve{\text{g}}$an and Sterman for Green functions, and develop the formalism for momentum space. A related analysis has been carried out by Collins for the Sudakov form factors, and is generalized here to any wide-angle kinematics with an arbitrary number of external momenta. We will first see that the approximations yield much richer IR structures than those of an original amplitude, then construct the forest formula and prove that all the singularities appearing in its subtraction terms cancel pairwise. With the help of the forest formula, the full amplitude can also be reorganized into a factorized expression, which helps to generalize the Sudakov form factor result to arbitrary numbers of external momenta. All our analysis will be on the amplitude level.}


\arxivnumber{1910.11304}

\begin{document}
\begin{flushright}
YITP-SB-19-34
\end{flushright}
\today
\maketitle
\flushbottom

\section{Introduction}
\label{introduction}

The use of forest-like structure of subtractions to remove singularities has inspired research since it was formulated from Bogoliubov's $R$-operation \cite{BglPrs57,Hepp66} for ultraviolet (UV) divergences by Zimmermann \cite{Zmm69}. The BPHZ formalism treats nested and overlapping divergences by a set of nested forests of subtractions. Later, the BPHZ theorem was generalized to include massless fields by Lowenstein and Zimmermann \cite{LwtZmm75_1,LwtZmm75_2}, and also to include Euclidean infrared (IR) singularities by Chetyrkin, Tkachov and Smirnov \cite{CtkTch82,CtkSmn84}. Beyond this, a Hopf algebraic structure has been discovered by Kreimer \cite{Krm97, BrkKrm15}, and its mathematical structures shed light on quantum field theory.

In comparison with the extensions mentioned above, our work concentrates on the forest-like treatment for IR divergences in Minkowski spacetime, with the subtraction terms motivated from the factorization theorems. This treatment remains under study because of the complex structures of pinch surfaces in Minkowski space \cite{Stm78I, LbyStm78}, on which much previous work has centered. Long ago, Humpert and van Neerven discussed the analogy between multiplicative BPHZ renormalization and mass factorization \cite{HptvNv81}, when they used a graphical method to achieve an alternative proof of the factorization of collinear singularities, with the factorized parts being the subtraction terms. Soon afterwards, Collins and Soper focused on the Drell-Yan process in the ``back-to-back'' limit \cite{ClsSpr81}. Working in axial gauge in that well-known paper, they used a ``botanical construction'' with concepts ``gardens'' and ``tulips'' to disentangle the nested and overlapping IR divergences. Later in Collins' book \cite{Cls11book}, he developed a forest formula in Feynman gauge for Drell-Yan and related processes, where there are two back-to-back external particles. An all-order factorization discussion has been given long ago in axial gauge for wide-angle scattering with color exchange by Sen \cite{Sen1983}, and more recently by Feige and Schwartz, using a ``factorization gauge'' \cite{FgeSwtz14}.

In a related work, Erdo$\breve{\text{g}}$an and Sterman have applied the forest formula to subtract the UV divergences for massless gauge theories in position space \cite{EdgStm15}, for arbitrary wide-angle kinematics. Based on these pioneering works, our paper aims to provide a generalization to multi-particle amplitudes. As we will see, to carry out the analysis in momentum space and Feynman gauge does not simply involve a Fourier transformation; rather, subtleties will arise due to the complexity of IR structure in the forest subtractions.

This complexity originates from the nontriviality of IR singularities of QCD amplitudes. They are described by the Landau equations, the solutions of which define pinch surfaces, a set of classical pictures with a combination of collinear and soft divergences. The pinch surfaces of an amplitude can be obtained from the Coleman-Norton interpretation \cite{ClmNtn65}. In more detail, each pinch surface consists of the hard, jet and soft subgraphs intertwining with each other. The short-distance interactions are encoded in the hard subgraph $H$, while the long-distance interactions are encoded in the jet subgraph $J$ and soft subgraph $S$. To evaluate the contribution of a pinch surface $\sigma$, we distinguish between its \emph{internal coordinates} and those transverse to $\sigma$, which are called \emph{normal coordinates}. By studying the behavior of the graph near $\sigma$ through the power counting technique of \cite{Stm78I, LbyStm78}, we can identify the IR divergent pinch surfaces. For the amplitudes studied here and many other QCD processes, the result is that the divergences are at worst logarithmic, when the following three requirements are satisfied on the pinch surface \cite{Stm78I, Cls11book, Stm95book}.
\begin{itemize}
\item[]{$\mathfrak{1}$. }A soft parton cannot be attached to the hard subgraph.
\end{itemize}
\begin{itemize}
\item[]{$\mathfrak{2}$. }A soft fermion or scalar cannot be attached to the jet subgraph.
\end{itemize}
\begin{itemize}
\item[]{$\mathfrak{3}$. }In each jet subgraph $J_I$, the full set of partons attached to a connected component of $H$ is made up of exactly one parton with physical polarization, and all others being scalar-polarized gauge bosons.
\end{itemize}
Strictly speaking, these requirements are not sufficient for an IR divergence. Imagine a pinch surface with these requirements satisfied, one of whose jets has only one internal vertex, to which two soft propagators are attached. Following the power counting procedure, we will find a suppression of the logarithmic divergence. On the other hand, a pinch surface would also be IR divergent without the third requirement satisfied. Namely, all the propagators of a jet that are attached to the hard subgraph are scalar-polarized gauge bosons \cite{Cls11book, LbtdStm85}. Such pinch surfaces are power divergent, giving a ``super-leading'' contribution \cite{Cls11book}. But if we sum over the graphs representing different attachments of the collinear gluons to the hard subgraph, they will vanish due to the Ward identity. So we do not treat them separately, and will regard the requirements $\mathfrak{1}-\mathfrak{3}$ as necessary conditions for an IR divergence.

\begin{figure}[t]
\centering
\includegraphics[width=14cm]{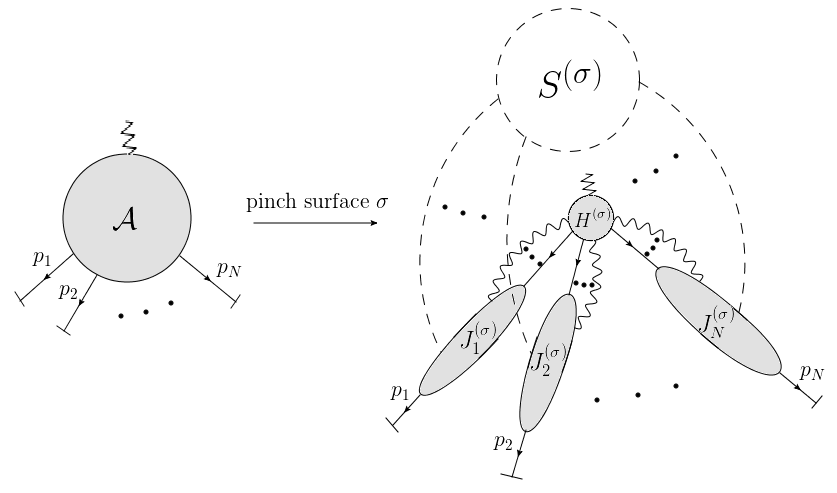}
\caption{Leading pinch surfaces of the decay process.}
\label{leading_pinch_surface}
\end{figure}

We call a pinch surface meeting the requirements $\mathfrak{1}-\mathfrak{3}$ above a \emph{leading pinch surface}. For a decay process with $n$ outgoing particles, for example, a leading pinch surface $\sigma$ can be graphically represented as the RHS of figure\ \ref{leading_pinch_surface}, where we have used $(\sigma)$ in superscripts to denote the subgraphs. The set of leading pinch surfaces of an amplitude $\mathcal{A}$ includes all its IR divergences, which are not cancelled in the sum over the gauge-invariant sets of graphs.

For each leading pinch surface $\sigma$, we will define an \emph{approximation operator} $t_\sigma$ such that $\mathcal{A} \mid_{\text{div.}\sigma} = t_\sigma\mathcal{A} \mid_{\text{div.}\sigma}$. That is, in region $\sigma$ the divergences of $t_\sigma\mathcal{A}$ are the same as these of $\mathcal{A}$. Approximation operators that correspond to nested pinch surfaces can also act on $\mathcal{A}$ repetitively as $t_{\sigma_n} ... t_{\sigma_1}\mathcal{A}$, where $\sigma_1 \subset...\subset \sigma_n$. With the help of the approximation operators, we will be able to construct the forest formula, which schematically reads:
\begin{eqnarray} \label{forest_formula_amplitude}
\left [ \sum_{F\in \mathcal{F}\left [ \mathcal{A} \right ]}^{ }\prod_{\sigma\in F}^{ }\left( -t_{\sigma} \right)\mathcal{A} \right ]_{\textup{div}}=0.
\end{eqnarray}
That is, after summing over all the ``forests'' $F$ in $\mathcal{F}\left [ \mathcal{A} \right ]$, each of which corresponds to a set of approximations acting repetitively on $\mathcal{A}$, all the IR divergences that may appear in any of the terms are cancelled. Note that these IR divergences include not only the ones from the original leading pinch surfaces, but also those of the subtraction terms, which are highly nontrivial and require a detailed discussion. But in the end, we will find that all these ``induced'' divergences form pairs to cancel each other, and are organized as the divergences along eikonal lines, which appear in a factorized expression of the full amplitude. The notations in eq.\ (\ref{forest_formula_amplitude}) will be explained in more detail in the following sections.

Besides the forest-like structure to subtract IR singularities, additional methods have been developed to separate the IR divergent parts from the finite parts in a Feynman integral, especially for the next-to-leading order (NLO) and the next-to-next-to-leading order (NNLO) cross sections. Some notable works define subtraction terms on the level of integrands, such as the Catani-Seymour method \cite{CtnSmr96, CtnSmr97}, the Nagy-Soper method \cite{NgySpr03}, the antenna method \cite{GdrGrmGlv05, GdrGrmGlvHrch07, CrrGlvWls13}, in addition to Refs. \cite{CtnGzn00, SmgTsyDDc05, SmgTsyDDc07, BznSmgTsy11}. Alternative ways are to define subtraction terms on the level of integration measures \cite{BnthHrch04, AntsMnkPtrl04, Czk10, ClMnkRsch17, Hzg18}, or the products of integrand and measure, like the Frixione-Kunszt-Signer method \cite{FrxKstSgn96, FrdrFrxMtnStz09}. These works mainly focus on the practical evaluations of multi-loop or multi-particle Feynman integrals up to certain orders, but suggest that local IR subtractions can regularize an arbitrary amplitude in momentum space. Our project here, therefore, aims to provide such an all-order IR subtraction procedure.

Most of our calculations and discussions in this paper center on eq.\ (\ref{forest_formula_amplitude}). Sections \ref{pinch_surface_amplitudes_approximations}--\ref{the_proof_of_cancellation} establish the validity of this formula. In section \ref{pinch_surface_amplitudes_approximations} we introduce the approximation operators, and study the IR singularities generated by them, which may not exist in the original amplitude $\mathcal{A}$. The approximation operators help to motivate the idea to cancel nested divergences. As for the cancellation of non-nested or ``overlapping'' divergences, the concept of enclosed pinch surfaces is introduced in section \ref{enclosed_pinch_surface}, which will be shown to be a leading pinch surface of $\mathcal{A}$. With the knowledge acquired, section \ref{the_proof_of_cancellation} then serves as a proof of the pairwise cancellations in eq.\ (\ref{forest_formula_amplitude}), by focusing on every IR divergent regions. Section \ref{factorization_subtraction_terms} is mainly concerned with the application of the forest formula. That is, we apply the factorization theorem to the subtraction terms, and show that with the help of the gauge theory Ward identity, the full amplitude can be written into a factorized expression. To visualize how these theories work, section \ref{NNLO_examples} offers examples on the $\mathcal{O}(\alpha^2)$-level. Finally, a summary and an outlook will be provided in section \ref{summary_outlook}.

Some detailed analyses are presented in the appendices. Explicitly, some details of the power counting evaluations in section \ref{divergences_are_logarithmic} are shown in appendix \ref{power_counting_repetitive_approximation}, some details of the analysis in section \ref{subgraphs_enclosed_pinch_surfaces} are shown in appendix \ref{enclosed_PS_details}, and a brief sketch of sections \ref{pinch_surface_amplitudes_approximations}--\ref{the_proof_of_cancellation}, from the position-space point of view, is provided in appendix \ref{interpretation_position_space}.

\section{Pinch surfaces of amplitudes and their approximations}
\label{pinch_surface_amplitudes_approximations}

A QCD amplitude $\mathcal{A}$ can be represented by an integral over its loop momenta, and we can use the Landau equations to identify all its pinch surfaces. For each of them, in order to study the asymptotic behavior in its neighborhood, it is natural to apply approximations to the integrand. Namely, we apply hard-collinear approximations on the jet momenta appearing the hard subgraph, and soft-collinear approximation on the soft momenta appearing in the jet subgraph. These approximations, as will be shown in eqs.\ (\ref{hard-collinear_approximation}) and (\ref{soft-collinear_approximation}), are close to those in \cite{Cls11book, ClsSprStm04}.\footnote{The approximations are similar to the those defined in Chapter 10.4.2 of \cite{Cls11book}, and equivalent to eqs.\ (147) and (158) in the review \cite{ClsSprStm04}.} They are also closely related to the expansions in Soft-Collinear Effective Theory (SCET) \cite{BurFlmLk00, BurPjlSwt02-1, BurPjlSwt02-2, BchBrgFrl15book}.

Since the integrand is changed while the integration measure is intact after the approximations, we expect the pinch surfaces of an approximated amplitude to be distinct from those of $\mathcal{A}$ in general. So a systematic enumeration of them is necessary.

This section is arranged as follows. In section\ \ref{neighborhoods & approximation_operators} we study the relations between pinch surfaces, their neighborhoods, and introduce approximation operators. Then the configurations that may appear in a pinch surface of $t_\sigma \mathcal{A}$ are discussed in section\ \ref{pinch_surfaces_from_single_approximation}. The results are generalized to $t_{\sigma_n}...t_{\sigma_1} \mathcal{A}$, after the rules for repetitive approximations are verified in section\ \ref{subtraction_terms}. With the whole zoo of IR divergences of an approximated amplitude in place, their degrees of divergence are evaluated in section\ \ref{divergences_are_logarithmic}, and are shown to be at worst logarithmic.

\subsection{Neighborhoods and approximation operators} \label{neighborhoods & approximation_operators}

Consider a graph with $N$ external lines of momentum $p_A$, with $A=1,...,N$. In this study we assume all the invariants $p_I\cdot p_K\ (I\neq K)$ are of the same order, $Q^2$. We start from formalizing the normal coordinates, which are normal to the specified pinch surface, and the intrinsic coordinates of a pinch surface $\sigma$, as are defined in \cite{Stm78I, Cls11book}. To be specific, suppose that $\left\{ q^\mu \right\}$ is the set of hard loop momenta, $\left\{ l^\mu \right\}$ the set of soft loop momenta, and $\left\{ k_A^\mu \right\}$ the set of a jet loop momenta in the direction of $\beta_A^\mu$ (a lightlike vector $\frac{1}{\sqrt{2}} (1, \widehat{\mathbf{v}}_A)$, with $A=1,...,N$), then the sets of normal and intrinsic coordinates of $\sigma$ are defined as
\begin{align} \label{normal_intrinsic_coordinates}
\begin{split}
\mathbb{N}_\sigma\equiv & \bigcup_{A=1}^{N}\left\{ k_A\cdot \beta_A,\ \frac{\left( k_A\cdot \beta_{A\perp} \right)^2}{k_A\cdot \overline{\beta}_A} \right\}\bigcup \left\{ l^\mu \right\}, \\
\mathbb{I}_\sigma\equiv & \bigcup_{A=1}^{N}\left\{ k_A\cdot \overline{\beta}_A \right\}\bigcup \left\{ q^\mu \right\}.
\end{split}
\end{align}
For each $\beta_A^\mu$, we define $\overline{\beta}_A^\mu \equiv \frac{1}{\sqrt{2}} (1, -\widehat{\mathbf{v}}_A)$, so that $\beta_A \cdot \overline{\beta}_A =1$.

To study the behavior of an amplitude $\mathcal{A}$ near the pinch surface, we scale the normal coordinates with $\lambda$ ($\ll 1$). Namely,
\begin{eqnarray} \label{momenta_scaling}
&&l^\mu \sim \left( \lambda, \lambda, \lambda, \lambda \right)Q, \nonumber\\
&&k_A^\mu = \left(k_A\cdot \overline{\beta}_A,\ k_A\cdot \beta_A,\ k_A\cdot \beta_{A\perp}\right) \sim \left(1,\lambda,\lambda^{1/2}\right)Q, \\
&&q^\mu \sim \left(1,1,1,1\right)Q. \nonumber
\end{eqnarray}
Note that by contour deformation, we can prove that there are no Glauber regions for wide-angle scatterings \cite{Sen1983, Cls11book, ClsStm81, BdwBskLpg81, Zeng15}, so eq.\ (\ref{momenta_scaling}) shows the unique way the soft momenta are scaled. We assume that the numerators and denominators of the integrand are polynomials in normal coordinates. As the normal coordinates are scaled as above near the pinch surface, each such polynomial can be approximated by keeping only the leading terms, which are isolated by the hard-collinear and soft-collinear approximations. These approximations act on the hard subgraph $H^{(\sigma)}$ and the jet subgraph $J_A^{(\sigma)}\ (A=1,...,N)$, as
\begin{eqnarray}
\label{hard-collinear_approximation}
H^{(\sigma)}\left( p_A^\mu-\sum_{i}^{ }k_i^\mu, \left \{ k_i^{\alpha_i} \right \} \right)^{\left \{ \mu_i \right \}}_\eta \xrightarrow[ ]{\text{hc}_A} && H^{(\sigma)} \left( \Big( ( p_A-\sum_{i}^{ }k_i ) \cdot\overline{\beta}_A \Big) \beta_A^\mu, \left \{ \left( k_i\cdot \overline{\beta}_A \right)\beta_A^{\alpha_i} \right \} \right)_{\left \{ \nu_i \right \},\eta}\nonumber\\
\cdot&& \prod_{j}^{ }\beta_A^{\nu_j}\overline{\beta}_A^{\mu_j} \cdot \left\{\begin{matrix}
\frac{1}{2}\left( \gamma\cdot\beta_A \right) \left( \gamma\cdot\overline{\beta}_A \right)\ \ \ \ \ \ \ \text{fermion line,}\\ 
\ \ \ \ \ \ \ \ \ \ \ 1\text{\ \ \ \ \ \ \ \ \ \ \ \ \ \ vector or scalar.}
\end{matrix}\right.\nonumber\\
\\
J_A^{(\sigma)} \left( \left\{ l_i^{\alpha_i} \right\} \right)_{\eta}^{ \left\{ \mu_i \right \}} \xrightarrow[ ]{\text{sc}_A} && J_A^{(\sigma)} \left( \left \{ \left (l_i\cdot\beta_A \right) \overline{\beta}_A^{\alpha_i} \right \} \right)_{\left \{ \nu_i \right \},\eta}\prod_{j}^{ }\overline{\beta}_A^{ \nu_j }\beta_A^{\mu_j}.
\label{soft-collinear_approximation}
\end{eqnarray}
Note that the hard function $H^{(\sigma)}$ is shown with only one jet's momenta, where the arguments of $H^{(\sigma)}$ are a set of momenta for each jet. In eqs.\ (\ref{hard-collinear_approximation}) and (\ref{soft-collinear_approximation}) we have $\left(p_A^\mu-\sum_{i}^{ }k_i^\mu\right)$ and $\eta$ as the momentum and the polarization index of the physical parton respectively, $k_i^\mu$'s as the momenta of the scalar-polarized gauge bosons and $l_i^\mu$'s as the momenta of the soft gauge bosons. The $\mu_i$'s are the polarization indices of the scalar-polarized gauge bosons in (\ref{hard-collinear_approximation}) and those of the soft gauge bosons in (\ref{soft-collinear_approximation}). In the hard-collinear approximation (denoted by $\text{hc}_A$ here), the jet momenta entering the hard subgraph are projected onto the directions of the jets, while the vector indices are projected onto the opposite directions of the jets. Moreover, for each fermion jet propagator attached to the hard subgraph, we insert the operator $\frac{1}{2}\left( \gamma\cdot\beta \right)\left( \gamma\cdot\overline{\beta} \right)$ where $\left( \gamma\cdot\beta \right)$ is next to the hard subgraph, to project on the spinor space which gives the leading power in the Dirac traces. In the soft-collinear approximations (denoted by $\text{sc}_A$ here), the projections on the momenta work in the reversed way compared to the hard-collinear approximations.

To be more specific, the typical denominators become under these approximations
\begin{eqnarray}
\label{approximated_denominators}
S: && l^2 \rightarrow l^2 \sim \mathcal{O}(\lambda^2), \nonumber \\ 
J_A: && \left( k_A+l \right)^2 \xrightarrow[ ]{\text{sc}_A} k_A^2+2\left( k_A\cdot \overline{\beta}_A \right)\left( l\cdot\beta_A \right) \sim \mathcal{O}(\lambda), \\ 
H: && \left( q+k_A \right)^2 \xrightarrow[ ]{\text{hc}_A} q^2+2\left( q\cdot\beta_A \right)\left( k_A\cdot \overline{\beta}_A \right) \sim \mathcal{O}(\lambda^0), \nonumber
\end{eqnarray}
where in the second line $k_A^\mu$ is a sum of jet loop momentum and $l^\mu$ is soft. In the third line $q^\mu$ is a hard momentum, of order $\lambda^0$ in all components.

The approximation operators are projections, so that
\begin{eqnarray}\label{projection_identity}
t_\sigma^2=t_\sigma
\end{eqnarray}
holds for any $\sigma$. Regarding these approximations, we shall introduce some convenient notation. First, we define
\begin{align} \label{hat_tilde_definition}
\begin{split}
&^{I}\widehat{k}^{\mu}\equiv \left (k\cdot\overline{\beta}_I \right) \beta_I^\mu,\\
&^{I}\widetilde{k}^{\mu}\equiv \left (k\cdot\beta_I \right) \overline{\beta}_I^{\mu},
\end{split}
\end{align}
representing the projected $I$-th jet momenta that appear in the hard part, and soft momenta in the $I$-th jet part, respectively. For simplicity, whenever the jet label $I$ is unambiguous, we shall use $\widehat{k}^\mu$ and $\widetilde{k}^\mu$.

\begin{figure}[t]
\centering
\includegraphics[width=10cm]{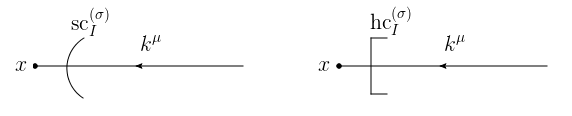}
\caption{Pictorial representations for the soft-collinear and hard-collinear approximations provided by $t_\sigma$. After being projected by the approximations, the momenta flowing into the vertex $x$ are $\left (k\cdot\beta_I \right) \overline{\beta}_I^{\mu}$ and $\left (k\cdot\overline{\beta}_I \right) \beta_I^\mu$ respectively.}
\label{convention_sc_hc}
\end{figure}

Since a propagator can belong to different subgraphs of different pinch surfaces (for example, a hard propagator in one pinch surface may be lightlike or soft in another), we will put the pinch surface in a bracket as an upper index of the subgraphs as in figure\ \ref{leading_pinch_surface}. For example, $H^{(\sigma)}$ refers to the hard subgraphs of $\sigma$, which may be no longer hard in other pinch surfaces. We will also use the notations $\text{sc}_I^{(\sigma)}$ and $\text{hc}_I^{(\sigma)}$ to denote soft-collinear or hard-collinear approximations in a given $t_\sigma$, with respect to the jet $J_I^{(\sigma)}$. For simplicity, when possible we will only use ``sc.'' and ``hc.'' if there are no ambiguities. Graphically we will draw round and square half-brackets to describe them, as in figure\ \ref{convention_sc_hc}, with the projected momenta or vector indices appearing outside the brackets.

To identify the regions where eqs.\ (\ref{hard-collinear_approximation}) and (\ref{soft-collinear_approximation}) are good approximations, we introduce the \emph{neighborhood of a pinch surface} $\sigma$ in terms of the coordinates in (\ref{normal_intrinsic_coordinates}). This is defined as a region containing $\sigma$, where the normal coordinates $\big\{ s_i^{(\sigma)} \big\}$ and the intrinsic coordinates $\big\{ r_j^{(\sigma)} \big\}$ satisfy \cite{EdgStm15}
\begin{align} \label{neighborhoods}
\begin{split}
& \sum_{i} \left | s_i^{(\sigma)} \right |^2\leqslant p_0^2,\ \ \ s_i^{(\sigma)}\in \mathbb{N}_\sigma, \\
& \left | r_j^{(\sigma)} \right |^2 \geqslant \left(\sum_{i} \left | s_i^{(\sigma)} \right |^2 \right)^{\delta_j}p_0^{2-2\delta_j},\ \ \ r_j^{(\sigma)}\in \mathbb{I}_\sigma,
\end{split}
\end{align}
where $p_0^2\ll Q^2$, and $0<\delta_j<1/2$ is fixed for each intrinsic coordinate. The reason for this range of $\delta_j$ is that if $\delta_j < 1/2$, we may always neglect $l^2$ and $k_A^2$ terms on the RHS of the soft- and hard-collinear approximations in eq.\ (\ref{approximated_denominators}), because they are relatively suppressed by $\mathcal{O}(\lambda^{1/2})$. This restricted region in (\ref{neighborhoods}), is denoted as $n\left[\sigma\right]$.

We now study the relations between pinch surfaces in momentum space. To do this, we define the $\emph{normal space}$ of a momentum $k^\mu$, $\mathcal{N}_\sigma \left(k\right)$, as the linear span of the sets of normal coordinates of momentum $k^\mu$ in $\sigma$, i.e.
\begin{eqnarray} \label{normal_space}
\mathcal{N}_\sigma\left( k^\mu \right)\equiv \begin{cases}
\varnothing\ \text{(empty)} & \text{ if } k^\mu \text{ is hard in }\sigma, \\ 
\text{span}\left \{ \overline{\beta}^\mu,\ \mathbf{\beta}_\perp^\mu \right \} & \text{ if } k^\mu \text{ is collinear to }\beta^\mu \text{ in }\sigma, \\ 
\text{the full 4-dim space} & \text{ if } k^\mu \text{ is soft in }\sigma.
\end{cases}
\end{eqnarray}
For any loop momentum $k_i^\mu$ of an amplitude $\mathcal{A}$ at a pinch surface $\sigma$, the larger the dimension of its normal space, the more it is constrained, and the smaller the dimension of $\sigma$ will be. For example, it would be most constrained if it is soft, since all its four components are zero. We use normal coordinates to define orderings of pinch surfaces. Given any two distinct pinch surfaces $\sigma_1$ and $\sigma_2$, we define that $\sigma_1\subset\sigma_2$ if and only if for any loop momentum $k_i^\mu$, its normal space in $\sigma_2$ is contained in (or equal to) that in $\sigma_1$ , i.e.
\begin{eqnarray} \label{containing_definition}
\sigma_1\subset\sigma_2 \Leftrightarrow \mathcal{N}_{\sigma_1}\left( k_i \right)\supseteq \mathcal{N}_{\sigma_2}\left( k_i \right),\ \forall\text{ loop momentum }k_i^\mu,
\end{eqnarray}
where the equal signs cannot be simultaneously taken for all the $k_i^\mu$. From this definition, we can deduce the relation between hard, jet and soft subgraphs. For $\sigma_1\subset\sigma_2$, we define $J^{(\sigma_i)}\equiv \bigcup_{I}^{ }J_I^{(\sigma_i)}$ ($i=1,2$), and then
\begin{align} \label{containing_subgraphs_relations}
\begin{split}
H^{\left( \sigma_1 \right)}&\subseteq \ H^{\left( \sigma_2 \right)}, \\
\left (H^{\left( \sigma_1 \right)}\ \cup J^{\left( \sigma_1 \right)}\ \right) &\subseteq \ \left (H^{\left( \sigma_2 \right)}\ \cup J^{\left( \sigma_2 \right)}\ \right), \\
\left (S^{\left( \sigma_1 \right)}\ \cup J^{\left( \sigma_1 \right)}\ \right) &\supseteq \ \left (S^{\left( \sigma_2 \right)}\ \cup J^{\left( \sigma_2 \right)}\ \right), \\
S^{\left( \sigma_1 \right)}&\supseteq \ S^{\left( \sigma_2 \right)},
\end{split}
\end{align}
where one can derive the full set of relations using any two of them. Again, the equal signs cannot be simultaneously taken. If neither $\sigma_1\subset\sigma_2$ nor $\sigma_2\subset\sigma_1$, and moreover,
\begin{eqnarray} \label{subgraphs_intersection_nonempty}
\bigcup_{A=1}^{N}\left( J_A^{(\sigma_1)}\cap J_A^{(\sigma_2)} \right)\bigcup \left( H^{(\sigma_1)}\cap H^{(\sigma_2)} \right)\bigcup \left( S^{(\sigma_1)}\cap S^{(\sigma_2)} \right)\neq \varnothing,
\end{eqnarray}
we say $\sigma_1$ and $\sigma_2$ are overlapping, denoted by the symbol $\sigma_1:o:\sigma_2$. If the left hand side of eq.\ (\ref{subgraphs_intersection_nonempty}) is empty, $\sigma_1$ and $\sigma_2$ are called disjoint pinch surfaces. Note that pinch surfaces of a lowest-order electroweak decay process can never be disjoint, since $\left( H^{(\sigma_1)}\cap H^{(\sigma_2)} \right)$ always includes the electroweak vertex, and thus is always nonempty. For others, like the scattering processes, the hard subgraphs of two pinch surfaces can be non-overlapping, but we will show in section\ \ref{extend_scattering_processes} that such configurations are not relevant in the forest formula.

Coming back to eq.\ (\ref{neighborhoods}), $\sigma_1 \subset \sigma_2$ does not imply $n\left[\sigma_1\right] \subset n\left[\sigma_2\right]$. Therefore, the neighborhoods of nested pinch surfaces may overlap each other. To avoid overcounting, we define the neighborhoods in the following ``reduced'' way:
\begin{eqnarray} \label{reduced_neighborhoods}
\mathfrak{n}\left [ \sigma \right ]\equiv n\left [ \sigma \right ]\setminus \bigcup_{\sigma' \subset\sigma}^{ }\left( n\left [ \sigma \right ]\bigcap n\left [ \sigma' \right ] \right).
\end{eqnarray}
Then the union of all the $\emph{reduced neighborhoods}$, as defined above, takes account of all the singularities of an amplitude without double-counting. Note that larger pinch surfaces correspond to smaller reduced graphs, and vice versa.

With these tools at hand, our next task is to study the action of a single approximation operator, and see how it changes pinch surfaces compared with those of the original amplitudes. This will be necessary for our analysis in the following sections.

\subsection{Pinch surfaces generated by a single approximation}
\label{pinch_surfaces_from_single_approximation}

In this subsection, we study the pinch surfaces of amplitudes with a single approximation operator $t_\sigma\mathcal{A}$, and our results will be generalized in section\ \ref{subtraction_terms}. The reason that the pinch surfaces of $t_\sigma\mathcal{A}$ are different from those of $\mathcal{A}$ is easy to see from the definitions eqs.\ (\ref{hard-collinear_approximation}) and (\ref{soft-collinear_approximation}), because after the action of the operator $t_\sigma$, only certain components of the momenta and numerator factors of $\mathcal{A}$ are kept in specified subgraphs. Naturally, we need to look at the effects of these approximations. Most of the reasoning in this subsection, as a result, will apply to lines that attach $S^{(\sigma)}$ to $J_A^{(\sigma)}$, or $J_A^{(\sigma)}$ to $H^{(\sigma)}$.

\begin{figure}[t]
\centering
\includegraphics[width=10cm]{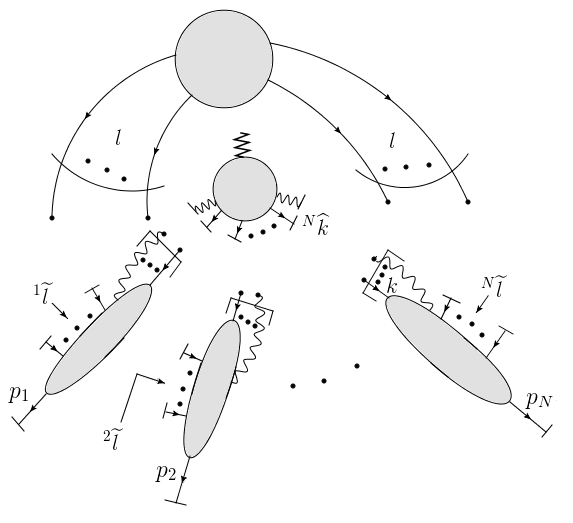}
\caption{A pictorial representation of the approximated amplitude $t_\sigma \mathcal{A}$, where $\sigma$ is shown in figure\ \ref{leading_pinch_surface}. The propagators with intact endpoints, denoted by dots, retain momenta before projections, while the truncated lines, denoted by bars, provide their projected momenta to the corresponding subgraphs. At each internal vertex of a bubble, the incoming and outgoing momenta are conserved.}
\label{approximated_amplitude}
\end{figure}

To be specific, we wish to classify all the pinch surfaces of $t_\sigma \mathcal{A}$. At pinch surface $\sigma$, the momenta are conserved to the leading order in the scaling variable $\lambda$ at each vertex. In region $\mathfrak{n}[\sigma]$, the action of $t_\sigma$ sets to zero only the components that are negligible in the momenta on which $t_\sigma$ acts. But the approximations still apply in other regions, where momentum conservation may not hold even to the leading order once these approximations have been made. A hard-collinear approximation provided by $t_\sigma$, for example, changes a jet momentum appearing in the hard subgraph, say $k^\mu$, into the form of $\left( k\cdot\overline{\beta}_I \right) \beta_I^\mu$. Of course these two momenta provide the same leading contribution in the region $\mathfrak{n}\left[\sigma\right]$. But when we consider another pinch surface where they are not necessarily identical in the leading contributions, this pinch surface may be different from any of the ones of $\mathcal{A}$. Similar considerations apply for the soft-collinear approximations. To synthesize all the approximations, we depict $t_\sigma \mathcal{A}$ in figure\ \ref{approximated_amplitude}. The approximation $t_\sigma$ defines hard, jet and soft bubbles. Inside each bubble, the Landau equation is applicable and the physical picture from Coleman-Norton interpretation still holds. However, between any two bubbles, the outgoing momenta of one bubble are generally not the same as the incoming momenta of the other one. Three comments regarding the momenta joining these bubbles summarize these new features.
\begin{itemize}
    \item [$\bullet$] The jet bubbles have external momenta only in the directions of $\beta_I^\mu$ and $\overline{\beta}_I^\mu$. We shall see that as a result, at pinch surfaces all their internal loops can only be soft, or hard, or collinear to $\beta_I^\mu$ or $\overline{\beta}_I^\mu$.
\end{itemize}
\begin{itemize}
    \item [$\bullet$] Loops joining the jet-$I$ bubble and the $H$ bubble depend only on the $\beta_I$-components of their momenta. These loops can have additional pinches at loops connecting $H$ and the jets, but they will only involve lines in the $\overline{\beta}_I$-direction for jet $I$.
\end{itemize}
\begin{itemize}
    \item [$\bullet$] Loops joining jets must flow through $S$, and can be pinched only in the jet directions.
\end{itemize}

Therefore, the original Coleman-Norton interpretation does not necessarily apply for the entire graph; we need a new analysis to see the formation of pinches for $t_\sigma \mathcal{A}$. To distinguish them from the pinch surfaces of $\mathcal{A}$, we will denote the pinch surfaces of $t_\sigma \mathcal{A}$ as $\rho^{ \left\{ \sigma \right\} }$. Most of the time we will keep the superscript, but during some specific discussions we may drop it for simplicity.

To enumerate the possible pinch surfaces of $t_\sigma \mathcal{A}$ in detail, we focus on any one of its propagators, whose momentum $k^\mu$ can be either soft, lightlike or hard before the projection. Denoting by $\left( t_\sigma k \right)^\mu$ the value that the projected momentum takes in $t_\sigma \mathcal{A}$, as a result of the approximations in eqs.\ (\ref{hard-collinear_approximation}) and (\ref{soft-collinear_approximation}), we then ask how the difference between $\left( t_\sigma k \right)^\mu$ and $k^\mu$ would change the pinch surface. Without loss of generality, we shall denote
\begin{eqnarray} \label{t_sigma_on_momentum}
\left(t_\sigma k\right)^\mu =\left( k\cdot\overline{v} \right)v^\mu
\end{eqnarray}
as the projected momentum after approximations are made. The vector $v^\mu$ here is a lightlike unit vector which is either in the same or opposite direction of a jet. That is, for a hard-collinear approximation with respect to the jet $J_I$, i.e. $k^\mu \rightarrow \left( k\cdot \overline{\beta}_I \right) \beta_I^\mu$, we have $v^\mu =\beta_I^\mu$; for a soft-collinear approximation $k^\mu \rightarrow \left( k\cdot \beta_I \right) \overline{\beta}_I^\mu$, we have $v^\mu= \overline{\beta}_I^\mu$. Both possibilities will be considered in terms of the examples in figures\ \ref{regular_pinch_surface1}--\ref{soft-exotic_example} below.

\begin{figure}[t]
\centering
\includegraphics[width=6cm]{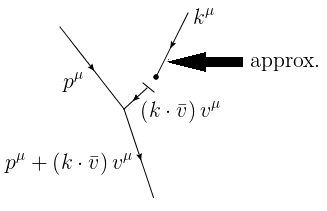}
\caption{The confluence of momenta $k^\mu$ and $p^\mu$ in $\rho^{ \left\{ \sigma \right\} }$, where the merged momentum is of the value $p^\mu+ \left( k\cdot\overline{v} \right)v^\mu$ because of the approximation $t_\sigma$. The approximation could be either soft-collinear or hard-collinear, for which we will show examples, and the discussions regarding to this figure are not restricted to 3-vertices.}
\label{confluence}
\end{figure}

Especially, we focus on a ``confluence'' of the projected momentum $\left( t_\sigma k \right)^\mu$ and another momentum $p^\mu$, resulting into a momentum $p^\mu+\left(k\cdot\overline{v}\right)v^\mu$, as is shown in figure\ \ref{confluence}. Note that $p^\mu$ is the momentum entering the confluence, which can be either the original momentum of a propagator or projected by $t_\sigma$.

We assume that these momenta are at a pinch surface of $t_\sigma \mathcal{A}$, say $\rho^{\left\{ \sigma \right\}}$, and will study how they relate to the pinch surfaces of $\mathcal{A}$ itself. In the paragraphs below, we will list all the possibilities by considering whether $k^\mu$ is soft, lightlike (in various directions) or hard in $\rho^{ \left\{ \sigma \right\} }$, and compare the obtained configurations of figure\ \ref{confluence} (corresponding to the pinch surfaces of $t_\sigma \mathcal{A}$) with the ones obtained by letting the original $k^\mu$ flow into the confluence (corresponding to the pinch surfaces of $\mathcal{A}$). In figure\ \ref{confluence} we exhibit a 3-point vertex, but the whole analysis also works in the presence of 4-point vertices.

\bigbreak
\centerline{\textbf{A. $k^\mu$ is soft in $\rho^{ \left\{ \sigma \right\} }$}}
This case is the simplest, since a soft momentum after any projections is still soft. Then the pinch surface at $k^\mu=0$ does not change if we replace $\left( t_\sigma k \right)^\mu =\left( k\cdot\overline{v} \right) v^\mu$ by $k^\mu$, meaning that the configuration in this case, being a subgraph of the pinch surface $\rho^{\left\{\sigma\right\}}$, also exists in a pinch surface $\rho$ of the original amplitude $\mathcal{A}$.

\bigbreak
\centerline{\textbf{B. $k^\mu$ is lightlike in $\rho^{ \left\{ \sigma \right\} }$ and not collinear to $\overline{v}^\mu$}}
Here $k^\mu$ is collinear to a certain lightlike vector, which is not necessarily $v^\mu$ but is not $\overline{v}^\mu$. Then the projected momentum will be collinear to $v^\mu$. To obtain the possible configuration of figure\ \ref{confluence}, the values of $p^\mu$ should be taken into account as well. We elaborate the discussion below, which involves a number of subcases.

\begin{itemize}
\item[(Bi)] $p^\mu$ is hard in $\rho^{ \left\{ \sigma \right\} }$. Then generally both $p^\mu +k^\mu$ and $p^\mu +\left( k\cdot\overline{v} \right) v^\mu$ are hard, so the configurations obtained by $k^\mu$ and $\left( k\cdot\overline{v} \right) v^\mu$ are identical.
\end{itemize}

\begin{itemize}
\item[(Bii)] $p^\mu$ is collinear to $v^\mu$ in $\rho^{ \left\{ \sigma \right\} }$. In other words, $p^\mu$ is pinched in alignment to the projected momentum $\left( k\cdot\overline{v} \right) v^\mu$. This configuration of momenta, as a subgraph of $\rho^{ \left\{ \sigma \right\} }$, may not exist in any pinch surface of the original amplitude $\mathcal{A}$.

In detail, if $k^\mu$ itself is also collinear to $v^\mu$, then $k^\mu$, $p^\mu$ and $p^\mu +\left( k\cdot\overline{v} \right) v^\mu$ are all collinear to $v^\mu$. Such a configuration can appear at a pinch surface of $\mathcal{A}$. But if $k^\mu$ is not in the direction of $v^\mu$, we will obtain a configuration where $k^\mu$ is collinear to one lightlike unit vector, while $p^\mu$ and $p^\mu +\left( k\cdot\overline{v} \right) v^\mu$ are collinear to another. In other words, the propagators in a connected jet subgraph are lightlike, but in different directions. Apparently, this never takes place in $\mathcal{A}$. To see how the pinches are formed, we examine the denominators in the expression of $t_\sigma \mathcal{A}$ that involve $p^\mu$, which are of the form:
\begin{eqnarray}
&&\left[ \big( \left( k\cdot\overline{v} \right)v+p \big)^2+i\epsilon \right] \left( p^2+i\epsilon \right)\nonumber\\
&&\hspace{0.5cm}=\left [ 2\left( p\cdot v \right)\left( p\cdot\overline{v}+k\cdot\overline{v} \right)-\left( p\cdot v_{\perp} \right)^2+i\epsilon \right ]\left [ 2\left( p\cdot v \right)\left( p\cdot\overline{v} \right)-\left( p\cdot v_{\perp} \right)^2+i\epsilon \right ]. \nonumber\\
&&
\end{eqnarray}
The solution that produces a pinch when both $p^\mu$ and $p^\mu +\left( k\cdot\overline{v} \right) v^\mu$ are lightlike is
\begin{eqnarray} \label{bii_result}
p\cdot\overline{v}=\alpha\left( k\cdot\overline{v} \right),\ \ -1<\alpha<0,\ \ p\cdot v=0,\ \ p\cdot v_\perp=0\ \text{ at }\ \rho^{ \left\{ \sigma \right\}}.
\end{eqnarray}
That is, when $p^\mu$ and $p^\mu +\left( k\cdot\overline{v} \right) v^\mu$ are both lightlike in $\rho^{ \left\{ \sigma \right\} }$, they can only be collinear to $(t_\sigma k)^\mu \propto v^\mu$, rather than the vector $k^\mu$ before approximations. The condition $-1<\alpha<0$ ensures a pinch. This analysis works for both hard-collinear and soft-collinear approximations, and examples are given for both cases in figure\ \ref{regular_pinch_surface1} below.
\begin{figure}[t]
\centering
\includegraphics[width=12cm]{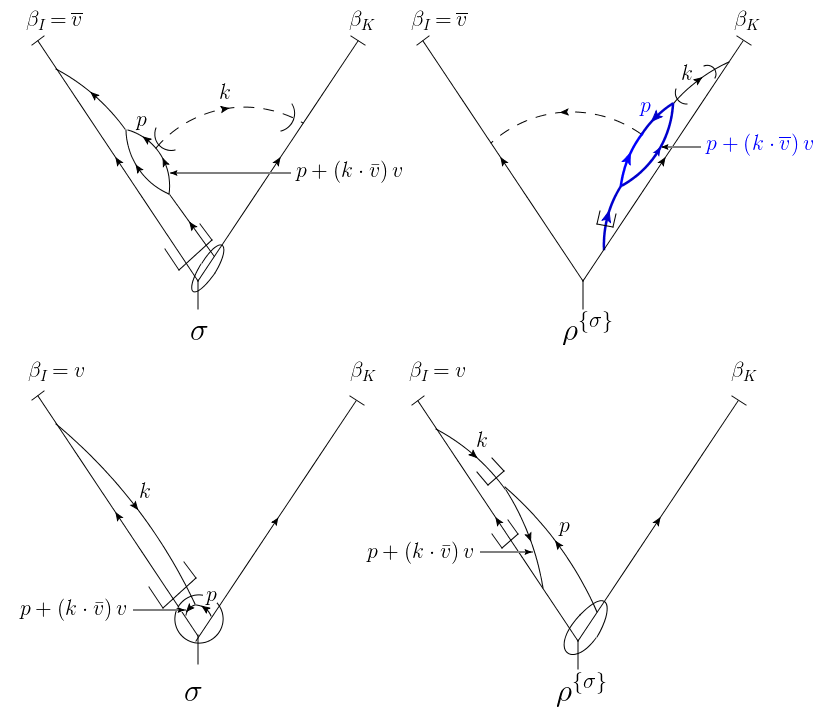}
\caption{Examples where $p^\mu$ in figure\ \ref{confluence} is pinched to be collinear to $v^\mu$ in $\rho^{ \left\{ \sigma \right\} }$. The pinch surface $\sigma$ of $\mathcal{A}$ is shown on the left, and the pinch surface $\rho^{ \left\{ \sigma \right\} }$ of $t_\sigma \mathcal{A}$ on the right. The upper row describes the case where $t_\sigma$ acts a soft-collinear approximation on $k^\mu$ and $v^\mu= \overline{\beta}_I^\mu$. The lower row describes a hard-collinear approximation for which $v^\mu= \beta_I^\mu$. Each of them forces the external momentum entering the $p^\mu$-loop to be in the direction of $v^\mu$ in $\rho^{\left\{ \sigma \right\}}$, and yields the configuration described in (Bii). The propagators marked bold and blue are for later use (to identify certain subgraphs) in section\ \ref{divergences_are_logarithmic}.}
\label{regular_pinch_surface1}
\end{figure}
\end{itemize}

\begin{itemize}
\item[(Biii)] $p^\mu$ is collinear to another vector $v'^\mu \left( \neq v^\mu \right)$ in $\rho^{ \left\{ \sigma \right\} }$. Then by construction, the confluence momentum $p^\mu+\left( k\cdot\overline{v} \right) v^\mu$ is hard, and both $\left( k\cdot\overline{v} \right) v^\mu$ and $k^\mu$ correspond to the same configuration of figure\ \ref{confluence}, i.e. two jet lines of different directions joining the hard subgraph together.
\end{itemize}

\begin{itemize}
\item[(Biv)] $p^\mu$ is soft in $\rho^{ \left\{ \sigma \right\} }$. Then $p^\mu+\left( k\cdot\overline{v} \right) v^\mu$ is in the same direction as $\left( k\cdot\overline{v} \right) v^\mu$. In the corresponding configuration of figure\ \ref{confluence}, a soft momentum $p^\mu$ is attached to a lightlike momentum $k^\mu$, which becomes $\left( k\cdot\overline{v} \right) v^\mu$ after the confluence. This configuration does not exist in any pinch surfaces of $\mathcal{A}$, unless $k^\mu$ is collinear to $v^\mu$.
\end{itemize}

\bigbreak
\centerline{\textbf{C. $k^\mu$ is collinear to $\overline{v}^\mu$ in $\rho^{ \left\{ \sigma \right\} }$}}
In this case $k^\mu$ is lightlike while $\left( t_\sigma k \right)^\mu = \left( k\cdot\overline{v} \right) v^\mu$ is soft. We again consider all the possible values of $p^\mu$, and discuss the configurations of figure\ \ref{confluence} in the following subcases.

\begin{itemize}
\item[(Ci)] $p^\mu$ is hard in $\rho^{ \left\{ \sigma \right\} }$. Then generally both $p^\mu +k^\mu$ and $p^\mu +\left( k\cdot\overline{v} \right) v^\mu$ are hard, so the configurations obtained by $k^\mu$ and $\left( k\cdot\overline{v} \right) v^\mu$ are identical.
\end{itemize}

\begin{itemize}
\item[(Cii)] $p^\mu$ is lightlike in $\rho^{ \left\{ \sigma \right\} }$. We start with a special case: $p^\mu$ is collinear to $\overline{v}^\mu$. Then $k^\mu$, $p^\mu$ and $p^\mu +\left( k\cdot\overline{v} \right) v^\mu$ are all collinear to $\overline{v}^\mu$. However, there is a difference between this configuration and that in an original amplitude $\mathcal{A}$, which we can observe from the following denominator factors:
\begin{eqnarray} \label{cii_denominator_factor}
&&\left[\left( p+(k\cdot\overline{v})v \right)^2 +i\epsilon\right] \left( k^2+i\epsilon \right)\nonumber\\
&&\hspace{0.5cm}=\left [ 2\left( p\cdot v \right)\left( p\cdot\overline{v} +k\cdot\overline{v} \right)-\left( p\cdot v_{\perp} \right)^2+i\epsilon \right ]\left [ 2\left( k\cdot v \right)\left( k\cdot\overline{v} \right)-\left( k\cdot v_{\perp} \right)^2+i\epsilon \right ]. \nonumber\\
\end{eqnarray}
The solution that produces a pinch when $k^\mu$ and $p^\mu$ are both in the direction of $\overline{v}^\mu$ is:
\begin{eqnarray} \label{cii_result}
k\cdot v=\alpha \left(p\cdot v\right),\ -\infty<\alpha<0,\ k\cdot \overline{v}=0,\ k\cdot v_\perp=0,\ \text{ at }\ \rho^{ \left\{ \sigma \right\}}.
\end{eqnarray}
We notice that the range of $\alpha$ that gives a pinch is unbounded, which is different from the configuration in an original amplitude $\mathcal{A}$, where $p^\mu$, $k^\mu$ and $p^\mu+k^\mu$ are all collinear to $\overline{v}^\mu$. This difference lies in the intrinsic coordinates.

In the general case, if $p^\mu$ is collinear to $v'^\mu \neq \overline{v}^\mu$, we still have eqs.\ (\ref{cii_denominator_factor}) and (\ref{cii_result}), and the obtained configuration is still different from any configuration in $\mathcal{A}$, due to the unbounded intrinsic variable $k\cdot v$. Meanwhile, since $v'^\mu \neq \overline{v}^\mu$, the two lightlike propagators carrying momenta $p^\mu$ and $k^\mu$ are in different directions, then join each other to form another propagator collinear to $p^\mu$. This is another difference from any configuration in $\mathcal{A}$, as we have encountered in (Bii) already.

With these differences in mind, we show in figure\ \ref{regular_pinch_surface2} two examples of case (Cii), with $t_\sigma$ as a hard- or soft-collinear approximation.
\begin{figure}[t]
\centering
\includegraphics[width=12cm]{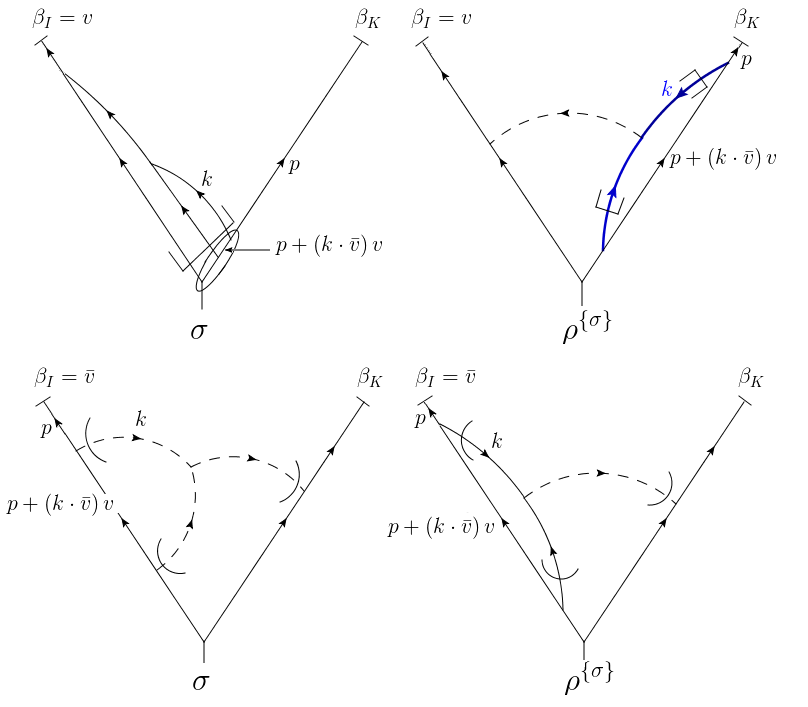}
\caption{Examples of case (Cii), where both $k^\mu$ and $p^\mu$ are lightlike in $\rho^{\left\{ \sigma \right\}}$, and specially, $k^\mu$ is in the direction of $\overline{v}^\mu$. The upper row describes the case where $t_\sigma$ acts as a hard-collinear approximation on $k^\mu$ with $v^\mu=\beta_I^\mu$, while the lower row describes a soft-collinear approximation with $v^\mu=\overline{\beta}_I^\mu$. Due to these approximations, the component of $k^\mu$ which joins $p^\mu$ is $\left( k\cdot\overline{v} \right)v^\mu$, so that $k^\mu$ is pinched in the direction of $\overline{v}^\mu$ from our analysis. Meanwhile, the propagator with momentum $p^\mu+\left( k\cdot\overline{v} \right)v^\mu$ is put on shell as a jet propagator. The propagators marked bold and blue are for later use in section\ \ref{divergences_are_logarithmic}.}
\label{regular_pinch_surface2}
\end{figure}
\end{itemize}

\begin{itemize}
\item[(Ciii)] $p^\mu$ is soft in $\rho^{ \left\{ \sigma \right\} }$. In this case the three incoming (outgoing) momenta at the confluence, $p^\mu$, $\left( k\cdot\overline{v} \right) v^\mu$ and $p^\mu+ \left( k\cdot\overline{v} \right) v^\mu$ are all soft, so they join at a soft vertex. This configuration does not exist in $\mathcal{A}$, because we have a jet propagator attached to two or more soft propagators. For this reason, we shall call such a jet propagator whose nonzero lightlike momentum becomes soft under $t_\sigma$, and is attached to a soft vertex as a \emph{soft-exotic propagator}. Two typical examples, where the $t_\sigma$ is either a hard-collinear or a soft-collinear approximation on $k^\mu$, are shown in figure\ \ref{soft-exotic_example}.
\begin{figure}[t]
\centering
\includegraphics[width=12cm]{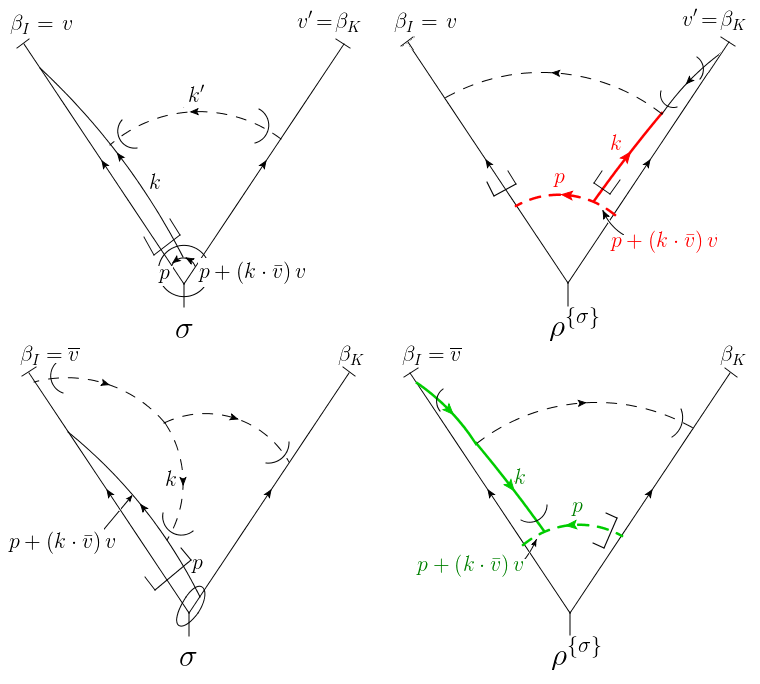}
\caption{Two examples where a jet propagator can end at a soft vertex in $\rho^{\left\{ \sigma \right\}}$. In the upper row, $t_\sigma$ acts as a hard-collinear approximation and only keeps the $v(=\beta_I)$-component of $k^\mu$ in $H^{(\sigma)}$. At the same time, it also acts as a soft-collinear approximation on the soft momentum entering $J_I^{(\sigma)}$. When this soft line in $\sigma$ becomes collinear to $v'^\mu (=\beta_K^\mu)$ in $\rho^{ \left\{ \sigma \right\} }$, for the same reasons as in (Bii) above, $k^\mu$ is pinched in the direction of $\overline{v}^\mu$. Then $\left( k\cdot\overline{v} \right) v^\mu$ vanishes at the pinch surface $\rho^{\left\{ \sigma \right\}}$, and if the internal momentum $p^\mu$ is soft as well, all the incoming momenta at the confluence of $p^\mu$ and $\left(t_\sigma k\right)^\mu$ will be soft. In the lower row, $t_\sigma$ acts as a soft-collinear approximation on $k^\mu$, and projects it onto its $v ( =\overline{\beta}_I)$-component. Subsequently, a soft vertex forms when $k^\mu$ is collinear to $\beta_I^\mu$. Some propagators are colored red or green for later uses in section\ \ref{divergences_are_logarithmic}.}
\label{soft-exotic_example}
\end{figure}

The two rows in figure\ \ref{soft-exotic_example} exhibit the lowest-order graphs, but in principle they can be the representatives of all-order graphs for the two cases, where the approximation on the momentum of the soft-exotic propagator ($k^\mu$) is hard-collinear or soft-collinear. To be specific, if the approximation is hard-collinear, $k^\mu$ must be lightlike in $\sigma$, and collinear to the opposite direction in $\rho^{ \left\{ \sigma \right\} }$ (this phenomenon will be explained below in Theorem 1). The projected momentum is then automatically soft. If the approximation is soft-collinear, the propagator with $k^\mu$ must be soft and attached to one jet in $\sigma$, and become part of that jet in $\rho^{ \left\{ \sigma \right\} }$. Under the soft-collinear approximation, only the component opposite to the jet's direction is kept, so the projected momentum will be automatically soft in $\rho^{ \left\{ \sigma \right\} }$ as well.
\end{itemize}

\bigbreak
\centerline{\textbf{D. $k^\mu$ is hard in $\rho^{ \left\{ \sigma \right\} }$}}

\begin{figure}[t]
\centering
\includegraphics[width=12cm]{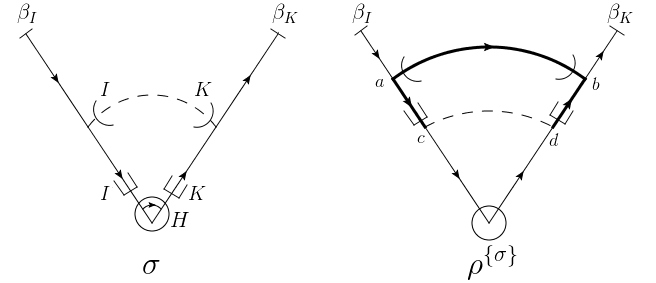}
\caption{A pinch surface $\sigma$ that yields a hard-exotic configuration in $\rho^{ \left\{ \sigma \right\} }$, where the hard subgraph is disconnected in the latter. Here we use bold lines and arcs to denote hard propagators, and dashed arcs to denote soft propagators. Due to the hard-collinear approximations from $t_\sigma$, only the $\beta_I$- and $\beta_K$-components of the hard momenta flow into vertices $c$ and $d$ in $\rho^{ \left\{ \sigma \right\} }$ separately. So IR structures can emerge in the subgraph with these lightlike momenta being the external momenta. By definition, the hard-exotic propagators are $ac$ and $bd$.}
\label{hard-exotic_PS_example}
\end{figure}

In this case, $\left( k\cdot\overline{v} \right) v^\mu$ is lightlike. Then if $p^\mu$ is neither collinear to $v^\mu$ nor soft, then $p^\mu+\left( k\cdot\overline{v} \right) v^\mu$ is hard, and we come up with a configuration where $p^\mu$ flows into the hard subgraph, as at a corresponding pinch surface of $\mathcal{A}$. If $p^\mu$ is collinear to $v^\mu$ or soft, the momentum $p^\mu+\left( k\cdot\overline{v} \right) v^\mu$ will be lightlike, and it is possible that some other collinear or soft subgraphs are pinched according to this lightlike momentum, and the hard subgraph may become disconnected (see figure\ \ref{hard-exotic_PS_example}). In other words, we have found a hard propagator attached to a jet vertex, all the other momenta flowing into which are collinear to a certain direction or soft. This can never happen in the pinch surfaces of $\mathcal{A}$. This is similar to the previously discussed case, where a jet propagator is attached at a soft vertex. In comparison, we call a propagator carrying hard momentum which becomes lightlike under $t_\sigma$ as a \emph{hard-exotic propagator}.

\bigbreak
\centerline{\textbf{Regular and exotic configurations}}

After the enumeration above, we can classify the configurations of figure\ \ref{confluence} into two types. To do so, we focus on a vertex of $t_\sigma \mathcal{A}$, say $x$, and identify the momentum flowing into (or out of) this vertex whose normal space in $\rho^{ \left\{ \sigma \right\} }$ has the smallest dimension. We say that $x$ is a \emph{soft (jet, hard) vertex}, if and only if the identified momentum is a soft (jet, hard) momentum. We then say \emph{the normal spaces are conserved at a given vertex} if one of the following statements is true:
\begin{itemize}
  \item [(1)] the vertex is a soft vertex, and all the propagators attached to this vertex are soft;
\end{itemize}
\begin{itemize}
  \item [(2)] the vertex is a jet vertex, and all the propagators attached to this vertex are either soft or lightlike;
\end{itemize}
\begin{itemize}
  \item [(3)] the vertex is a hard vertex.
\end{itemize}
Otherwise, we say that the normal spaces are not conserved at this vertex.

This concept helps us to classify the configurations of figure\ \ref{confluence}. That is, for a given vertex $x$ of an approximated amplitude, the subgraph composed by $x$ and its attached propagators is called a \emph{regular configuration} if and only if the normal spaces are conserved at $x$. Otherwise, it is called an \emph{exotic configuration}.\footnote{An exotic configuration reflects a pinch for internal loop momenta in a jet or hard subdiagram, induced by the action of the approximations, which set its lines on shell. This is a general feature of nested subtractions, and we expect all such singularities to be cancelled in the full sum over forests. This will turn out to be the case.} Specifically, a jet propagator attached to a soft vertex corresponds to a \emph{soft-exotic configuration}, while a hard propagator attached to a jet vertex corresponds to a \emph{hard-exotic configuration}. A pinch surface with soft- or hard-exotic configurations is called an \emph{exotic pinch surface}, otherwise it is called a \emph{regular pinch surface}. These two types of pinch surfaces of $t_\sigma\mathcal{A}$ are thus denoted as $\rho_{\text{exo}}^{ \left\{ \sigma \right\} }$ and $\rho_{\text{reg}}^{ \left\{ \sigma \right\} }$, and the divergences near their neighborhoods both should be considered in detail.

\begin{table}[t]
\captionsetup{justification=centering,margin=0cm}
\caption{Summary of the possible configurations of figure\ \ref{confluence}, which depend on $p^\mu$, $k^\mu$ in $\rho^{ \left\{ \sigma \right\} }$ and $p^\mu+\left(t_\sigma k \right)^\mu= p^\mu+ \left( k\cdot \overline{v} \right) v^\mu$. The abbreviation ``col.'' represents ``collinear''.}
\begin{center}
\begin{tabular}{ |c||c|c|c|c|c| } 
\hline
 \multirow{1}{*}{$k^\mu$} & \multirow{1}{*}{$p^\mu$} & $p^\mu+\left(t_\sigma k \right)^\mu$ & Description & Classification \\ 
 \hline
 \hline
 Soft & No con- & \multirow{2}{*}{$p^\mu$} & A soft propagator & \multirow{2}{*}{Regular}\\ 
 (A) & straints & & joining $p^\mu$ & \\
 \hline
 & \multirow{2}{*}{(i) Hard} & \multirow{2}{*}{Hard} & A lightlike propagator joining & \multirow{2}{*}{Regular} \\
 & & & the hard subgraph $H^{(\rho^{\left\{ \sigma \right\}})}$ & \\ \cline{2-5}
 Col. to & & \multirow{4}{*}{Col. to $v^\mu$} & $k^\mu \parallel v^\mu:$ $p^\mu$ and $k^\mu$ are & \multirow{2}{*}{Regular} \\
 any & (ii) Col. to & & lightlike in the same direction & \\ \cline{4-5}
 vector & vector $v^\mu$ & & $k^\mu \nparallel v^\mu:$ $p^\mu$ and $k^\mu$ are & \multirow{2}{*}{Regular} \\
 except & & & lightlike in different directions & \\ \cline{2-5}
 $\overline{v}^\mu$ & (iii) Col. to & \multirow{2}{*}{Hard} & $p^\mu$ and $k^\mu$ lightlike, and & \multirow{2}{*}{Regular} \\
 (B) & $v'^\mu$ ($\neq v^\mu$) & & join the hard subgraph $H^{(\rho^{\left\{ \sigma \right\}})}$ & \\ \cline{2-5}
 & \multirow{2}{*}{(iv) Soft} & \multirow{2}{*}{Col. to $v^\mu$} & A lightlike propagator attached & \multirow{2}{*}{Regular}\\
 & & & by a soft momentum $p^\mu$ & \\
 \hline
 & \multirow{2}{*}{(i) Hard} & \multirow{8}{*}{$p^\mu$} & A lightlike propagator joining & \multirow{2}{*}{Regular}\\
 & & & the hard subgraph $H^{(\rho^{\left\{ \sigma \right\}})}$ & \\ \cline{2-2}\cline{4-5}
 \multirow{2}{*}{Col.} & (ii) Col. to & & $p^\mu$ and $k^\mu$ are lightlike & \multirow{2}{*}{Regular}\\
 \multirow{2}{*}{to $\overline{v}^\mu$} & vector $\overline{v}^\mu$ & & in the same direction & \\ \cline{2-2}\cline{4-5}
 \multirow{2}{*}{(C)} & (ii) Col. to & & $p^\mu$ and $k^\mu$ are lightlike & \multirow{2}{*}{Regular}\\
 & $v'^\mu (\neq \overline{v}^\mu) $ & & in different directions & \\ \cline{2-2}\cline{4-5}
 & \multirow{2}{*}{(iii) Soft} & & A lightlike propagator is & \multirow{2}{*}{Soft-exotic} \\
 & & & attached to a soft vertex &\\
 \hline
 \multirow{2}{*}{Hard} & Col. to $v^\mu$ & \multirow{2}{*}{Col. to $v^\mu$} & A hard propagator is attached & \multirow{2}{*}{Hard-exotic} \\
 \multirow{2}{*}{(D)}& or soft & & to soft or lightlike propagators &\\ \cline{2-5}
 & (Otherwise) & Hard & $p^\mu$ joining the hard subgraph & Regular \\ 
 \hline
\end{tabular}
\end{center}
\label{single_approximated_PS_summary}
\end{table}

We summarize our discussions in this subsection in table\ \ref{single_approximated_PS_summary}, giving all the possibilities of figure\ \ref{confluence} that serve as configurations of sub pinch surfaces of $t_\sigma\mathcal{A}$. All the information in the table is given in our analysis and definitions above.

\bigbreak
\centerline{\textbf{General approximated subgraphs of $\rho^{\left\{ \sigma \right\}}$}}

With these preparations, we now study the approximated subgraphs of $\rho^{\left\{ \sigma \right\}}$. For each $\rho^{\left\{ \sigma \right\}}$, its general picture can be obtained by combining the configurations in table\ \ref{single_approximated_PS_summary}. To make it clearer, we focus on the propagators of subgraph $J_I^{(\sigma)}$ that are also lightlike in $\rho^{\left\{ \sigma \right\}}$, and make an observation that will be quite useful in the upcoming sections. We formalize it in bold as follows:

\textbf{Theorem 1: In the pinch surface $\rho^{ \left\{ \sigma \right\} }$, all the propagators of $J_I^{(\sigma)}$ that are lightlike can only be collinear to $\beta_I^\mu$ or $\overline{\beta}_I^\mu$}.

\textit{Proof of Theorem 1:} Consider all the propagators of $J_I^{(\sigma)}$ that are lightlike in any directions but $\beta_I^\mu$ in $\rho^{ \left\{ \sigma \right\} }$. We denote the set of these propagators (as a subgraph of $J_I^{(\sigma)}$) by $\gamma$, and aim to prove that the propagators of $\gamma$ can only be collinear to $\overline{\beta}_I^\mu$ in $\rho^{ \left\{ \sigma \right\} }$.

\begin{figure}[t]
\centering
\includegraphics[width=6cm]{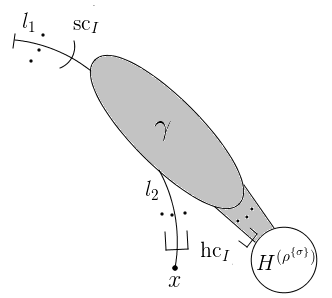}
 \caption{The subgraph $\gamma$, which includes the propagators marked by momentum $l_2^\mu$ as well as the shaded area. The hard-collinear and soft-collinear approximations are from $t_\sigma$.}
\label{two_jets_intersection}
\end{figure}

Taking account of the associated approximations, we can depict $\gamma$ in $\rho^{ \left\{ \sigma \right\} }$, as is shown in figure\ \ref{two_jets_intersection}. The whole subgraph includes the shaded area as well as those propagators whose momenta are denoted by $l_2^\mu$, where the vertices denoted by $x$ are arbitrary jet vertices in $\rho^{\left\{ \sigma \right\}}$. The approximations are from $t_\sigma$: in $\sigma$ the momenta $l_1^\mu$ are soft external momenta of $J_I^{(\sigma)}$ while the $l_2^\mu$ are lightlike (in the $\beta_I^\mu$-direction) external momenta of $H^{(\sigma)}$ in $\sigma$. Some propagators of $\gamma$ may be attached to the hard part of $\rho^{ \left\{ \sigma \right\} }$, either approximated by $t_\sigma$ or not.

The directions of the propagators in $\gamma$ are determined only by the momenta $l_1^\mu$ and $l_2^\mu$. On one hand, the momenta $l_1^\mu$ are projected and become $\left( l_1 \cdot \beta_I \right) \overline{\beta}_I^\mu$ by the soft-collinear approximation of $t_\sigma$. The momenta $\left( l_1 \cdot \beta_I \right) \overline{\beta}_I^\mu$ are lightlike, meaning all the external jet momenta from $l_1$ that enter the shaded area are parallel to $\overline{\beta}_I^\mu$, and only momenta parallel to $\overline{\beta}_I^\mu$ can satisfy Landau equations for the internal loops of $\gamma$.

On the other hand, only the $\beta_I$-components of $l_2^\mu$ enter the subgraph $H^{(\sigma)}$ as shown by the hard-collinear approximation. Recalling that any vertex represented by $x$ is a jet vertex, we identify the lightlike momentum that enters $x$ (which is not $\left(l_2\cdot \overline{\beta}_I\right) \beta_I^\mu$), denote it by $p^\mu$, and assume it to be parallel to some jet direction $\beta_K^\mu$. We may have $K=I$ or not. If $K=I$, then no matter what directions the $l_2^\mu$ are in, all the momenta entering $x$ are soft or collinear to $\beta_I^\mu$. In this case the $l_2^\mu$ do not fix the direction of the propagators in $\gamma$. If $K\neq I$, the propagators that contain the momenta $l_2^\mu$ have denominators of the form $2\left( p\cdot\overline{\beta}_K \right) \left (\beta_I\cdot\beta_K \right) \left( l_2 \cdot \overline{\beta}_I \right)+i\epsilon$ if they are in the $K$-jet, and $l_2^2+i\epsilon$ if they are in $\gamma$. In order that these propagators are both lightlike at $\rho^{ \left\{ \sigma \right\} }$, $l_2^\mu$ has to be pinched in the direction of $\overline{\beta}_I^\mu$ since the only candidates for normal coordinates are their $\beta_I$-components $\left( l_2 \cdot \overline{\beta}_I \right)$.

From these two aspects we see the whole shaded area can only be collinear to $\overline{\beta}_I^\mu$ rather than any other directions, due to the the effects of $t_\sigma$. In conclusion, Theorem 1 is proved.

\begin{figure}[t]
\centering
\includegraphics[width=12cm]{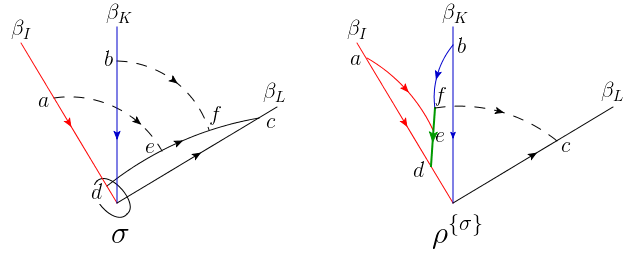}
\caption{The example where two jet subgraphs have a nontrivial overlap. In $\rho^{\left\{\sigma\right\}}$, the red propagators are parallel to $\beta_I^\mu$; the blue propagators are parallel to $\beta_K^\mu$; the green propagators are parallel to $\overline{\beta}_L^\mu$. The fact that these propagators are fixed to certain directions has been explained in this subsection. From our definition of jet subgraphs above, $J_I^{(\rho^{\left\{\sigma\right\}})}$ includes all the red propagators and $ed$, and $J_K^{(\rho^{\left\{\sigma\right\}})}$ includes all the blue propagators, as well as $fe$ and $ed$. As a result, $ed$ is the propagator shared by both $J_I^{(\rho^{\left\{\sigma\right\}})}$ and $J_K^{(\rho^{\left\{\sigma\right\}})}$.}
\label{two_jets_merge}
\end{figure}

\bigbreak
With the help of Theorem 1, we now define the jet subgraphs in $\rho^{\left\{ \sigma \right\}}$. We imagine a flow starting from the $I$-th external momentum and going inward to the hard subgraph. The flow only covers lightlike propagators, including those collinear to $\beta_I^\mu$ in $\rho^{\left\{ \sigma \right\}}$, and those lightlike in another direction $\beta_K^\mu\ (\neq \beta_I^\mu)$ in $\sigma$, and become collinear to $\overline{\beta}_K^\mu$ in $\rho^{\left\{ \sigma \right\}}$. The set of lightlike propagators that carry this flow, is defined as $J_I^{(\rho^{\left\{ \sigma \right\}})}$. For example, in the upper row of figure\ \ref{regular_pinch_surface1} the subgraph $J_K^{(\rho^{\left\{ \sigma \right\}})}$ constructed in this way also contains the bold and blue lines, although they are parallel to $\overline{\beta}_I^\mu$ rather than $\beta_K^\mu$. Under this definition, each jet subgraph in $\rho^{\left\{ \sigma \right\}}$ may contain lightlike propagators in several different directions, and two jet subgraphs may even have a nontrivial overlap.\footnote{By saying that two jets have a trivial overlap, we mean they share a vertex in the hard part as opposed to sharing a line.} An example is shown in figure\ \ref{two_jets_merge}.

In this figure, both $fe$ and $ed$ are lightlike in the direction of $\overline{\beta}_L^\mu$. From our previous discussions to obtain table\ \ref{single_approximated_PS_summary}, the confluences at $e$ and $f$ correspond to case (Bii), and the confluences at $a$ and $b$ correspond to case (Cii).

\begin{figure}[t]
\centering
\includegraphics[width=13cm]{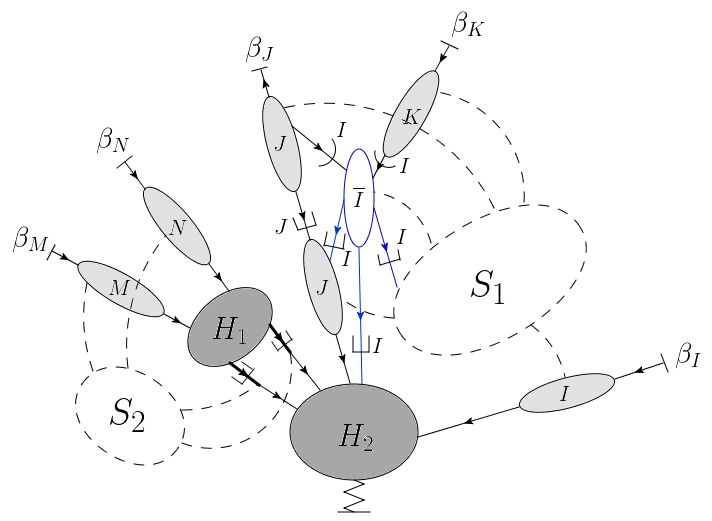}
\caption{A general picture of the pinch surfaces of $t_\sigma \mathcal{A}$, which contains all the configurations in table\ \ref{single_approximated_PS_summary}. There are five external jet momenta in total, separately in the directions of $\beta_I^\mu$, $\beta_J^\mu$, $\beta_K^\mu$, $\beta_M^\mu$ and $\beta_N^\mu$. The blob and lines marked blue are in the direction of $\overline{\beta}_I^\mu$.}
\label{general_PS_t_sigma_A}
\end{figure}

To end this subsection, we depict a general picture of $\rho^{\left\{ \sigma \right\}}$, figure\ \ref{general_PS_t_sigma_A}, which contains all the configurations displayed in table\ \ref{single_approximated_PS_summary}, as well as the ``merging of jets''. To be specific, case (A) can occur inside the soft subgraphs $S_1$ and $S_2$. Case (Bi) occurs at the connection between each jet and $H_1$ (or $H_2$); case (Bii) occurs where the propagators collinear to $\beta_J^\mu$ or $\beta_K^\mu$ flow into the blob in the direction of $\overline{\beta}_I^\mu$; case (Biv) occurs where the soft propagators of $S_1$ is attached to the blob in the $\overline{\beta}_I$-direction. Cases (Ci), (Cii) and (Ciii) separately occurs at the connections between the propagators collinear to $\overline{\beta}_I^\mu$ and projected by the approximation $\text{hc}_I^{(\sigma)}$, and other subgraphs: $H_2$, the blob in the direction of $\beta_J^\mu$, and $S_1$. Finally, the hard-exotic configuration in case (D) occurs at the vertices where $H_1$ meets $S_2$; the remaining case in case (D) can occur inside the hard subgraphs $H_1$ and $H_2$.

\subsection{Subtraction terms with repetitive approximations}
\label{subtraction_terms}

In this subsection we study the approximated amplitudes with repetitive approximations. The reason to introduce them, taking $t_{\sigma_2} t_{\sigma_1} \mathcal{A}$ for example, is to subtract double-counted and unphysical divergences from terms (approximation amplitudes) with fewer approximations, in other words $t_{\sigma_1} \mathcal{A}$ and $t_{\sigma_2} \mathcal{A}$. The extra divergences of $t_{\sigma_2} t_{\sigma_1} \mathcal{A}$, are cancelled by terms with more approximations. In order to obtain the subtractions that eliminate the infrared divergences properly, we should be clear about the rules such operators obey when transforming momenta and vector indices under repetitive approximations.

The rules themselves, should meet the following requirements. First, we only consider the ``nested'' repetitive approximations, namely, $\sigma_1\subset ... \subset\sigma_n$, and denote the relevant operator as $t_{\sigma_n} ... t_{\sigma_1}$. We will define the action of $t_{\sigma_n}$ on $t_{\sigma_{n-1}} ... t_{\sigma_1} \mathcal{A}$ in terms of its projections on the momenta and vector indices. Of course, momenta and vector indices may respond differently. For any two nested pinch surfaces $\sigma_1\subset\sigma_2$, we further require:
\begin{eqnarray} \label{requirement1}
t_{\sigma_2}\left( 1-t_{\sigma_1} \right)\mathcal{A}\mid _{\text{div }\mathfrak{n}\left [ \sigma_1^{ \left\{ \sigma_2 \right\} } \right ]}=0,
\end{eqnarray}
and
\begin{eqnarray} \label{requirement2}
\left( 1-t_{\sigma_2} \right)t_{\sigma_1}\mathcal{A}\mid _{\text{div }\mathfrak{n}\left [ \sigma_2^{ \left\{ \sigma_1 \right\} } \right ]}=0.
\end{eqnarray}
Both these relations imply two aspects. Taking eq.\ (\ref{requirement1}) for example, it implies the coincidence of pinch surfaces, i.e. $\sigma_1^{ \left\{ \sigma_2 \right\} } = \sigma_1^{ \left\{ \sigma_2\sigma_1 \right\} }$, as well as the exactness of $t_{\sigma_1}$ in the neighborhood $\mathfrak{n}\big [ \sigma_1^{ \left\{ \sigma_2 \right\} } \big ]$. In this subsection, we will give the rule for repetitive approximations, and only show that the rule is compatible with the exactness of relevant approximation operators. In section\ \ref{pairwise_cancellation_regular: nested} later, we will see that $\sigma_1^{ \left\{ \sigma_2 \right\} } = \sigma_1^{ \left\{ \sigma_2\sigma_1 \right\} }$, whose loop momenta have the same normal spaces as those of $\sigma_1$; similarly, $\sigma_2^{ \left\{ \sigma_1 \right\} } = \sigma_2^{ \left\{ \sigma_2\sigma_1 \right\} }$, with identical normal spaces as $\sigma_2$. These results are within our assumption in this subsection.

Before we work on the exactness of $t_{\sigma_1}$ in eq.\ (\ref{requirement1}) and $t_{\sigma_2}$ in (\ref{requirement2}), we make the following crucial observation: \textbf{For any subtraction term $t_{\sigma_n} ... t_{\sigma_1} \mathcal{A}$ that is not vanishing, each momentum or vector index of $\mathcal{A}$ is projected at most twice. More precisely, the projection is given by a soft-collinear approximation with respect to some $\beta_I^\mu$ followed by a hard-collinear approximation with respect to $\beta_K^\mu(\neq \beta_I^\mu)$.} We see this as follows. First, when we increase the index $i$ of $\sigma_i$, going from smaller to larger regions, by eq.\ (\ref{containing_subgraphs_relations}) a soft line may move into jets, and then become hard. So the approximations $t_{\sigma_n} ... t_{\sigma_1}$ can be seen as several $\text{sc}_I$'s followed by several $\text{hc}_{I'}$'s, where $I'$ need not be different from $I$. Next, because both sc. and hc. act as projections, $t_{\sigma_n} ... t_{\sigma_1}$ for any given line, is indeed an $\text{sc}_I$ followed by an $\text{hc}_{I'}$. Finally, if $I'=I$, we will see at the end of this subsection (at eq.\ (\ref{generation_of_beta_square})) that a $\beta_I^2=0$ factor will be produced. By neglecting those vanishing terms, the only nontrivial case we should consider is $I'\neq I$.

\begin{figure}[t]
\centering
\includegraphics[width=12cm]{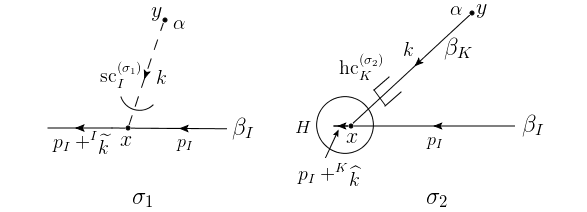}
\caption{Subgraphs of two leading pinch surfaces $\sigma_1$ and $\sigma_2$ of $\mathcal{A}$. The propagator with momentum $k^\mu$ is soft in $\sigma_1$, while collinear to $\beta_K^\mu\ (\neq \beta_I^\mu)$ in $\sigma_2$. In the approximated amplitude $t_{\sigma_1} \mathcal{A}$, the momentum $k^\mu$ appearing in $\left(p_I+k\right)^\mu$ is replaced by $^{I}\widetilde{k}^\mu= \left( k\cdot\beta_I \right) \overline{\beta}_I^\mu$; while in $t_{\sigma_2} \mathcal{A}$, $k^\mu$ is replaced by $^{K}\widehat{k}^\mu= \left( k\cdot\overline{\beta}_K \right) \beta_K^\mu$.}
\label{requirement_nontrivial_structure}
\end{figure}

From the observations above, in order to verify eqs.\ (\ref{requirement1}) and (\ref{requirement2}), the only nontrivial case of $\sigma_1$ and $\sigma_2$ we need to consider is shown in figure\ \ref{requirement_nontrivial_structure} (as well as their corresponding approximations $t_{\sigma_1}$ and $t_{\sigma_2}$). Note that if $\beta_I^\mu$ and $\beta_K^\mu$ are back-to-back ($\beta_I^\mu= \overline{\beta}_K^\mu$), the proof of the two equations will be trivial. So for generality, we do not assume that.

Now we can write down the rule of repetitive approximations on a momentum, and verify it is compatible with eqs.\ (\ref{requirement1}) and (\ref{requirement2}) by considering how the denominator factor $\left(p_I+k\right)^2$ changes according to the approximations $t_{\sigma_1}$, $t_{\sigma_2}$ and $t_{\sigma_2} t_{\sigma_1}$. According to (\ref{approximated_denominators}), we have:
\begin{align}
\begin{split}
t_{\sigma_1} \left( p_I+k \right)^2 &=\left( p_I+\left( k\cdot\beta_I \right)\overline{\beta}_I \right)^2=p_I^2+2\left( p_I\cdot\overline{\beta}_I \right)\left( k\cdot\beta_I \right),\\
t_{\sigma_2} \left( p_I+k \right)^2 &=\left( \left( p_I\cdot\overline{\beta}_I \right)\beta_I+\left( k\cdot\overline{\beta}_K \right)\beta_K \right)^2=2\left( p_I\cdot\overline{\beta}_I \right)\left( k\cdot\overline{\beta}_K \right)\left( \beta_I\cdot\beta_K \right).
\end{split}
\end{align}
The rule of repetitive approximations is as follows:
\begin{eqnarray} \label{denominator_change}
t_{\sigma_2}t_{\sigma_1} \left(p_I+k\right)^2 &&=t_{\sigma_2} \left( p_I+ \left( k\cdot\beta_I \right) \overline{\beta}_I \right)^2 \nonumber\\
&&=\left( \left( p_I\cdot\overline{\beta}_I \right) \beta_I+ \left( k\cdot\overline{\beta}_K \right) \left(\beta_K\cdot\beta_I \right) \overline{\beta}_I \right)^2 \nonumber\\
&&=2\left( p_I\cdot\overline{\beta}_I \right) \left( k\cdot\overline{\beta}_K \right) \left( \beta_I\cdot\beta_K \right) \nonumber\\
&&= t_{\sigma_2} \left(p_I+k\right)^2.
\end{eqnarray}

Apparently, eq.\ (\ref{requirement1}) is automatically satisfied. Eq.\ (\ref{requirement2}) is also satisfied if we retain the leading behaviors of the normal coordinates, because in the neighborhood of $\sigma_2^{\left\{ \sigma_1 \right\}}$ we have $p_I^\mu \approx \left( p_I\cdot \overline{\beta}_I \right) \beta_I^\mu$ and $k^\mu \approx \left( k\cdot \overline{\beta}_K \right)\beta_K^\mu$. This implies that (\ref{denominator_change}) describes the correct rule for repetitive approximation on a momentum.

Next, we show and verify the rule of repetitive approximations on a vector index. In order to do this, we consider a propagator carrying momentum $k^\mu$ and a vertex to which the gauge boson attaches with index $\alpha$ in figure\ \ref{requirement_nontrivial_structure}, and denote their product as $V^\alpha$. $V^\alpha$ corresponds to either a $\overline{\psi} \psi A^\alpha$ vertex or a $\partial\phi \phi A^\alpha$ vertex, which can be generalized to a 3-gluon vertex. Again, it becomes trivial when $\beta_I^\mu= \overline{\beta}_K^\mu$, so we do not assume this below.

If the vertex is $\overline{\psi} \psi A^\alpha$, then $V^\alpha=(\slashed{p}_I+\slashed{k} )\gamma^\alpha$ and under the approximations,
\begin{align}
\begin{split}
t_{\sigma_1} V^\alpha &=\left [ \slashed{p}_I+ \left( k\cdot\beta_I \right)\overline{\slashed{\beta}}_I \right ]\left( \gamma\cdot\overline{\beta}_I \right)\beta_I^\alpha=\slashed{p}_I\overline{\slashed{\beta}}_I\beta_I^\alpha, \\
t_{\sigma_2} V^\alpha &=\left [ \left( p_I\cdot\overline{\beta}_I \right)\slashed{\beta}_I+ \left( k\cdot\overline{\beta}_K \right)\slashed{\beta}_K \right ]\left( \gamma\cdot\beta_K \right)\overline{\beta}_K^\alpha=\left( p_I\cdot\overline{\beta}_I \right)\slashed{\beta}_I\slashed{\beta}_K\overline{\beta}_K^\alpha.
\end{split}
\end{align}
Then the rule gives
\begin{eqnarray} \label{fermion_numerator_change}
t_{\sigma_2} t_{\sigma_1} V^\alpha &&=\left [ \left( p_I\cdot\overline{\beta}_I \right)\slashed{\beta}_I+ \left( k\cdot\overline{\beta}_K \right)\left( \beta_I\cdot\beta_K \right)\overline{\slashed{\beta}}_I \right ]\left( \gamma\cdot\overline{\beta}_I \right)\left( \beta_I\cdot\beta_K \right)\overline{\beta}_K^\alpha\nonumber\\
&&=\left( p_I\cdot\overline{\beta}_I \right)\slashed{\beta}_I\overline{\slashed{\beta}}_I\left( \beta_I\cdot\beta_K \right)\overline{\beta}_K^\alpha.
\end{eqnarray}

To check $t_{\sigma_2}\left( 1-t_{\sigma_1} \right)\mathcal{A}\mid _{\text{div }\mathfrak{n}\big [ \sigma_1^{\left\{ \sigma_2 \right\}} \big ]}=0$, we see that
\begin{eqnarray} \label{fermion_numerator_change_check}
t_{\sigma_2}\left( 1-t_{\sigma_1} \right) V^\alpha=\left( p_I\cdot\overline{\beta}_I \right)\slashed{\beta}_I\left[\slashed{\beta}_K-\overline{\slashed{\beta}}_I\left( \beta_I\cdot\beta_K \right)\right]\overline{\beta}_K^\alpha.
\end{eqnarray}
In region $\mathfrak{n}\big[ \sigma_1^{\left\{ \sigma_2 \right\}}\big]$, $V^\alpha$ always appears in the combination for $V^\alpha \slashed{\beta}_I$. Then eq.\ (\ref{fermion_numerator_change_check}) contributes $O\left( \lambda \right)$ to $t_{\sigma_2} \left( 1-t_{\sigma_1} \right) \mathcal{A}$ in $\mathfrak{n}\big[ \sigma_1^{\left\{ \sigma_2 \right\}}\big]$ because we can decompose
\begin{eqnarray}
\slashed{\beta}_K=\overline{\slashed{\beta}}_I\left( \beta_I\cdot\beta_K \right)+\slashed{\beta}_I\left( \overline{\beta}_I\cdot\beta_K \right)+\slashed{\beta}_{I\perp}\left( \beta_{I\perp}\cdot\beta_K \right),
\end{eqnarray}
where only the first term in this expression is leading, and is cancelled in (\ref{fermion_numerator_change_check}).

To check $\left( 1-t_{\sigma_2} \right)t_{\sigma_1}\mathcal{A}\mid _{\text{div }\mathfrak{n}\big [ \sigma_2^{\left\{ \sigma_1 \right\}} \big ]}=0$ is similar
\begin{eqnarray}
\left( 1-t_{\sigma_2} \right)t_{\sigma_1} V^\alpha &&=\slashed{p}_I\overline{\slashed{\beta}}_I\beta_I^\alpha-\left( p_I\cdot\overline{\beta}_I \right)\slashed{\beta}_I\overline{\slashed{\beta}}_I\left( \beta_I\cdot\beta_K \right)\overline{\beta}_K^\alpha\nonumber\\
&&\approx \left( p_I\cdot\overline{\beta}_I \right)\slashed{\beta}_I\overline{\slashed{\beta}}_I \left [ \beta_I^\alpha-\left( \beta_I\cdot\beta_K \right)\overline{\beta}_K^\alpha \right ],
\end{eqnarray}
where we have only kept the leading terms in the second line. This result contributes $O\left(\lambda\right)$ to $\left( 1-t_{\sigma_2} \right)t_{\sigma_1}\mathcal{A}$ because for the leading term in $\mathfrak{n}\Big[ {\sigma_2^{\left\{ \sigma_1 \right\}}} \Big]$, $k^\alpha$ is scalar-polarized in the $K$-direction, so we can insert $\beta_{K \alpha '}\overline{\beta}_K^\alpha $ in the first term without changing the leading behavior, which then cancels the second term exactly.

If the vertex is $\partial\phi \phi A^\alpha$, then $V^\alpha=\left(2p_I+k\right)^\alpha$ and the rule is
\begin{eqnarray} \label{scalar_numerator_change}
t_{\sigma_1} V^\alpha &&=\beta_I^\alpha\overline{\beta}_{I\alpha '}\left (2p_I+ \left( k\cdot\beta_I \right)\overline{\beta}_I \right)^{\alpha '}=2\left(p_I\cdot\overline{\beta}_I\right)\beta_I^\alpha, \nonumber\\
t_{\sigma_2} V^\alpha &&=\overline{\beta}_K^\alpha\beta_{K\alpha '}\left (2p_I+\left( k\cdot\overline{\beta}_K \right)\beta_K \right)^{\alpha '}=2\left(p_I\cdot\beta_K\right)\overline{\beta}_K^\alpha, \nonumber\\
t_{\sigma_2} t_{\sigma_1} V^\alpha &&=\overline{\beta}_K^\alpha\beta_{K\alpha '}\beta_{I}^{\alpha '}\overline{\beta}_{I\alpha ''}\left (2p_I+\left( k\cdot\overline{\beta}_K \right)\left (\beta_K\cdot\beta_I \right)\overline{\beta}_I \right)^{\alpha ''}. \nonumber\\
&&=2\left(p_I\cdot\overline{\beta}_I\right)\left(\beta_I\cdot\beta_K\right)\overline{\beta}_K^\alpha
\end{eqnarray}
As above we easily verify that
\begin{eqnarray}
t_{\sigma_2}\left( 1-t_{\sigma_1} \right) V^\alpha\mid _{\text{div }\mathfrak{n}\big[ \sigma_1^{\left\{ \sigma_2 \right\}} \big]}=0=\left( 1-t_{\sigma_2} \right) t_{\sigma_1} V^\alpha\mid _{\text{div }\mathfrak{n}\big[ \sigma_2^{\left\{ \sigma_1 \right\}} \big]}.
\end{eqnarray}
The first equality is due to the expansion of $p_I$ in $\mathfrak{n}\big[ \sigma_1^{\left\{ \sigma_2 \right\}} \big]$, while the second equality is due to treating $k^\alpha$ as scalar-polarized in $\mathfrak{n}\big[ \sigma_2^{\left\{ \sigma_1 \right\}} \big]$, so we can insert a $\overline{\beta}_K^\alpha \beta_{K\alpha '}$ in $t_{\sigma_1} V^\alpha$. Note that this explanation can be generalized in the same way to a 3-gluon vertex, implying that eqs.\ (\ref{fermion_numerator_change}) and (\ref{scalar_numerator_change}) describe the correct rules for repetitive approximations on a vector index.

Having constructed the rules for repetitive approximations, eqs.\ (\ref{denominator_change}), (\ref{fermion_numerator_change}) and (\ref{scalar_numerator_change}), on an amplitude $\mathcal{A}$ from the requirements (\ref{requirement1}) and (\ref{requirement2}), we emphasize again that in order to completely verify the requirements, we also need to show the coincidence of pinch surfaces. This will be done in section\ \ref{pairwise_cancellation_regular: nested}, when we deal with a stronger relation, eq.\ (\ref{pairwise_cancellation_nested_conclusion}) there. The terms of the form $t_{\sigma_n} ... t_{\sigma_1} \mathcal{A}$, will be seen as the proper subtractions to remove IR divergences.

\begin{figure}[t]
\centering
\includegraphics[width=4cm]{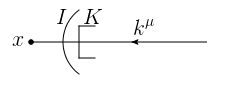}
\caption{The graphical description for the repetitive approximations. The momentum $k^\mu$ after the projections, which flows into the vertex $x$ is $\left( k\cdot\overline{\beta}_K \right)\left( \beta_K\cdot\beta_I \right)\overline{\beta}_I^\mu$.}
\label{convention_sc_followed_by_hc}
\end{figure}

For convenience, we add one more notation besides those introduced in eq.\ (\ref{hat_tilde_definition}). As we have argued at the beginning of this subsection, the only nontrivial combination of approximations on a given momentum is a soft-collinear with respect to $\beta_I^\mu$ followed by a hard-collinear with respect to $\beta_K^\mu$. In the upcoming text, especially in some of the figures, we will abbreviate a momentum projected by such a combination as $^{IK}\widetilde{\widehat{k}}$ (where $k^\mu$ is the original momentum) for simplicity. Explicitly,
\begin{eqnarray} \label{hat_plus_tilde_definition}
^{IK}\widetilde{\widehat{k}}^\mu \equiv \left( k\cdot\overline{\beta}_K \right)\left( \beta_K\cdot\beta_I \right)\overline{\beta}_I^\mu.
\end{eqnarray}
In terms of graphs, they are represented by figure\ \ref{convention_sc_followed_by_hc}.

\begin{figure}[t]
\centering
\includegraphics[width=4.5cm]{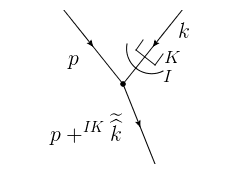}
\caption{The confluence of momenta $k^\mu$ and $p^\mu$ in $\rho^{ \left\{ \sigma \right\} }$, where the merged momentum is of the value $\Big( p+ ^{IK}\widetilde{\widehat{k}} \Big)^\mu \equiv p^\mu+ \left( k\cdot\overline{\beta}_K \right) \left( \beta_K\cdot\beta_I \right) \overline{\beta}_I^\mu$, due to the approximations $t_{\sigma_1}\ (\text{sc}_I)$ and $t_{\sigma_2}\ (\text{hc}_K)$.}
\label{confluence_repetitive}
\end{figure}

\bigbreak
Given these results for repetitive approximations, we should generalize the analysis in section\ \ref{pinch_surfaces_from_single_approximation} to an amplitude acted on by repetitive projections: $t_{\sigma_n} ... t_{\sigma_1} \mathcal{A}$ ($\sigma_1 \subset ... \subset \sigma_n$). Compared with the graphical representation of $t_\sigma \mathcal{A}$ in figure\ \ref{approximated_amplitude}, now we have more subgraphs whose external momenta have been modified for $t_{\sigma_n} ... t_{\sigma_1} \mathcal{A}$. Inside each of them there is a classical picture from the Landau equations at any pinch surface. To study the pinch surfaces, which are denoted by $\rho^{ \left\{ \sigma_n...\sigma_1 \right\} }$, we again study the configuration in figure\ \ref{confluence}, but with repetitive approximations taken into consideration. In other words, we focus on the ``confluence'' of the double-projected momentum $\left( t_{\sigma_2} t_{\sigma_1} k \right)^\mu$ with another momentum $p^\mu$ in figure\ \ref{confluence_repetitive}, using our notation for repetitive approximations introduced above.

In the figure $I$ and $K$ can be either equal or not. Let's assume $I\neq K$ first. Then we can analyze the configurations of figure\ \ref{confluence_repetitive} as we have done for figure\ \ref{confluence}, and one can check that everything follows similarly. The results are summarized in table\ \ref{repetitive_approximated_PS_summary}, which can be seen as a generalization of table\ \ref{single_approximated_PS_summary}: in the former there are two vectors of reference ($\beta_I^\mu$ and $\beta_K^\mu$) while in the latter there is only one. If we assume that $\overline{\beta}_I^\mu = \beta_K^\mu \equiv v^\mu$, table\ \ref{repetitive_approximated_PS_summary} then becomes exactly table\ \ref{single_approximated_PS_summary}.
\begin{table}[t]
\captionsetup{justification=centering,margin=1.7cm}
\caption{Summary of the possible configurations of figure\ \ref{confluence}, which depend on $\Big( p+ ^{IK}\widetilde{\widehat{k}} \Big)^\mu \equiv p^\mu+ \left( k\cdot\overline{\beta}_K \right) \left( \beta_K\cdot\beta_I \right) \overline{\beta}_I^\mu\ (I\neq K)$.}
\begin{center}
\begin{tabular}{ |c||c|c|c|c|c| } 
\hline
 \multirow{1}{*}{$k^\mu$} & \multirow{1}{*}{$p^\mu$} & $p^\mu+\left(t_\sigma k \right)^\mu$ & Description & Classification \\ 
 \hline
 \hline
 \multirow{2}{*}{Soft} & No con- & \multirow{2}{*}{$p^\mu$} & A soft propagator & \multirow{2}{*}{Regular}\\ 
 & straints & & joining $p^\mu$ & \\
 \hline
 & \multirow{2}{*}{Hard} & \multirow{2}{*}{Hard} & A lightlike propagator joining & \multirow{2}{*}{Regular} \\
 & & & the hard subgraph $H^{(\rho^{\left\{ \sigma \right\}})}$ & \\ \cline{2-5}
 \multirow{2}{*}{Col. to} & & \multirow{4}{*}{Col. to $\overline{\beta}_I^\mu$} & $k^\mu \parallel \overline{\beta}_I^\mu:$ $p^\mu$ and $k^\mu$ are & \multirow{2}{*}{Regular} \\
 \multirow{2}{*}{any} & \multirow{2}{*}{Col. to $\overline{\beta}_I^\mu$} & & lightlike in the same direction & \\ \cline{4-5}
 \multirow{2}{*}{vector} & & & $k^\mu \nparallel \overline{\beta}_I^\mu:$ $p^\mu$ and $k^\mu$ are & \multirow{2}{*}{Regular} \\
 \multirow{2}{*}{except} & & & lightlike in different directions & \\ \cline{2-5}
 \multirow{2}{*}{$\overline{\beta}_K^\mu$} & Col. to & \multirow{2}{*}{Hard} & $p^\mu$ and $k^\mu$ lightlike, and & \multirow{2}{*}{Regular} \\
 & $\beta_L^\mu$ ($\neq \overline{\beta}_I^\mu$) & & join the hard subgraph $H^{(\rho^{\left\{ \sigma \right\}})}$ & \\ \cline{2-5}
 & \multirow{2}{*}{Soft} & \multirow{2}{*}{Col. to $\overline{\beta}_I^\mu$} & A lightlike propagator attached & \multirow{2}{*}{Regular}\\
 & & & by a soft momentum $p^\mu$ & \\
 \hline
 & \multirow{2}{*}{Hard} & \multirow{8}{*}{$p^\mu$} & A lightlike propagator joining & \multirow{2}{*}{Regular}\\
 & & & the hard subgraph $H^{(\rho^{\left\{ \sigma \right\}})}$ & \\ \cline{2-2}\cline{4-5}
 & Col. & & $p^\mu$ and $k^\mu$ are lightlike & \multirow{2}{*}{Regular}\\
 Col. & to $\overline{\beta}_K^\mu$ & & in the same direction & \\ \cline{2-2}\cline{4-5}
 to $\overline{\beta}_K^\mu$ & Col. to & & $p^\mu$ and $k^\mu$ are lightlike & \multirow{2}{*}{Regular}\\
 & $\beta_L^\mu (\neq \overline{\beta}_K^\mu) $ & & in different directions & \\ \cline{2-2}\cline{4-5}
 & \multirow{2}{*}{Soft} & & A lightlike propagator is & \multirow{2}{*}{Soft-exotic} \\
 & & & attached to a soft vertex &\\
 \hline
 \multirow{3}{*}{Hard} & Col. to $\overline{\beta}_I^\mu$ & \multirow{2}{*}{Col. to $\overline{\beta}_I^\mu$} & A hard propagator is attached & \multirow{2}{*}{Hard-exotic} \\
 & or soft & & to soft or lightlike propagators &\\ \cline{2-5}
 & (Otherwise) & Hard & $p^\mu$ joining the hard subgraph & Regular \\ 
 \hline
\end{tabular}
\end{center}
\label{repetitive_approximated_PS_summary}
\end{table}

If $I=K$, since $\beta_I\cdot\beta_K=0$, the confluence momentum $\Big( p+ ^{IK}\widetilde{\widehat{k}} \Big)^\mu = p^\mu$. Now we claim that whatever the configuration of figure\ \ref{confluence_repetitive} is, there is always a zero appearing as the overall factor in the approximated amplitude, which then will not contribute. This is because the soft-collinear approximations can only act on gauge bosons. In the presence of a hard-collinear approximation acting on the same line, this gauge boson must be scalar-polarized. From the definitions in eqs.\ (\ref{hard-collinear_approximation}) and (\ref{soft-collinear_approximation}), the vector index of the gauge boson is projected and we obtain
\begin{eqnarray} \label{generation_of_beta_square}
V^{...\alpha} \xrightarrow[ ]{\text{approx.}} V^{...\alpha''} \overline{\beta}_{\alpha''} \beta^{\alpha'} \beta_{\alpha'} \overline{\beta}^\alpha=0,
\end{eqnarray}
for some jet velocity $\beta$. This vanishes because $\beta^2=0$. This observation implies that we do not need to consider the case $I=K$ whenever we study the pinch surfaces of a repetitively-approximated amplitude and require them to be IR divergent, as we will see in Theorems 2, 4 and 6. Nevertheless, these terms will still be included in the analysis of section\ \ref{the_proof_of_cancellation} so as to manifest the IR cancellation in a more direct way.

With the knowledge of table\ \ref{repetitive_approximated_PS_summary}, we classify the various configurations of a pinch surface $\rho^{ \left\{ \sigma_n...\sigma_1 \right\} }$ into the types of regular and exotic, just as we have done for a $\rho^{\left\{ \sigma \right\}}$. Namely, the normal spaces are conserved at regular configurations and not conserved at exotic configurations. As one studies the divergences of $t_{\sigma_n} ... t_{\sigma_1} \mathcal{A}$, all such pinch surfaces need to be taken into consideration. Theorem 1, given at the end of section\ \ref{pinch_surfaces_from_single_approximation}, can also be generalized from the knowledge of table\ \ref{repetitive_approximated_PS_summary}.

\textbf{Theorem 2: In the pinch surface $\rho^{ \left\{ \sigma_n...\sigma_1 \right\} }$, all the propagators of $J_I^{(\sigma_i)}\ (\sigma_1 \subseteq \sigma_i \subseteq \sigma_n)$ that are lightlike can only be collinear to $\beta_I^\mu$ or $\overline{\beta}_I^\mu$.}

\textit{Proof of Theorem 2:} As in the proof of Theorem 1, we denote the set of lightlike propagators (in $\rho^{ \left\{ \sigma_n...\sigma_1 \right\} }$) of $J_I^{(\sigma_i)}$ that are not collinear to $\beta_I^\mu$, as $\gamma$.

\begin{figure}[t]
\centering
\includegraphics[width=6.5cm]{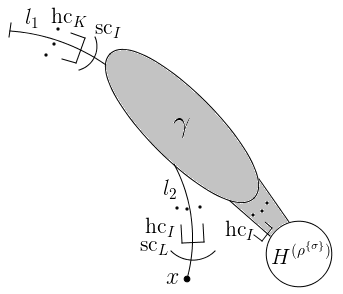}
\caption{The subgraph $\gamma \equiv J_I^{(\sigma_i)} \bigcap J_K^{(\rho^{ \left\{ \sigma_n...\sigma_1 \right\} })}$, which includes the propagators marked by momentum $l_2^\mu$ as well as the shaded area. The combinations of hard-collinear and soft-collinear approximations are given by $t_{\sigma_n} ... t_{\sigma_1}$.}
\label{two_jets_intersection_generalize}
\end{figure}

We depict $\gamma$ in figure\ \ref{two_jets_intersection_generalize}, which includes the shaded area as well as those propagators whose momenta are denoted as $l_2^\mu$. The vertices $x$ are arbitrary jet vertices in $\rho^{ \left\{ \sigma_n...\sigma_1 \right\} }$. Since the propagators with a single approximation from $t_{\sigma_n} ...t_{\sigma_1}$ have been taken into account in the proof of Theorem 1, we only consider those with repetitive approximations. Each momentum denoted by $l_1$ is collinear to some $\beta_K^\mu$ in $\rho^{ \left\{ \sigma_n...\sigma_1 \right\} }$ and attached to $H$ in some $\sigma_{k_1}$, while soft and attached to $J_I$ in some $\sigma_{i_1}$ ($\subset \sigma_{k_1}$). The momenta denoted by $l_2$ are collinear to $\beta_I^\mu$ and attached to $H$ in some $\sigma_{i_2}$, while soft and attached to $J_L$ in some $\sigma_{j_2}$ ($\subset \sigma_{i_2}$). By construction, there are hard-collinear approximations $\text{hc}_K$ acting on $l_1^\mu$, and soft-collinear approximations $\text{sc}_L$ on $l_2^\mu$.

Similarly to the proof of Theorem 1, first we focus on the external propagators with momenta $l_1^\mu$. Due to the repetitive approximations, the external momenta that enter the shaded area will be of the form $\left(l_1\cdot\overline{\beta}_K\right) \left(\beta_K\cdot\beta_I\right) \overline{\beta}_I^\mu$, which is always in the direction of $\overline{\beta}_I^\mu$, ensuring that the propagators of $\gamma$ that contains $l_1^\mu$ can only be collinear to $\overline{\beta}_I^\mu$ in $\rho^{ \left\{ \sigma_n...\sigma_1 \right\} }$ as well. Then we focus on the propagators with momenta $l_2^\mu$. After being acted on by the approximations, all the momenta entering the subgraph $J_K^{(\rho^{ \left\{ \sigma_n...\sigma_1 \right\} })} \setminus J_I^{(\sigma_i)}$ through the jet vertices $x$ are of the form $\left(l_2\cdot\overline{\beta}_I\right) \left(\beta_I\cdot\beta_L\right) \overline{\beta}_L^\mu$, rather than $l_2^\mu$. In other words, the propagators that contain the momenta $l_2^\mu$ have denominators of the form $2\left( p\cdot\overline{\beta}_L \right) \left (\beta_I\cdot\beta_L \right) \left( l_2 \cdot \overline{\beta}_I \right)+i\epsilon$ if they are in the subgraph $J_K^{(\rho^{ \left\{ \sigma_n...\sigma_1 \right\} })} \setminus J_I^{(\sigma_i)}$, and $l_2^2+i\epsilon$ if they are in $\gamma$. In order that these types of propagators are both lightlike at the pinch surface, the normal coordinates can only be $\left( l_2\cdot \overline{\beta}_I \right)$, meaning that the propagators marked by momentum $l_2^\mu$ are also parallel to $\overline{\beta}_I^\mu$ at $\rho^{ \left\{ \sigma_n...\sigma_1 \right\} }$. In conclusion, Theorem 2 is proved.

\subsection{Divergences are logarithmic}
\label{divergences_are_logarithmic}

Another natural question on the effects of the approximation operators is whether they preserve the degree of divergences of the leading term near a pinch surface, which is logarithmic from power counting. We expect so, since in the process of showing IR finiteness, the IR divergences in an approximated amplitude need to be cancelled by some other subtraction terms, and if all the divergences are still logarithmic, we only need to show the coincidence of their leading terms (differing by a minus sign). Otherwise we will need the cancellations for next-to-leading terms, etc., which would be more difficult.

We shall take an arbitrary pinch surface $\rho^{ \left\{ \sigma_n ... \sigma_1 \right\} }$ and discuss all its possible configurations in table\ \ref{single_approximated_PS_summary}, and relations with the forest $\left\{ \sigma_1, ..., \sigma_n \right\}$. We classify the possibilities into four cases: ``nested \& regular'', ``overlapping \& regular'', ``soft-exotic'' and ``hard-exotic''. We will explain their meanings, and analyze them one by one. For the ``overlapping \& regular'' and ``soft-exotic'' cases, we only consider a single approximation in this subsection, and put the generalizations to repetitive approximations in appendix \ref{power_counting_repetitive_approximation}. After these analyses, we discuss whether the three features of a leading pinch surface of $\mathcal{A}$, as introduced in section\ \ref{introduction}, still hold for a pinch surface $\rho^{\left\{ \sigma_1, ..., \sigma_n \right\}}$ with logarithmic divergence.

\bigbreak
\centerline{\textbf{Nested \& Regular}}
By ``nested \& regular'' we mean that $\rho^{ \left\{ \sigma_n ... \sigma_1 \right\} }$ is a regular pinch surface, and nested with every $\sigma_i$ ($1 \leqslant i \leqslant n$, and one may recall the definition in eq.\ (\ref{containing_definition})). Without loss of generality, we assume $\sigma_1 \subset...\subset \sigma_m \subseteq \rho^{ \left\{ \sigma_n ... \sigma_1 \right\} } \subset \sigma_{m+1} \subset...\subset \sigma_n$. We argue below that though $\rho^{ \left\{ \sigma_n ... \sigma_1 \right\} }$ may differ from any pinch surface of $\mathcal{A}$, we can always find a corresponding pinch surface of $\mathcal{A}$, say $\rho_{\mathcal{A}}$, which has the same set of normal coordinates with $\rho^{ \left\{ \sigma_n ... \sigma_1 \right\} }$. In other words, the power counting procedures for $\rho_{\mathcal{A}}$ and $\rho^{ \left\{ \sigma_n ... \sigma_1 \right\} }$ are identical, assuring the degree of divergence near $\rho^{ \left\{ \sigma_n ... \sigma_1 \right\} }$ is at worst logarithmic.

The approximations acting inside $S^{(\rho^{ \left\{ \sigma_n ... \sigma_1 \right\} })}$ are provided by $t_{\sigma_{m+1}}, ..., t_{\sigma_n}$, because they must correspond to the pinch surfaces that contain $\rho^{ \left\{ \sigma_n ... \sigma_1 \right\} }$. But since a soft momentum remains soft after any projections, we can simply remove these approximations without changing the configuration of $\rho^{ \left\{ \sigma_n ... \sigma_1 \right\} }$ or its degree of IR divergence (though the value of the leading term may vary). Similarly, the approximations inside $H^{(\rho^{ \left\{ \sigma_n ... \sigma_1 \right\} })}$ are provided by $t_{\sigma_1}, ..., t_{\sigma_{m-1}}$. But since the momenta in $H^{(\rho^{ \left\{ \sigma_n ... \sigma_1 \right\} })}$ are hard and hence do not contribute to the degree of divergence, we can simply remove the approximations without changing the configuration of $\rho^{ \left\{ \sigma_n ... \sigma_1 \right\} }$ or its degree of divergence.

Now we examine the approximations acting inside $J_I^{(\rho^{ \left\{ \sigma_n ... \sigma_1 \right\} })}$. They come from the following two sources: (1) the hard-collinear approximations of $t_{\sigma_{m+1}}, ..., t_{\sigma_n}$ which project the jet momenta attached to $H^{(\sigma_j)} \bigcap J_I^{(\rho^{ \left\{ \sigma_n ... \sigma_1 \right\} })}\ (j=m+1, ..., n)$ in $\sigma_j$, and carry the momenta $^I\widehat{k}^\mu$ in the $\beta_I$-direction into $H^{(\sigma_j)}$; (2) the soft-collinear approximations of $t_{\sigma_1}, ..., t_{\sigma_{m-1}}$, which project the soft momenta of $S^{(\sigma_k)} \bigcap J_I^{(\rho^{ \left\{ \sigma_n ... \sigma_1 \right\} })}\ ( k= 1, ..., m-1 )$ in $\sigma_k$, and carry the momenta $^I\widetilde{k}^\mu$ in the $\overline{\beta}_I$-direction into $J_I^{(\rho^{ \left\{ \sigma_n ... \sigma_1 \right\} })}$. Both these types of approximations act on the momenta of $J_I^{(\rho^{ \left\{ \sigma_n ... \sigma_1 \right\} })}$, and assure that they can only be pinched collinear to $\beta_I^\mu$ at $\rho^{ \left\{ \sigma_n ... \sigma_1 \right\} }$. The only difference between $\rho_\mathcal{A}^{ }$ and $\rho^{ \left\{ \sigma_n ... \sigma_1 \right\} }$ is due to the soft-collinear approximations. That is, as is explained in the paragraph below eq.\ (\ref{cii_result}), the $\beta_I^\mu$-component of the momenta of $S^{(\sigma_k)} \bigcap J_I^{(\rho^{ \left\{ \sigma_n ... \sigma_1 \right\} })}$ is either positive with no upper bound or negative with no lower bound. But this difference is only for the intrinsic coordinates. As a result, neither the configuration of $\rho^{ \left\{ \sigma_n ... \sigma_1 \right\} }$ or the degree of divergence of the leading term will be changed, if we simply remove the approximations inside $J_I^{(\rho^{ \left\{ \sigma_n ... \sigma_1 \right\} })}$.

In conclusion, each $\rho^{ \left\{ \sigma_n ... \sigma_1 \right\} }$ that is nested with every $\sigma_i\ (i=1,...,n)$ corresponds to a $\rho_\mathcal{A}$ --- a pinch surface of $\mathcal{A}$ --- with their degrees of divergence being identical. Since $\rho_\mathcal{A}$ is at worst logarithmically divergent, we conclude that the IR divergence of $\rho^{ \left\{ \sigma_n ... \sigma_1 \right\} }$, is at worst logarithmic.

\bigbreak
\centerline{\textbf{Overlapping \& Regular}}
By ``overlapping \& regular'' we mean the case where $\rho^{ \left\{ \sigma_n ... \sigma_1 \right\} }$ is regular, and overlaps with some $\sigma_i (1\leqslant i\leqslant n)$. To evaluate its degree of divergence, we consider the pinch surfaces with only a single approximation operator here, i.e. $\rho^{\left\{ \sigma \right\}}$, and include the discussion on repetitive approximations in appendix \ref{overlapping_regular_repetitive}. For simplicity, we will drop the superscript and use $\rho$ instead, until the end of the power counting evaluation.

According to the definition, eq.\ (\ref{subgraphs_intersection_nonempty}), overlapping implies one of the two possibilities. First, a hard, jet or soft subgraph of $t_\sigma\mathcal{A}$ at $\rho$ contains the corresponding subgraph at $\sigma$ while another hard, jet or soft subgraph of $\mathcal{A}$ at $\sigma$ contains the corresponding subgraph at $\rho$. Second, some jet subgraphs overlap, i.e. $J_I^{(\sigma)}\bigcap J_K^{(\rho)}\neq \varnothing\ (I\neq K)$.

The first case is simpler, since as in the ``nested \& regular'' case, the action of $t_\sigma$ does not change the power counting procedure at $\rho$. We immediately come to the conclusion that the divergence near $\rho$ is at worst logarithmic.

\begin{figure}[t]
\centering
\includegraphics[width=9cm]{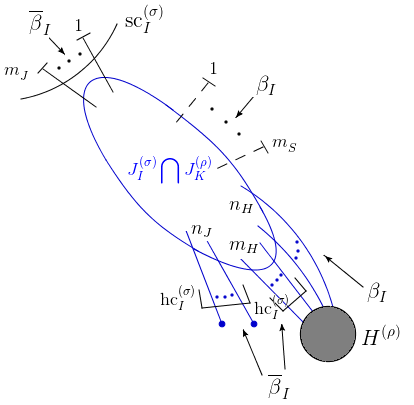}
\caption{The subgraph $\gamma \equiv J_I^{(\sigma)} \bigcap J_K^{(\rho)}$ in $\rho$, which includes all the blue lines. There are $\left(m_J^{ }+m_S^{ }\right)$ external lines whose propagators are not included, and $\left(n_J^{ }+m_H^{ }\right)$ internal lines whose propagators are included. The set of the solid lines represents $J_I^{(\rho)}$, and the black subgraph is $H^{(\rho)}$.}
\label{structure_for_overlapping_pc}
\end{figure}

Turning to the second case, the subtleties originate from the fact that $J_I^{(\sigma)}\bigcap J_K^{(\rho)}$ is in the direction of $\overline{\beta}_I^\mu$, which may not be the same as the other parts of $J_K^{(\rho)}$, as is explained in Theorem 1 of section\ \ref{pinch_surfaces_from_single_approximation}. We draw the subgraph $\gamma\equiv J_I^{(\sigma)}\bigcap J_K^{(\rho)}$ in $\rho$ together with the approximations given by $t_\sigma$, and mark it blue in figure\ \ref{structure_for_overlapping_pc} below as a generalization of our previous examples in (the upper rows of) figures\ \ref{regular_pinch_surface1} and \ref{regular_pinch_surface2}.

In detail, the external and internal propagators of $\gamma$ with approximations on their momenta, are from different sources. (1) Some of them are soft and attached to $J_I^{(\sigma)}$ in $\sigma$, while lightlike in the direction of $\beta_K^\mu$ in $\rho$. These are external propagators of $\gamma$. (2) Some are collinear to $\beta_I^\mu$ in $\sigma$ and attached to $H^{(\sigma)}$, while becoming collinear to the opposite direction ($\overline{\beta}_I^\mu$) in $\rho$, being internal propagators of $\gamma$, either attached to $H^{(\rho)}$ or not. The two types (1) and (2) are respectively associated with soft-collinear and hard-collinear approximations from $t_\sigma$. In comparison, some other propagators of $\gamma$ are internal propagators of $J_I^{(\sigma)}$ in $\sigma$, and are soft or attached to $H^{(\rho)}$ in $\rho$. The latter are not associated with approximations. All these types are shown in figure\ \ref{structure_for_overlapping_pc}.

In the figure, $m_J^{ }$ is the number of external lines of $\gamma$ that belong to the subgraph $S^{(\sigma)} \bigcap J_K^{(\rho)}$; $m_S^{ }$ is the number of external lines that belong to $J_I^{(\sigma)}\bigcap S^{(\rho)}$; $n_J^{ }$ is the number of internal lines that are attached to $H^{(\sigma)}$ but not to $H^{(\rho)}$; $n_H^{ }$ is the number of internal lines that are attached to $H^{(\rho)}$ but not to $H^{(\sigma)}$; finally, there are $m_H^{ }$ internal lines, each one having an endpoint attached to both $H^{(\rho)}$ and $H^{(\sigma)}$. These carry polarization $\overline{\beta}_I^\mu$ into $\gamma$. Due to the operator $t_\sigma$ and the ranges of momenta in $\rho$, vectors $\beta_I^\mu$ and $\overline{\beta}_I^\mu$ together with momenta from vertices in $\gamma$ form invariants in the leading term, as is shown in the figure.

Now we undertake the power counting for this leading term. Suppose the degree of divergence is $p^{(\rho)}\left( \gamma \right)$, then by definition,
\begin{eqnarray} \label{overlapping_regular_pc1}
p^{(\rho)}\left( \gamma \right)=2L-N+n_{\text{num}}^{(\rho)},
\end{eqnarray}
where $L$ is the number of loops, $N$ is the number of propagators, $V$ is the number of vertices, and $n_{\text{num}}^{(\rho)}$ represents the numerator contribution. The number of independent loops in $\gamma$ as well as those formed by the $n_J^{ }$ propagators, can be expressed in terms of $N$ and $V$ by Euler's formula,
\begin{eqnarray} \label{euler_formula1}
L=N-\left(V_3+V_4+1\right)+1=N-V_3-V_4,
\end{eqnarray}
where the +1 in the bracket corresponds to the external vertices of the $n_J^{ }$ propagators. Combining with the identity counting half-edges,
\begin{eqnarray} 
2N=3V_3+4V_4+\left( n_J^{ }+m_H^{ }+n_H^{ } \right)- \left( m_J^{ }+m_S^{ } \right),
\end{eqnarray}
we have
\begin{eqnarray} \label{overlapping_regular_pc2}
p^{(\rho)}\left( \gamma \right) =-\frac{1}{2}V_3 +\frac{1}{2} \left( n_J^{ }+m_H^{ }+n_H^{ }-m_J^{ }-m_S^{ } \right) +n_{\text{num}}^{(\rho)}.
\end{eqnarray}

We now calculate $n_{\text{num}}^{(\rho)}$. In $\rho$, let $\alpha_{ \overline{I} I}$ be the number of the invariants $\left(\beta_I\cdot \overline{\beta}_I\right)$ appearing in the expression of $\gamma$, $\alpha_{\overline{I}l}$ be the number of the invariants $\left(l\cdot \overline{\beta}_I\right)$, and so on. Since from eq.\ (\ref{momenta_scaling}) every momentum of the propagators in $\gamma$ can be expressed as
\begin{eqnarray} 
l^\mu \sim \mathcal{O}\left(Q\right)\overline{\beta}_I^\mu+\mathcal{O}\left(\lambda Q\right)\beta_I^\mu+\mathcal{O}\left(\lambda^{1/2}Q\right)\beta_{I\perp}^\mu,
\end{eqnarray}
the numerator contribution can then be rewritten as
\begin{eqnarray} \label{numerator_contribution}
n_{\text{num}}^{(\rho)}=\alpha_{\overline{I}l}+\alpha_{ll},
\end{eqnarray}
in which each invariant counted in $\alpha_{\overline{I}l}$ and $\alpha_{ll}$ contributes a factor $\lambda$, while the other invariants counted in $\alpha_{\overline{I}I}$ and $\alpha_{Il}$ contribute orders $\lambda^0$.

On one hand, the uppermost $m_J^{ }$ propagators in figure\ \ref{structure_for_overlapping_pc} are external propagators of $\gamma$, and projected onto the $\overline{\beta}_I$-component by the soft-collinear approximation of $t_\sigma$. So each of them provides a $\overline{\beta}_I^\mu$ to the subgraph $\gamma$. On the other hand, the lowest $\left( n_J^{ }+m_H^{ } \right)$ propagators are internal propagators of $\gamma $, and only the $\beta_I^\mu$-component remains after the hard-collinear approximation. So equivalently, each of them is contracted with a $\overline{\beta}_I^\mu$. These lightlike vectors are generated by $t_\sigma$, and there are also some generated when we focus on the leading behavior near $\rho^{\left\{ \sigma \right\}}$. For example, the $m_S^{ }$ external propagators are soft in $\rho$, so we can impose an $\text{sc}_{\bar{I}}$ (soft-collinear approximation with respect to $\overline{\beta}_I^\mu$) on each of them without changing the leading behavior, and a $\beta_I^\mu$ is automatically provided to $\gamma$. Similarly, the $n_H^{ }$ internal propagators are collinear to $\overline{\beta}_I^\mu$ and attached to $H^{(\rho)}$ in $\rho$, so we can impose an $\text{hc}_{\overline{I}}^{(\rho)}$ on each of them, and a $\beta_I^\mu$ is automatically provided.

Now we can relate the numbers of different invariants by counting the vectors $\overline{\beta}_I^\mu$, $\beta_I^\mu$ and $l^\mu$. Explicitly, we have
\begin{eqnarray} \label{alpha_relations_1}
\alpha_{\overline{I}I}+\alpha_{\overline{I}l}=m_J^{ }+m_H^{ }+n_J^{ }\ \ \ (\text{to count }\overline{\beta}_I^\mu),
\end{eqnarray}
\begin{eqnarray} \label{alpha_relations_2}
\alpha_{\overline{I}I}+\alpha_{Il}=m_S^{ }+n_H^{ }\ \ \ (\text{to count }\beta_I^\mu),
\end{eqnarray}
\begin{eqnarray} \label{alpha_relations_3}
\alpha_{\overline{I}l}+\alpha_{Il}+2\alpha_{ll}=V_3\ \ \ (\text{to count }l^\mu).
\end{eqnarray}
We can combine these relations and solve for $\alpha_{ll}$ as
\begin{eqnarray} \label{overlapping_regular_pc3}
\alpha_{ll}=\frac{1}{2}\left( V_3-m_J^{ }-m_H^{ }-n_J^{ }-m_S^{ }-n_H^{ }+2\alpha_{\overline{I}I} \right).
\end{eqnarray}
Then eqs.\ (\ref{numerator_contribution}), (\ref{alpha_relations_1}) and (\ref{overlapping_regular_pc3}) in (\ref{overlapping_regular_pc2}) give the final result
\begin{eqnarray} \label{overlapping_regular_pc4}
p^{(\rho)} \left(\gamma\right)=m_H^{ }+n_J^{ }-m_S^{ }.
\end{eqnarray}

The power counting carried out above is for the subgraph $\gamma$. If we consider the IR divergence of the entire graph $t_\sigma\mathcal{A}$, we also need to study how the external propagators of $\gamma$ can affect the power counting for $t_\sigma\mathcal{A} \setminus\gamma$. First, lines counted in $m_H^{ }$ attach to $H^{(\rho)}$ and hence produce no other contributions to power counting. Second, each line counted in $n_J^{ }$ can produce a $-1$ in power counting by attaching to a line in some jet $K\ (\neq I)$. Such lines would be in $H^{(\sigma)} \bigcap J_K^{(\rho)}$. Finally, each line in the set labelled $m_S^{ }$ produces a $-1$ in (\ref{overlapping_regular_pc4}) from the power counting of $\gamma$. As we shall see, this is necessary to produce logarithmic divergences. Explicitly, for any regular $\rho$ that overlaps with $\sigma$ through a set of nonempty subgraphs $J_I^{(\sigma)} \bigcap J_K^{(\rho)}$, we have
\begin{eqnarray} \label{overlapping_regular_pc5}
p^{(\rho)}\left( t_\sigma\mathcal{A} \right) &&= \sum_{A=1}^N p_A^{(\rho)} +p^{(\rho)}\left( S \right)\nonumber \\
&&= \sum_{A=1}^N \Big[ p^{(\rho)}\big( J_A^{[A]} \big) +p^{(\rho)}\left( \gamma_A \right) \Big]+p^{(\rho)}\left(S\right),
\end{eqnarray}
where $p_A^{(\rho)}$ is the contribution to $p^{(\rho)}\left( t_\sigma\mathcal{A} \right)$ from the jet subgraph in $\rho$. We decomposed $p_A^{(\rho)}$ further in the second line above: $\gamma_A^{ }$ is the subgraph of $J_A^{(\sigma)}$ whose propagators are in $J_A^{(\sigma)} \bigcap \left( \bigcup_{B\neq A} J_B^{(\rho)} \right)$, whose lines are collinear to $\overline{\beta}_A^\mu$ in $\rho$, and $J_A^{[A]}$ is the subgraph of $J_A^{(\rho)}$ whose propagators are collinear to $\beta_A^\mu$ in $\rho$. Lines in $J_A^{[A]}$ may be either from $J_A^{(\sigma)}$, $H^{(\sigma)}$ or $S^{(\sigma)}$. That is, the contribution from the jet subgraph in $\rho$ can be rewritten as the sum over those from $J_A^{[A]}$ and $\gamma_A^{ }$, as is shown above. The lower bound of the first term can be evaluated as
\begin{eqnarray} \label{overlapping_regular_pc5_term1}
p^{(\rho)}\big( J_A^{[A]} \big) \geqslant -v_{A}^{[A]} -\overline{n}_{J}^{[A]} +\text{num}_A^{ }(n_J),
\end{eqnarray}
where $v_{A}^{[A]}$ is the number of soft gauge bosons attached to $J_{A}^{[A]}$, and $\overline{n}_J^{[A]}$ is the number of the $n_J$ propagators in figure\ \ref{structure_for_overlapping_pc} that are attached to $J_{A}^{[A]}$ from subgraphs $\gamma_B^{ }\ (B\neq A)$. The first term in eq.\ (\ref{overlapping_regular_pc5_term1}) is from the standard power counting of a jet subgraph, which is obtained by removing the attachments of the $n_J^{[A]}$ propagators from $J_A^{[A]}$, and the second (third) term is the extra denominator (numerator) contribution that is generated by these attachments.

The evaluation result of $p^{(\rho)}\left( \gamma_A^{ } \right)$ can be directly read from eq.\ (\ref{overlapping_regular_pc4}), and the leading contribution of $p^{(\rho)}\left(S\right)$ can be obtained from a standard power counting. That is,
\begin{eqnarray}
p^{(\rho)}\left( \gamma_A^{ } \right)&&= m_{H}^{[A]} +n_{J}^{[A]} -m_{S}^{[A]}; \label{overlapping_regular_pc5_term2}\\
p^{(\rho)}\left(S\right) &&\geqslant v_{A}^{[A]} +m_{S}^{[A]} \label{overlapping_regular_pc5_term3}.
\end{eqnarray}
In (\ref{overlapping_regular_pc5_term2}), the symbols $m_{H}^{[A]}$, $n_{J}^{[A]}$ and $m_{S}^{[A]}$ are the numbers of specific lines of $\gamma_A^{ }$, which separately correspond to $m_H$, $n_J$ and $m_S$ in figure\ \ref{structure_for_overlapping_pc}. (Notice that $n_{J}^{[A]}$ is different from $\overline{n}_J^{[A]}$.) With this construction, $\left( v_{A}^{[A]} +m_{S}^{[A]} \right)$ in (\ref{overlapping_regular_pc5_term3}) is the number of soft gauge bosons attached to $J_A$, which is equal to the leading contribution of $p^{(\rho)}\left(S\right)$. Using eqs.\ (\ref{overlapping_regular_pc5_term1})--(\ref{overlapping_regular_pc5_term3}) in (\ref{overlapping_regular_pc5}), and that $\sum_A \overline{n}_J^{[A]} = \sum_A n_J^{[A]}$, we have
\begin{eqnarray} \label{overlapping_regular_pc6}
p^{(\rho)}\left( t_\sigma\mathcal{A} \right)= \sum_{A=1}^N \big( m_{H}^{[A]}+\text{num}_A^{ }(n_J) \big) = m_H^{ } +\text{num}(n_J) \geqslant 0.
\end{eqnarray}
In order that the pinch surface $\rho$ is divergent, we now have two additional requirements:
\begin{itemize}
\item[]{$\textit{1.}$} $m_H^{[A]}=0$ for all jets $J_A^{(\rho)}$. This means that the propagators of $J_I^{(\sigma)} \bigcap J_K^{(\rho)}$ are not attached to $H^{(\sigma)}\bigcap H^{(\rho)}$, a result that will be revisited in section\ \ref{leading_enclosed_pinch_surfaces}.
\end{itemize}
\begin{itemize}
\item[]{$\textit{2.}$} $\text{num}(n_J)=0$, which means that at the external vertices of the $n_{J}^{ }$ propagators (see figure\ \ref{structure_for_overlapping_pc}), the numerator contribution must be of $\mathcal{O}(1)$. This property will be very helpful, and we will revisit it several times in the later analysis of section\ \ref{pairwise_cancellation_regular: overlapping}.
\end{itemize}

In figure\ \ref{structure_for_overlapping_pc}, we have stated that for the leading contribution, each of the $n_H^{ }$ propagators provides a $\beta_I^\mu$ to contract with the other vectors in $\gamma$. If some of these $n_H^{ }$ propagators provide transverse polarizations $\beta_{I\perp}^\mu$ instead, we can carry out a calculation similar to that from eq.\ (\ref{overlapping_regular_pc1}) to (\ref{overlapping_regular_pc6}), and find that each of such propagators gives a $\lambda^{1/2}$-suppression. Therefore, for an IR-divergent $\rho^{\left\{ \sigma \right\}}$, none of the $n_H^{ }$ propagators can be transversely polarized gauge bosons. The same conclusion holds for scalars and fermions.\footnote{This argument can also be justified in Item $\mathfrak{2}$ of section\ \ref{leading_enclosed_pinch_surfaces}.} As a result, the subgraph $\gamma$ can only be attached to $H^{(\rho^{\left\{ \sigma \right\}})}$ through scalar-polarized gauge bosons.

The analysis above is for a single approximation, and that for repetitive approximations is in appendix\ \ref{overlapping_regular_repetitive}. We restore the superscripts of $\rho$ and draw the conclusion: the IR divergence at a regular pinch surface $\rho^{ \left\{ \sigma_n...\sigma_1 \right\} }$, which overlaps with some of the $\left\{ \sigma_i \right\}$, is at worst logarithmic.

To end the discussion under the title of ``overlapping \& regular'', we comment that for each $\rho^{ \left\{ \sigma_n...\sigma_1 \right\} }$ as a pinch surface of $t_{\sigma_n} ...t_{\sigma_1} \mathcal{A}$, we can find a corresponding pinch surface of $\mathcal{A}$, say $\rho_\mathcal{A}$, as long as the jets do not have nontrivial overlaps (see figure\ \ref{two_jets_merge} for example). To obtain the corresponding $\rho_\mathcal{A}$, we simply remove the approximation operators inside the solution. For example, after we do this for $\rho^{ \left\{ \sigma_n...\sigma_1 \right\} }$ in the upper row of figure\ \ref{regular_pinch_surface1}, the subgraph $J_I^{(\sigma)} \bigcap J_K^{(\rho)}$ (blue lines) become collinear to $\beta_K^\mu$. Similar procedures can be implemented for the repetitive-approximation case. The change from $\rho^{ \left\{ \sigma_n...\sigma_1 \right\} }$ to $\rho_\mathcal{A}$ preserves the propagator types, i.e. a lightlike (hard, soft) propagator remains lightlike (hard, soft), though its direction may change. In other words, the set of normal coordinates of $\rho^{ \left\{ \sigma_n...\sigma_1 \right\} }$ may differ from that of $\rho_\mathcal{A}$, as long as $\beta_I^\mu$ and $\beta_K^\mu$ are not back-to-back, but their elements are in one-to-one correspondence.

\bigbreak
\centerline{\textbf{Soft-Exotic}}
Having discussed the regular pinch surfaces, now we study the power counting when there are soft-exotic configurations. Again, we consider a single approximation $t_\sigma$ here, and the case of repetitive approximations will be discussed later in appendix\ \ref{soft-exotic_repetitive}.

\begin{figure}[t]
\centering
\includegraphics[width=15cm]{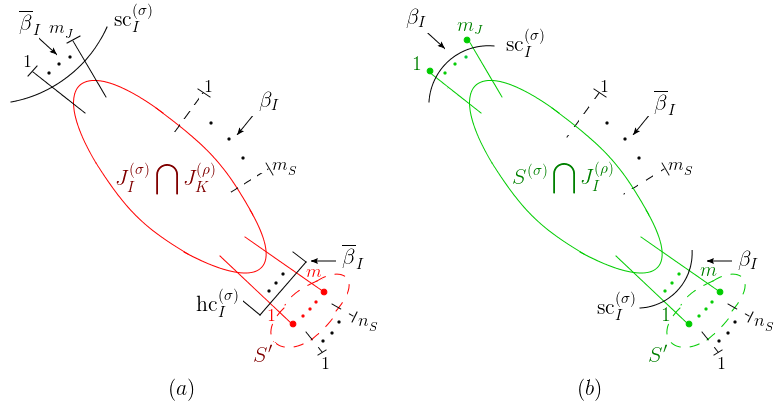}
\caption{The subgraph $\gamma_{(a)} \equiv
\left( J_I^{(\sigma)}\bigcap J_K^{(\rho)} \right) \bigcup S'$ that is marked red in $(a)$, and $\gamma_{(b)}\equiv \left( S^{(\sigma)}\bigcap J_I^{(\rho)} \right)\bigcup S'$ that is marked green in $(b)$. Both subgraphs are depicted in $\rho$. The intermediate $m$ lightlike propagators are soft-exotic propagators (defined in case (Ciii) of section\ \ref{pinch_surfaces_from_single_approximation}), whose projected momenta are soft. The projection is due to hard-collinear approximations in $(a)$, and soft-collinear approximations in $(b)$. In both $(a)$ and $(b)$, $m_J^{ }$ is the number of propagators in $S^{(\sigma)} \bigcap J_K^{(\rho)}$, and $n_S^{ }$ and $m_S^{ }$ are separately the number of external soft propagators of $S'$ and $J_I^{(\sigma)}\bigcap J_K^{(\rho)}$.}
\label{structure_for_soft-exotic_pc}
\end{figure}

As is indicated before, a soft-exotic configuration can be induced by both hard-collinear and soft-collinear approximations. The first case is encountered when lines in a jet $J_I^{(\sigma)}$ become part of another jet in $\rho^{ \left\{ \sigma \right\} }$. As discussed in Theorem 1 of section\ \ref{pinch_surfaces_from_single_approximation}, these momenta are pinched only in the direction of $\overline{\beta}_I^\mu$. The hard-collinear approximation then forces the projected momenta that flow into $H^{(\sigma)}$ to be soft, producing a pinch surface where a jet line appears to ``decay'' into soft lines (see figure\ \ref{structure_for_soft-exotic_pc}$(a)$). The second case is encountered when a subgraph of $S^{(\sigma)}$, which contains lines attached to $J_I^{(\sigma)}$, becomes part of $J_I^{(\rho)}$. Since $t_\sigma$ provides us with the soft-collinear approximation projecting certain momenta onto their $\overline{\beta}_I$-component, the projected values that flow back into $J_I^{(\sigma)}$ are soft in $\rho^{ \left\{ \sigma \right\} }$. This can again produce a pinch surface where a jet line ``decays'' into soft lines (see figure\ \ref{structure_for_soft-exotic_pc}$(b)$). We should consider both these cases, and shall start from the first one. The analysis of the second case will follow similarly.

For the hard-collinear case, the subgraph whose degree of divergence we shall calculate is $J_I^{(\sigma)}\bigcap J_K^{(\rho)}$ as well as a soft subgraph $S'$, where the $m$ vertices of the exotic propagators are internal. In the following discussion, we define $\gamma\equiv \left (J_I^{(\sigma)}\bigcap J_K^{(\rho)} \right)\bigcup S'$. We have marked this structure red in figure\ \ref{structure_for_soft-exotic_pc}$(a)$ below, as well as our previous example in (the upper row of) figure\ \ref{soft-exotic_example}. The idea is similar to that in the calculations for the ``overlapping \& regular'' case.

First, by dimensional analysis, the degree of divergence of the soft subgraph $S'$, with external propagators removed is
\begin{eqnarray} \label{soft_degree_of_divergence}
4-\#\begin{pmatrix}
\text{external}\\ 
\text{bosons}
\end{pmatrix}
-\frac{3}{2}\cdot \#\begin{pmatrix}
\text{external}\\ 
\text{fermions}
\end{pmatrix}.
\end{eqnarray}
The degree of divergence $p^{(\rho)} \left( \gamma \right)$ is then
\begin{eqnarray} \label{soft-exotic_pc1}
p^{(\rho)}\left( \gamma \right)&&=2L_J- N_J+ n_{\text{num},J}^{(\rho)}+ \left( 4-m-b-\frac{3}{2}f \right) \nonumber\\
&&=N_J-2V_J+2+n_{\text{num},J}^{(\rho)}+ 4-m-b-\frac{3}{2}f\nonumber\\
&&=-\frac{1}{2}V_{3,J}-\frac{1}{2}m-\frac{1}{2}m_J^{ }-\frac{1}{2}m_S^{ } +4-b-\frac{3}{2}f+n_{\text{num},J}^{(\rho)},
\end{eqnarray}
where $n_{\text{num},J}^{(\rho)}$ is the numerator contribution from the subgraph $J_I^{(\sigma)}\bigcap J_K^{(\rho)}$, and $b$ and $f$ are the external bosonic and fermionic propagators of $S'$ that are not included in $\gamma$, so in the figure $n_S^{ } =b+f$. In the second and third equalities, we have separately used Euler's formula and counted the half-edges of $J_I^{(\sigma)}\bigcap J_K^{(\rho)}$. For the $m_S^{ }$ soft lines, we keep only the leading $\beta_I^\mu$ numerator projection. All other external lines of $J_I^{(\sigma)}\bigcap J_K^{(\rho)}$ are projected with $\overline{\beta}_I^\mu$, as is shown in the figure. The numerator contribution is evaluated as above, in terms of the power of invariants, in the same notations as eqs.\ (\ref{alpha_relations_1})--(\ref{alpha_relations_3}). We have,
\begin{eqnarray} \label{soft-exotic_numerator_contribution}
n_{\text{num},J}^{(\rho)}=\alpha_{\overline{I}l}+\alpha_{ll},
\end{eqnarray}
where the $\alpha$'s satisfy
\begin{eqnarray} \label{soft-exotic_alpha_relations}
&&\alpha_{\overline{I}l}+\alpha_{\overline{I}I}=m+m_J^{ }, \nonumber\\
&&\alpha_{Il}+\alpha_{\overline{I}I}=m_S^{ }, \\
&&\alpha_{\overline{I}l}+\alpha_{Il}+2\alpha_{ll}=V_{3,J}. \nonumber
\end{eqnarray}

From these, we can solve for $n_{\text{num},J}^{(\rho)}$ and hence $p^{(\rho)}\left( \gamma \right)$, with the result
\begin{eqnarray}
n_{\text{num},J}^{(\rho)}= \frac{1}{2}\left( V_{3J} +m +m_J^{ } -m_S^{ } \right),
\end{eqnarray}
\begin{eqnarray} \label{soft-exotic_pc2}
p^{(\rho)}\left( \gamma \right)=4-b-\frac{3}{2}f-m_S^{ },
\end{eqnarray}
which shows the divergence is at worst logarithmic, because by referring to eq.\ (\ref{soft_degree_of_divergence}), the first part $\left( 4-b-\frac{3}{2}f \right)$ is the contribution from a normal soft subgraph $S'$, and the other term $\left( -m_S^{ } \right)$ fits the contribution from the soft propagators attached to the jets, as we explained in the ``Overlapping \& Regular'' discussion. So in this case the degree of divergence is unchanged.

Now we consider the case where the projection from $t_\sigma$ that acts on the lightlike propagators is a soft-collinear approximation. This time, the structure we shall study is $\left( S^{(\sigma)} \bigcap J_I^{(\rho)} \right) \bigcup S'$, where $S'$ is again a soft subgraph containing the $m$ vertices of the soft-exotic propagators. Pictorially it can be expressed as figure\ \ref{structure_for_soft-exotic_pc}$(b)$.

Comparing figure\ \ref{structure_for_soft-exotic_pc}$(b)$ with \ref{structure_for_soft-exotic_pc}$(a)$, we see that this configuration can be treated in the same way as the previous case, since we can simply exchange $\beta_I$ and $\overline{\beta}_I$. So it is indicated that the degree of divergence should be the same as in eq.\ (\ref{soft-exotic_pc2}). In conclusion, we have verified that the soft-exotic configuration preserves the logarithmic degree of divergence.

\bigbreak
\centerline{\textbf{Hard-Exotic}}
Finally we study the degree of divergence of a pinch surface $\rho^{\left\{ \sigma_n...\sigma_1 \right\}}$ with hard-exotic configurations. Denoting the graph of $\rho^{\left\{ \sigma_n...\sigma_1 \right\}}$ as $\mathcal{G}$, we can find a general procedure to decompose it into several subgraphs, whose degrees of divergence are known results, or easy to evaluate. The method separates contributions from the disjoint hard subgraphs that occur at generic hard-exotic pinch surfaces. This is achieved from the following recursive steps:
\begin{itemize}
  \item[]{\textit{Step 1. }}Imagine a flow along the lightlike propagators of $\mathcal{G}$ in $\rho^{\left\{ \sigma_n...\sigma_1 \right\}}$, which starts from the external propagators and points towards the origin. Whenever a branch of the flow hits a vertex, it streams into the other propagators at this vertex, whose momenta are collinear to the same direction. A branch comes to its endpoint when it encounters a hard subgraph of $\rho^{\left\{ \sigma_n...\sigma_1 \right\}}$, and the whole flow is stopped when every branch comes to an end. We consider the union of the constructed flow, and contract all the endpoints together as a hard vertex, and denote the obtained subgraph as $\mathcal{M}_1$.
\end{itemize}
\begin{itemize}
  \item[]{\textit{Step 2. }}Next we focus on the set of the hard propagators of $\mathcal{G}/\mathcal{M}_1$ ``coterminous with'' $\mathcal{M}_1$: those attached to the flow endpoints. We enlarge this set by including all the hard propagators which can join them through a series of other hard propagators, and denote this enlarged set as $\mathcal{N}_1$. The momenta of $\mathcal{M}_1$ that flow into $\mathcal{N}_1$ can be regarded as the external momenta of the truncated propagators of $\mathcal{N}_1$.
\end{itemize}
\begin{itemize}
  \item[]{\textit{Step 3. }}If $\mathcal{M}_1 \bigcup \mathcal{N}_1$ includes all the hard and jet propagators of $\mathcal{G}$, there are no hard-exotic configurations, and we can jump to \textit{Step 4}. Otherwise, some momenta of the propagators in $\mathcal{N}_1$ must be projected to become lightlike, and combine with the normal coordinates of the loop momenta in $\mathcal{G}_1\equiv \mathcal{G} \setminus \left( \mathcal{M}_1 \bigcup \mathcal{N}_1 \right)$ to form pinches. From the definition in case (D) of section\ \ref{pinch_surfaces_from_single_approximation}, these projected hard propagators are called hard-exotic propagators. Treating the momenta of these projected hard-exotic propagators as the external momenta of $\mathcal{G}_1$, we can recursively follow the same routine of \textit{Step 1} and \textit{2} to decompose $\mathcal{G}_1$, into the new subgraphs $\mathcal{M}_i$ and $\mathcal{N}_i$ ($i$=2, 3...), until $\bigcup_{i=1}^{ }\left( \mathcal{M}_i \bigcup \mathcal{N}_i \right)$ has covered the whole of the hard and jet subgraphs of $\mathcal{G}$. Note that in the $\mathcal{M}_i$'s, the ``nested \& regular'' and ``overlapping \& regular'' configurations can occur, which we have analyzed above.
\end{itemize}
\begin{itemize}
  \item[]{\textit{Step 4. }}Finally we consider the subgraph of $\mathcal{G}$, which is soft in $\rho^{\left\{ \sigma_n...\sigma_1 \right\}}$, and denote it as $\mathcal{S}$. Technically $\mathcal{S}$ can be attached to both $\mathcal{M}_i$ and $\mathcal{N}_i$. Given a vertex of $\bigcup_{i=1}^{ }\left( \mathcal{M}_i \bigcup \mathcal{N}_i \right)$ where some propagators of $\mathcal{S}$ are attached, if it is a jet or soft vertex (defined in the ``regular and exotic configurations'' part of section\ \ref{pinch_surfaces_from_single_approximation}), we say that $\mathcal{S}$ is attached to $\mathcal{M}$ at this vertex; if it is a hard vertex, we say that $\mathcal{S}$ is attached to $\mathcal{N}$. For example, in figure\ \ref{hard-exotic_PS_example}, the soft subgraph $\mathcal{S}$ is attached to $\mathcal{M}$ at vertices $a$ and $b$. Note that $\mathcal{S}$ can also combine with $\mathcal{M}$ to form soft-exotic configurations, which we have analyzed above.
\end{itemize}

Following \textit{Steps 1-4}, the decomposition of figure\ \ref{hard-exotic_PS_example} is shown in figure\ \ref{{hard-exotic_decomposition}}. Note that there are two external soft propagators attached at $a$ and $b$ in $\mathcal{M}_2$, and there is one soft loop hidden in $\mathcal{S}$.
\begin{figure}[t]
\centering
\includegraphics[width=15cm]{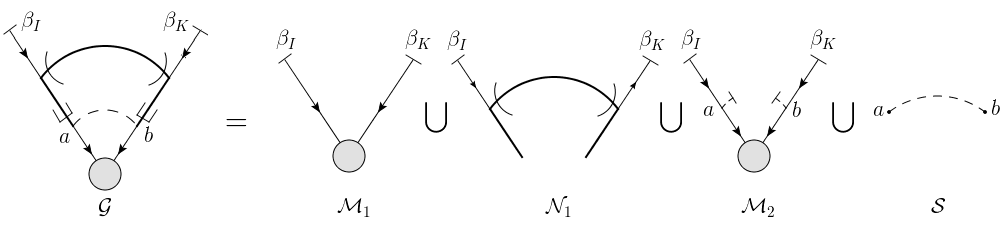}
\caption{The decomposition of a pinch surface with hard-exotic configurations (represented by $\mathcal{G}$) into several subgraphs $\mathcal{M}_1$, $\mathcal{N}_1$, $\mathcal{M}_2$ and $\mathcal{S}$.}
\label{{hard-exotic_decomposition}}
\end{figure}

The result of the procedure above is that the degree of divergence of $\mathcal{G}$ can be regarded as the sum of the contributions from $\left\{ \mathcal{M}_i \right\}$, $\left\{ \mathcal{N}_i \right\}$, and $\mathcal{S}$. For each $i$, $\mathcal{N}_i$ contributes zero to the degree of divergence, because it is made up of propagators with off-shell momenta in $\rho^{\left\{ \sigma_n...\sigma_1 \right\}}$. In other words,
\begin{eqnarray} \label{hard_exotic_pc1}
p^{(\rho)}\left( t_{\sigma_n}...t_{\sigma_1} \mathcal{A} \right)=\sum_{i}^{ } p^{(\rho)}\left( \mathcal{M}_i \right) + p^{(\rho)}\left( \mathcal{S} \right).
\end{eqnarray}
Here we have again dropped the superscript of $\rho$ for simplicity.

Each subgraph $\mathcal{M}_i$ can be seen as a set of jet subgraphs with external soft propagators, whose ``external states'' are the real external states of $\mathcal{G}$ (for $i=1$), or the projected hard momenta from $\mathcal{N}_i$ (for $i>1$). Given a jet subgraph $J_{Ki}$ of $\mathcal{M}_i$, suppose $b_{Ki}$ is the number of soft bosons attached to it, $f_{Ki}$ is the number of attached soft fermions, $v_{Ki}$ is the number of its vertices to which a soft gauge boson is attached, and $h_{Ki}$ is the number of its physical partons attached to $\mathcal{N}_i$, then the degree of divergence of $J_{Ki}$ is
\begin{eqnarray} \label{hard_exotic_pc2}
p^{(\rho)}\left( J_{Ki} \right)\geqslant \frac{1}{2}\left( h_{Ki}-1 \right)- \frac{1}{2}\left( b_{Ki}+f_{Ki}+v_{Ki} \right),
\end{eqnarray}
by standard power counting methods like those used above \cite{Cls11book, Stm78I, Stm95book, Stm96lectures}. The contribution from $\mathcal{S}$ is
\begin{eqnarray} \label{hard_exotic_pc3}
p^{(\rho)}\left( \mathcal{S} \right)=&& \sum_{i,K}^{ } \left[ \left( 4-b_{Ki}- \frac{3}{2} f_{Ki} \right)+ 4\left( b_{Ki}+f_{Ki}-1 \right) -2b_{Ki} -f_{Ki} \right] \nonumber\\
=&&\sum_{i,j}^{ } \left( b_{Ki}+ \frac{3}{2}f_{Ki} \right),
\end{eqnarray}
where $\left( b_{Ki}+f_{Ki}-1 \right)$ represents the number of soft loops generated by $b_{Ki}$ and $f_{Ki}$. Inserting eqs.\ (\ref{hard_exotic_pc2}) and (\ref{hard_exotic_pc3}) into (\ref{hard_exotic_pc1}), we have
\begin{eqnarray} \label{hard_exotic_pc4}
p^{(\rho)}\left( t_{\sigma_n}...t_{\sigma_1} \mathcal{A} \right)\geqslant \sum_{i,K}^{ } \left[ \frac{1}{2}\left( h_{Ki}-1 \right)+ f_{Ki}+ \frac{1}{2}\left( b_{Ki}-v_{Ki} \right) \right].
\end{eqnarray}
By definition, $f_{Ki}\geqslant 0$, $b_{Ki}\geqslant v_{Ki}$ and $h_{Ki}\geqslant 1$ (otherwise the divergence could be cancelled by the Ward identity). $p^{(\rho)}\left( t_{\sigma_n}...t_{\sigma_1} \mathcal{A} \right)\geqslant 0$ is then obvious. For example, the degree of divergence of figure\ \ref{hard-exotic_PS_example} can then be easily evaluated. According to figure\ \ref{{hard-exotic_decomposition}}, the contributions from $\mathcal{M}_1$, $\mathcal{N}_1$, $\mathcal{M}_2$ and $\mathcal{S}$ are separately $0,\ 0,\ -2$ and $+2$. So we have $p^{(\rho)}\left( t_{\sigma_n}...t_{\sigma_1} \mathcal{A} \right)=0$: the approximated amplitude $t_{\sigma_n}...t_{\sigma_1} \mathcal{A}$ has a logarithmic divergence at $\rho$.

\bigbreak
From all the analyses for these four cases (nested \& regular, overlapping \& regular, soft-exotic and hard-exotic), we have verified that the approximation operators preserve the degree of IR divergence, which are hence still at worst logarithmic.

\bigbreak
\centerline{\textbf{Features of Leading $\rho^{\left\{\sigma_1...\sigma_n\right\}}$}}

As stated in the introduction, a leading pinch surface of $\mathcal{A}$ possesses three features: no soft lines attached to the hard subgraph, no soft fermions or scalars attached to the jet subgraph, and at most one line with physical polarization in each jet subgraph attached to the hard subgraph. However, due to the complex structures, as well as the breakdown of normal space conservation (defined in the ``regular and exotic configurations'' part of section\ \ref{pinch_surfaces_from_single_approximation}), we are not guaranteed that these features are still present in a given $t_{\sigma_n}...t_{\sigma_1} \mathcal{A}$ at an arbitrary $\rho^{\left\{ \sigma_1...\sigma_n \right\}}$, even though the integral is (logarithmically) divergent there.

For example, in our power counting analysis for the soft-exotic configuration in figure\ \ref{structure_for_soft-exotic_pc}, a soft fermion or scalar can join one of the $m$ jet propagators without suppressing the logarithmic divergence, because they join each other at a soft vertex, which does not exert any constraints on the types of the soft partons. For hard-exotic configurations that yield logarithmic divergence, figure\ \ref{hard-exotic_PS_example} for example, a soft parton can be attached to a hard propagator, and we can have more than one jet propagator with physical polarization attached to the (union of the connected components of the) hard subgraph.

Nevertheless, the basic features are still present for a regular pinch surface (more precisely, at a regular configuration). The reason is simple. As we have seen in the discussion of ``nested \& regular'' and ``overlapping \& regular'' in this subsection, as long as the jets do not have a nontrivial overlap in $\rho^{\left\{ \sigma_1...\sigma_n \right\}}$, each regular pinch surface of $\rho^{\left\{ \sigma_1...\sigma_n \right\}}$ can be ``mapped'' to a pinch surface of $\mathcal{A}$ ($\rho_\mathcal{A}$) by removing the approximations, and they have one-to-one corresponding normal coordinates. So when we carry out the power counting procedure for $\rho^{\left\{ \sigma_1...\sigma_n \right\}}$, the factors that suppress its degree of divergences are exactly those that suppress the degree of divergence of $\rho_\mathcal{A}$. Namely, whenever there is more than one physical jet parton in the same direction, or a soft parton attached to the hard subgraph, or a soft fermion or scalar attached to the jet subgraph, a power suppression emerges.

This conclusion also holds when the jets in $\rho^{\left\{ \sigma_1...\sigma_n \right\}}$ do have a nontrivial overlap (see figure\ \ref{two_jets_merge} for example), because from our definition, the propagators of any given $J_I^{(\sigma)}$ are from two sources: 1, those in the direction of $\beta_I^\mu$; 2, those from $J_K^{(\sigma)}\ (K\neq I)$ that are collinear to $\overline{\beta}_K^\mu$. As we have mentioned in the discussion after eq.\ (\ref{overlapping_regular_pc6}), those propagators from the second source can only be attached to $H^{\rho^{\left\{ \sigma \right\}}}$ through scalar-polarized gauge bosons. Therefore, there is still exactly one physical parton in each $J_I^{(\left\{ \sigma \right\})}$. The other two features of leading pinch surfaces follow identically as above.

\bigbreak

The whole of this section has centered on the approximation operators extensively. To summarize, we have figured out all the possible configurations in a pinch surface of $t_\sigma \mathcal{A}$, from which we deduced those in $t_{\sigma_n}...t_{\sigma_1} \mathcal{A}$, and verified that all the pinch surfaces appearing in eq.\ (\ref{forest_formula_amplitude}), though various, still lead to logarithmic divergences. These results are fundamental in the upcoming study of IR cancellations in the forest formula.

\section{Enclosed pinch surfaces}\label{enclosed_pinch_surface}

In section\ \ref{pinch_surface_amplitudes_approximations} we have studied the IR singularities of the approximated amplitude $t_{\sigma_n} ...t_{\sigma_1} \mathcal{A}$. An approximation operator in the approximated amplitude, say $t_{\sigma_i}$, describes the asymptotic behavior where the normal coordinates of $\sigma_i$ approach zero. The two terms $t_{\sigma_n} ...t_{\sigma_i} ...t_{\sigma_1} \mathcal{A}$ and $t_{\sigma_n} ...t_{\sigma_{i+1}} t_{\sigma_{i-1}} ...t_{\sigma_1} \mathcal{A}$, then cancel in $\sigma_i$ for any choice of the other $\sigma$'s. This motivates the idea of pairwise cancellations of the divergences near nested pinch surfaces. We must also, however, consider the cancellations near overlapping pinch surfaces.

More precisely, given a pinch surface $\rho^{\left\{ \sigma_n ...\sigma_1 \right\}}$ that overlaps with one or more $\sigma_i \in \left\{ \sigma_1,...,\sigma_n \right\}$, how does one find the pairwise cancellation for its divergence? The way to do this, as we will explain in this subsection, is to study the maximal region simultaneously contained in $\sigma_i$ and $\rho^{\left\{ \sigma_n ...\sigma_1 \right\}}$, which we call the ``enclosed pinch surface of $\sigma_i$ and $\rho^{\left\{ \sigma_n ...\sigma_1 \right\}}$'', and denote as ``$\text{enc}\left [ \sigma_i, \rho^{\left\{ \sigma_n ...\sigma_1 \right\}} \right ]$'' \cite{EdgStm15}. The formal construction of $\text{enc}\left [ \sigma_i, \rho^{\left\{ \sigma_n ...\sigma_1 \right\}} \right ]$ follows from the definition of ordering in eq.\ (\ref{containing_definition}), by defining the normal space (see (\ref{normal_space})) of the loop momenta $l^\mu$ at $\text{enc}\left [ \sigma, \rho^{\left\{\sigma\right\}} \right ]$ as
\begin{eqnarray} \label{enc_definition}
\mathcal{N}_{\text{enc}\left [ \sigma_i, \rho^{\left\{ \sigma_n ...\sigma_1 \right\}} \right ]} \left( l^\mu \right) = \mathcal{N}_{\sigma_i} \left( l^\mu \right) \oplus \mathcal{N}_{\rho^{\left\{ \sigma_n ...\sigma_1 \right\}}} \left( l^\mu \right).
\end{eqnarray}
The action of the direct sum symbol $\oplus$ is given in table\ \ref{oplus_definition}. In the table, we use the notation, as in eq.\ (\ref{normal_space}),
\begin{eqnarray}
\label{N_soft_definition}
\mathcal{N}^{(\text{soft})}\equiv &&\text{the full 4-dim space}, \\
\label{N_I_definition}
\mathcal{N}^{(I)}\equiv &&\text{span}\left\{\overline{\beta}_I^\mu, \beta_{I\perp}^\mu \right\},
\end{eqnarray}
with $\mathcal{N}^{(I)}$ the normal space of momenta in the direction of $\beta_I^\mu$.
In the special case of a single approximation, (\ref{enc_definition}) becomes
\begin{eqnarray} \label{enc_definition_single}
\mathcal{N}_{\text{enc}\left [ \sigma, \rho^{\left\{\sigma\right\}} \right ]} \left( l^\mu \right) =\mathcal{N}_\sigma \left( l^\mu \right) \oplus \mathcal{N}_{\rho^ {\left\{\sigma\right\}}} \left( l^\mu \right).
\end{eqnarray}

\begin{table}[t]
\captionsetup{justification=centering,margin=2.65cm}
\caption{Table for the operation $\oplus$, in the notations of eqs.\ (\ref{N_soft_definition}) and (\ref{N_I_definition}), with $I$ and $K$ labelling different jets.} \label{oplus_definition}
\centering
\begin{tabular}{ | c || c| c | c | c | }
\hline
 \multirow{1}{*}{$\oplus$} & $\mathcal{N}^{(\text{soft})}$ & $\mathcal{N}^{(I)}$ & $\mathcal{N}^{(K)}$ & $\varnothing$ \\ 
\hline
\hline
$\mathcal{N}^{(\text{soft})}$ & $\mathcal{N}^{(\text{soft})}$ & $\mathcal{N}^{(\text{soft})}$ & $\mathcal{N}^{(\text{soft})}$ & $\mathcal{N}^{(\text{soft})}$ \\ 
\hline
$\ \mathcal{N}^{(I)}\ $ & $\mathcal{N}^{(\text{soft})}$ & $\mathcal{N}^{(I)}$ & $\mathcal{N}^{(\text{soft})}$ & $\mathcal{N}^{(I)}$ \\ 
\hline
$\mathcal{N}^{(K)}$ & $\mathcal{N}^{(\text{soft})}$ & $\mathcal{N}^{(\text{soft})}$ & $\mathcal{N}^{(K)}$ & $\mathcal{N}^{(K)}$ \\ 
\hline
$\varnothing$ & $\mathcal{N}^{(\text{soft})}$ & $\mathcal{N}^{(I)}$ & $\mathcal{N}^{(K)}$ & $\ \ \varnothing\ \ $ \\ 
\hline
\end{tabular}
\end{table}
This definition is natural because a larger normal space implies a smaller pinch surface (and a larger reduced graph). A direct sum of two normal spaces corresponds to a pinch surface simultaneously enclosed by the two pinch surfaces.

In the following, we will show that enclosed pinch surfaces are leading pinch surfaces of $\mathcal{A}$. Once this is demonstrated, we are assured to find pairs of the subtraction terms from the forest formula (\ref{forest_formula_amplitude}), which contain the approximation operator associated with this enclosed pinch surface or not. This will result in the cancellation of overlapping divergences. We should note that if this were not the case, either the cancellation of overlapping divergences would not be this simple, or we would need to design a corresponding approximation operator for this enclosed pinch surface, as well as the subtraction terms containing this operator, then add them into the forest formula. This would turn out to enlarge the workload greatly, or even endlessly, which would be disastrous.

For this reason, it is necessary to study the definition, eq.\ (\ref{enc_definition}) further. The first thing to do is to study the relations between the soft, jet and hard subgraphs of $\sigma_i$, $\rho^{\left\{ \sigma_n ...\sigma_1 \right\}}$ and $\text{enc}\left[ \sigma_i, \rho^{\left\{ \sigma_n ...\sigma_1 \right\}} \right]$. We will do this by developing a new algebra of normal spaces of pinch surfaces in section\ \ref{subgraphs_enclosed_pinch_surfaces}. We then turn to the question of whether $\text{enc}\left[ \sigma_i, \rho^{\left\{ \sigma_n ...\sigma_1 \right\}} \right]$ is a leading pinch surface of $\mathcal{A}$. In section\ \ref{leading_enclosed_pinch_surfaces}, we first assume $\rho^{\left\{ \sigma_n ...\sigma_1 \right\}}$ to be a regular pinch surface (defined in section\ \ref{pinch_surfaces_from_single_approximation}), and then show that under this assumption, $\text{enc}\left[ \sigma_i, \rho^{\left\{ \sigma_n ...\sigma_1 \right\}} \right]$ possesses the three features of a leading pinch surface of $\mathcal{A}$ given in the introduction. After that, we consider the soft- and hard-exotic configurations of $\rho^{\left\{ \sigma_n ...\sigma_1 \right\}}$ in section\ \ref{extension_exotic_structures}, and confirm that they do not produce any anomalous structures of $\text{enc}\left[ \sigma_i, \rho^{\left\{ \sigma_n ...\sigma_1 \right\}} \right]$ that violate the three features. For convenience, the analysis in sections\ \ref{leading_enclosed_pinch_surfaces} and \ref{extension_exotic_structures} is implemented within the framework of decay processes. Later in section\ \ref{extend_scattering_processes}, we explain why this analysis extends to wide-angle scatterings.

We note that section\ \ref{subgraphs_enclosed_pinch_surfaces} shows the detailed analysis for a single approximation, and the corresponding details for repetitive approximations are provided in appendix\ \ref{enclosed_PS_details}. In sections\ \ref{leading_enclosed_pinch_surfaces} and \ref{extension_exotic_structures} it is sufficient to only work on the single-approximated amplitudes because the analysis for repetitive approximations will follow identically. Throughout these sections, we will denote $\text{enc}\left[ \sigma, \rho^{\left\{\sigma\right\}} \right] \equiv \tau$ for convenience.

\subsection{Soft, jet and hard subgraphs of enclosed pinch surfaces}
\label{subgraphs_enclosed_pinch_surfaces}

From the definition eq.\ (\ref{enc_definition}) we can easily obtain the relation between the loop momentum $l^\mu$ in $\sigma$, $\rho^{ \left\{ \sigma \right\}}$ and $\tau$. If $l^\mu$ is hard in $\tau$, then it does not have any normal coordinates, i.e. $\mathcal{N}_{\tau} \left( p^\mu \right)$ is empty, so it must be hard in $\sigma$ and $\rho^{\left\{ \sigma \right\}}$ as well. If $l^\mu$ is collinear in $\tau$, to $\beta^\mu$ for example, then it is either collinear to $\beta^\mu$ in $\sigma$ and hard in $\rho^{\left\{ \sigma \right\}}$, or vice versa, or collinear to $\beta^\mu$ in both $\sigma$ and $\rho^{\left\{ \sigma \right\}}$. If it is soft, then either it is also soft in $\sigma$ and/or $\rho^{\left\{ \sigma \right\}}$, or it is collinear in both $\sigma$ and $\rho^{\left\{ \sigma \right\}}$, but to different lightlike vectors. 

It is not yet adequate, however, to only relate the loop momenta in $\tau$ to those of $\sigma$ and $\rho^{\left\{ \sigma \right\}}$. Rather, we need to extend such relations to line momenta. In other words, we aim to prove the following, extending eq.\ (\ref{enc_definition}) for loop momenta to all line momenta.

\textbf{Theorem 3: For the momentum of any propagator of $\mathcal{A}$, say $p^\mu$, we have}
\begin{eqnarray} \label{enc_to_prove}
\mathcal{N}_{\tau} \left( p^\mu \right) =\mathcal{N}_\sigma \left( p^\mu \right) \oplus \mathcal{N}_{\rho^ {\left\{\sigma\right\}}} \left( t_\sigma p^\mu \right),
\end{eqnarray}
where $\mathcal{N}_{\tau} \left( p^\mu \right)$ is the same normal space defined by eq.\ (\ref{enc_definition}). Once Theorem 3 is proved, we can immediately relate the subgraphs of $\tau$ to those of $\sigma$ and $\rho^{\left\{ \sigma \right\}}$. That is, applying the very reasoning given in the first paragraph of this subsection to each propagator, we have
\begin{align} \label{enc_subgraph_relations}
\begin{split}
H^{\left (\tau \right)}=&H^{\left( \sigma \right)}\bigcap H^{\left( \rho^{ \left\{ \sigma \right\} } \right)},\\
J_I^{\left (\tau \right)}=&\left (J_I^{\left( \sigma \right)}\bigcap H^{\left( \rho^{ \left\{ \sigma \right\} } \right)} \right)\bigcup \left( H^{\left( \sigma \right)}\bigcap J_I^{\left( \rho^{ \left\{ \sigma \right\} } \right)} \right) \bigcup \left( J_I^{\left( \sigma \right)}\bigcap J_I^{\left( \rho^{ \left\{ \sigma \right\} } \right)} \right),\\
S^{\left (\tau \right)}=&S^{\left( \sigma \right)}\bigcup S^{\left( \rho^{ \left\{ \sigma \right\} } \right)}\bigcup \left( \bigcup_{K\neq I}^{ }\left (J_I^{\left( \sigma \right)}\bigcap J_K^{\left( \rho^{ \left\{ \sigma \right\} } \right)} \right) \right),
\end{split}
\end{align}
where, for example, every line in $J_I^{(\tau)}$ carrying momentum $p^\mu$ has $\mathcal{N}_\tau (p^\mu)= \mathcal{N}^{(I)}$. Eq.\ (\ref{enc_subgraph_relations}) will be a powerful tool in constructing and understanding the subgraphs of an enclosed pinch surface.

\begin{figure}[t]
\centering
\includegraphics[width=6cm]{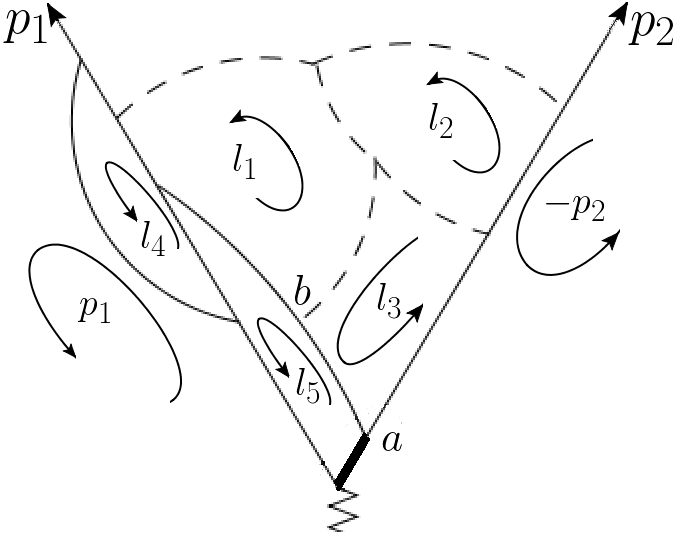}
\caption{The way we mark the loop momenta of $\mathcal{A}$.}
\label{mark_loop_momenta}
\end{figure}

To make the relation between the loop momenta and line momenta specific, we begin with planar graphs. We assign the loop momenta of $\mathcal{A}$ as the counter-clockwise momenta going around its loops as shown in figure\ \ref{mark_loop_momenta}. In order to match the values of the external momenta, we mark the ``external loop momenta'' $\pm p_1^\mu$, ..., $\pm p_N^\mu$, which are outside the graph and lightlike.

In this notation, the momentum of a propagator can be easily obtained from those of the loops. For planar graphs, it is simply the difference between two of its loop momenta that flow through the propagator. In figure\ \ref{mark_loop_momenta}, the momentum of propagator $ab$ (from $a$ to $b$) is thus $l_5^\mu-l_3^\mu$. By comparison, in a nonplanar graph we may have a linear combination (with coefficients $\pm 1$) of three or more loop momenta. We will first prove that a momentum of the form $p_{ij}^\mu \equiv \pm l_i^\mu \pm l_j^\mu$ also satisfies eq.\ (\ref{enc_to_prove}), and then directly generalize the conclusion to nonplanar graphs. After showing these results, Theorem 3 is proved, and we have the relations between subgraphs, eq.\ (\ref{enc_subgraph_relations}).

As a preliminary to the proof, we introduce an operator denoted by $\star:\ \left (\mathcal{N},\mathcal{N} \right) \mapsto \mathcal{N}$. The action of $\star$ is defined in table\ \ref{star_definition}, where the notations $\mathcal{N}^{(\text{soft})}$ and $\mathcal{N}^{(I)}$ have been introduced in eqs.\ (\ref{N_soft_definition}) and (\ref{N_I_definition}).\footnote{With $\oplus$, the operation $\star$ provides an algebra for normal spaces that will be very useful below.}

\begin{table}[t]
\caption{Multiplication table for the operation $\star$.} \label{star_definition}
\centering
\begin{tabular}{ | c || c| c | c | c | }
\hline
 \multirow{1}{*}{$\star$} & $\mathcal{N}^{(\text{soft})}$ & $\mathcal{N}^{(I)}$ & $\mathcal{N}^{(K)}$ & $\varnothing$ \\ 
\hline
\hline
$\mathcal{N}^{(\text{soft})}$ & $\mathcal{N}^{(\text{soft})}$ & $\ \mathcal{N}^{(I)}\ $ & $\mathcal{N}^{(K)}$ & $\varnothing$ \\ 
\hline
$\ \mathcal{N}^{(I)}\ $ & $\ \mathcal{N}^{(I)}\ $ & $\mathcal{N}^{(I)}$ & $\varnothing$ & $\varnothing$ \\ 
\hline
$\mathcal{N}^{(K)}$ & $\mathcal{N}^{(K)}$ & $\varnothing$ & $\mathcal{N}^{(K)}$ & $\varnothing$ \\ 
\hline
$\varnothing$ & $\varnothing$ & $\varnothing$ & $\varnothing$ & $\ \ \varnothing\ \ $ \\ 
\hline
\end{tabular}
\end{table}

The motivation for this operation is to obtain the normal space of a momentum that is the linear combination of two independent momenta, whose normal spaces are known. In other words, given any independent momenta $l_i^\mu$ and $l_j^\mu$ together with a pinch surface $\lambda$ ($=\sigma,\ \rho^{\left\{ \sigma \right\}}\text{ or }\tau$), the linear combination of $l_i^\mu$ and $l_j^\mu$ satisfies
\begin{eqnarray}
\mathcal{N}_\lambda \left( \pm l_i \pm l_j \right)&&\equiv \begin{cases}
\varnothing\ \text{(empty)} & \text{ if } l_i^\mu \text{ or } l_j^\mu \text{ is hard in }\lambda, \\ 
\varnothing\ \text{(empty)} & \text{ if } l_i^\mu \text{ and } l_j^\mu \text{ are collinear to different directions in }\lambda, \\ 
\mathcal{N}_\lambda\left( l_i \right) & \text{ if } l_j^\mu \text{ is soft in }\lambda \text{, and vice versa}, \\
\mathcal{N}_\lambda\left( l_i \right) & \text{ if } l_i^\mu \text{ and } l_j^\mu \text{ are collinear to the same direction in }\lambda.
\end{cases}\nonumber \\
&&= \mathcal{N}_\lambda \left( l_i \right) \star \mathcal{N}_\lambda \left( l_j \right).
\end{eqnarray}
Clearly, this star symbol relates the loop momenta with the propagator momenta in our construction, though $l_i^\mu$ and $l_j^\mu$ can be either projected by $t_\sigma$ or not. In more detail, suppose $p_{ij}^\mu =\pm l_i^\mu \pm l_j^\mu$ in $\mathcal{A}$, then $p_{ij}^\mu =\pm t_\sigma l_i^\mu \pm t_\sigma l_j^\mu$ in $t_\sigma \mathcal{A}$, where for some lines, $t_\sigma l^\mu= l^\mu$. The normal space of $p_{ij}$ in $\sigma$ and $\rho^{\left\{ \sigma \right\}}$ then, separately satisfies
\begin{eqnarray} \label{star_property1}
\mathcal{N}_\sigma \left( p_{ij} \right) &&= \mathcal{N}_\sigma\left( l_i \right) \star \mathcal{N}_\sigma\left( l_j \right), \nonumber\\
\mathcal{N}_{\rho^{\left\{ \sigma \right\}}} \left( p_{ij} \right) &&= \mathcal{N}_{\rho^{\left\{ \sigma \right\}}}\left( t_\sigma l_i \right) \star \mathcal{N}_{\rho^{\left\{ \sigma \right\}}}\left( t_\sigma l_j \right).
\end{eqnarray}
Now we can go on with the proof of Theorem 3.

\textit{Proof of Theorem 3:} Using eq.\ (\ref{star_property1}) with momentum $p_{ij}^\mu$, the result we wish to prove, eq.\ (\ref{enc_to_prove}), becomes
\begin{eqnarray} \label{enc_to_prove_before'}
\mathcal{N}_\tau \left(l_i\right) \star \mathcal{N}_\tau \left(l_j\right) = \left (\mathcal{N}_\sigma\left( l_i \right) \star \mathcal{N}_\sigma\left( l_j \right) \right) \oplus \left (\mathcal{N}_{\rho^{\left\{ \sigma \right\}}}\left( t_\sigma l_i \right) \star \mathcal{N}_{\rho^{\left\{ \sigma \right\}}}\left( t_\sigma l_j \right) \right).
\end{eqnarray}
Next, using the defining property of the normal space for a single loop momentum, eq.\ (\ref{enc_definition}), we can rewrite (\ref{enc_to_prove_before'}) as
\begin{eqnarray} \label{enc_to_prove'}
&&\left (\mathcal{N}_\sigma \left(l_i\right) \oplus \mathcal{N}_{\rho^{\left\{\sigma\right\}}} \left(l_i\right) \right) \star \left (\mathcal{N}_\sigma \left(l_j\right) \oplus \mathcal{N}_{\rho^{\left\{\sigma\right\}}} \left(l_j\right) \right)= \nonumber\\
&&\hspace{2cm} \left (\mathcal{N}_\sigma\left( l_i \right) \star \mathcal{N}_\sigma\left( l_j \right) \right) \oplus \left (\mathcal{N}_{\rho^{\left\{ \sigma \right\}}}\left( t_\sigma l_i \right) \star \mathcal{N}_{\rho^{\left\{ \sigma \right\}}}\left( t_\sigma l_j \right) \right).
\end{eqnarray}
We shall prove this relation, which is equivalent to (\ref{enc_definition}) and hence Theorem 3. The method of proving eq.\ (\ref{enc_to_prove'}) is to find all the cases of $\mathcal{N}_{\rho^{\left\{ \sigma \right\}}} \left(p^\mu\right)$ that appear in $t_\sigma \mathcal{A}$, given the pinch surface $\sigma$, which may differ from $\mathcal{N}_{\rho} \left(p^\mu\right)$ where $\rho$ is a another pinch surface of $\mathcal{A}$. The proof depends on the action of $t_\sigma$ on the momenta in eq.\ (\ref{enc_to_prove'}). This action can be an identity operator, or exert a hard-collinear or soft-collinear approximation on $l_i^\mu$ or $l_j^\mu$.

\textit{1, $t_\sigma=1$ in eq.\ (\ref{enc_to_prove'}).} First we analyze the case where $t_\sigma=1$ on $l_i^\mu$ and $l_j^\mu$. This happens when the confluence of $l_i^\mu$ and $l_j^\mu$ is at an internal vertex of $H^{(\sigma)}$, $J^{(\sigma)}$ or $S^{(\sigma)}$. Equivalently, $l_i^\mu$ and $l_j^\mu$ are simultaneously hard, soft or collinear to a given direction in $\sigma$. For the case where they are both hard in $\sigma$, we have $\mathcal{N}_\sigma \left( l_i \right)=\mathcal{N}_\sigma \left( l_j \right)= \varnothing$, and (\ref{enc_to_prove'}) reduces to the identity
\begin{eqnarray}
\left (\mathcal{N}_{\rho^{\left\{ \sigma \right\}}}\left( l_i \right) \star \mathcal{N}_{\rho^{\left\{ \sigma \right\}}}\left( l_j \right) \right) =\left (\mathcal{N}_{\rho^{\left\{ \sigma \right\}}}\left( l_i \right) \star \mathcal{N}_{\rho^{\left\{ \sigma \right\}}}\left( l_j \right) \right).
\end{eqnarray}
Similarly, for cases where they are both soft in $\sigma$, it is obvious that both sides of (\ref{enc_to_prove'}) equal a 4-dim space.

For the case where $l_i^\mu$ and $l_j^\mu$ are both collinear to, say $\beta_I^\mu$ in $\sigma$, $\left (\mathcal{N}_\sigma\left( l_i \right) \star \mathcal{N}_\sigma\left( l_j \right) \right)$ is then equal to $\text{span}\left\{ \overline{\beta}_I^\mu, \beta_{I\perp}^\mu \right\}$. Suppose first that $\beta_I^\mu$ is contained in $\left (\mathcal{N}_{\rho^{\left\{ \sigma \right\}}}\left( l_i \right) \star \mathcal{N}_{\rho^{\left\{ \sigma \right\}}}\left( l_j \right) \right)$. If it is, then the RHS of eq.\ (\ref{enc_to_prove'}) equals the whole space, because $\beta_I^\mu$ is not contained in $\mathcal{N}^{(I)}$. Moreover, $\beta_I^\mu$ must be in both $\mathcal{N}_{\rho^{\left\{ \sigma \right\}}}\left( l_i \right)$ and $\mathcal{N}_{\rho^{\left\{ \sigma \right\}}}\left( l_j \right)$. So on the LHS of (\ref{enc_to_prove'}), using table\ \ref{star_definition} we have
\begin{eqnarray}
\left (\mathcal{N}_\sigma \left(l_i\right) \oplus \mathcal{N}_{\rho^{\left\{\sigma\right\}}} \left(l_i\right) \right) \star \left (\mathcal{N}_\sigma \left(l_j\right) \oplus \mathcal{N}_{\rho^{\left\{\sigma\right\}}} \left(l_j\right) \right)= \mathcal{N}^{(\text{soft})}\star \mathcal{N}^{(\text{soft})}= \mathcal{N}^{(\text{soft})},
\end{eqnarray}
and the two sides of (\ref{enc_to_prove'}) match.

If $\beta_I^\mu \notin \left (\mathcal{N}_{\rho^{\left\{ \sigma \right\}}}\left( l_i \right) \star \mathcal{N}_{\rho^{\left\{ \sigma \right\}}}\left( l_j \right) \right)$ on the other hand, the RHS of eq.\ (\ref{enc_to_prove'}) equals the $\text{span}\left\{ \overline{\beta}_I^\mu, \beta_{I\perp}^\mu \right\}$. Moreover, either $\beta_I^\mu \notin \mathcal{N}_{\rho^{\left\{ \sigma \right\}}}\left( l_i \right)$ or $\beta_I^\mu \notin \mathcal{N}_{\rho^{\left\{ \sigma \right\}}}\left( l_j \right)$ (or both). Let's say $\beta_I^\mu \notin \mathcal{N}_{\rho^{\left\{ \sigma \right\}}}\left( l_i \right)$, then, on the left of (\ref{enc_to_prove'}),
\begin{align}
\begin{split}
\Big (\mathcal{N}_{\sigma}\left( l_i \right) &\oplus \mathcal{N}_{\rho^{\left\{ \sigma \right\}}}\left( l_i \right) \Big)= \text{span} \left\{\overline{\beta}_I^\mu, \beta_{I\perp}^\mu\right\} = \mathcal{N}^{(I)},\\
&\left (\mathcal{N}_{\sigma}\left( l_j \right) \oplus \mathcal{N}_{\rho^{\left\{ \sigma \right\}}}\left( l_j \right) \right) \supseteq \mathcal{N}^{(I)}.
\end{split}
\end{align}
So the LHS of (\ref{enc_to_prove'}) equals $\mathcal{N}^{(I)}$ as well.

\begin{figure}[t]
\centering
\includegraphics[width=12cm]{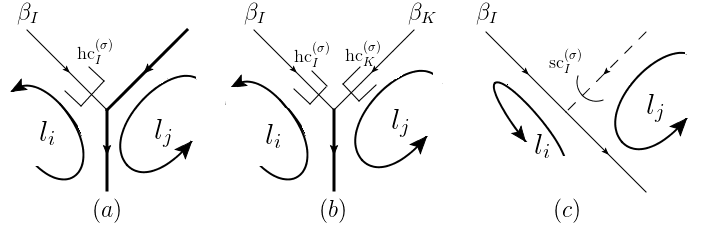}
\caption{The three cases where one of the loop momenta $l_i^\mu$ or $l_j^\mu$ (or both) is projected nontrivially by the approximations of $t_\sigma$.}
\label{momenta_confluence}
\end{figure}

\textit{2, $t_\sigma \neq 1$ in eq.\ (\ref{enc_to_prove'}).} For cases where $t_\sigma$ provides nontrivial approximations on $l_i^\mu$ and $l_j^\mu$, figure\ \ref{momenta_confluence} shows how these arise.

In figure\ \ref{momenta_confluence}, $(a)$ describes a jet momentum collinear to $\beta_I$ joining the hard part, so a hard-collinear approximation is applied on $l_i^\mu$; $(b)$ describes the case of two jet momenta of different directions ($\beta_I$ and $\beta_K$) merging as a hard momenta together, so two hard-collinear approximations are applied separately on $l_i^\mu$ and $l_j^\mu$; $(c)$ describes a soft momentum joining a jet momentum which is in the direction of $\beta_I$, so a soft-collinear approximation is applied on $l_j^\mu$. The considerations below apply when the line of momentum $l_i^\mu \pm l_j^\mu$ is internal to $H^{(\sigma)}$ (for $(a)$) or a $J_I^{(\sigma)}$ (for $(c)$), while $(b)$ can only appear as tree-level configurations shown in the figure.

For figure\ \ref{momenta_confluence}$(a)$, we combine the definition of the $\star$-symbol in table\ \ref{star_definition} with the rules of $\text{hc}_I^{(\sigma)}$, i.e. eq.\ (\ref{hard-collinear_approximation}), to rewrite (\ref{enc_to_prove'}) as
\begin{eqnarray} \label{enclosed_normal_coordinates_requirement1}
\left (\mathcal{N}^{(I)} \oplus \mathcal{N}_{\rho^{\left\{\sigma\right\}}} \left(l_i\right) \right) \star \mathcal{N}_{\rho^{\left\{\sigma\right\}}} \left(l_j\right) = \mathcal{N}_{\rho^{ \left\{ \sigma \right\} }}\left( ^{I}\widehat{l}_i \right) \star \mathcal{N}_{\rho^{ \left\{ \sigma \right\} }} \left( l_j \right),
\end{eqnarray}
with $^{I}\widehat{l}_i$ defined in (\ref{hat_tilde_definition}). The RHS of (\ref{enclosed_normal_coordinates_requirement1}) indicates that $l_i^\mu$ is replaced by $^{I}\widehat{l}_i^\mu\ (=t_\sigma l_i^\mu)$ inside $H^{(\sigma)}$. Also, note that $\mathcal{N}_{\rho^{ \left\{ \sigma \right\} }}\left( ^{I}\widehat{l}_i \right)$ may not be equal to $\mathcal{N}^{(I)}$ because $l_i^\mu$ can be soft or parallel to $\overline{\beta}_I^\mu$ in $\rho^{ \left\{ \sigma \right\} }$.

Similarly, for figure\ \ref{momenta_confluence}$(b)$, eq.\ (\ref{enc_to_prove'}) becomes
\begin{eqnarray} \label{enclosed_normal_coordinates_requirement1'}
\left (\mathcal{N}^{(I)} \oplus \mathcal{N}_{\rho^{\left\{\sigma\right\}}} \left(l_i\right) \right) \star \left (\mathcal{N}^{(K)} \oplus \mathcal{N}_{\rho^{\left\{\sigma\right\}}} \left(l_j\right) \right) = \mathcal{N}_{\rho^{ \left\{ \sigma \right\} }}\left( ^{I}\widehat{l}_i \right) \star \mathcal{N}_{\rho^{ \left\{ \sigma \right\} }} \left( ^{K}\widehat{l}_j \right).
\end{eqnarray}
To verify eqs.\ (\ref{enclosed_normal_coordinates_requirement1}) and (\ref{enclosed_normal_coordinates_requirement1'}), we show first
\begin{eqnarray} \label{enclosed_normal_coordinates_requirement1_lemma}
\left (\mathcal{N}^{(I)} \oplus \mathcal{N}_{\rho^{\left\{\sigma\right\}}} \left(l_i\right) \right)= \mathcal{N}_{\rho^{ \left\{ \sigma \right\} }}\left( ^{I}\widehat{l}_i \right).
\end{eqnarray}
This can be achieved by recalling Theorem 1 in section\ \ref{pinch_surfaces_from_single_approximation} that in $\rho^{\left\{\sigma\right\}}$, $l_i^\mu$ can be of only four types: hard, collinear to $\beta_I^\mu$, collinear to $\overline{\beta}_I^\mu$ and soft. Then it is direct to verify the answer: for the first two types, both sides are $\text{span}\left\{ \overline{\beta}_I^\mu, \beta_{I\perp}^\mu \right\}$; for the other two types, both sides are the 4-dim entire space. Given (\ref{enclosed_normal_coordinates_requirement1_lemma}), eq.\ (\ref{enclosed_normal_coordinates_requirement1}) is immediate. Using (\ref{enclosed_normal_coordinates_requirement1_lemma}) and Theorem 1 for $^K\widehat{l}_j^\mu$, we also verify (\ref{enclosed_normal_coordinates_requirement1'}). With (\ref{enclosed_normal_coordinates_requirement1}) and (\ref{enclosed_normal_coordinates_requirement1'}) proved, (\ref{enc_to_prove'}) holds for the configurations in figures\ \ref{momenta_confluence}$(a)$ and $(b)$.

For the configuration in figure\ \ref{momenta_confluence}$(c)$, we use that $\mathcal{N}_\sigma \left(l_j\right)= \text{full space}$, and $\mathcal{N}_\sigma \left(l_i\right) \star \mathcal{N}_\sigma \left(l_j\right)= \mathcal{N}^{(I)}$ to show that eq.\ (\ref{enc_to_prove'}) is equivalent to
\begin{eqnarray} \label{enclosed_normal_coordinates_requirement2}
\mathcal{N}^{(I)} \oplus \mathcal{N}_{\rho^{\left\{\sigma\right\}}} \left(l_i\right) = \mathcal{N}^{(I)} \oplus \left( \mathcal{N}_{\rho^{ \left\{ \sigma \right\} }}\left( l_i \right) \star \mathcal{N}_{\rho^{ \left\{ \sigma \right\} }} \left( ^{I}\widetilde{l}_j \right) \right).
\end{eqnarray}
To verify this relation, we make the following observation. Because the approximation $\text{sc}_I$ projects the momentum $l_j^\mu$ onto the direction of $\overline{\beta}_I^\mu$, the coordinate $\beta_I^\mu$ must be included in $\mathcal{N}_{\rho^{ \left\{ \sigma \right\} }} \left( ^{I}\widetilde{l}_j \right)$. After this observation the idea is the same as that in the last paragraph. By noticing that in $\rho^{\left\{ \sigma \right\}}$, $l_i^\mu$ can only be of four types (hard, collinear to $\beta_I^\mu$, collinear to $\overline{\beta}_I^\mu$ and soft), we study them one by one. For the case of being hard or collinear to $\beta_I^\mu$, both sides of (\ref{enclosed_normal_coordinates_requirement2}) are equal to $\text{span}\left\{ \overline{\beta}_I^\mu, \beta_{I\perp}^\mu \right\} = \mathcal{N}^{(I)}$. When $l_i^\mu$ is collinear to $\overline{\beta}_I^\mu$ or soft in $\rho^{\left\{ \sigma \right\}}$, the LHS of (\ref{enclosed_normal_coordinates_requirement2}) is the full 4-dim space. Also, we have $\beta_I^\mu \in \mathcal{N}_{\rho^{ \left\{ \sigma \right\} }}\left( l_i \right) \star \mathcal{N}_{\rho^{ \left\{ \sigma \right\} }} \left( ^{I}\widetilde{l}_j \right)$, which means the RHS is the full 4-dim space as well. Therefore, we have proved eq.\ (\ref{enclosed_normal_coordinates_requirement2}), and (\ref{enc_to_prove'}) holds for figure\ \ref{momenta_confluence}$(c)$ as well.

Finally, we comment that the proof above is also sufficient for non-planar graphs, or other loop assignments in a planar graph, where the propagator momenta are of the form $p^\mu= {\sum}_{i=1}^L a_i l_i^\mu$, with $L$ being the number of loops, and $a_i= \pm 1, 0$. Taking the case where $p^\mu=l_i^\mu+l_j^\mu+l_k^\mu$ as an example, we use the associativity of $\star$ and aim to prove the following relation:
\begin{eqnarray}
\mathcal{N}_\tau \left(l_{ijk}\right) = \mathcal{N}_\tau \left(l_i\right) \star \mathcal{N}_\tau \left(l_j\right) \star \mathcal{N}_\tau \left(l_k\right) =\mathcal{N}_\sigma \left(l_{ijk}\right) \oplus \mathcal{N}_{\rho^{\left\{ \sigma \right\}}} \left(l_{ijk}\right),
\end{eqnarray}
which is equivalent to the following analogue of eq.\ (\ref{enc_to_prove'}):
\begin{eqnarray} \label{enc_to_prove'_nonplanar}
&&\left (\mathcal{N}_\sigma \left(l_i\right) \oplus \mathcal{N}_{\rho^{\left\{\sigma\right\}}} \left(l_i\right) \right) \star \left (\mathcal{N}_\sigma \left(l_j\right) \oplus \mathcal{N}_{\rho^{\left\{\sigma\right\}}} \left(l_j\right) \right) \star \left (\mathcal{N}_\sigma \left(l_k\right) \oplus \mathcal{N}_{\rho^{\left\{\sigma\right\}}} \left(l_k\right) \right)= \nonumber\\
&&\hspace{0.9cm} \left (\mathcal{N}_\sigma\left( l_i \right) \star \mathcal{N}_\sigma\left( l_j \right) \star \mathcal{N}_\sigma\left( l_k \right) \right) \oplus \left (\mathcal{N}_{\rho^{\left\{ \sigma \right\}}} \left( t_\sigma l_i \right) \star \mathcal{N}_{\rho^{\left\{ \sigma \right\}}}\left( t_\sigma l_j \right) \star \mathcal{N}_{\rho^{\left\{ \sigma \right\}}}\left( t_\sigma l_k \right) \right).
\end{eqnarray}
Define $l_{ij}^\mu \equiv l_i^\mu+l_j^\mu$. Then from the proof of (\ref{enc_to_prove'}) we have
\begin{eqnarray}
\mathcal{N}_\tau \left(l_{i}\right) \star \mathcal{N}_\tau \left(l_{j}\right)= \mathcal{N}_\tau \left(l_{ij}\right)= \mathcal{N}_\sigma \left(l_{ij}\right) \oplus \mathcal{N}_{\rho^{\left\{ \sigma \right\}}} \left(l_{ij}\right). 
\end{eqnarray}
Inserting this into (\ref{enc_to_prove'_nonplanar}) and using (\ref{star_property1}) again, we only need to show
\begin{eqnarray} \label{enc_to_prove'_nonplanar_end}
&&\left (\mathcal{N}_\sigma \left(l_{ij}\right) \oplus \mathcal{N}_{\rho^{\left\{\sigma\right\}}} \left(l_{ij}\right) \right) \star \left (\mathcal{N}_\sigma \left(l_k\right) \oplus \mathcal{N}_{\rho^{\left\{\sigma\right\}}} \left(l_k\right) \right)= \nonumber\\
&&\hspace{2cm}\left (\mathcal{N}_\sigma\left( l_{ij} \right) \star \mathcal{N}_\sigma\left( l_k \right) \right) \oplus \left (\mathcal{N}_{\rho^{\left\{ \sigma \right\}}}\left( t_\sigma l_{ij} \right) \star \mathcal{N}_{\rho^{\left\{ \sigma \right\}}}\left( t_\sigma l_k \right) \right),
\end{eqnarray}
which is of the same form as (\ref{enc_to_prove'}), with $l_i\rightarrow l_{ij},\ l_j\rightarrow l_k$. Therefore, we see that the same reasoning works for non-planar graphs as well. This completes the proof of Theorem 3.

\bigbreak
Now we have verified the correctness of eq.\ (\ref{enc_to_prove}), the normal space relation for arbitrary lines, which then implies the graphical relation (\ref{enc_subgraph_relations}) for the enclosed pinch surface. Relating the subgraphs of different pinch surfaces, (\ref{enc_subgraph_relations}) is very helpful for our understanding of the structures of an arbitrary enclosed pinch surface.

We emphasize that to derive the relation between subgraphs, the approximation operator $t_\sigma$ is indispensable in eq.\ (\ref{enc_definition}). That is, if we rewrite (\ref{enc_definition}) by replacing $\rho^{ \left\{ \sigma \right\} }$ by $\rho$ (a pinch surface of $\mathcal{A}$) in the definition of enclosed pinch surfaces, no relations between the subgraphs of $\sigma$, $\rho$ and $\text{enc}\left[ \sigma,\rho \right]$, like (\ref{enc_subgraph_relations}), will still hold in general. To be specific, suppose we define the enclosed pinch surface of $\sigma$ and $\rho$, which are two pinch surfaces of $\mathcal{A}$, as
\begin{eqnarray} \label{enc_definition_trial}
\mathcal{N}'_{\text{enc}\left [ \sigma, \rho \right ]} \left( l^\mu \right) =\mathcal{N}_\sigma \left( l^\mu \right) \oplus \mathcal{N}_{\rho} \left( l^\mu \right).
\end{eqnarray}
Then we can still construct the pinch surface $\text{enc}\left [ \sigma, \rho \right ]$ from the loop momenta. But we would not have any line-momentum relations that are as simple as eq.\ (\ref{enc_subgraph_relations}). An example to illustrate this is provided in appendix \ref{enclosed_PS_details_approximations_importance}.

Finally we generalize Theorem 3 by taking repetitive approximations into account.

\textbf{Theorem 4: Suppose $\sigma_1\subset... \subset \sigma_n$ are a series of nested leading pinch surfaces of $\mathcal{A}$. For the momentum of any propagator of $\mathcal{A}$, say $p^\mu$, we have}
\begin{eqnarray} \label{enc_to_prove_generalize}
\mathcal{N}_{\tau} \left( p^\mu \right) =\mathcal{N}_{\sigma_m} \left( p^\mu \right) \oplus \mathcal{N}_{\rho^ {\left\{\sigma_n...\sigma_1\right\}}} \left( p^\mu \right),
\end{eqnarray}
\textbf{where $\tau=\text{enc} \left[ \sigma_m, \rho^ {\left\{\sigma_n...\sigma_1\right\}} \right]$ and $\sigma_m$ ($1\leqslant m\leqslant n$) is the smallest one of all the pinch surfaces in $\left\{\sigma_1,...,\sigma_n\right\}$ that are not contained in $\rho^ {\left\{\sigma_n...\sigma_1\right\}}$.}

The proof is similar to that of Theorem 3, but it requires a more extensive discussion. We give the detailed proof in appendix \ref{enclosed_PS_details_theorem_4_proof}, and only provide a sketch here.

The case where every $\sigma_i$ is contained in $\rho^ {\left\{\sigma_n...\sigma_1\right\}}$ is automatic, because both sides of eq.\ (\ref{enc_to_prove_generalize}) are $\mathcal{N}_{\sigma_n} \left( p^\mu \right)$. Otherwise, we can make use of the defining property (\ref{enc_definition}) and table\ \ref{star_definition}, to rewrite (\ref{enc_to_prove_generalize}) into a form similar to (\ref{enc_to_prove'}). The result is,
\begin{eqnarray} \label{enc_to_prove'_generalize}
&&\left (\mathcal{N}_{\sigma_m} \left(l_i\right) \oplus \mathcal{N}_{\rho^{\left\{\sigma_n...\sigma_1\right\}}} \left(l_i\right) \right) \star \left (\mathcal{N}_{\sigma_m} \left(l_j\right) \oplus \mathcal{N}_{\rho^{\left\{\sigma_n...\sigma_1\right\}}} \left(l_j\right) \right)= \nonumber\\
&&\hspace{0.9cm} \left (\mathcal{N}_{\sigma_m} \left( l_i \right) \star \mathcal{N}_{\sigma_m} \left( l_j \right) \right) \oplus \left (\mathcal{N}_{\rho^{\left\{ \sigma_n...\sigma_1 \right\}}}\left( t_{\sigma_n}...t_{\sigma_1} l_i \right) \star \mathcal{N}_{\rho^{\left\{ \sigma_n...\sigma_1 \right\}}}\left( t_{\sigma_n}...t_{\sigma_1} l_j \right) \right),
\end{eqnarray}
which depends upon the combination of approximations, $t_{\sigma_n}...t_{\sigma_1}$.

We classify the explicit expressions of eq.\ (\ref{enc_to_prove'_generalize}) into two types: those where $t_{\sigma_m}$ is a identity operator for $l_i^\mu$ and $l_j^\mu$, and those where $t_{\sigma_m}$ is not the identity for $l_i^\mu$ or $l_j^\mu$. In the case where $t_{\sigma_m}=1$, $l_i^\mu$ and $l_j^\mu$ must be simultaneously hard, soft or collinear to a certain direction in $\sigma_m$. In the case where $t_{\sigma_m}\neq 1$, its actions on $l_i^\mu$ and $l_j^\mu$ are sufficiently described by the three cases in figure\ \ref{momenta_confluence}. These are in common with our arguments in proving Theorem 3. Meanwhile, approximations from other pinch surfaces, i.e. $\sigma_i$ where $i\neq m$ may also act on $l_i^\mu$ and $l_j^\mu$. We then need to consider all the possibilities, and take them into account to check the relation (\ref{enc_to_prove'_generalize}) is true in each of them. Throughout the proof, we will see the information that $\sigma_m$ is the smallest pinch surface that contains or overlaps with $\rho^ {\left\{\sigma_n...\sigma_1\right\}}$, plays a key role.

Once Theorem 4 is proved, we can immediately come to the following relations among the subgraphs of $\tau$, $\sigma$ and $\rho^{ \left\{ \sigma_n...\sigma_1 \right\} }$, which generalizes eq.\ (\ref{enc_subgraph_relations}) without changing its algebra:
\begin{eqnarray} \label{enc_subgraph_relations_generalize}
H^{\left (\tau \right)}&=&H^{\left( \sigma_m \right)}\bigcap H^{\left( \rho^{ \left\{ \sigma_n...\sigma_1 \right\} } \right)}, \nonumber\\
J_I^{\left (\tau \right)}&=&\left (J_I^{\left( \sigma_m \right)}\bigcap H^{\left( \rho^{ \left\{ \sigma_n...\sigma_1 \right\} } \right)} \right)\bigcup \left( H^{\left( \sigma_m \right)}\bigcap J_I^{\left( \rho^{ \left\{ \sigma_n...\sigma_1 \right\} } \right)} \right) \bigcup \left( J_I^{\left( \sigma_m \right)}\bigcap J_I^{\left( \rho^{ \left\{ \sigma_n...\sigma_1 \right\} } \right)} \right), \nonumber\\
S^{\left (\tau \right)}&=&S^{\left( \sigma_m \right)}\bigcup S^{\left( \rho^{ \left\{ \sigma_n...\sigma_1 \right\} } \right)}\bigcup \left( \bigcup_{K\neq I}^{ }\left (J_I^{\left( \sigma_m \right)}\bigcap J_K^{\left( \rho^{ \left\{ \sigma_n...\sigma_1 \right\} } \right)} \right) \right).
\end{eqnarray}

\subsection{Enclosed pinch surfaces are leading}
\label{leading_enclosed_pinch_surfaces}

In the previous subsection we verified eq.\ (\ref{enc_subgraph_relations}), the relation between the subgraphs of an enclosed pinch surface $\tau\equiv \text{enc}\left[ \sigma, \rho^{\left\{ \sigma \right\}} \right]$ and those of $\sigma $ and $\rho^{\left\{ \sigma \right\}}$. We also generalized it to repetitive approximations in (\ref{enc_subgraph_relations_generalize}). If we construct an enclosed pinch surface by means of such relations, the result should be a well-defined pinch surface of $\mathcal{A}$. Our next goal is to confirm that it is leading, when both $\sigma$ and $\rho^{ \left\{ \sigma \right\} }$ are IR-divergent pinch surfaces of $\mathcal{A}$ and $t_\sigma \mathcal{A}$ respectively. This is formulated as Theorem 5.

\textbf{Theorem 5: If $\sigma$ is a set of nested leading pinch surfaces of $\mathcal{A}$, and $\rho^{\left\{ \sigma \right\}}$ is a pinch surface of $t_\sigma \mathcal{A}$ such that $\sigma:o:\rho^{\left\{ \sigma \right\}}$, and}
\begin{eqnarray}
\left( t_\sigma \mathcal{A}\right)_{\text{div }\mathfrak{n}\left[\rho^{\left\{ \sigma \right\}}\right]}\neq 0,
\end{eqnarray}
\textbf{then }$\tau\equiv\text{enc} \left[\sigma,\rho^{\left\{ \sigma \right\}}\right]$\textbf{ is a leading pinch surface of $\mathcal{A}$.} The notation of overlapping is defined in Sec\ \ref{neighborhoods & approximation_operators}.

If $\rho^{\left\{ \sigma \right\}}$ is an IR-divergent regular pinch surface, then from the discussion at the end of section\ \ref{divergences_are_logarithmic}, it possesses the three features of a leading pinch surface of $\mathcal{A}$ described in the introduction. A natural idea for the proof of Theorem 5 then, is to proceed by contradiction. Namely, suppose some of the three features are violated for our constructed $\tau\equiv \text{enc}\left[ \sigma, \rho^{\left\{ \sigma \right\}} \right]$ from eq.\ (\ref{enc_subgraph_relations}). Then we aim to find the contradiction by proving that either $\sigma$ or $\rho^{\left\{ \sigma \right\}}$ does not preserve these three features, which implies that $\left( \mathcal{A} \right)_{\text{div }\mathfrak{n} \left[ \sigma \right]}= 0$, or $\left( t_\sigma \mathcal{A}\right)_{\text{div }\mathfrak{n}\left[ \rho^{\left\{ \sigma \right\}} \right]}= 0$. This is what we shall do in this subsection, which can be directly generalized to repetitive approximations, as will be discussed at the end.

For convenience, we again drop the superscript of $\rho ^{\left\{ \sigma \right\}}$ and denote it by $\rho$ within this subsection. Some notations to be used are in table\ \ref{leading_enclosed_PS_notations}.

\begin{table}[t]
\caption{Some notations in section\ \ref{leading_enclosed_pinch_surfaces}}
 \begin{tabular}{| c | c |} 
 \hline
 $J_I^{(H^\lambda)}$ ($\lambda=\sigma,\rho,\tau$) & The subgraph of $J_I^{(\lambda)}$ whose propagators are attached to $H^{(\lambda)}$ \\ 
 \hline
 $j_{I,\text{phys}}^{(H^\mu)}$ ($\mu=\sigma,\rho$) & The (unique) physical propagator of $J_I^{(H^\mu)}$ \\
 \hline
 \multirow{2}{*}{$J_{I,\text{phys}}^{(H^\tau)}$} & The subgraph consisting of the physical propagators of $J_{I}^{(H^\tau)}$ \\
 & (which we will show also contains only one propagator) \\
 \hline 
 \end{tabular}
 \label{leading_enclosed_PS_notations}
\end{table}

\bigbreak
\centerline{$\mathfrak{1.}$ \textbf{Connections between $S^{(\tau)}$ and $H^{(\tau)}$}}
We begin by showing that no soft propagators can join to the hard subgraph in $\tau$.

We only need to focus on one connected component of $H^{\left(\tau\right)}$, and will prove the claim by contradiction: we suppose there exists a soft line attached to the hard subgraph in $\tau$. It is easy to see that the specific soft propagator cannot be from $S^{\left(\sigma\right)}$ or $S^{\left(\rho \right)}$, because otherwise either $\sigma$ or $\rho$ will have a soft line connected directly to its hard subgraph, and would then not be a leading pinch surface by power counting. So the propagator can only be from $J_I^{\left(H^\sigma\right)}\bigcap J_K^{\left( H^\rho\right)}\ (I\neq K)$ which implies that $\sigma:o:\rho$. In addition, by eq.\ (\ref{enc_subgraph_relations}), one of its endpoints should be in $H^{\left(\tau\right)}= H^{\left(\sigma\right)} \bigcap H^{\left(\rho\right)}$.

Then we recall Theorem 1, which states that in $\rho$, any line of the subjet $J_I^{\left(\sigma\right)}\bigcap J_K^{\left(\rho\right)}$ is lightlike in the $\overline{\beta}_I$-direction. Then from the ``overlapping \& regular'' discussion in section\ \ref{divergences_are_logarithmic}, one of the conclusions from eq.\ (\ref{overlapping_regular_pc6}) which is the power counting result of $J_I^{\left(\sigma\right)}\bigcap J_K^{\left(\rho\right)}$, states that the number of these lines ($m_H$) must vanish to avoid a power suppression. In other words, such a propagator of $J_I^{\left(H^\sigma\right)}\bigcap J_K^{\left( H^\rho\right)}$ cannot exist when $\left( t_\sigma \mathcal{A}\right)_{\text{div }\mathfrak{n}\left[\rho^{\left\{ \sigma \right\}}\right]}\neq 0$. As a result $S^{(\tau)}$ and $H^{(\tau)}$ are disconnected.

\bigbreak
\centerline{$\mathfrak{2.}$ \textbf{Soft fermions and scalars attached to $J^{(\tau)}$}}
We next show that no soft fermions or scalars can be attached to the jet subgraphs in the pinch surface $\tau$.

We shall prove this by contradiction: in the presence of a soft fermion or scalar attached to $J^{\left(\tau\right)}$, the IR divergence of $\mathcal{A}$ at $\sigma$, or that of $t_\sigma \mathcal{A}$ at $\rho$, would be suppressed. Suppose such a soft fermion or scalar propagator is labelled $ab$, attaching vertices $a$ and $b$ of the graph. Similarly to the arguments above, we observe that $ab$ can neither be from $S^{(\sigma)}$ nor $S^{(\rho)}$, because otherwise, from $S^{(\sigma)}$ for example, it would then be attached to either $J^{(\sigma)}$ or $H^{(\sigma)}$. Neither case is allowed since $\sigma$ is a leading pinch surface of $\mathcal{A}$.

To see that $ab$ is not from $J_I^{\left(\sigma\right)}\bigcap J_K^{\left(\rho\right)}$ ($I \neq K$ and $\sigma :o: \rho$) either, we need to analyze several subcases. First it cannot be an internal jet propagator of both $J_I^{\left(\sigma\right)}$ and $J_K^{\left(\rho\right)}$.\footnote{By saying ``internal jet propagators of $J_I$'', we refer to those propagators of $J_I$, both of whose endpoints are jet vertices (defined in the ``regular and exotic configurations'' part of section\ \ref{pinch_surfaces_from_single_approximation}).} Otherwise not only $ab$ is soft in $\tau$, but all its attached propagators are soft as well. Then it would be an internal propagator of $S^{(\tau)}$, which is not involved in our consideration.

So only two subcases are left for $ab$. (i) $ab= j_{I,\text{phys}}^{\left(H^\sigma\right)}$ in $\sigma$ while $ab= j_{K,\text{phys}}^{\left(H^\rho\right)}$ in $\rho$. (ii) $ab= j_{I, \text{phys}}^{\left( H^\sigma \right)}$ in $\sigma$ while it is an internal propagator of $J_K^{\left(\rho\right)}$ in $\rho$, or vice versa. We explain below why these subcases are not possible either.

\begin{itemize}
\item[(i)]{} By construction, we suppose that in the pinch surface $\sigma$, one of the endpoints $a$ is an internal jet vertex of $J_I^{(\sigma)}$, and the other endpoint $b$ is in $H^{(\sigma)}$. Now consider $a$ and $b$ in $\rho$. If $b$ is in $H^{(\rho)}$, then it is in $H^{(\tau)}$ as well, and we have a soft propagator attached to the hard subgraph in $\tau$, which is suppressed from the arguments in Part $\mathfrak{1}$ above. So with the requirement that $\left( t_\sigma \mathcal{A}\right)_{\text{div }\mathfrak{n} \left[ \rho \right]}\neq 0$, we only need to consider the case where $a \in H^{(\rho)} \bigcap J_I^{(\sigma)}$ while $b \in H^{(\sigma)} \bigcap J_K^{(\rho)}$.

Then we focus on the subgraph $\gamma \equiv H^{(\sigma)}\bigcap H^{(\rho)}$. If line $ab$ is a fermion or scalar that carries charge into $H^{(\sigma)}$ and out of $H^{(\rho)}$, it cannot flow into $\gamma$. As a result, for $\left( t_\sigma \mathcal{A}\right)_{\text{div }\mathfrak{n} \left[ \rho \right]}\neq 0$, $\gamma$ cannot have external scalars or fermions. Instead, all its external propagators are from the following two sources: (1) gauge bosons of $\bigcup_I J_I^{(H^\sigma)}$ that are scalar-polarized in $\sigma$ and acted on by $\text{hc}_I^{(\sigma)}$; (2) gauge bosons of $\bigcup_K J_K^{(H^\rho)}$, which are scalar-polarized in $\rho$. These gauge bosons are illustrated by figure\ \ref{soft_fermion_scalar_case_i}. According to the hard-collinear approximation provided by $t_\sigma$, those gauge bosons of type (1) are projected to be scalar-polarized in both regions; those of type (2) are scalar-polarized in region $\rho$. So in the leading term of $t_\sigma \mathcal{A}$ near $\rho$, all the external propagators of $\gamma$ are scalar-polarized gauge bosons. Such a configuration may lead to an IR divergence, but vanishes in the sum over all the gauge-invariant sets of graphs of $\gamma$ due to the Ward identity. As a result, the case of $ab \in J_{I,\text{phys}}^{\left(H^\sigma\right)} \bigcap J_{K,\text{phys}}^{\left(H^\rho\right)}$ is not consistent with $\rho$ being a leading pinch surface.
\begin{figure}[t]
\centering
\includegraphics[width=12cm]{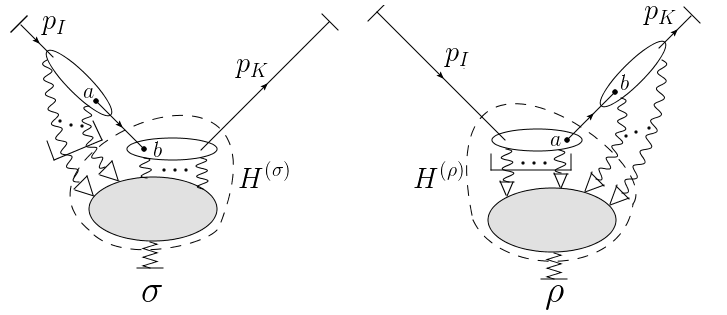}
\caption{The case where $ab$ belongs to $J_{I,\text{phys}}^{\left(H^\sigma\right)} \bigcap J_{K,\text{phys}}^{\left(H^\rho\right)}$. The shadowed area is the subgraph $\gamma\equiv H^{(\sigma)}\bigcap H^{(\rho)}$, whose external propagators are all gauge bosons in this case. As is indicated by the arrows on the right, all these gauge bosons are scalar-polarized in $\rho$.}
\label{soft_fermion_scalar_case_i}
\end{figure}
\end{itemize}

\begin{itemize}
\item[(ii)]{} We now treat the possibility that $ab= j_{I,\text{phys}}^{\left(H^\sigma\right)}$ and is internal in $J_K^{(\rho)}$. Let us define a ``chain'' as the path-connected component of the fermion or scalar propagators that describes the flow of charge carried by line $ab$. From the observation we have just made, it can either be a closed loop, or be open and extending into the final or initial states. Since there is exactly one such chain element ($ab$) that is in $J_I^{\left(H^\sigma\right)}$, the entire chain must extend into the external line of $J_I^{(\sigma)}$, which then can only be a scalar or a fermion.

Now we consider the picture in $\rho$, where $ab$ goes to another direction, as an internal propagator of $J_K^{(\rho)}$. The whole chain is then ``distorted'': it either passes through $S^{(\rho)}$ or $H^{(\rho)}$. There is an obvious suppression on the IR divergence in the former case, because we will obtain soft scalars or fermions attached to the jets in $\rho$. The latter case, on the other hand, can be depicted almost identically as figure\ \ref{soft_fermion_scalar_case_i}, except that the vertex $a$ is not included in $H^{(\rho)}$ (enclosed by the dashed curve) in $\rho$. The argument that we have given in case (i) above, still applies, because all the external propagators of $\gamma \equiv H^{(\sigma)}\bigcap H^{(\rho)}$ are scalar-polarized gauge bosons. Such configurations vanish in the sum over all gauge-invariant set of graphs, and again region $\rho$ is not leading.

Similarly, we rule out the possibility that $ab= j_{I,\text{phys}}^{\left(H^\rho\right)}$ in $\rho$ while being an internal propagator of $J_K^{\left(\sigma\right)}\ (K\neq I)$ in $\sigma$.
\end{itemize}

In conclusion, no soft fermions or scalars can be attached to the jets in $\tau$.

\bigbreak
\centerline{$\mathfrak{3.}$ \textbf{Number of propagators in $J_{I,\text{phys}}^{(H^\tau)}$}}
The final step in verifying that $\tau$ is a leading pinch surface of $\mathcal{A}$ is to prove that in each jet $J_I^{\left (\tau \right)}$, there is at most one physical jet parton attached to any connected component of the hard part $H^{\left(\tau\right)}$. The physical parton may be a fermion, scalar or transversely polarized gauge boson.

To see this, recall the result for $J_I^{\left (\tau \right)}$ in eq.\ (\ref{enc_subgraph_relations}). Each propagator in $J_I^{\left(\tau\right)}$ must be either from $J_I^{\left(\sigma\right)}$ or $J_I^{\left(\rho\right)}$ (or both). To begin with, we consider the subgraph $H^{\left(\tau\right)}$. Since it is the intersection of the hard subgraphs in $\sigma$ and $\rho$, all jet propagators attached to $H^{(\tau)}$ are then attached to either $H^{(\sigma)}$ or $H^{(\rho)}$, from which we can observe that $j_{I,\text{phys}}^{(H^\sigma)}$ and $j_{I,\text{phys}}^{(H^\rho)}$ are the only possible elements of $J_{I,\text{phys}}^{(H^{\tau})}$. As a result, we only need to show that $j_{I,\text{phys}}^{(H^\sigma)}$ and $j_{I,\text{phys}}^{(H^\rho)}$ do not contribute two or zero to the number of propagators in $J_{I,\text{phys}}^{(H^{\tau})}$. This can be done by examining first the possible positions of $j_{I,\text{phys}}^{(H^\rho)}$ in the pinch surface $\sigma$.

We start by proving that the contribution is at most one. We notice that $j_{I,\text{phys}}^{(H^\rho)}$ cannot be from $S^{(\sigma)}$ or $J_{K}^{\left(\sigma\right)}$ where $K \neq I$, because from eq.\ (\ref{enc_subgraph_relations}) it would then be a part of $S^{\left(\tau\right)}$ rather than $J_I^{\left(\tau\right)}$. We have treated these cases above, so all we will need to consider is $j_{I,\text{phys}}^{(H^\rho)} \subseteq J_I^{(\sigma)}$, or $j_{I,\text{phys}}^{(H^\rho)} \subseteq H^{(\sigma)}$.

First suppose $j_{I,\text{phys}}^{(H^\rho)} \subseteq J_I^{(\sigma)}$. If it is an element of $J_{I,\text{phys}}^{(H^{\tau})}$, it must be attached to $H^{\left(\tau\right)}$, hence to $H^{\left(\sigma\right)}$. Then we see that $j_{I,\text{phys}}^{(H^\rho)}$ and $j_{I,\text{phys}}^{(H^\sigma)}$ must be identical, because $J_{I,\text{phys}}^{(H^{\sigma})}$ has only a single element. If $j_{I,\text{phys}}^{\left(H^\rho\right)}$ is not an element of $J_{I,\text{phys}}^{(H^{\tau})}$, it can only be an internal propagator of $J_I^{(\sigma)}$, and $j_{I,\text{phys}}^{(H^\sigma)}$ itself is then the only possible element of $J_{I,\text{phys}}^{(H^{\tau})}$. Under either circumstance, $j_{I,\text{phys}}^{(H^\sigma)}$ and $j_{I,\text{phys}}^{(H^\rho)}$ contribute at most one to the number of propagators in $J_{I,\text{phys}}^{(H^{\tau})}$.

\begin{figure}[t]
\centering
\includegraphics[width=10cm]{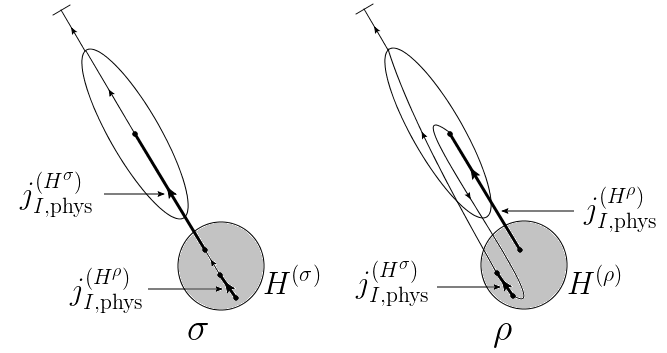}
\caption{The physical chain containing $j_{I,\text{phys}}^{(H^\sigma)}$ and $j_{I,\text{phys}}^{(H^\rho)}$, which are marked bold. This figure implies that if $j_{I,\text{phys}}^{(H^\sigma)}$ and $j_{I,\text{phys}}^{(H^\rho)}$ are different, then we will have more than one propagator in $J_{I,\text{phys}}^{(H^\rho)}$, which suppresses the IR divergence.}
\label{physical_chain_cross}
\end{figure}

For the case of $j_{I,\text{phys}}^{(H^\rho)} \subseteq H^{(\sigma)}$, we consider the position of $j_{I,\text{phys}}^{(H^\sigma)}$ in $\rho$. The arguments in the paragraph above show that if $j_{I,\text{phys}}^{(H^\sigma)} \subseteq J_I^{(\rho)}$, $j_{I,\text{phys}}^{(H^\sigma)}$ and $j_{I,\text{phys}}^{(H^\rho)}$ contribute only one to the number of propagators in $J_{I,\text{phys}}^{(H^{\tau})}$, simply by reversing the roles of $\sigma$ and $\rho$. The only possibility that there are two elements in $J_{I,\text{phys}}^{(H^{\tau})}$ is that $j_{I,\text{phys}}^{(H^\sigma)} \subseteq H^{(\rho)}$ while at the same time $j_{I,\text{phys}}^{(H^\rho)} \subseteq H^{(\sigma)}$. This is not possible because as we will see below, the two physical parton lines $j_{I,\text{phys}}^{(H^\sigma)}$ and $j_{I,\text{phys}}^{(H^\rho)}$ must be linked by a ``chain of physical partons'', which extends to the $I$-th external particle. Graphically, such a chain must cross the boundary between $J_I^{(\rho)}$ and $H^{(\rho)}$ at least three times: it enters $H^{(\rho)}$, goes through $j_{I,\text{phys}}^{(H^\sigma)}$, exits $H^{(\rho)}$, and enters $H^{(\rho)}$ by going through $j_{I,\text{phys}}^{(H^\rho)}$. See figure\ \ref{physical_chain_cross}. But this is not the case for $\rho$ leading.

The reason for the existence of a physical chain is as follows. If the $I$-th external particle is a scalar or a fermion, then as noted in case (ii) of the discussions on soft fermions and scalars attached to $J^{(\tau)}$ above, such a chain describes the flow of charge. Next we consider the case where the external particle is a transversely polarized gauge boson. For any graph $\gamma$ with internal propagators lightlike in the direction of $\beta_I^\mu$, and gauge bosons being its external propagators, we want to show that if one of these gauge bosons is transversely polarized, there must be another one transversely polarized. Denoting the external momenta of $\gamma$ by $p^\mu$, and the loop momenta by $l^\mu$, then we have generic factors in the expression of $\gamma$ of the type:
\begin{eqnarray}
\int d^4l \left( \epsilon_{\perp\mu_1} g^{\mu_1\mu_2} \right) l^{\mu_3}...l^{\mu_m} p^{\nu_1}...p^{\nu_n}, \hspace{0.5cm} \int d^4l \left( \epsilon_\perp\cdot l \right)l^{\mu_1}...l^{\mu_m} p^{\nu_1}...p^{\nu_n},
\end{eqnarray}
where $\mu_1,...,\mu_m, \nu_1,...,\nu_n$ are the vector indices of the external gauge bosons. Apparently in the first factor, $\mu_2= \perp$, and in order that the second factor is nonzero, one of the $\mu_i$'s $(1\leqslant i\leqslant m)$ must be $\perp$. In other words, a gauge boson with transverse polarization must get out after going into $\gamma$. As a result, transversely polarized gauge bosons must also form a physical chain.

We end by explaining why the number of lines in $J_{I,\text{phys}}^{(H^\tau)}$ is not zero. Suppose it is zero, we focus on the subgraph $H^{(\sigma)} \bigcap H^{(\rho)}$, and observe that in $\rho$ every attached parton plays the role of a scalar-polarized gauge boson. Figure\ \ref{soft_fermion_scalar_case_i} again shows the general case. Though such pinch surfaces may be IR divergent, the sum over all similar configurations vanishes due to the Ward identity \cite{Cls11book,LbtdStm85} in region $\rho$, so that $\rho$ is not leading.

\bigbreak

In summary, we have shown that if $\rho^{\left\{ \sigma \right\}}$ is a regular pinch surface that is divergent, (1) $S^{(\tau)}$ and $H^{(\tau)}$ are disjoint; (2) soft fermions or scalars of $S^{(\tau)}$ cannot be attached to $J^{(\tau)}$; (3) there is exactly one physical parton in each $J_I^{(\tau)}$ that is attached to $H^{(\tau)}$. Also, we note that our arguments can be directly generalized to the repetitive case, i.e. $\tau\equiv \text{enc}\left[ \sigma_m, \rho^{\left\{ \sigma_n...\sigma_1 \right\}} \right]$ with $\sigma_m$ the smallest pinch surface of $\left\{ \sigma_1,...,\sigma_n \right\}$ that overlaps with $\rho^{\left\{ \sigma_n ... \sigma_1 \right\}}$. Every step above will follow identically.

To finish the proof of Theorem 5, we also need to consider the exotic configurations of $\rho^{\left\{ \sigma \right\}}$, and show that any enclosed pinch surface produced by them does not violate the three features of a leading pinch surface of $\mathcal{A}$. We do this in the next subsection.

\subsection{Extension to exotic configurations}
\label{extension_exotic_structures}

In section\ \ref{leading_enclosed_pinch_surfaces}, we have assumed $\rho^{\left\{ \sigma \right\}}$ as an IR-divergent regular pinch surface, so that we can apply the three features of $\mathcal{A}$ introduced in section\ \ref{introduction} to $\rho^{\left\{ \sigma \right\}}$. That is, a soft parton cannot be attached to the hard subgraph; a soft fermion or scalar cannot be attached to he jet subgraph; in each jet there is exactly one physical parton attached to the hard subgraph. We ``endowed'' $\rho^{\left\{ \sigma \right\}}$ with these features, with which $\tau$ can be proved leading.

However, as is pointed out at the end of section\ \ref{divergences_are_logarithmic}, these three features no longer hold at exotic configurations. As a result, we need to remove this limitation by showing that whatever exotic configurations $\rho^{\left\{ \sigma \right\}}$ has, the corresponding configuration of $\tau\equiv \text{enc}\left[ \sigma, \rho^{\left\{ \sigma \right\}} \right]$ never violates the three features above. This is what we shall do in this subsection.

Before we start, we recall that the jet propagators in $\rho^{ \left\{ \sigma \right\} }$, whose projected momenta are soft and are attached to other soft propagators, have been denoted as ``soft-exotic propagators'' in case (Ciii) of section\ \ref{pinch_surfaces_from_single_approximation} (see figure\ \ref{soft-exotic_example} and table\ \ref{single_approximated_PS_summary}). Similarly, the hard propagators in $\rho^{ \left\{ \sigma \right\} }$ whose projected momenta are collinear to certain directions, and make some other momenta pinched in alignment with them, have been denoted as ``hard-exotic propagators'' (figure\ \ref{hard-exotic_PS_example} and table\ \ref{single_approximated_PS_summary}).

\begin{figure}[t]
\centering
\includegraphics[width=10cm]{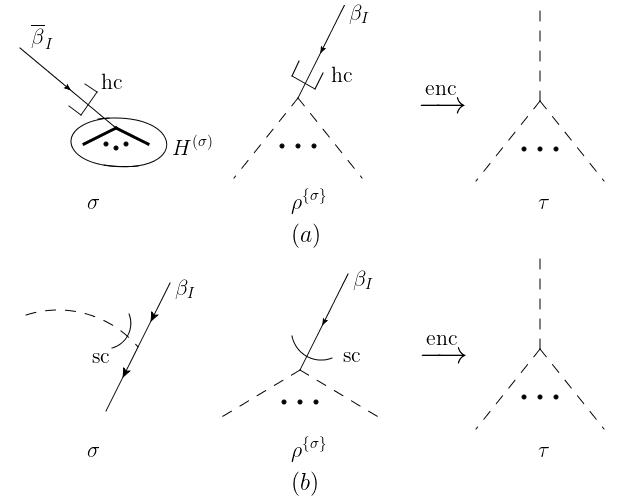}
\caption{The two generic soft-exotic configurations. The upper row is from the hard-collinear approximation, while the lower is from the soft-collinear. The configurations corresponding to $\tau\equiv \text{enc}\left[\sigma, \rho^{ \left\{ \sigma \right\}} \right]$ are all-soft, as are drawn at the rightmost. The intermediate dots indicates that there can be one or two lines, but at each vertex the total number of lines attached does not exceed four.}
\label{all_soft-exotic_structures}
\end{figure}

\begin{figure}[t]
\centering
\includegraphics[width=10.5cm]{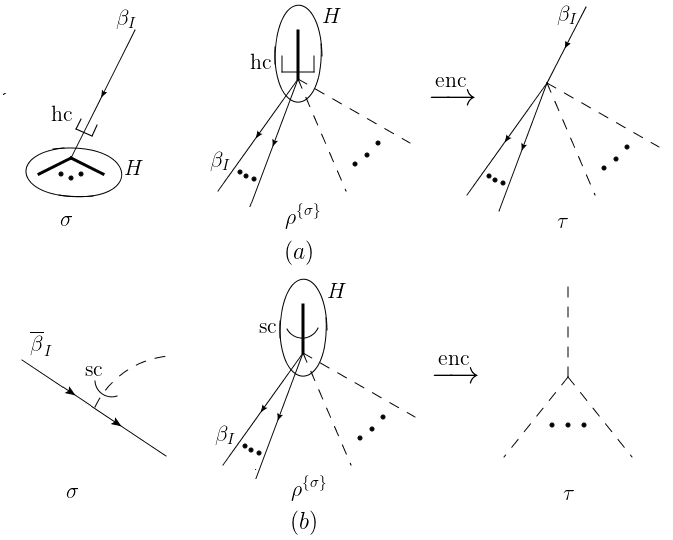}
\caption{The two generic hard-exotic configurations. The upper row is from the hard-collinear approximation, while the lower is from the soft-collinear. The configurations corresponding to $\tau\equiv \text{enc}\left[\sigma, \rho^{ \left\{ \sigma \right\}} \right]$ are drawn at the rightmost. The intermediate dots indicates that there can be one or two lines, but at each vertex the total number of lines attached does not exceed four.}
\label{all_hard-exotic_structures}
\end{figure}

We depict possible exotic configurations in $\rho^{ \left\{ \sigma \right\} }$ in figures\ \ref{all_soft-exotic_structures} (soft-exotic) and \ref{all_hard-exotic_structures} (hard-exotic) below. For a soft-exotic propagator that is collinear to, say $\beta_I^\mu$, in $\rho^{ \left\{ \sigma \right\} }$, $t_\sigma$ must project it onto the direction of $\overline{\beta}_I^\mu$ in order to make it soft. If $t_\sigma$ acts as a hard-collinear approximation, then by definition, the soft-exotic propagator must be collinear to $\overline{\beta}_I^\mu$ in $\sigma$ and thus belong to $J_{\bar{I}}^{(\sigma)} \bigcap J_K^{(\rho^{ \left\{ \sigma \right\} })}\ (K\neq \bar{I})$, as is shown in figure\ \ref{all_soft-exotic_structures}$(a)$. If $t_\sigma$ acts as a soft-collinear approximation, then the soft-exotic propagator is a soft propagator in $\sigma$ and attached to $J^{(\sigma)}$, and becomes a lightlike propagator attached to a soft vertex in $\rho^{ \left\{ \sigma \right\}}$, as is shown in figure\ \ref{all_soft-exotic_structures}$(b)$.

The analysis for hard-exotic configurations is similar. For a given hard-exotic propagator, let's suppose $t_\sigma$ projects it onto the direction of $\beta_I^\mu$. If $t_\sigma$ acts as a hard-collinear approximation, then by definition, the hard-exotic propagator must be collinear to $\beta_I^\mu$ in $\sigma$ and attached to $H^{(\sigma)}$. From the point of view of $\rho^{\left\{ \sigma \right\}}$, the propagators in $H^{(\sigma)}$ attached to this hard-exotic propagator can only be soft or collinear to $\beta_I^\mu$, as is shown in figure\ \ref{all_hard-exotic_structures}$(a)$. If $t_\sigma$ acts as a soft-collinear approximation, then the hard-exotic propagator is soft in $\sigma$ and attached to jet lines in the direction of $\overline{\beta}_I^\mu$. In $\rho^{\left\{ \sigma \right\}}$, all the propagators that are attached to the hard-exotic propagator are soft or collinear to $\beta_I^\mu$, as is shown in figure\ \ref{all_hard-exotic_structures}$(b)$.

These described configurations show all the possibilities of an exotic configuration in $\rho^{\left\{ \sigma \right\}}$. Now we study the corresponding configurations in the enclosed pinch surface $\tau$.

From figures\ \ref{all_soft-exotic_structures} and \ref{all_hard-exotic_structures}, we see that $\rho^{\left\{ \sigma \right\}}$ and $\sigma$ always overlap if $\rho^{\left\{ \sigma \right\}}$ is an exotic pinch surface. To be specific, in figure\ \ref{all_soft-exotic_structures}$(a)$ some jet propagators change directions; in figure\ \ref{all_soft-exotic_structures}$(b)$ some soft propagators become lightlike while some jet propagators become soft; in figure\ \ref{all_hard-exotic_structures}$(a)$ some hard propagators become lightlike and some jet propagators become soft; in figure\ \ref{all_hard-exotic_structures}$(b)$ some jet propagators change directions.

Moreover, the corresponding configurations in $\tau$ are relatively simpler. In figures\ \ref{all_soft-exotic_structures} and \ref{all_hard-exotic_structures}$(b)$, the enclosed pinch surface includes a soft vertex and the (soft) propagators attached to it, which is compatible with any $\tau$ that is leading. In figure\ \ref{all_hard-exotic_structures}$(a)$, region $\tau$ has a jet subgraph in a certain direction and one or two soft partons entering it. In order that it is compatible with a leading pinch surface $\tau$, we need to show that these soft partons are gauge bosons. This is direct from the requirement that $\left( t_\sigma \mathcal{A}\right)_{\text{div }\mathfrak{n} \left[\rho^{\left\{ \sigma \right\}}\right]}\neq 0$, because otherwise there will be a suppression compared to the logarithmic divergence of $\rho^{\left\{ \sigma \right\}}$, making it IR finite. Therefore, all the configurations of $\tau$ in figures\ \ref{all_soft-exotic_structures} and \ref{all_hard-exotic_structures} are compatible with a leading pinch surface of the original amplitude $\mathcal{A}$.

We comment on generalizing our argument above to repetitive approximations, i.e. configurations of the pinch surfaces $\rho^{ \left\{ \sigma_n ... \sigma_1 \right\} }$. Given a soft-exotic or hard-exotic propagator, if there is only one approximation acting on it, then everything follows identically to the single-approximation case. If there are two approximations, then from our explanations in section\ \ref{subtraction_terms}, these approximations must be an $\text{hc}_K^{(\sigma_p)}$ and an $\text{sc}_I^{(\sigma_q)}$ ($I\neq K,\ \sigma_q \subset \sigma_p$). Still, every step of our arguments above applies. The conclusions then become as follows: with an exotic configuration, $\rho^{ \left\{ \sigma_n ... \sigma_1 \right\} }$ always overlaps with a certain pinch surface $\sigma_m$ ($1\leqslant m \leqslant n$), which is the smallest pinch surface of the subset $\left\{ \sigma_m,...,\sigma_n \right\}\subset \left\{ \sigma_1,...,\sigma_n \right\}$ that is not contained in $\rho^{ \left\{ \sigma_n ... \sigma_1 \right\} }$. Any subgraph of $\tau\equiv \text{enc}\left[ \sigma_m, \rho^{\left\{ \sigma_1...\sigma_n \right\}} \right]$ from such a $\rho^{ \left\{ \sigma_n ... \sigma_1 \right\} }$, is compatible with a leading pinch surface of $\mathcal{A}$.

After we combine these conclusions with those obtained in section\ \ref{leading_enclosed_pinch_surfaces}, Theorem 5, which is introduced at the beginning of section\ \ref{leading_enclosed_pinch_surfaces}, is proved. Moreover, we are able to generalize it to repetitive approximations. In summary, we have

\textbf{Theorem 6: If $\left\{ \sigma_1,...,\sigma_n \right\}$ is a set of nested leading pinch surfaces of $\mathcal{A}$, with the relation $\sigma_1 \subset ... \subset \sigma_n$, and $\rho^{\left\{ \sigma_n ... \sigma_1 \right\}}$ is a pinch surface of $t_{\sigma_n}...t_{\sigma_1} \mathcal{A}$ that overlaps with some of the $\sigma_i$'s, and}
\begin{eqnarray}
\left( t_{\sigma_n}...t_{\sigma_1} \mathcal{A}\right)_{\text{div }\mathfrak{n}\left[\rho^{\left\{ \sigma_n ... \sigma_1 \right\}}\right]}\neq 0,
\end{eqnarray}
\textbf{then }$\tau\equiv\text{enc} \left[\sigma_m,\rho^{\left\{ \sigma_n ... \sigma_1 \right\}}\right]$\textbf{ is a leading pinch surface of $\mathcal{A}$. Here $\sigma_m$ is the smallest pinch surface of $\left\{ \sigma_1,...,\sigma_n \right\}$ that overlaps with $\rho^{\left\{ \sigma_n ... \sigma_1 \right\}}$.}

We emphasize again that this conclusion is of a great significance in the pairwise cancellations of the divergences in the forest formula, eq.\ (\ref{forest_formula_amplitude}). To be specific, each approximated amplitude $t_{\sigma_n}...t_{\sigma_1} \mathcal{A}$ corresponds to a series of nested pinch surfaces $\sigma_1,...,\sigma_n$. If an IR-divergent pinch surface $\rho^{\left\{ \sigma_n...\sigma_1 \right\}}$ is overlapping with some of them, we find the smallest one among them, say $\sigma_m$ ($1\leqslant m \leqslant n$). Then from Theorem 6, $\tau\equiv \text{enc}\left[ \sigma_m, \rho^{\left\{ \sigma_n...\sigma_1 \right\}} \right]$ is also a leading pinch surface of $\mathcal{A}$ ($\sigma_m \subset \tau \subseteq \sigma_{m-1}$), so the terms with $t_\tau$ appear in the forest formula. What we will find is, by adding or eliminating $t_\tau$ in $t_{\sigma_n}...t_{\sigma_1} \mathcal{A}$, we will obtain two terms whose divergences near $\rho^{\left\{ \sigma_n...\sigma_1 \right\}}$ are cancelled by each other. Such a pairwise cancellation will be discussed in more detail in section\ \ref{pairwise_cancellation_regular: overlapping}.

So far the discussions in sections\ \ref{leading_enclosed_pinch_surfaces} and \ref{extension_exotic_structures} apply to electroweak induced decay processes, for which $H^{(\sigma)} \bigcap H^{(\rho)}$ is always non-empty, since it must contain the electroweak vertex (or other external current). In fact, we can generalize the analysis to wide-angle scatterings.

\subsection{The case of wide-angle scatterings}
\label{extend_scattering_processes}

\begin{figure}[t]
\centering
\includegraphics[width=7.5cm]{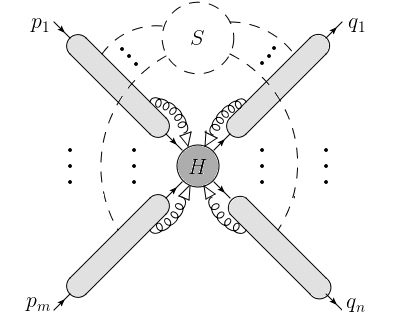}
\caption{The leading pinch surface of the $m \rightarrow n$ wide-angle scattering.}
\label{scattering_process}
\end{figure}

We briefly recap what we have now. Given an amplitude $\mathcal{A}$ of the decay process, we can list its leading pinch surfaces $\left\{ \sigma_i \right\}$, and obtain approximated amplitudes of the form $t_{\sigma_n}... t_{\sigma_1} \mathcal{A}$. For each approximated amplitude, though its IR-divergent pinch surfaces may not be leading pinch surfaces of the original amplitude $\mathcal{A}$, we find that the enclosed pinch surface $\tau\equiv \text{enc}\left[ \sigma, \rho^{\left\{ \sigma \right\}} \right]$ must be, which can be generalized to repetitive approximations. Now we extend our conclusion to the $m \rightarrow n$ wide-angle scattering amplitudes, namely, $m$ external particles of the initial state are scattered into $n$ external particles of the final state, with all the particle momenta in different directions. A leading pinch surface is shown in figure\ \ref{scattering_process}, with the external momenta being $\left\{ p_i \right\}_{i=1}^m$ and $\left\{ q_j \right\}_{j=1}^n$.

Compared with the decay processes, the only subtlety in wide-angle scatterings lies in the intersection of hard subgraphs. That is, unless $H^{(\sigma)}\bigcap H^{(\rho^{\left\{ \sigma \right\}})} = \varnothing$ the pinch surfaces of wide-angle scattering can be seen as identical to those of decay processes, and our analyses in sections\ \ref{leading_enclosed_pinch_surfaces} and \ref{extension_exotic_structures} apply. Now we prove by contradiction that the case $H^{(\sigma)}\bigcap H^{(\rho^{\left\{ \sigma \right\}})} = \varnothing$ never occurs for $\rho^{\left\{ \sigma \right\}}$. Note if $H^{(\sigma)}\bigcap H^{(\rho^{\left\{ \sigma \right\}})} = \varnothing$, $H^{(\rho^{\left\{ \sigma \right\}})}$ must transfer momentum by lines from $S^{(\sigma)}$. But according to the soft-collinear approximations of $t_\sigma$, momenta flowing out of $S^{(\sigma)}$ are always in the direction of $\overline{\beta}_I^\mu$, and hence never scatter the jet lines carrying momenta $p_1^\mu,...,p_m^\mu$ into external momenta $q_1^\mu,...,q_n^\mu$; instead, they are only taken off-shell.

Figure\ \ref{hard-exotic_PS_example} in section\ \ref{pinch_surfaces_from_single_approximation} serves as an example for the explanations above. We can consider figure\ \ref{hard-exotic_PS_example} as part of a scattering process, with $\beta_I$ and $\beta_K$ labelling final-state lines, resulting from a one-loop hard scattering $H^{(\sigma)}$. In $\rho^{\left\{ \sigma \right\}}$ the propagator of $S^{(\sigma)}$ is hard, which, after projection by the soft-collinear approximations, becomes lightlike in the directions of $\overline{\beta}_I^\mu$ and $\overline{\beta}_K^\mu$. These lightlike momenta join the flows of the external momenta, and take them off-shell. These off-shell lines of $H^{(\rho^{\left\{ \sigma \right\}})}$ attach to $H^{(\sigma)}$ at vertices labelled $c$ and $d$ in the figure, so that $H^{(\sigma)}$ and $H^{(\rho^{\left\{ \sigma \right\}})}$ are not disjoint. (The exotic pinch surface internal to $H^{(\sigma)}$ does not affect this result.)

In conclusion, Theorems 5 and 6 not only work for decay processes, but for hard scattering processes as well. We are now ready to work on the IR cancellations in the forest formula, eq.\ (\ref{forest_formula_amplitude}).

\section{The proof of cancellations}
\label{the_proof_of_cancellation}

In this section, we confirm that the full set of forest subtractions, eq.\ (\ref{forest_formula_amplitude}) eliminates all singularities. For convenience, we rewrite (\ref{forest_formula_amplitude}) and explain its notations in more detail:
\begin{eqnarray} \label{forest_formula_amplitude_rewrite}
\left [ \sum_{F\in \mathcal{F}\left [ \mathcal{A} \right ]}^{ } \left( -t_{\sigma_n} \right) \left( -t_{\sigma_{n-1}} \right)... \left( -t_{\sigma_1} \right) \mathcal{A} \right ]_{\textup{div}}=0.
\end{eqnarray}
Here $\mathcal{F}\left[ \mathcal{A} \right]$ refers to the set of forests of $\mathcal{A}$, in which each forest $F$ is defined by a set of leading pinch surfaces (LPS):
\begin{eqnarray}\label{forest_definition}
F\equiv \left \{ \sigma_1,...,\sigma_n \mid \begin{matrix}
\sigma_i\text{'s are LPSs of }\mathcal{A}; \\ 
\sigma_1 \subset\sigma_2\subset...\subset \sigma_n .
\end{matrix} \right \}.
\end{eqnarray}
For each $F$, we have a corresponding series of $\left\{ \sigma_1,...,\sigma_n \right\}$, and the $t_{\sigma_i}$ products are ordered. (The value of $n$ depends implicitly on each forest $F$.) Namely, the approximation operator with a smaller pinch surface appears to the right of that with a larger pinch surface, as discussed in section\ \ref{subtraction_terms}. Finally, the lower notation ``div'' refers to the IR divergences from the whole sum over forests, except for singularities that are cancelled by the Ward identity in the sum over all hard subgraphs of the same order. Then eq.\ (\ref{forest_formula_amplitude_rewrite}) means that in the whole sum, the IR divergences are thoroughly cancelled. In other words, the remainder of $\mathcal{A}$ after all the subtractions is finite. If we set the on-shell particle masses to be small rather than zero, then all the large logarithms disappear in the remainder.

As noted in the introduction, the idea of a forest formula originates from the BPHZ renormalization scheme as a subtraction method of UV divergences in Feynman graphs \cite{BglPrs57,Hepp66,Zmm69}. In the UV case, for the integrand of a general Feynman graph, with overlapping and nested subgraphs whose degrees of UV divergence are non-negative, the prescription for the renormalized integrand is
\begin{eqnarray} \label{uv_forest_formula}
R_\mathcal{A}\left( p,k \right)=\sum_{U\in\mathcal{U}\left( \mathcal{A} \right)}^{ }\prod_{\gamma\in U}^{ }\left( -t_{p\left( \gamma \right)}^{d\left( \gamma \right)} \right)I_\mathcal{A}\left( p,k \right),
\end{eqnarray}
where $t_{p\left( \gamma \right)}^{d\left( \gamma \right)}$ is the operator on the subgraph $\gamma$, which acts by performing a Taylor expansion in its external momenta $p\left( \gamma \right)$ up to the degree of UV divergence $d\left( \gamma \right)$. For the remaining subgraph $\mathcal{A} \setminus \gamma$, the operator acts as an identity. The $t_{p\left( \gamma \right)}^{d\left( \gamma \right)}$ products are ordered, so that the operator with a smaller subgraph $\gamma$ appears on the right. The integral after subtractions, $\int \prod_{i}^{ }dk_i R_\mathcal{A}\left( p,k \right)$, is then absolutely convergent.

The subtraction terms in eq.\ (\ref{uv_forest_formula}) result from replacing the $\gamma$'s by local counterterms, while in comparison, the subtraction terms in our forest formula (\ref{forest_formula_amplitude_rewrite}) result from expanding the integrand near all the pinch surfaces of $\mathcal{A}$. Due to the complicated IR structures (compared with UV), different constructions are needed for the proof of (\ref{forest_formula_amplitude_rewrite}), as we have seen in the previous sections.

Our method is to focus on any IR-divergent pinch surface $\rho^{\left\{ ... \right\}}$ of an arbitrary term in eq.\ (\ref{forest_formula_amplitude_rewrite}), to be specific, $\rho^{\left\{ \sigma_n...\sigma_1 \right\}}$ of $t_{\sigma_n} ...t_{\sigma_1} \mathcal{A}$. We aim to find a unique other term $t_{\sigma_m'} ...t_{\sigma_1'} \mathcal{A}$ in (\ref{forest_formula_amplitude_rewrite}) that cancels the specific divergence near $\rho^{\left\{ \sigma_n...\sigma_1 \right\}}$,
\begin{eqnarray}\label{pairwise_cancellation_form}
\Big [ t_{\sigma_n} ...t_{\sigma_1} \mathcal{A} \Big ]_{\text{div }\mathfrak{n} \left [ \rho \right ]}+\Big [ t_{\sigma_m'} ...t_{\sigma_1'} \mathcal{A} \Big ]_{\text{div }\mathfrak{n}\left [ \rho \right ]}=0.
\end{eqnarray}
We will use the defining properties of approximation operators and enclosed pinch surfaces to identify these pairs. Note that the assignments of the pairs are unique in both ways: each divergence of $t_{\sigma_n} ...t_{\sigma_1} \mathcal{A}$ corresponds to that of $t_{\sigma_m'} ...t_{\sigma_1'} \mathcal{A}$, and vice versa. Once this is done, we need two more steps in order to show this pairwise cancellation. They are:
\begin{itemize}
    \item [($\mathfrak{a}$)]{} to assure that $\rho^{\left\{ \sigma_n...\sigma_1 \right\}}$ is a pinch surface of $t_{\sigma_m'} ...t_{\sigma_1'} \mathcal{A}$ as well, by verifying that the denominators of the integrands of $t_{\sigma_n} ...t_{\sigma_1} \mathcal{A}$ and $t_{\sigma_m'} ...t_{\sigma_1'} \mathcal{A}$ match;
\end{itemize}
\begin{itemize}
    \item [($\mathfrak{b}$)]{} to show that the approximations $t_{\sigma_n} ...t_{\sigma_1}$ and $t_{\sigma_m'} ...t_{\sigma_1'}$ are equivalent at $\rho^{\left\{ \sigma_n...\sigma_1 \right\}}$, by verifying that the numerators of the integrands of $t_{\sigma_n} ...t_{\sigma_1} \mathcal{A}$ and $t_{\sigma_m'} ...t_{\sigma_1'} \mathcal{A}$ match.
\end{itemize}
These two steps verify the IR cancellation of each pair of terms, eq.\ (\ref{forest_formula_amplitude_rewrite}).

To understand how the pairs are assigned, we need to discuss the relations between $\rho^{\left\{ \sigma_n...\sigma_1 \right\}}$ and the elements of $\left\{ \sigma_1, ..., \sigma_n \right\}$. This is how we organize this section. In section\ \ref{pairwise_cancellation_regular: nested} we study the case where $\rho^{\left\{ \sigma_n...\sigma_1 \right\}}$ is nested with all the $\sigma_i$'s, and in section\ \ref{pairwise_cancellation_regular: overlapping} we deal with the case where $\rho^{\left\{ \sigma_n...\sigma_1 \right\}}$ overlaps with certain elements of $\left\{ \sigma_1, ..., \sigma_n \right\}$. In both these subsections, $\rho^{\left\{ \sigma_n...\sigma_1 \right\}}$ is assumed to be a regular pinch surface (defined in section\ \ref{subtraction_terms}), and in each subcase we shall discuss, the two steps above to show IR cancellation introduced in the previous paragraph will be applied. After that, we include exotic configurations into our analysis in section\ \ref{pairwise_cancellation_exotic}. Some comments on the proof are added in section\ \ref{discussion_forest_formula}.

\subsection{Divergences at nested pinch surfaces}
\label{pairwise_cancellation_regular: nested}

We first consider the case where $\rho^{\left\{ \sigma_n...\sigma_1 \right\}}$ is nested with every $\sigma_i$ appearing in $t_{\sigma_n} ...t_{\sigma_1} \mathcal{A}$. In other words, there exists an $m\in \left\{ 1,...,n \right\}$, such that $\rho^{\left\{ \sigma_n...\sigma_1 \right\}}$ is ``smaller than'' or ``the same as'' $\sigma_m, ..., \sigma_n$, while ``larger than'' $\sigma_1, ..., \sigma_{m-1}$. In terms of our definition in eq.\ (\ref{containing_definition}), for each loop momentum in $\rho^{\left\{ \sigma_n...\sigma_1 \right\}}$, its normal space contains the corresponding ones of $\sigma_m, ..., \sigma_n$, and is contained in the corresponding ones of $\sigma_1, ..., \sigma_{m-1}$. We then identify a certain leading pinch surface of $\mathcal{A}$, the normal spaces of whose loop momenta are identical to those of $\rho^{\left\{ \sigma_n...\sigma_1 \right\}}$, and denote it as $\sigma_\rho$. Note that $\sigma_\rho$ can be different from $\rho^{\left\{ \sigma_n...\sigma_1 \right\}}$, because they are pinch surfaces of different integrals, $\mathcal{A}$ for $\sigma_\rho$ and $t_{\sigma_n}... t_{\sigma_1} \mathcal{A}$ for $\rho^{\left\{ \sigma_n...\sigma_1 \right\}}$. Moreover, $\sigma_\rho$ may be included in the forest $F \equiv \left\{ \sigma_1, ..., \sigma_n \right\}$ or not, but without loss of generality, we assume $\sigma_\rho \in F$, and explicitly, $\sigma_\rho = \sigma_m$.

We claim that the following two terms have the same divergence at $\rho$ up to a minus sign, so they cancel in the sum. In other words, eq.\ (\ref{pairwise_cancellation_form}) is rewritten as:
\begin{eqnarray} \label{pairwise_cancellation_nested_conclusion}
\Bigg[ \prod_{j=m+1}^{n} \left( -t_{\sigma_j} \right)\left( 1-t_{\sigma_m} \right) \prod_{i=1}^{m-1} \left( -t_{\sigma_i} \right)\mathcal{A} \Bigg] _{\text{div}\ \mathfrak{n}\left [ \rho \right ]}=0.
\end{eqnarray}
As is indicated by the two steps ($\mathfrak{a}$) and ($\mathfrak{b}$) above, in order to show this relation, we need to first ascertain that $t_{\sigma_n}...t_{\sigma_{m+1}} t_{\sigma_{m-1}}...t_{\sigma_1} \mathcal{A}$ is also divergent at $\rho^{\left\{ \sigma_n...\sigma_1 \right\}}$, and then prove that $t_{\sigma_\rho}$ is exact for $t_{\sigma_n}...t_{\sigma_{m+1}} t_{\sigma_{m-1}}...t_{\sigma_1} \mathcal{A}$ at $\rho^{\left\{ \sigma_n...\sigma_1 \right\}}$. We carry out these two steps below.

\begin{figure}[t]
\centering
\includegraphics[width=10cm]{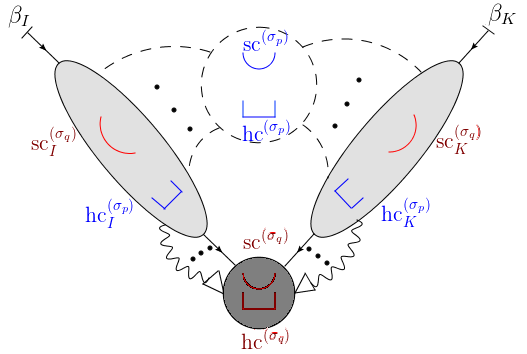}
\caption{The pinch surfaces $\rho^{\left\{ \sigma_n...\sigma_1 \right\}}$ of a back-to-back decay process as an example. As is indicated in the text, the approximations from $t_{\sigma_p}\ (m\leqslant p\leqslant n)$ reside in the soft and jet subgraphs, and we mark them blue. In comparison, the approximations from $t_{\sigma_q}\ (1\leqslant q\leqslant m)$ reside in the jet and hard subgraphs, which we mark dark red.}
\label{nested_PS}
\end{figure}

$(\mathfrak{a})$ We confirm the divergence by studying the difference between the two pinch surfaces $\sigma_\rho( =\sigma_m )$ and $\rho^{\left\{ \sigma_n...\sigma_1 \right\}}$. By definition, $\sigma_\rho$ and $\rho^{\left\{ \sigma_n...\sigma_1 \right\}}$ have the same set of normal coordinates, so their differences can only lie in the ranges of intrinsic coordinates. It is then relatively easy to see that the approximation operators $t_{\sigma_n}, ..., t_{\sigma_m}$ do not lead to any differences between $\sigma_\rho$ and $\rho^{\left\{ \sigma_n...\sigma_1 \right\}}$. Any given loop-momentum projection from these $t$'s acts either on $S^{(\sigma_\rho)}$ or $J^{(\sigma_\rho)}$. On one hand, at $\sigma_\rho$ there are no intrinsic coordinates in $S^{(\sigma_\rho)}$. On the other hand, as shown in figure\ \ref{nested_PS} it follows from the ordering of the trees that these approximations in $J_I^{(\sigma_\rho)}$ can only be $\text{hc}_I$, and do not make any approximations on the intrinsic coordinates ($\beta_I-$components of the jet momenta). So the ranges of intrinsic coordinates are unaffected by $t_{\sigma_n}, ..., t_{\sigma_m}$.

The approximations $t_{\sigma_1}, ..., t_{\sigma_{m-1}}$ apply projections on the loop momenta in $H^{(\sigma_\rho)}$ and $J^{(\sigma_\rho)}$. For $H^{(\sigma_\rho)}$, since the hard loop momenta are always integrated over the entire 4-dim space, the projections do not make any difference between $\sigma_\rho$ and $\rho^{\left\{ \sigma_n...\sigma_1 \right\}}$. For the subgraph $J_I^{(\sigma_\rho)}$, the approximations can only be $\text{sc}_I$, which projects the jet momenta onto their $\overline{\beta}_I-$component. As we have shown in eq.\ (\ref{cii_result}), this leads to a pinch surface where the ranges of the $\beta_I-$components of the projected jet momenta (which are intrinsic coordinates) are unbounded. This is the only difference between $\sigma_\rho$ and $\rho^{\left\{ \sigma_n...\sigma_1 \right\}}$. However, this difference does not affect eq.\ (\ref{pairwise_cancellation_nested_conclusion}), because the same $t_{\sigma_1}, ..., t_{\sigma_{m-1}}$ are present in both terms.

From these two paragraphs above, it is then obvious that both terms in eq.\ (\ref{pairwise_cancellation_nested_conclusion}) are pinched at $\rho^{\left\{ \sigma_n...\sigma_1 \right\}}$.

$(\mathfrak{b})$ From our previous analysis on the products of approximation operators in section\ \ref{subtraction_terms}, more precisely, eqs.\ (\ref{requirement1}) and (\ref{requirement2}), it is also direct that $t_{\sigma_m}$ is exact at $\rho^{\left\{ \sigma_n...\sigma_1 \right\}}$, which implies the two terms in (\ref{pairwise_cancellation_nested_conclusion}), which differ by a minus sign, cancel in the leading order. As we have analyzed in section\ \ref{divergences_are_logarithmic}, the divergences in (\ref{pairwise_cancellation_nested_conclusion}) are at worst logarithmic, so the cancellation in the leading terms is sufficient to show IR finiteness.

\subsection{Divergences at overlapping pinch surfaces}
\label{pairwise_cancellation_regular: overlapping}

In this subsection we consider the case where $\rho^{\left\{ \sigma_n...\sigma_1 \right\}}$ overlaps with certain $\sigma_i$'s appearing in $t_{\sigma_n} ...t_{\sigma_1} \mathcal{A}$. Then we consider the pinch surface $\tau \equiv \text{enc}\left[ \sigma_m, \rho^{\left\{ \sigma_n...\sigma_1 \right\}} \right]$, where $\sigma_m$ is the smallest of all the pinch surfaces in the forest $F\equiv \left\{ \sigma_1, ..., \sigma_n \right\}$ that overlap with $\rho^{\left\{ \sigma_n...\sigma_1 \right\}}$. By definition, $\tau$ is nested with the $\sigma_i$'s. From sections\ \ref{subgraphs_enclosed_pinch_surfaces}--\ref{extension_exotic_structures}, we know $\tau$ is a leading pinch surface of $\mathcal{A}$, which may have been included in $F$ or not. Without loss of generality, we assume $\tau \notin F$. We will confirm that the IR divergences at $\rho^{\left\{ \sigma_n...\sigma_1 \right\}}$, are cancelled between the following two terms:
\begin{eqnarray} \label{pairwise_cancellation_overlapping_conclusion}
\bigg[ \left( -t_{\sigma_n} \right)... \left( -t_{\sigma_m} \right)... \left( -t_{\sigma_1} \right) \mathcal{A} + \left( -t_{\sigma_n} \right)... \left( -t_{\sigma_m} \right) \left( -t_{\tau} \right)... \left( -t_{\sigma_1} \right) \mathcal{A} \bigg] _{\text{div}\ \mathfrak{n}\left [ \rho \right ]}=0,
\end{eqnarray}
which differ by a minus sign and the operator $t_\tau$. To prove eq.\ (\ref{pairwise_cancellation_overlapping_conclusion}), we can first work on a simpler version with the minimum number of approximation operators:
\begin{eqnarray} \label{center_formula}
t_{\sigma}\left( 1-t_\tau \right) \mathcal{A} \mid _{\text{div}\ \mathfrak{n}\left [ \rho^{\left\{ \sigma \right\}} \right ]}=0.
\end{eqnarray}
Here $\tau\equiv \text{enc}\left [ \sigma, \rho^{\left\{ \sigma \right\}} \right ]$, and we have both $\left( \mathcal{A} \right)_{\text{div }\mathfrak{n}\left[\rho\right]}\neq 0$ and $\left( t_\sigma \mathcal{A} \right)_{\text{div }\mathfrak{n}\left[\rho\right]}\neq 0$. Eq.\ (\ref{center_formula}) is what we aim to prove in this subsection. This is a special case of (\ref{pairwise_cancellation_overlapping_conclusion}), in which $n=m=1$. In fact the proof of (\ref{center_formula}) is sufficient to deduce (\ref{pairwise_cancellation_overlapping_conclusion}), as we will explain at the end of this subsection.

As above, we need to show two things in order to prove eq.\ (\ref{center_formula}): ($\mathfrak{a}$) $t_\tau t_\sigma \mathcal{A}$ is also divergent at $\rho^{\left\{ \sigma \right\}}$, namely, $\rho^{ \left\{ \sigma \right\} }=\rho^{ \left\{ \sigma\tau \right\} }$; ($\mathfrak{b}$) $t_\tau$ is exact for $t_\sigma \mathcal{A}$ there. Our method is to focus on an arbitrary configuration of $\rho^{\left\{ \sigma \right\}}$, go through all the ways in which $t_\tau$ may act on line momenta and vector indices, and verify the two aspects above for each case. Throughout this subsection, $\rho^{\left\{ \sigma \right\}}$ is assumed to be regular. Exotic configurations in $\rho^{\left\{ \sigma \right\}}$ will be discussed in section\ \ref{pairwise_cancellation_exotic}.

Suppose $t_\tau$ acts as a hard-collinear approximation, say $\text{hc}_I$, on some momentum $k^\mu$ of a propagator. Then by definition this propagator is in the set $J_I^{(H^\tau)}$ (see the notation in table\ \ref{leading_enclosed_PS_notations}). From eq.\ (\ref{enc_subgraph_relations}), $k^\mu$ either belongs to $J_I^{\left(H^\sigma\right)}$ or $J_I^{\left(H^\rho\right)}$ (or both). Whenever $k^\mu \in J_I^{\left(H^\sigma\right)}$, from the operator identity $t_\sigma^2=t_\sigma$ restricted to momentum $k^\mu$, it is immediate to see that the action of $t_\tau$ on line $k^\mu$ is consistent with both $\rho^{ \left\{ \sigma \right\} }=\rho^{ \left\{ \sigma\tau \right\} }$ and $t_{\sigma}\left( 1-t_\tau \right) =0$ for this line. Then we only need to consider the case where the propagator belongs to $J_I^{\left(H^\rho\right)}$ but not $J_I^{\left(H^\sigma\right)}$. Then by construction it is an internal propagator of $H^{(\sigma)}$. For this case the argument reduces to that of the nested case in section\ \ref{pairwise_cancellation_regular: nested}, and we can make use of (\ref{pairwise_cancellation_nested_conclusion}). Therefore, we have verified (\ref{center_formula}) when $t_\tau$ acts as a hard-collinear approximation.

Then suppose $t_\tau$ acts as a soft-collinear approximation on the momentum $k^\mu$ of a propagator $ab$. Again from eq.\ (\ref{enc_subgraph_relations}), $ab$ is in the following three subgraphs: $S^{(\sigma)}$, $S^{(\rho)}$ and/or $J_I^{\left( \sigma \right)} \bigcap J_K^{\left( \rho \right)}\ (I\neq K)$. For the cases of $S^{(\sigma)}$ and $S^{(\rho)}$, the arguments in the previous paragraph can be applied straightforwardly, after replacing the hard subgraphs there by jet subgraphs here, and jet subgraphs there by soft subgraphs here. We find that the action of $t_\tau$ is exact and consistent with $\rho^{ \left\{ \sigma \right\} }=\rho^{ \left\{ \sigma\tau \right\} }$.

The case of $ab\in J_I^{\left( \sigma \right)} \bigcap J_K^{\left( \rho \right)} \subset S^{(\tau)}$ is more complicated: although line $ab$ is acted on by a soft-collinear approximation in $t_\tau$, it is also acted upon by a hard-collinear approximation from $t_\sigma$, if it is attached to $H^{(\sigma)}$. We can classify all the relations between $ab$ and the hard subgraphs, $H^{(\sigma)}$ and $H^{(\rho)}$, into four types: (I) $ab$ is attached to neither $H^{(\sigma)}$ nor $H^{(\rho)}$; (II) $ab$ is attached to both $H^{(\sigma)}$ and $H^{(\rho)}$; (III) $ab$ is attached to $H^{(\sigma)}$ but not $H^{(\rho)}$; (IV) $ab$ is attached to $H^{(\rho)}$ but not $H^{(\sigma)}$. These subcases are considered one by one to verify eq.\ (\ref{center_formula}). Again, in each of them we must verify $(\mathfrak{a})$ $\rho^{ \left\{ \sigma \right\} }=\rho^{ \left\{ \sigma\tau \right\} }$, and $(\mathfrak{b})$ $t_\tau$ is exact for $t_\sigma \mathcal{A}$ in $\rho^{\left\{ \sigma \right\}} (=\rho^{ \left\{ \sigma\tau \right\} })$.

\begin{itemize}
\item[(I)]{} The case $ab \notin J_I^{(H^\sigma)}\text{ or }J_K^{(H^\rho)}$ is simple, because it implies that $ab$ is an internal soft propagator in $\tau$. Then $t_\tau$ only acts as the identity operator on its momenta and vector indices, and we immediately have $\rho^{ \left\{ \sigma \right\} } = \rho^{ \left\{ \sigma\tau \right\} }$ and $t_{\sigma} \left( 1-t_\tau \right)=0$ in this case.
\end{itemize}

\begin{itemize}
\item[(II)]{} $ab\in J_I^{(H^\sigma)} \bigcap J_K^{(H^\rho)}$. As we have shown in section\ \ref{leading_enclosed_pinch_surfaces}, $S^{(\tau)}$ cannot be attached to $H^{(\tau)}= H^{(\sigma)} \bigcap H^{(\rho)}$ directly. So if $ab$ is attached to $H^{(\sigma)}$ at $a$, then it can only be attached to $H^{(\rho)}$ at $b$, and vice versa.

\begin{figure}[t]
\centering
\includegraphics[width=13cm]{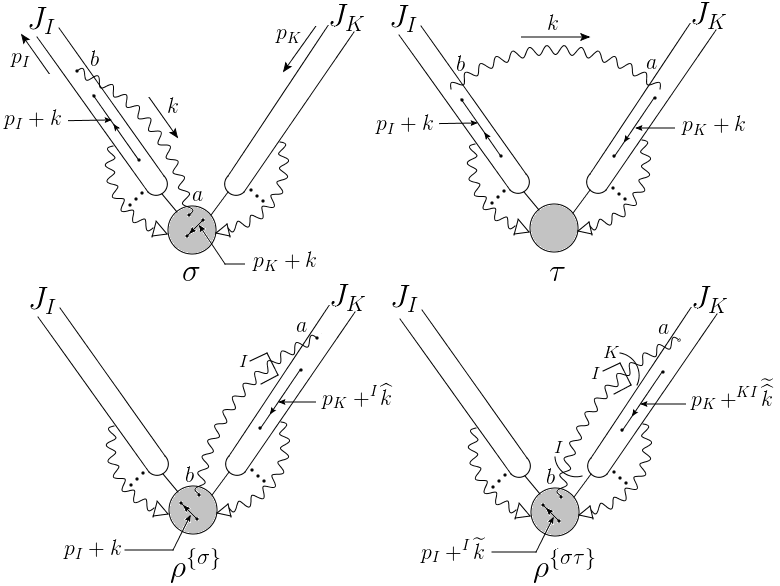}
\caption{A pictorial illustration of pinch surfaces $\sigma$ and $\tau$ of $\mathcal{A}$, $\rho^{ \left\{ \sigma \right\} }$ of $t_\sigma \mathcal{A}$ and $\rho^{ \left\{ \sigma\tau \right\} }$ of $t_\sigma t_\tau \mathcal{A}$ in case (II), where $ab$ is an element of $J_I^{\left( \sigma \right)} \bigcap J_K^{\left( \rho \right)} \subset S^{(\tau)}$, whose endpoint $a$ is in $H^{(\sigma)}\bigcap J_K^{\left( \rho \right)}$, and $b$ is in $H^{(\rho)}\bigcap J_I^{\left( \sigma \right)}$. The shaded areas represent the hard subgraphs in each region (and are not all the same). Denote the momentum of $ab$ as $k^\mu$. All the propagator momenta in $\mathcal{A}$ that contain $k$ are of the forms $p_I+k$ and $p_K+k$, where $p_I$ ($p_K$) is the momentum of a propagator in $J_I$ ($J_K$). In the approximated amplitude $\rho^{ \left\{ \sigma \right\} }$, $p_K+k$ becomes $p_K+^I\widehat{k}$, while in $\rho^{ \left\{ \sigma\tau \right\} }$, $p_I+k$ becomes $p_I+^I\widetilde{k}$ and $p_K+k$ becomes $p_K+^{KI}\widetilde{\widehat{k}}$.}
\label{effects_on_overlapping_PS_hh}
\end{figure}
($\mathfrak{a}$) Figure\ \ref{effects_on_overlapping_PS_hh} shows how the line $ab$ appears in regions $\sigma$, $\rho^{ \left\{ \sigma \right\} }$ and $\rho^{ \left\{ \sigma\tau \right\} }$. From Theorems 1 and 2 in section\ \ref{pinch_surface_amplitudes_approximations}, in both $\rho^{ \left\{ \sigma \right\} }$ and $\rho^{ \left\{ \sigma\tau \right\} }$ the lightlike momentum $k^\mu$ is pinched in the direction of $\overline{\beta}_I^\mu$. To make sure that $\rho^{ \left\{ \sigma \right\} }$ and $\rho^{ \left\{ \sigma\tau \right\} }$ coincide, we examine the differences between the denominators of $t_\sigma \mathcal{A}$ and $t_\sigma t_\tau \mathcal{A}$. The figure shows two possible differences: momentum $p_I+k$ in $\rho^{ \left\{ \sigma \right\} }$ becomes $p_I+^I\widetilde{k}$ in $\rho^{ \left\{ \sigma\tau \right\} }$; $p_K+^I\widehat{k}$ in $\rho^{ \left\{ \sigma \right\} }$ becomes $p_K+^{KI}\widetilde{\widehat{k}}$ in $\rho^{ \left\{ \sigma\tau \right\} }$, with the notations for projected momenta in eqs.\ (\ref{hat_tilde_definition}) and (\ref{hat_plus_tilde_definition}). But when $k^\mu$ is parallel to $\overline{\beta}_I^\mu$ in $\rho^{ \left\{ \sigma \right\} }$ and $\rho^{ \left\{ \sigma\tau \right\} }$, $k^\mu\sim \left( k\cdot\beta_I \right) \overline{\beta}_I^\mu= ^I\widetilde{k}^\mu$, and then
\begin{align}
\begin{split}
&\Big( p_I+k \Big)^2 = \left( p_I+^I\widetilde{k} \right)^2 + \mathcal{O}(\lambda); \\
&\Big( p_K+^I\widehat{k} \Big)^2 = p_K^2+ 2\left( p_K\cdot\beta_I \right) \left( k\cdot\overline{\beta}_I \right)= \left( p_K+^{KI}\widetilde{\widehat{k}} \right)^2+ \mathcal{O}(\lambda^2).
\end{split}
\end{align}
So the action of $t_\tau$ leaves the denominators unchanged, and the two IR-divergent pinch surfaces $\rho^{ \left\{ \sigma \right\} }$ and $\rho^{ \left\{ \sigma\tau \right\} }$ are identical.

($\mathfrak{b}$) It remains to check the agreement in the numerators of $t_{\sigma}\mathcal{A} \mid _{\mathfrak{n}\left [ \rho \right ]}$ and $t_\sigma t_\tau \mathcal{A} \mid _{\mathfrak{n}\left [ \rho \right ]}$ in order to verify eq.\ (\ref{center_formula}). The only possible difference is from the current at vertex $a$ in the figure, say $v^\mu$. Respectively, we have $\left( v\cdot\beta_I \right)\overline{\beta}_I^\mu$ in $t_\sigma \mathcal{A}$ while $\left( v\cdot \overline{\beta}_K \right) \left( \beta_K \cdot \beta_I \right) \overline{\beta}_I^\mu$ in $t_\sigma t_\tau \mathcal{A}$. However, at the pinch surface $\rho^{\left\{ \sigma \right\}}$ that we study, their contributions are both $\left( v\cdot\overline{\beta}_K \right)\left( \beta_K\cdot\beta_I \right)\overline{\beta}_I^\mu$. The reason is that the component of $v^\mu$ that appears, in the leading term of $t_{\sigma}\mathcal{A} \mid _{\mathfrak{n}\left [ \rho \right ]}$, is its $\beta_K-$component, i.e. $\left( v\cdot\overline{\beta}_K \right)\beta_K^\mu$, because $a$ is a jet vertex with jet momenta parallel to $\beta_K^\mu$, while the other endpoint $b$ is in $H^{(\rho)}$. Therefore, (\ref{center_formula}) holds when $ab$ is attached to both $H^{(\sigma)}$ and $H^{(\rho)}$.
\end{itemize}

\begin{itemize}
\item[$\begin{matrix} \hspace{0.3cm}(\text{III})\\ \& (\text{IV}) \end{matrix}$]{} The two cases where $ab\in J_I^{(H^\sigma)}\text{ but }ab\notin J_K^{(H^\rho)}$ in (III), and $ab\notin J_I^{(H^\sigma)}\text{ but }ab\in J_K^{(H^\rho)}$ in (IV), can be simultaneously represented by figure\ \ref{effects_on_overlapping_PS_jh}. In general, there can be four types of propagators in each figure, which are denoted by their momenta $k_i\ (i=1,2,3,4)$. Explicitly, the propagators labeled by $k_1$ are internal lines of $J_I^{(\sigma)}$, and soft lines attached to $J_I^{(\tau)}$, because in $\rho^{\left\{ \sigma \right\}}$ they are soft and attached to both $J_I^{(\sigma)}$ and $J_K^{(\sigma)}$. The propagators $a_4b_4$, which are $ab$ in case (IV), are labeled by $k_2$; the propagators $a_3b_3$, which are $ab$ in case (III), are labeled by $k_3$. The propagators labeled by $k_4$ are soft in $\sigma$, and become internal jet lines of $J_K^{(\rho)}$ in $\rho^{\left\{ \sigma \right\}}$ and $\rho^{\left\{ \sigma\tau \right\}}$. For each type of propagator, its configurations in $\sigma$ and $\rho$ determine the approximation operators acting on it in $\rho^{\left\{ \sigma \right\}}$ and $\rho^{\left\{ \sigma\tau \right\}}$. Note that the number of propagators of each type can be arbitrary, as long as momentum conservation is satisfied. For example, it is possible that the number of $k_3$- or $k_4$-propagators is zero, but they cannot be simultaneously zero in order to keep momentum conservation in $\rho^{\left\{ \sigma \right\}}$ and $\rho^{\left\{ \sigma\tau \right\}}$. In each figure, there can also be an arbitrary number of scalar-polarized jet gauge bosons attached to the hard part, as are shown in figure\ \ref{effects_on_overlapping_PS_hh}. But they do not affect the reasoning, so for simplicity we do not exhibit them here.
\begin{figure}[t]
\centering
\includegraphics[width=15.1cm]{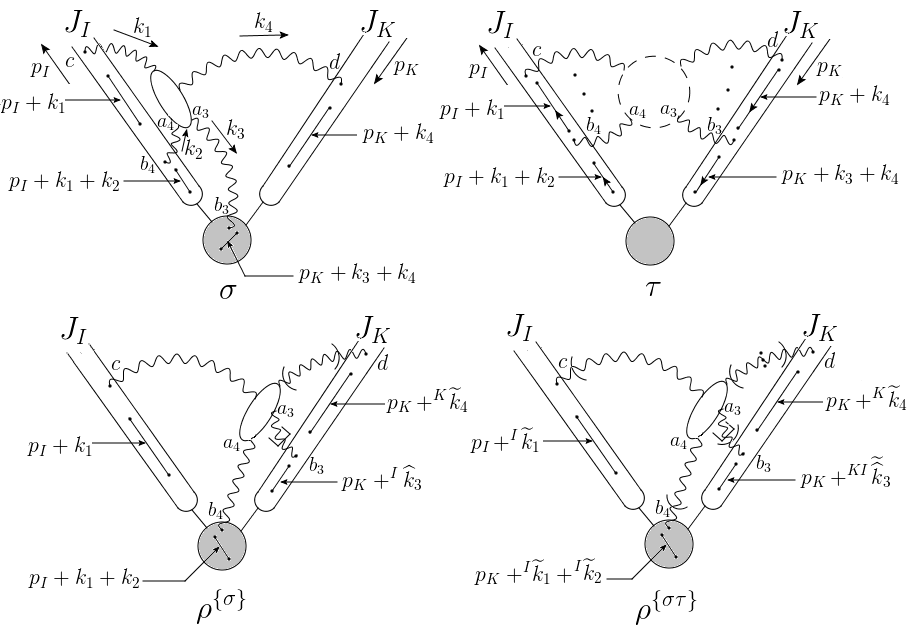}
\caption{A pictorial illustration of pinch surfaces $\sigma$ and $\tau$ of $\mathcal{A}$, $\rho^{ \left\{ \sigma \right\} }$ of $t_\sigma \mathcal{A}$ and $\rho^{ \left\{ \sigma\tau \right\} }$ of $t_\sigma t_\tau \mathcal{A}$, where $a_3b_3$ and $a_4b_4$ play the role of $ab$ in cases (III) and (IV), respectively. The shaded areas represent the hard subgraphs in each region. The white blob in the middle is parallel to $\beta_I^\mu$ in $\sigma$, $\overline{\beta}_I^\mu$ in $\rho^{ \left\{ \sigma \right\} }$ and $\rho^{ \left\{ \sigma\tau \right\} }$, and is therefore soft in $\tau$. After we label the four types of momenta attached the blob by $k_i^\mu\ (i=1,2,3,4)$, all the line momenta in $\mathcal{A}$ that contain $k_i$ are of the form $p_I+k_1$, $p_I+k_1+k_2$, $p_K+k_3+k_4$ and $p_K+k_4$. Their values change as a result of the approximations in $t_\sigma \mathcal{A}$ and $t_\sigma t_\tau \mathcal{A}$, as have been marked in $\rho^{ \left\{ \sigma \right\} }$ and $\rho^{ \left\{ \sigma\tau \right\} }$. Notice that $k_4^\mu$ does not enter the propagator momenta that are labelled by $p_K+^I\widehat{k}_3$ (in $\rho^{ \left\{ \sigma \right\} }$) and $p_K+^{KI}\widetilde{\widehat{k}}_3$ (in $\rho^{ \left\{ \sigma\tau \right\} }$) according to the hard-collinear approximation $\text{hc}_K^{(\sigma)}$ (not shown in the figure).}
\label{effects_on_overlapping_PS_jh}
\end{figure}

We need to show for figure\ \ref{effects_on_overlapping_PS_jh} that $(\mathfrak{a})$ $\rho^{ \left\{ \sigma \right\} }=\rho^{ \left\{ \sigma\tau \right\} }$, and $(\mathfrak{b})$ $t_\tau$ is exact for $t_\sigma \mathcal{A}$ in $\rho^{\left\{ \sigma \right\}}(=\rho^{ \left\{ \sigma\tau \right\} })$.

($\mathfrak{a}$) In both $\rho^{ \left\{ \sigma \right\} }$ and $\rho^{ \left\{ \sigma\tau \right\} }$, $k_1^\mu$ is soft, $k_2^\mu$ and $k_3^\mu$ are collinear to $\overline{\beta}_I^\mu$, and $k_4^\mu$ is collinear to $\beta_K^\mu$. With these values of the $k_i^\mu$'s, all the denominators of $t_\sigma \mathcal{A}$ in $\rho^{ \left\{ \sigma \right\} }$ coincide with those of $t_\sigma t_\tau \mathcal{A}$ in $\rho^{ \left\{ \sigma\tau \right\} }$, because the only possibly different denominators are identical to the leading term:
\begin{align}
\begin{split}
\left( p_I+^I\widetilde{k}_1 \right)^2 &= p_I^2+ 2\left( p_I\cdot \overline{\beta}_I \right) \left( k_1\cdot \beta_I\right)= \left( p_I+k_1 \right)^2 +\mathcal{O}(\lambda^2), \\
\left( p_I+ ^I\widetilde{k}_1+ ^I\widetilde{k}_2 \right)^2 &= p_I^2+ 2\left( p_I\cdot \overline{\beta}_I \right) \left( (k_1+k_2)\cdot \beta_I\right)= \left( p_I+k_1+k_2 \right)^2 +\mathcal{O}(\lambda), \\
\left( p_K+^{KI}\widetilde{\widehat{k}}_3 \right)^2 &= p_K^2+ 2\left( p_K\cdot \overline{\beta}_K \right) \left( \beta_I\cdot\beta_K \right) \left( k_3\cdot \overline{\beta}_I \right) = \left( p_K+ ^I\widehat{k}_3 \right)^2+ \mathcal{O}\left( \lambda^2 \right).
\end{split}
\end{align}
Meanwhile other denominator factors are exactly the same in $\rho^{ \left\{ \sigma \right\} }$ and $\rho^{ \left\{ \sigma\tau \right\} }$. This implies that $\rho^{ \left\{ \sigma \right\} }$ and $\rho^{ \left\{ \sigma\tau \right\} }$ are identical.

($\mathfrak{b}$) Now we check the agreement in the numerators of $t_{\sigma}\mathcal{A} \mid _{\mathfrak{n}\left [ \rho \right ]}$ and $t_\sigma t_\tau \mathcal{A} \mid _{\mathfrak{n}\left [ \rho \right ]}$, by verifying the coincidence of currents at each vertex (at the leading order).\footnote{It suffices to only consider the currents, because all the other numerator contributions are from fermion propagators. At $\rho^{ \left\{ \sigma \right\} }$ and $\rho^{ \left\{ \sigma\tau \right\} }$, the momenta $k_3$ and $k_4$ do not enter the numerators of the propagators according to the approximations, and $^I\widetilde{k}_1$ and $^I\widetilde{k}_2$ are good approximations of $k_1$ and $k_2$ separately.} It is relatively easy for the currents at vertices $a_3$, $a_4$, $c$ and $d$ in the figure. In detail, $t_\tau=1$ for the currents at $a_3$ and $a_4$, $t_\tau= t_\sigma$ for the currents at $d$ (therefore $t_\sigma t_\tau= t_\sigma$), meanwhile $t_\tau$ is a good approximation for the currents at $c$ in the neighborhood of the pinch surface $\rho^{ \left\{ \sigma \right\} } (=\rho^{ \left\{ \sigma\tau \right\} })$.

Similarly, the two currents at $b_4$ in $\rho^{ \left\{ \sigma \right\} }$ and $\rho^{ \left\{ \sigma\tau \right\} }$ also agree, because $a_4b_4$ is collinear to $\overline{\beta}_I^\mu$ there. Then the soft-collinear approximation in the $\beta_I-$direction, which is equal to the hard-collinear approximation in the $\overline{\beta}_I-$direction, is a good approximation.

It is relatively more complicated for the vector indices of the currents $v^\mu$ at vertices $b_3$. They belong to the scalar-polarized gauge boson in $J_I^{(H^\sigma)}$, and are marked by $k_3^\mu$. By definition, any such vertex is projected differently by $t_\sigma$ and $t_\sigma t_\tau$: it appears as $\left( v\cdot\beta_I \right) \overline{\beta}_I^\mu$ in $t_\sigma\mathcal{A}$ and $\left( v\cdot\overline{\beta}_K \right) \left (\beta_I\cdot\beta_K \right)\overline{\beta}_I^\mu$ in $t_\sigma t_\tau\mathcal{A}$. In order to see that their contributions to the leading terms are the same near the pinch surface $\rho^{ \left\{ \sigma \right\} } (=\rho^{ \left\{ \sigma\tau \right\} })$, we recall the second conclusion we have drawn from the power counting result eq.\ (\ref{overlapping_regular_pc6}), that the numerators combined with $v^\mu$ should offer an $\mathcal{O}(1)$-contribution to the leading term, otherwise it will be suppressed. Then it is clear from the expressions of the three-point vertices in figure\ \ref{physical_vertices} that if the propagators marked by momentum $k_3^\mu$ are attached to scalars or fermions at $b_3$, the only component of $v^\mu$ that leads to $\mathcal{O}(1)$ is $\left( v\cdot \overline{\beta}_I \right)$. Therefore, in the leading term we actually have $\left( v\cdot\beta_I \right) \overline{\beta}_I^\mu\rightarrow \left( v\cdot\overline{\beta}_K \right) \left (\beta_I\cdot\beta_K \right)\overline{\beta}_I^\mu$, and the coincidence is automatic.
\begin{figure}[t]
\centering
\includegraphics[width=8cm]{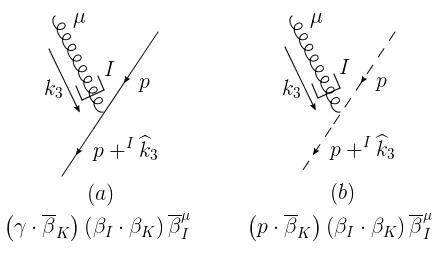}
\caption{Vertices of $t_\sigma \mathcal{A}$ where the propagators with momentum $p^\mu$ are $(a)$ fermions or $(b)$ scalars, depicted at pinch surface $\rho^{ \left\{ \sigma \right\} } (=\rho^{ \left\{ \sigma\tau \right\} })$. At the bottom of each figure we show the leading contribution of this vertex in the neighborhood. To obtain such values, note that $k_3^\mu$ does not appear in the expressions because it is parallel to $\overline{\beta}_I^\mu$ in $\rho^{ \left\{ \sigma \right\} }$, and after the hard-collinear approximation $\text{hc}_I$ only its $\beta_I$-component, which is $\mathcal{O}(\lambda)$, enters the vertex. Based on the expressions under both figures, it is clear that the contribution from $v^\mu$ to the leading term is in fact $\left( v\cdot\overline{\beta}_K \right) \left (\beta_I\cdot\beta_K \right)\overline{\beta}_I^\mu$.}
\label{physical_vertices}
\end{figure}

It remains to analyze the case where the scalar-polarized gluons marked by $k_3^\mu$ are attached to other gluons through three- or four-gluon vertices at $b_3$. We verify that the numerators of $t_{\sigma}\mathcal{A} \mid _{\mathfrak{n}\left [ \rho \right ]}$ and $t_\sigma t_\tau \mathcal{A} \mid _{\mathfrak{n}\left [ \rho \right ]}$ contribute identically to the leading term as follows.

Consider first that the junction is a three-gluon vertex $V^{\alpha \beta \gamma} \left( p,\ ^I\widehat{k}_3 \right)$, with $\gamma$ being the vector index associated with $k_3^\mu$ in $\mathcal{A}$, as is shown in figure\ \ref{3g_vertices}.
\begin{figure}[t]
\centering
\includegraphics[width=3.6cm]{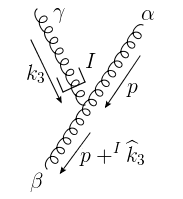}
\caption{A three-gluon vertex $V^{\alpha \beta \gamma'} \left( p,\ ^I\widehat{k}_3 \right) \cdot \beta_{I\gamma'} \overline{\beta}_I^\gamma$ that appears in $t_\sigma \mathcal{A}$, where only the $\beta_I$-component of the momentum $k_3^\mu$ joins the gluon with momentum $p^\mu$. Also, the vector index $\gamma$ is also projected onto its $\overline{\beta}_I$-component according to the same hard-collinear approximation.}
\label{3g_vertices}
\end{figure}
Then in $t_\sigma \mathcal{A}$ it reads:
\begin{eqnarray} \label{unphysical_1}
&&V^{\alpha \beta \gamma'} \left( p,\ ^I\widehat{k}_3 \right) \cdot \beta_{I\gamma'} \overline{\beta}_I^\gamma \nonumber\\
= &&gf^{abc} \left[ 2\left( p\cdot\beta_I \right) g^{\alpha\beta} \overline{\beta}_I^\gamma -\left( k_3\cdot \overline{\beta}_I\right) \beta_I^\alpha \beta_I^\beta \overline{\beta}_I^\gamma -p^\alpha \beta_I^\beta \overline{\beta}_I^\gamma -\beta_I^\alpha p^\beta \overline{\beta}_I^\gamma \right]\nonumber\\
= &&gf^{abc} \left[ 2\left( p\cdot\beta_I \right) g^{\alpha\beta} -p^\alpha \beta_I^\beta -\beta_I^\alpha p^\beta \right] \overline{\beta}_I^\gamma +\mathcal{O}(\lambda),
\end{eqnarray}
where the terms with $k_3^\mu$ are of $\mathcal{O}(\lambda)$ because $k_3^\mu$ is collinear to $\overline{\beta}_I^\mu$ in $\rho^{ \left\{ \sigma \right\} } (=\rho^{ \left\{ \tau\sigma \right\} })$. Similarly, using $\overline{\beta}_K\cdot ^{KI}\widetilde{\widehat{k}}_3 =0$, with $^{KI}\widetilde{\widehat{k}}$ defined by eq.\ (\ref{hat_plus_tilde_definition}), we have in $t_\sigma t_\tau \mathcal{A}$:
\begin{eqnarray} \label{unphysical_2}
&&V^{\alpha \beta \gamma''} \left( p,\ ^{KI}\widetilde{\widehat{k}}_3 \right) \cdot \overline{\beta}_{K\gamma''} \beta_K^{\gamma'} \beta_{I\gamma'} \overline{\beta}_I^\gamma \nonumber\\
= &&gf^{abc} \left[ 2\left( p\cdot\overline{\beta}_K \right) \left( \beta_I\cdot\beta_K \right) g^{\alpha\beta} -p^\alpha \left( \beta_K\cdot\beta_I \right) \overline{\beta}_K^\beta -\left( \beta_I\cdot\beta_K \right)\overline{\beta}_K^\alpha p^\beta \right] \overline{\beta}_I^\gamma +\mathcal{O}(\lambda).\nonumber\\
&&
\end{eqnarray}
According to one of the conclusions in section\ \ref{divergences_are_logarithmic} (more precisely, the second proposition following eq.\ (\ref{overlapping_regular_pc6})), in order to prove that eqs.\ (\ref{unphysical_1}) and (\ref{unphysical_2}) contribute identically to the leading terms of $t_\sigma \mathcal{A}$ and $t_\sigma t_\tau \mathcal{A}$, it is equivalent to verify that (\ref{unphysical_1}) and (\ref{unphysical_2}) agree at $\mathcal{O}(1)$. Obviously, their first terms in the square bracket agree, because $p^\mu$ is in the $\beta_K$-direction in $\rho$, so their contributions are the same. As for their second and third terms, notice that the only vectors appearing in the square bracket of (\ref{unphysical_1}) are $p^\mu$ and $\beta_I^\mu$. Since $p^\mu = \left( p\cdot \overline{\beta}_K \right) \beta_K^\mu$ at $\rho^{ \left\{ \sigma \right\} }$, every $\beta_I^\mu$ in the whole expression of the leading term must form an invariant with $\beta_K^\mu$, i.e. $\left( \beta_I\cdot \beta_K \right)$. Equivalently, every $\beta_I^\mu$ is projected onto its $\overline{\beta}_K$-component. Therefore, the leading contributions from (\ref{unphysical_1}) and (\ref{unphysical_2}) near $\rho^{ \left\{ \sigma \right\} } (=\rho^{ \left\{ \tau\sigma \right\} })$ are automatically identical.

This argument also works for four-gluon vertices, because then we have the following factors in $t_\sigma \mathcal{A}$:
\begin{eqnarray} \label{unphysical_3}
V^{\alpha \beta \gamma \delta'} \cdot \beta_{I\delta'} \overline{\beta}_I^{\delta} = \left( c_1 g^{\alpha\beta} \beta_I^\gamma +c_2 g^{\alpha\gamma} \beta_I^\beta +c_3 \beta_I^\alpha g^{\beta\gamma} \right)\overline{\beta}_I^\delta,
\end{eqnarray}
for a single line in $J_I^{(\sigma)}$, or
\begin{eqnarray} \label{unphysical_4}
V^{\alpha \beta \gamma' \delta'} \cdot \beta_{I\gamma'} \overline{\beta}_I^{\gamma} \beta_{I\delta'} \overline{\beta}_I^{\delta} = c_4 \left( \beta_I^\alpha \beta_I^\beta \right) \overline{\beta}_I^\gamma \overline{\beta}_I^\delta,
\end{eqnarray}
for two $J_I^{(\sigma)}$ lines, where $c_i\ (i=1,2,3,4)$ are color factors. For both cases above, the only vector appearing in the bracket is $\beta_I$. Similarly to the analysis of the three-point vertices, a $\beta_I^\mu$ must contract with a $\beta_K^\mu$ from other three-gluon vertices to form invariants of the form $\left( \beta_I\cdot\beta_K \right)$ in the leading term, which exactly appears in $t_\sigma t_\tau \mathcal{A}$.

In conclusion, we have proved that eq.\ (\ref{center_formula}) also holds when $ab$ is attached to $H^{(\sigma)}$ but not $H^{(\rho)}$.
\end{itemize}

Now that we have finished the proof of eq.\ (\ref{center_formula}), we next show that it is equivalent to (\ref{pairwise_cancellation_overlapping_conclusion}), which can be rewritten as
\begin{eqnarray} \label{pairwise_cancellation_overlapping_conclusion_rewrite}
\Big[ t_{\sigma_n}... t_{\sigma_m} \left( 1-t_{\tau} \right) t_{\sigma_{m-1}}... t_{\sigma_1} \mathcal{A} \Big] _{\text{div}\ \mathfrak{n}\left [ \rho \right ]}=0.
\end{eqnarray}

We focus on any line momentum of $t_{\sigma_{m-1}}... t_{\sigma_1} \mathcal{A}$, say $k^\mu$, and examine how $t_\tau$ may project it. For a vector index that is contracted in $t_{\sigma_{m-1}}... t_{\sigma_1} \mathcal{A}$, the analysis below follows in the same way. First, eq.\ (\ref{pairwise_cancellation_overlapping_conclusion_rewrite}) would be trivial if $t_\tau$ acts as an identity operator on $k^\mu$, so we assume that $t_\tau$ is either a hard-collinear or a soft-collinear approximation. Then we recall our observation that for both $t_{\sigma_n}... t_{\sigma_1} \mathcal{A}$ and $t_{\sigma_n}...t_{\sigma_m} t_\tau t_{\sigma_{m-1}}... t_{\sigma_1} \mathcal{A}$, $k^\mu$ is projected at most twice. Thus if $t_{\sigma_{m-1}}... t_{\sigma_1}$ acts as an identity operator on $k^\mu$, the operator $t_{\sigma_n}... t_{\sigma_m} \left( 1-t_{\tau} \right) t_{\sigma_{m-1}}... t_{\sigma_1}$ can be rephrased into $t_\sigma \left( 1-t_\tau \right)$, where $t_\sigma$ is the net projection on $k^\mu$ from $t_{\sigma_n}... t_{\sigma_m}$. The result follows immediately from (\ref{center_formula}).

So we only need to consider the case where both $t_\tau$ and $t_{\sigma_{m-1}}... t_{\sigma_1}$ are nontrivial (and not identical) on $k^\mu$. The only case is when $t_\tau$ is hard-collinear and $t_{\sigma_{m-1}}... t_{\sigma_1}$ is soft-collinear. We now consider the action of $t_{\sigma_n}... t_{\sigma_m}$, from which there are two possibilities. If $t_{\sigma_n}... t_{\sigma_m}$ is the same as $t_\tau$ on $k^\mu$, being a hard-collinear approximation, then from $t_\tau^2= t_\tau$ we see that eq.\ (\ref{pairwise_cancellation_overlapping_conclusion_rewrite}) is also trivial. Otherwise $t_{\sigma_n}... t_{\sigma_m}=1$ on $k^\mu$, which means that the propagator with momentum $k^\mu$ can only be an internal hard propagator in the pinch surfaces $\sigma_m,..., \sigma_n$. Since $\tau\equiv \text{enc}\left[ \sigma_m, \rho^{\left\{ \sigma_n...\sigma_1 \right\}} \right]$, in order that $t_\tau$ offers a hard-collinear approximation on $k^\mu$, the propagator must be lightlike and attached to $H^{(\rho^{\left\{ \sigma_n...\sigma_1 \right\}})}$ in $\rho^{\left\{ \sigma_n...\sigma_1 \right\}}$. In this case, $t_\tau$ is a good approximation at $\rho^{\left\{ \sigma_n...\sigma_1 \right\}}$, and the cancellation of IR divergences in (\ref{pairwise_cancellation_overlapping_conclusion_rewrite}) is immediate.

In conclusion, eqs.\ (\ref{pairwise_cancellation_overlapping_conclusion}) and (\ref{center_formula}) are equivalent, which indicates that for any approximated amplitude with an overlapping divergence, as long as it corresponds to a regular pinch surface, we can always find a counterterm to cancel it. This cancellation is pairwise, and in each pair one term has a $t_\tau$ while the other does not. To extend our analysis to all types of overlapping divergences, we will check for the exotic configurations in the next subsection.

\subsection{Divergences at exotic pinch surfaces}
\label{pairwise_cancellation_exotic}

The analysis in the last subsection is based on the assumption that $\rho^{\left\{ \sigma \right\}}$ is a regular pinch surface. For example, when we discussed the case where $t_\tau$ acts as a soft-collinear approximation on the momentum of a propagator, and that propagator is from $S^{(\rho)}$, we then deduced that it must be attached to a jet subgraph in $\rho^{\left\{ \sigma \right\}}$, where $t_\tau$ is a good approximation. However, in the presence of exotic configurations, it is possible that a soft propagator is attached to the hard subgraph in region $\rho^{\left\{ \sigma \right\}}$. So if we take such configurations into account, the analysis in section\ \ref{pairwise_cancellation_regular: overlapping} does not immediately apply.

Fortunately, in section\ \ref{extension_exotic_structures} we have enumerated all the possible exotic configurations in $\rho^{\left\{ \sigma \right\}}$, as well as the corresponding pinch surfaces $\sigma$ that provide approximations, for which $t_\sigma \mathcal{A}$ has these configurations. From $\sigma$ and $\rho^{\left\{ \sigma \right\}}$ we can derive $\tau$, as are shown in figures\ \ref{all_soft-exotic_structures} and \ref{all_hard-exotic_structures}. For figures\ \ref{all_soft-exotic_structures}$(a)$, $(b)$ and \ref{all_hard-exotic_structures}$(b)$, the exotic configurations in $\rho^{\left\{ \sigma \right\}}$ correspond to an internal soft vertex in $\tau$, which only contributes identity operators to $t_\tau$. So for these cases, eqs.\ (\ref{pairwise_cancellation_overlapping_conclusion}) and (\ref{center_formula}) are automatic.

As for the case of figure\ \ref{all_hard-exotic_structures}$(a)$, $t_\tau$ contains a soft-collinear approximation $\text{sc}_I$ on the soft propagator attached to the hard subgraph at $\rho^{\left\{ \sigma \right\}}$, and we claim that it is a good approximation. The reason is simple: the vertex in $\rho^{\left\{ \sigma \right\}}$, to which the soft lines are attached, is a jet vertex, because all the lightlike momenta entering it are parallel to $\beta_I^\mu$. As a result, all the invariants formed by the jet momenta and the soft momenta in the leading term at $\rho^{\left\{ \sigma \right\}}$, can only involve the $\overline{\beta}_I$-component of the soft momenta.

After all these discussions, we can assert that $t_\tau$ is always exact in eqs.\ (\ref{pairwise_cancellation_overlapping_conclusion}) and (\ref{center_formula}), with or without exotic configurations. Sections\ \ref{pairwise_cancellation_regular: nested}--\ref{pairwise_cancellation_exotic} altogether constitutes our proof of the forest formula, eq.\ (\ref{forest_formula_amplitude}).

\subsection{Discussion}\label{discussion_forest_formula}

Our discussions below treat four topics relevant to the arguments and results of this section. In Item $\mathit{1}$ below we explain why the proof is not graph-by-graph, and in Items $\mathit{2}$-$\mathit{4}$ we relate our forest formula, eq.\ (\ref{forest_formula_amplitude}), to other subtraction methods formulated in forest-structural expressions.

\begin{itemize}
\item[$\mathit{1. }$]{} The fact that eq.\ (\ref{forest_formula_amplitude}) is not graph-by-graph is due to the possibility of unphysical pinch surfaces, namely, the solutions of the Landau equation where all the lightlike propagators of one or more jets that are attached to the hard subgraph are scalar-polarized gauge bosons. These pinch surfaces are not the leading pinch surfaces by definition, but the IR behaviors in their neighborhoods can be power divergent (see eq.\ (\ref{hard_exotic_pc4}), when one or more $h_{Ki}=0$). Nevertheless, these divergences are cancelled by the Ward identity in the sum over all the attachments between the scalar-polarized gauge bosons and the hard subgraph. But given a single Feynman graph, $\mathcal{A}$, the remainder after all our subtractions is still divergent near these ``super-leading'' pinch surfaces \cite{Cls11book, LbtdStm85}, because the approximation operators appearing in the forest formula match only physical pinch surfaces. Therefore, the IR finiteness in (\ref{forest_formula_amplitude}) is not graph-by-graph.
\end{itemize}

\begin{itemize}
\item[$\mathit{2. }$]{} Our forest formula sums over all the forests of $\mathcal{A}$, which are defined in eq.\ (\ref{forest_definition}). Another way to formulate the forest formula, as is done by Collins and Soper in \cite{ClsSpr81}, is to sum over only the ``inequivalent forests''. Two forests of $\mathcal{A}$, say $\left\{ \sigma_{a(1)},...,\sigma_{a(m)} \right\}$ and $\left\{ \sigma_{b(1)},...,\sigma_{b(n)} \right\}$, are called equivalent if the two approximations $t_{\sigma_{a(m)}}...t_{\sigma_{a(1)}}$ and $t_{\sigma_{b(n)}}...t_{\sigma_{b(1)}}$ are the same. Given a series of equivalent forests, only one subtraction term is needed, and the overall sign is $(-1)^\mathcal{T}$, where $\mathcal{T}$ is the maximum number of trees in any forest of this class. For example, in the upcoming example in section\ \ref{2_loop_example_forest_formula}, we will see that $t_{\sigma_6}= t_{\sigma_7}t_{\sigma_6}= t_{\sigma_8}t_{\sigma_6}$, so $\left\{ \sigma_6 \right\}$, $\left\{ \sigma_6,\ \sigma_7 \right\}$ and $\left\{ \sigma_6,\ \sigma_8 \right\}$ are three equivalent forests. As a result, we only need to include a single term, for example $\left( t_{\sigma_6} \mathcal{A} \right)$, rather than the whole combination $\left( -t_{\sigma_6} \mathcal{A} +t_{\sigma_7} t_{\sigma_6} \mathcal{A} +t_{\sigma_7} t_{\sigma_6} \mathcal{A} \right)$, in our forest formula (\ref{forest_formula_amplitude}). We believe that this equivalence of using forests and inequivalent forests can be generalized to arbitrary orders, but a rigorous proof is left for future research.
\end{itemize}

\begin{itemize}
\item[$\mathit{3. }$]{} The whole of our analysis of sections\ \ref{pinch_surface_amplitudes_approximations}--\ref{the_proof_of_cancellation} can also be interpreted in position space. Previous work has already been carried out by Erdo$\breve{\text{g}}$an and Sterman in \cite{EdgStm15}, where they focused on UV divergences of massless gauge theories in position space. In light of scale invariance, such UV structures of an original amplitude $\mathcal{A}$ are very similar to the IR structures in our momentum-space study. The work of Erdo$\breve{\text{g}}$an and Sterman offers the precedent for this project, and we have provided, in the previous sections, a detailed illustration of what these singularities are like in momentum space, and how they are cancelled in the forest formula. We will also provide a sketch of the position-space version, especially for the pinch surfaces of $t_\sigma \mathcal{A}$, in appendix\ \ref{interpretation_position_space}.
\end{itemize}

\begin{itemize}
\item[$\mathit{4. }$]{} Finally we compare our work with some other works that use similar IR subtractions. Collins in his book \cite{Cls11book}, constructed the forest formula for color-singlet hard scattering with subtraction terms from hard-collinear and soft-collinear approximations as well. He proved the forest formula using an inductive strategy.

A recent work by Anastasiou and Sterman \cite{AntsStm18}, studies the IR behaviors of fixed-angle scatterings from an iterative perspective, illustrating the idea at two loops. In contrast to the latter treatment, the forest formula method we take here offers a viewpoint of the IR singularities of an all-order amplitude, with or without approximation. This treatment, though much more laborious, enables us to generalize to arbitrary orders and numbers of external momenta, and observe a number of general principles of IR cancellations.

The forest-like subtraction also appears in another recent work, which is from a slicing approach by Herzog \cite{Hzg18}. In the paper he promotes the subtraction method by employing suitable phase space mappings. This method is based on the geometry of IR regions, and is carried out explicitly at NLO and NNLO. Although his construction of subtraction terms is different from ours, the formula that summarizes the combinatorics of various counterterms is still forest-like.
\end{itemize}

\section{Factorization of the subtraction terms} \label{factorization_subtraction_terms}

As has been mentioned, our subtraction method implies a factorization structure. In more detail, for a QCD hard process with external momenta $p_1^\mu,...,p_N^\mu$, the forest formula holds for every Feynman graph $\mathcal{A}^{(n)} \left( p_1,...,p_N \right)$, where $n$ represents the order $\mathcal{O}(\alpha^n)$. After we sum over all the $\mathcal{A}^{(n)}$'s as well as the orders, we obtain the full amplitude $\mathcal{M} \left( p_1,...,p_N \right)$,
\begin{eqnarray} \label{def_graph_amplitude}
\mathcal{M}\equiv \sum_{n=0}^\infty \mathcal{M}^{(n)}\equiv \sum_{n=0}^\infty \sum_{\mathcal{A}^{(n)}} \mathcal{A}^{(n)}.
\end{eqnarray}
After we replace each $\mathcal{A}^{(n)}$ by its subtraction terms, the sum over graphs will lead to a factorized expression of $\mathcal{M}$, to obtain which is our aim in this section.

In the derivation, we will use the symbols $\gamma_H^{ }$, $\gamma_J^{ }$ and $\gamma_S^{ }$ to denote the hard, jet and soft subgraphs, in order to emphasize that their loop momenta are integrated over the full 4-dim space, rather than certain restricted ranges. In section\ \ref{factorization_presence_approximations} we show eq.\ (\ref{factorization_formula_hard&jet_approximations}), a key result that is subsidiary to the factorization in the presence of repetitive approximations. In section\ \ref{partonic_amplitudes_factorized_forms}, we use this result to derive the factorized expression for $\mathcal{M}$. A sketch of the argument is as follows.
\begin{itemize}
    \item[]{$\textit{Step 1}$. }We use the forest formula to rewrite the $\mathcal{A}^{(n)}$ in eq.\ (\ref{def_graph_amplitude}) as the sum over forests. For each forest, we identify a specific pinch surface $\sigma_0^*$: it has the largest ``reduced hard subgraph'' (explained below), and is the smallest among all the other pinch surfaces in this forest that have the same reduced hard subgraph. We denote the hard (jet, soft) subgraph of $\sigma_0^*$ as $\gamma_{H_0}$ ($\gamma_{J_0}$, $\gamma_{S_0}$).
\end{itemize}
\begin{itemize}
    \item[]{$\textit{Step 2}$. }In the sum over $\mathcal{A}^{(n)}$, the soft subgraph $\gamma_{S_0}^{ }$ is factorized from the hard-and-jet subgraph $\gamma_{H_0J_0}^{ }$. After we take the approximations inside $\gamma_{S_0}^{ }$ into account, the soft part contributes to a factor, which we denote $\gamma_{S_0,\text{eik}}^{ }/ \mathcal{J}_{0,\text{eik}}$.
\end{itemize}
\begin{itemize}
    \item[]{$\textit{Step 3}$. }From our analysis in section\ \ref{factorization_presence_approximations}, $\gamma_{H_0J_0}^{ }$ can be further factorized into the reduced subgraphs $\overline{\gamma}_{H_0}$ and $\overline{\gamma}_{J_0}$. The approximations inside $\overline{\gamma}_{J_0}$, which are soft-collinear, help us rewrite $\overline{\gamma}_{J_0}$ as the factor $\mathcal{J}_{0,\text{part}}$.
\end{itemize}
\begin{itemize}
    \item[]{$\textit{Step 4}$. }We combine the remaining part, which involves $\overline{\gamma}_{H_0}$ and its subtraction terms, together with the factors obtained in Steps 2 and 3. The final result is eq.\ (\ref{full_amplitude_factorization_result}).
\end{itemize}

\subsection{Factorization in the presence of approximations}
\label{factorization_presence_approximations}

To illustrate that QCD factorization can be achieved in the presence of repetitive approximations, we need to set up the following concept, which relates the pinch surfaces of different Feynman graphs. Given any two $\mathcal{O}\left( \alpha^n \right)$ Feynman graphs from the whole set $\left\{ \mathcal{A}^{(n)} \right\}$, say $\mathcal{A}_1^{(n)}$ and $\mathcal{A}_2^{(n)}$, we say that two pinch surfaces $\sigma'$ (of $\mathcal{A}_1^{(n)}$) and $\sigma''$ (of $\mathcal{A}_2^{(n)}$) are \emph{normal-space-equivalent} ($\mathcal{N}$-equivalent), if and only if there exists a one-to-one correspondence between their loop momenta, such that both momenta in the pair share the same normal space. In the text below, we will associate all the $\mathcal{O}\left( \alpha^n \right)$ Feynman graphs sharing a given pinch surface, say $\sigma^*$, and denote it as $\sigma^*\big[ \mathcal{A}^{(n)} \big]$ to represent each element of the $\mathcal{N}$-equivalent class. Note that the symbol $\sigma^*$ contains the information about the normal spaces of the loop momenta, i.e. the orders of the soft, hard, and jet subgraphs.

Following this convention, we will use $t_{\sigma^*[ \mathcal{A}^{(n)} ]}$ to denote the approximation operator that is associated with $\sigma^*$. Note that $\sigma^*\big[ \mathcal{A}^{(n)} \big]$ may not be a leading pinch surface of every $\mathcal{A}^{(n)}$, and if so, $t_{\sigma^*[ \mathcal{A}^{(n)} ]}$ serves to annihilate $\mathcal{A}^{(n)}$, i.e. we define
\begin{eqnarray}
t_{\sigma^*\left[ \mathcal{A}^{(n)} \right]}\mathcal{A}^{(n)} \equiv 0\ \ \ \text{if}\ \sigma^*\big[ \mathcal{A}^{(n)} \big]\ \text{is not a LPS of}\ \mathcal{A}^{(n)}.
\end{eqnarray}
Otherwise, the rules of such operators are exactly identical to those we have given in section\ \ref{neighborhoods & approximation_operators}, in which case we say that $\mathcal{A}^{(n)}$ \emph{is compatible with} (the loop assignments in) $\sigma^*$.

With the help of the Ward identity applied to soft-collinear and hard-collinear attachments \cite{ClsSprStm04}, we obtain the following factorization relation for each $\mathcal{N}$-equivalent class $\left\{ \sigma^*\big[ \mathcal{A}^{(n)} \big] \right\}$:
\begin{eqnarray} \label{factorization_formula}
\sum_{\mathcal{A}^{(n)}} t_{\sigma^*\left[ \mathcal{A}^{(n)} \right]} \mathcal{A}^{(n)} = \sum_{\gamma_{S_*,\text{eik}}^{ }} \gamma_{S_*,\text{eik}}^{(a_*)} \cdot \prod_{A=1}^N \sum_{\overline{\gamma}_{J_{A*}}^{ }} \overline{\gamma}_{J_{A*}}^{(b_{A*})} \cdot \sum_{\overline{\gamma}_{H_*}^{ }} \overline{\gamma}_{H_*}^{(n-a_*-b_*)},
\end{eqnarray}
where $b_*\equiv \sum_{A=1}^N b_{A*}$. In this expression, $\gamma_{S_*,\text{eik}}^{(a_*)} \equiv \gamma_{S_*,\text{eik}}^{(a_*)} \left( \beta_1,...,\beta_N \right)$ is obtained from any $\mathcal{O} \left( \alpha^{a_*} \right)$ soft subgraph $\gamma_{S_*}^{(a_*)}$, with its external lines attached to eikonal lines in the directions of $\beta_1^\mu,...,\beta_N^\mu$. The subgraph $\overline{\gamma}_{J_{A*}}^{(b_{A*})} \equiv \overline{\gamma}_{J_{A*}}^{(b_{A*})} \left( p_A,\overline{\beta}_A \right)$ is obtained by starting from the $\mathcal{O} \left( \alpha^{b_{A*}} \right)$ jet subgraph $\gamma_{J_{A_*}}^{(b_{A*})}$, deleting the soft lines attached to it, and attaching its lines that were previously attached to the hard subgraph to an eikonal line in the $\overline{\beta}_A^\mu$ direction \cite{ClsSprStm04}. Similarly, $\overline{\gamma}_{H_*}^{(n-a_*-b_*)}$ is a subgraph dependent on $p_1^\mu,...,p_N^\mu$, obtained from the hard subgraph $\gamma_{H_*}^{ }$ by deleting the unphysically polarized jet gauge bosons attached to it. The subgraph orders, $a_*$, $b_{A*}$'s and $(n-a_*-b_*)$, are determined by specifying $\sigma^*$. We note that eq.\ (\ref{factorization_formula}) not only holds for the full Feynman graphs, but also for subgraphs. For example, in the upcoming analysis of section\ \ref{partonic_amplitudes_factorized_forms}, we will apply this relation to the graphs with their soft subgraphs factorized out, say $\gamma_{H_*J_*}$. In this case, $a_*=0$, and we have
\begin{eqnarray} \label{factorization_formula_hard&jet}
\sum_{\gamma_{H_*J_*}^{ }} t_{\sigma^*[ \gamma_{H_*J_*}^{(n)} ]} \gamma_{H_*J_*}^{(n)} = \prod_{A=1}^N \sum_{\overline{\gamma}_{J_{A*}}^{ }} \overline{\gamma}_{J_{A*}}^{(b_{A*})} \cdot \sum_{\overline{\gamma}_{H_*}^{ }} \overline{\gamma}_{H_*}^{(n-b_*)},
\end{eqnarray}
where the $\sigma^*\big[ \gamma_{H_*J_*}^{(n)} \big]$'s are the pinch surfaces of $\gamma_{H_*J_*}^{(n)}$, where all the loop momenta are either hard or lightlike.

For each subtraction term of the forest formula, there can be more than one approximation acting repetitively on the amplitude. In order to generalize eq.\ (\ref{factorization_formula_hard&jet}) to fit such cases, we aim to prove the following statement. Given $\left\{ \gamma_{H_*J_*}^{(n)} \right\}$ and $\left\{ \sigma^*\big[ \gamma_{H_*J_*}^{(n)} \big] \right\}$ introduced above, we focus on another $\mathcal{N}$-equivalent class $\left\{ \sigma_<\big[ \gamma_{H_*J_*}^{(n)} \big] \right\}$, whose elements are the pinch surfaces contained in the $\sigma^*\big[ \gamma_{H_*J_*}^{(n)} \big]$'s, and we find
\begin{eqnarray} \label{factorization_formula_hard&jet_approximations}
\sum_{\gamma_{H_*J_*}^{ }} t_{\sigma^*[ \gamma_{H_*J_*}^{(n)} ]} t_{\sigma_<[ \gamma_{H_*J_*}^{(n)} ]} \gamma_{H_*J_*}^{(n)} = \prod_{A=1}^N \sum_{\overline{\gamma}_{J_{A*}}^{ }} t_{\sigma_<[ \overline{\gamma}_{J_{A*}} ]} \overline{\gamma}_{J_{A*}}^{(b_{A*})} \cdot \sum_{\overline{\gamma}_{H_*}^{ }} t_{\sigma_<[ \overline{\gamma}_{H_*} ]} \overline{\gamma}_{H_*}^{(n-b_*)},
\end{eqnarray}
where $\sigma_<[ \overline{\gamma}_{J_{A*}} ]$ and $\sigma_<[ \overline{\gamma}_{H_*} ]$ are the pinch surfaces of the factorized graphs $\overline{\gamma}_{J_{A*}}^{ }$ and $\overline{\gamma}_{H_*}^{ }$, whose loop momenta are identical to those of $\sigma_<[ \gamma_{H_*J_*}^{(n)} ]$. The subtlety of (\ref{factorization_formula_hard&jet_approximations}) is as follows: on the LHS of the equation, only the graphs ($\gamma_{H_*J_*}$) that are compatible with the loop assignments in both $\sigma^*$ and $\sigma_<$ contribute. In other words, there are fewer terms on the LHS of (\ref{factorization_formula_hard&jet_approximations}) than those on the LHS of (\ref{factorization_formula_hard&jet}), which are what we need for factorization. Nevertheless, we will prove (\ref{factorization_formula_hard&jet_approximations}) as follows.

\begin{figure}[t]
\centering
\includegraphics[width=12cm]{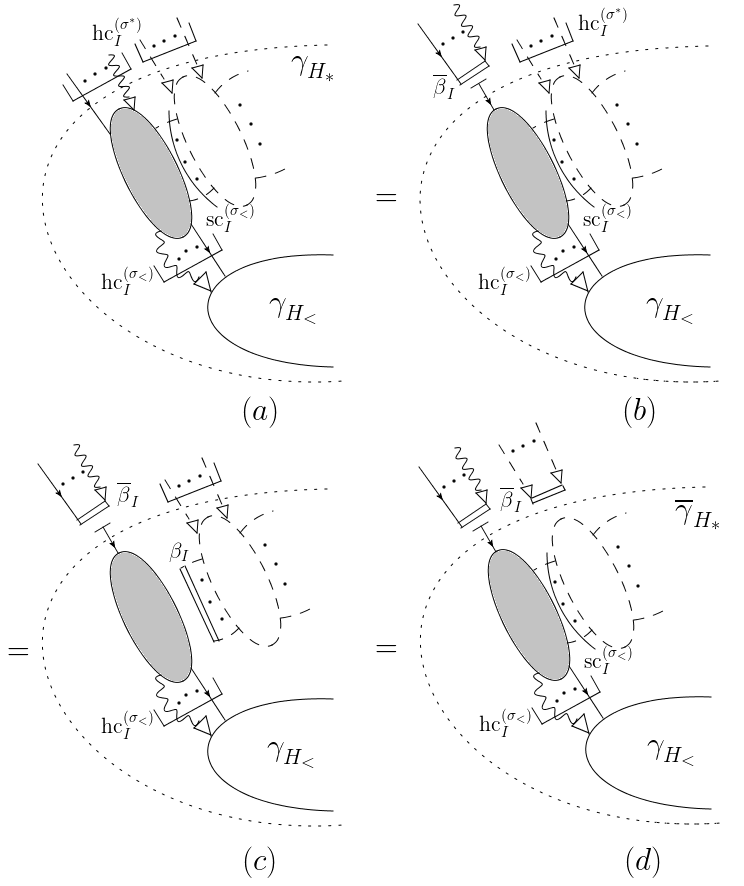}
\caption{The graphic procedure to prove eq.\ (\ref{factorization_formula_hard&jet_approximations}). The figures $(b)$ and $(c)$ are the intermediate steps from $(a)$ to $(d)$, to obtain which we have repeatedly applied the Ward identity for factorization. The subgraph enclosed by the dotted curve is $\gamma_{H_*}^{ }$ in $(a)$, and becomes the reduced hard subgraph $\overline{\gamma}_{H_*}^{ }$ in $(d)$ after factorization. At $\sigma_<$, some of its propagators become soft (represented by dashed lines), some become lightlike (represented by the shadowed region), and some remains hard, which is $\gamma_{H_<}^{ }$.}
\label{factorization_approximations_procedure}
\end{figure}

The most direct way to prove eq.\ (\ref{factorization_formula_hard&jet_approximations}) is to rewrite both its sides graphically. The LHS of (\ref{factorization_formula_hard&jet_approximations}), which is pre-factorization, is illustrated by figure\ \ref{factorization_approximations_procedure}$(a)$. In this figure, some of the jet loop momenta in $\sigma^*$ become soft in $\sigma_<$ while the others remain lightlike, and the hard loop momenta in $\sigma^*$ can become either lightlike or soft, or remain hard in $\sigma_<$. We arrive at figure\ \ref{factorization_approximations_procedure}$(b)$ by applying the Ward identity on the scalar-polarized gauge bosons that are lightlike in both $\sigma^*$ and $\sigma_<$. Then we can factorize the soft lines from the jets in $\gamma_{H_*}^{ }$ to obtain figure\ \ref{factorization_approximations_procedure}$(c)$. Finally, we factorize the remaining scalar-polarized gauge bosons attached to $\gamma_{H_*}^{ }$ (which are soft in $\sigma_<$) and restore the soft-collinear approximations inside $\gamma_{H_*}^{ }$ to get figure\ \ref{factorization_approximations_procedure}$(d)$, which is exactly the RHS of (\ref{factorization_formula_hard&jet_approximations}). Two explicit examples will also be provided later in figures\ \ref{an_example_on_factorization_cancellation_1st} and \ref{an_example_on_factorization_cancellation_2nd}.

In summary, eq.\ (\ref{factorization_formula_hard&jet_approximations}) shows that in the sum over hard-collinear attachments, the jet subgraph can still be factorized from the hard subgraph in the presence of repetitive approximations.

\subsection{Partonic amplitudes in factorized forms}
\label{partonic_amplitudes_factorized_forms}

In this subsection we derive the factorized expression for $\mathcal{M}$. We have shown that for each graph $\mathcal{A}^{(n)}$, the forest formula subtracts all its IR divergences from its leading pinch surfaces, and renders the remainder $R\left[ \mathcal{A}^{(n)} \right]$ finite (up to terms that can be cancelled in the sum over graphs of $\mathcal{A}^{(n)}$ with $\mathcal{N}$-equivalent pinch surfaces by the Ward identity):
\begin{eqnarray} \label{graph_forest_formula1}
\mathcal{A}^{(n)} = -\sum_{\substack{F\in \mathcal{F}[\mathcal{A}^{(n)}] \\ F\neq \varnothing}} \prod_{\sigma\in F} \left( -t_\sigma \right) \mathcal{A}^{(n)} + R\left[ \mathcal{A}^{(n)} \right].
\end{eqnarray}

We now rewrite $R\left[ \mathcal{A}^{(n)} \right]$ into a sum over subtraction terms. Consider the trivial ``pinch surface'' of $\mathcal{A}^{(n)}$ where the soft and jet subgraphs are trivial; in other words, all the loop momenta are off-shell. We denote this region by $\eta$. By definition, for any leading pinch surface of $\mathcal{A}^{(n)}$, say $\sigma$, we have $\sigma \subset \eta$. Defining the approximation operator $t_\eta$ as an identity operator on $\mathcal{A}^{(n)}$, we can rewrite $R\left[ \mathcal{A}^{(n)} \right]$ as:
\begin{eqnarray} \label{remaining_term_definition}
R\left[ \mathcal{A}^{(n)} \right]= \mathcal{A}^{(n)}+ \sum_{\substack{F\in \mathcal{F}[\mathcal{A}^{(n)}] \\ F\neq \varnothing}} \prod_{\sigma\in F} \left( -t_\sigma \right) \mathcal{A}^{(n)} = -\sum_{\substack{F\in \mathcal{F}[\mathcal{A}^{(n)}] \\ F\neq \varnothing}} \left( -t_\eta \right) \prod_{\sigma\in F} \left( -t_\sigma \right) \mathcal{A}^{(n)}.
\end{eqnarray}
We define the set of the \emph{extended leading pinch surfaces} of $\mathcal{A}^{(n)}$ to include all the leading pinch surfaces of $\mathcal{A}^{(n)}$ as well as $\eta$, and the \emph{extended forests} of $\mathcal{A}^{(n)}$ as the sets of nested extended leading pinch surfaces. For a given extended forest, we denote it as $\overline{F}$, which may contain $\eta$ or not. Using this notation, we substitute eq.\ (\ref{remaining_term_definition}) into (\ref{graph_forest_formula1}) and get
\begin{eqnarray} \label{graph_forest_formula2}
\mathcal{A}^{(n)} = -\sum_{\substack{\overline{F}\in \overline{\mathcal{F}}[\mathcal{A}^{(n)}] \\ \overline{F}\neq \varnothing}} \prod_{\sigma\in \overline{F}} \left( -t_\sigma \right) \mathcal{A}^{(n)}.
\end{eqnarray}

Given an arbitrary nonempty extended forest $\overline{F}$, we first focus on the subset of its extended pinch surfaces that have the largest \emph{reduced hard subgraph}. A reduced hard subgraph is obtained from the original hard subgraph by removing all the unphysical jet lines (scalar-polarized gauge bosons) from it. Then we select the smallest extended pinch surface from this set, namely the one having the largest soft subgraph (smallest jet subgraph). We denote this pinch surface by $\sigma_0^*$, and its corresponding hard (jet, soft) subgraph by $\gamma_{H_0}^{ }$ ($\gamma_{J_0}^{ }$, $\gamma_{S_0}^{ }$). For the special case where $\eta\in \overline{F}$, we have $\sigma_0^*=\eta$, hence $\gamma_{H_0}^{ }=\mathcal{A}^{(n)}$ and $\gamma_{J_0}^{ }= \gamma_{S_0}^{ }= \varnothing$. Otherwise $\sigma_0^*$ is a leading pinch surface of $\mathcal{A}^{(n)}$.

With this construction we can reorganize the sum over forests as:
\begin{eqnarray} \label{graph_forest_formula3}
\mathcal{A}^{(n)}= \sum_{\sigma_0^* [\mathcal{A}^{(n)}]} \sum_{F_>\in \mathcal{F}_>[\sigma_0^*]} \prod_{\sigma_>\in F_>} \left( -t_{\sigma_>} \right) t_{\sigma_0^*} \sum_{F_<\in \mathcal{F}_<[\sigma_0^*]} \prod_{\sigma_<\in F_<} \left( -t_{\sigma_<} \right) \mathcal{A}^{(n)}.
\end{eqnarray}
In this expression, $\sigma_>$ are pinch surfaces with the same hard subgraphs $\gamma_{H_0}^{ }$, but with smaller soft subgraphs. In comparison, the pinch surfaces denoted as $\sigma_<$ have smaller hard subgraphs than $\gamma_{H_0}^{ }$, and are contained in $\sigma_0^*$. Note that the overall minus sign of the first term in eq.\ (\ref{graph_forest_formula1}) has cancelled the minus sign in $(-t_{\sigma_0^*})$.

This reorganized sum enables us to arrive at a preliminary factorized form of $\mathcal{M}^{(n)}$ after we sum over all the $\mathcal{A}^{(n)}$'s. That is,
\begin{eqnarray} \label{amplitude_forest_formula1}
\mathcal{M}^{(n)}= \sum_{\mathcal{A}^{(n)}} \mathcal{A}^{(n)}= \sum_{i=0}^n \sum_{\gamma_{S_0,\text{eik}}^{(i)}} \sum_{F_>\in \mathcal{F}_> \left[ \gamma_{S_0,\text{eik}}^{(i)} \right]} \prod_{\sigma_>\in F_>} (-t_{\sigma_>}) \gamma_{S_0,\text{eik}}^{(i)} \cdot \Gamma_{H_0J_0}^{(n-i)},
\end{eqnarray}
where the $t_{\sigma_>}$'s only act on $\gamma_{S_0,\text{eik}}^{(i)}$ because the $\sigma_>$'s have the same reduced hard subgraph as that of $\sigma_0^*$, and soft subgraphs contained in $\gamma_{S_0}^{ }$. The hard-and-jet function
\begin{eqnarray} \label{hard&jet_forest_formula1}
\Gamma_{H_0J_0}^{(n-i)} \equiv \sum_{\gamma_{H_0J_0}^{(n-i)}} \sum_{\sigma_0^*} t_{\sigma_0^*} \sum_{F_{HJ}\in \mathcal{F}[\gamma_{H_0J_0}^{(n-i)}]} \prod_{\sigma_{HJ}\in F_{HJ}} \left( -t_{\sigma_{HJ}} \right) \gamma_{H_0J_0}^{(n-i)},
\end{eqnarray}
which is a function of $p_1^\mu,...,p_N^\mu$.

We describe the idea to arrive at eqs.\ (\ref{amplitude_forest_formula1}) and (\ref{hard&jet_forest_formula1}) before interpreting the details of the notation. We write the full $\mathcal{M}^{(n)}$ as the sum over all $\mathcal{A}^{(n)}$'s. To perform this sum, we first group the terms with identical $\gamma_{H_0}^{ }$, $\gamma_{J_0}^{ }$ and the approximations inside these subgraphs, but with different $\gamma_{S_0}^{ }$'s. In the sum over all the elements in each group, the factor involving the loop momenta of $\gamma_{S_0}^{ }$ can be reformed into a multi-eikonal graph $\gamma_{S_0,\text{eik}}^{ }$ due to the soft-collinear approximation from $t_{\sigma_0^*}$ and the Ward identity. In the obtained $\gamma_{S_0,\text{eik}}^{ }$, the external lines of the $\gamma_{S_0}^{ }$'s are attached to eikonal lines in the directions of $\beta_A^\mu$'s. In the spirit of the factorization theorem \cite{ClsSprStm04, Cls11book}, this graph is decoupled from the rest part of $\mathcal{A}^{(n)}$. Then we sum over the results from different groups of $\gamma_{H_0}^{ }$ and $\gamma_{J_0}^{ }$, from which all the possible hard-and-jet subgraphs are automatically included. Suppose the soft part is $\mathcal{O}(\alpha^i)$, then the sum over $\mathcal{A}^{(n)}$ can be rewritten as the three-fold sum over $i$ (from $0$ to $n$), the $\mathcal{O}(\alpha^i)$ soft subgraphs, and the $\mathcal{O}(\alpha^{n-i})$ hard-and-jet subgraphs. In this way, the sum over $\sigma_0^*\left[ \mathcal{A}^{(n)} \right]$ only remains in the hard-and-jet part. By definition, the approximations from the forests $F_>$ only act inside the soft part, while those from $F_<$ and the hard-collinear branch of $t_{\sigma_0^*}$ only act inside the hard-and-jet part. After this we arrive at the RHS of (\ref{amplitude_forest_formula1}) and (\ref{hard&jet_forest_formula1}).

With the idea explained, the notations in eqs.\ (\ref{amplitude_forest_formula1}) and (\ref{hard&jet_forest_formula1}) are natural. The symbol $\gamma_{S_0,\text{eik}}^{ }$ denotes the graph $\gamma_{S_0}^{ }$ with its external gauge boson lines attached to eikonal lines in the directions of $\beta_1,...,\beta_N$, as we have explained in the paragraph above. In comparison, $\gamma_{H_0J_0}^{ }$ denotes the remaining part of the Feynman graph, which is the union of $\gamma_{H_0}^{ }$ and $\gamma_{J_0}^{ }$, whose corresponding set of forests is denoted by $\mathcal{F}[\gamma_{H_0J_0}]$. The forests $F_{HJ}$ are sets of nested pinch surfaces $\sigma_{HJ}$, which determine approximation operators acting on $\gamma_{H_0J_0}^{ }$. Besides these, there are also hard-collinear approximations from $t_{\sigma_0^*}$ acting on $\gamma_{H_0J_0}^{ }$. For the special case where $\sigma_0^*= \eta$, we have $i=0$, so the factor involving $\gamma_{S_0,\text{eik}}^{ }$ is simply $1$, and $\gamma_{H_0J_0}^{(n-i)}= \mathcal{A}^{(n)}$.

The full amplitude $\mathcal{M}=\sum_{n=0}^\infty \mathcal{M}^{(n)}$ then reads:
\begin{eqnarray} \label{amplitude_forest_formula2}
\mathcal{M}\left( p_1,...,p_N \right)= \left( \sum_{i=0}^\infty \sum_{\gamma_{S_0,\text{eik}}^{(i)}} \sum_{F_>\in \mathcal{F}_> \left[ \gamma_{S_0,\text{eik}}^{(i)} \right]} \prod_{\sigma_>\in F_>} (-t_{\sigma_>}) \gamma_{S_0,\text{eik}}^{(i)} \right)\cdot \left( \sum_{j=0}^\infty \Gamma_{H_0J_0}^{(j)} \right).
\end{eqnarray}
In the paragraphs below, we shall separately study the two factors in the brackets.

\begin{figure}[t]
\centering
\includegraphics[width=13cm]{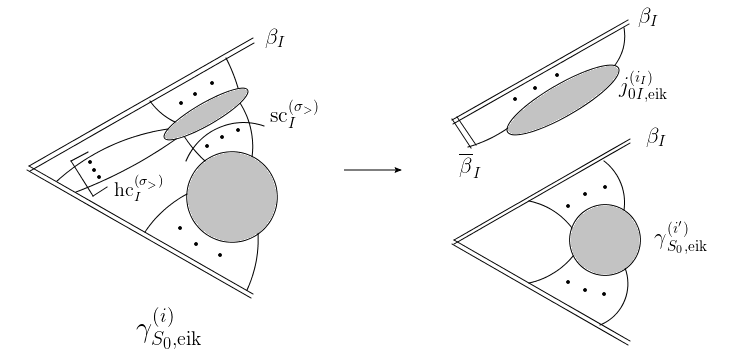}
\caption{A pictorial illustration of ${\gamma}_{S_0,\text{eik}}^{(i')}\left( \beta_1,...,\beta_N \right)$ and $j_{0I,\text{eik}}^{(i_I)}\left( \beta_I,\overline{\beta}_I \right)$ for any given direction $\beta_I^\mu$. The subgraphs ${\gamma}_{S_0,\text{eik}}^{(i')}$ and $j_{0I,\text{eik}}^{(i_I)}$ are disjoint due to the factorization from the approximations $\text{hc}_I^{(\sigma_>)}$ and $\text{sc}_I^{(\sigma_>)}$ inside $\gamma_{S_0,\text{eik}}^{(i)}$.}
\label{sc_factorize_soft}
\end{figure}

\bigbreak
\centerline{\textbf{The soft factor}}
First we study the factor involving $\gamma_{S_0,\text{eik}}^{(i)}$, which we call the ``soft factor''. As is shown in eq.\ (\ref{amplitude_forest_formula2}), all the approximations acting on $\gamma_{S_0,\text{eik}}^{(i)}$ are from $t_{\sigma_>}$. Moreover, they are of two types: (a) some of them are hard-collinear, because some soft lines in $\sigma_0^*$ can become lightlike and attached to $\gamma_{H_0}^{ }$ in $\sigma_>$; (b) some of them are soft-collinear, because some soft lines in $\sigma_0^*$ can become lightlike, which the other soft lines can be attached to in $\sigma_>$. The two types are depicted on the LHS of figure\ \ref{sc_factorize_soft}. After we perform the sum over all the $\gamma_{S_0}^{(i)}$'s and the approximations acting on them, $\gamma_{S_0}^{(i)}$ will be further factorized into disjoint cusps. They are depicted on the RHS of figure\ \ref{sc_factorize_soft}. Here we have denoted its subgraph that is from the soft part of $\sigma_>$ as ${\gamma}_{S_0,\text{eik}}^{(i')}$, and that from the $A$-th jet as $j_{0A,\text{eik}}^{(i_A)}$. The superscripts $i'$ and $i_A$'s are the orders of the graphs; with special cases, ${\gamma}_{S_0,\text{eik}}^{(0)}= 1= j_{0A,\text{eik}}^{(0)}$.

In more detail, according to the Ward identity, we have
\begin{eqnarray} \label{sum_graphs_factorization_soft&jet}
t_{\sigma_>} \sum_{\gamma_{S_0,\text{eik}}^{(i)}} \gamma_{S_0,\text{eik}}^{(i)} \left( \beta_1,...,\beta_N \right)= && \prod_{A=1}^{N} \sum_{i_A=0}^\infty \sum_{j_{0A}^{(i_A)}} j_{0A,\text{eik}}^{(i_A)}\left( \beta_A,\overline{\beta}_A \right) \nonumber\\
&&\times \sum_{i'=0}^\infty \sum_{{\gamma}_{S_0}^{(i')}} {\gamma}_{S_0,\text{eik}}^{(i')}\left( \beta_1,...,\beta_N \right)\cdot \delta_{\sum_{A=1}^N i_A,i-i'} \nonumber\\
= && \sum_{i'=1}^i \mathcal{J}_{0,\text{eik}}^{(i-i')}\left( \left\{\beta_A,\overline{\beta}_A\right\}_{A=1}^N \right) \cdot \sum_{{\gamma}_{S_0}^{(i')}} {\gamma}_{S_0,\text{eik}}^{(i')}\left( \beta_1,...,\beta_N \right), \nonumber\\
&&
\end{eqnarray}
where the Kronecker delta factor controls the orders on both sides to be $\mathcal{O} \left( \alpha^i \right)$. The subgraphs ${\gamma}_{S_0}^{ }$ and $j_{0A}$ are attached to eikonal lines so we have added ``eik'' to their subscripts. For clarity, we have defined the union of $j_{0A,\text{eik}}^{ }$ ($A=1,...,N$) as $\mathcal{J}_{0,\text{eik}}^{ }$, i.e.
\begin{eqnarray} \label{script_J_eik_definition}
\mathcal{J}_{0,\text{eik}}^{(j)}\left( \left\{\beta_A,\overline{\beta}_A\right\}_{A=1}^N \right)\equiv \prod_{A=1}^{N} \sum_{i_A=0}^\infty \sum_{j_{0A}^{(i_A)}} j_{0A,\text{eik}}^{(i_A)}\left( \beta_A,\overline{\beta}_A \right)\cdot \delta_{j,\sum_{A=1}^N i_A}.
\end{eqnarray}
From our construction, it is apparent that $\mathcal{J}_{0,\text{eik}}^{(0)}=1$ and $\mathcal{J}_{0,\text{eik}}^{(1)}= \sum_{A=1}^N \sum_{j_A^{(1)}} j_{A,\text{eik}}^{(1)}$. Each operator $t_{\sigma_>}$ in (\ref{amplitude_forest_formula2}) leads to a factor $\mathcal{J}_{0,\text{eik}}^{(j)}$ with $j>0$. (Note that $j\neq 0$ because different $t_{\sigma_>}$'s provide different soft-collinear approximations.) We also define
\begin{eqnarray} \label{order_sum_abbreviations}
\mathcal{J}_{0,\text{eik}}\equiv \sum_{i=0}^\infty \mathcal{J}_{0,\text{eik}}^{(i)}, \hspace{1cm} \gamma_{S_0,\text{eik}}^{ }\equiv \sum_{i=0}^\infty \sum_{\gamma_{S_0,\text{eik}}^{(i)}} \gamma_{S_0,\text{eik}}^{(i)},
\end{eqnarray}
for convenience.

Combining eqs.\ (\ref{sum_graphs_factorization_soft&jet})--(\ref{order_sum_abbreviations}) together, we rewrite the soft factor of (\ref{amplitude_forest_formula2}) as
\begin{eqnarray} \label{soft_forest_formula1}
&&\sum_{i=0}^\infty \sum_{\gamma_{S_0,\text{eik}}^{(i)}} \sum_{F_>\in \mathcal{F}_> \left[ \gamma_{S_0,\text{eik}}^{(i)} \right]} \prod_{\sigma_>\in F_>} (-t_{\sigma_>}) \gamma_{S_0,\text{eik}}^{(i)}\nonumber\\
&&\hspace{2cm} = \left[ 1+\sum_{n_{\text{sc}}=1}^\infty \left( \sum_{j=1}^\infty \left( -\mathcal{J}_{0,\text{eik}}^{(j)} \right) \right)^{n_{\text{sc}}} \right]\cdot \gamma_{S_0,\text{eik}}^{ },
\end{eqnarray}
where $n_{\text{sc}}$ is the number of different soft-collinear approximations that act inside $\gamma_{S_0,\text{eik}}^{(i)}$. The factor in the bracket can be further simplified, i.e.
\begin{eqnarray} \label{soft_factor_computation}
1+\sum_{n_{\text{sc}}=1}^\infty \left( \sum_{j=1}^\infty \left( -\mathcal{J}_{0,\text{eik}}^{(j)} \right) \right)^{n_{\text{sc}}}= 1+\sum_{n_{\text{sc}}=1}^\infty \left( 1-\mathcal{J}_{0,\text{eik}} \right)^{n_{\text{sc}}}= \frac{1}{\mathcal{J}_{0,\text{eik}}}.
\end{eqnarray}
Therefore, the factor involving $\gamma_{S_0,\text{eik}}^{(i)}$ in eq.\ (\ref{amplitude_forest_formula2}) is
\begin{eqnarray} \label{soft_forest_formula2}
\sum_{i=0}^\infty \sum_{\gamma_{S_0,\text{eik}}^{(i)}} \sum_{F_>\in \mathcal{F}_> \left[ \gamma_{S_0,\text{eik}}^{(i)} \right]} \prod_{\sigma_>\in F_>} (-t_{\sigma_>}) \gamma_{S_0,\text{eik}}^{(i)}= \frac{\gamma_{S_0,\text{eik}}^{ }\left( \beta_1,...,\beta_N \right) }{\mathcal{J}_{0,\text{eik}} \left( \left\{ \beta_A,\overline{\beta}_A \right\}_{A=1}^N \right)}.
\end{eqnarray}

\bigbreak
\centerline{\textbf{The hard-and-jet factor}}
Now we focus on the hard-and-jet factor $\sum_{j=0}^\infty \Gamma_{H_0J_0}^{(j)}$, whose expression is in eq.\ (\ref{hard&jet_forest_formula1}). In light of our discussions in section\ \ref{factorization_presence_approximations}, we perform the sum over $\gamma_{H_0J_0}^{(n-i)}$ to factorize the hard part from the jets in $\sigma_0^*$, according to (\ref{factorization_formula_hard&jet_approximations}). The result is,
\begin{eqnarray} \label{sum_graphs_factorization_hard&jet}
&&\Gamma_{H_0J_0}\left( p_1,...,p_N \right)= \sum_{i=0}^\infty \Gamma_{H_0J_0}^{(i)} \left( p_1,...,p_N \right)\nonumber\\
&&\hspace{0.5cm}= \sum_{h=0}^\infty \sum_{\overline{\gamma}_{H_0}^{(h)}} \sum_{j=0}^\infty \sum_{\overline{\gamma}_{J_0}^{(j)}} \sum_{F_{HJ}\in \mathcal{F}[\gamma_{H_0J_0}^{(h+j)}]} \prod_{\sigma_{HJ}\in F_{HJ}} \left( -t_{\sigma_{HJ}} \right) \left( \overline{\gamma}_{H_0}^{(h)} \cdot\overline{\gamma}_{J_0}^{(j)} \right).
\end{eqnarray}
In this factorized expression, $\overline{\gamma}_{H_0}^{(h)}= \overline{\gamma}_{H_0}^{(h)}\left( p_1,...,p_N \right)$ is the reduced hard subgraph, which we introduced at the beginning of this subsection. Similarly, $\overline{\gamma}_{J_0}^{(j)} =\overline{\gamma}_{J_0}^{(j)}\left( \left\{ p_A, \overline{\beta}_A \right\}_{A=1}^N \right)$ is obtained from $\gamma_{J_0}^{(j)}$ by attaching its scalar-polarized gauge bosons, which are decoupled from $\overline{\gamma}_{H_0}^{(h)}$, to eikonal lines that are in the directions of $\overline{\beta}_A^\mu$ ($A=1,...,N$). Note that in eq.\ (\ref{sum_graphs_factorization_hard&jet}), $\gamma_{H_0J_0}^{(h+j)}$ appears as $\overline{\gamma}_{H_0}^{(h)} \cdot\overline{\gamma}_{J_0}^{(j)}$ due to the hard-collinear approximations in $t_{\sigma_0^*}$. For the special case where $\sigma_0^*=\eta$, we have $j=0$, and $\gamma_{H_0J_0}^{(h+j)}= \gamma_{H_0}^{(h)}= \overline{\gamma}_{H_0}^{(h)}$.

In general, the approximation operators $t_{\sigma_{HJ}}$ act on both $\overline{\gamma}_{H_0}^{(h)}$ and $\overline{\gamma}_{J_0}^{(j)}$, which prevents us from factorizing eq.\ (\ref{sum_graphs_factorization_hard&jet}) immediately. But from our construction, the hard subgraph of any $\sigma_{HJ}^{ }$ is included in $\overline{\gamma}_{H_0}^{ }$, so all the approximations from the operators $\left( -t_{\sigma_{HJ}} \right)$ that are inside $\overline{\gamma}_{J_0}$, can only be soft-collinear. This enables us to extract the factors generated by these approximations from the remaining part of the $\overline{\gamma}_{J_0}^{ }$ function, which is similar to what we have done for the soft factor above.

\begin{figure}[t]
\centering
\includegraphics[width=13cm]{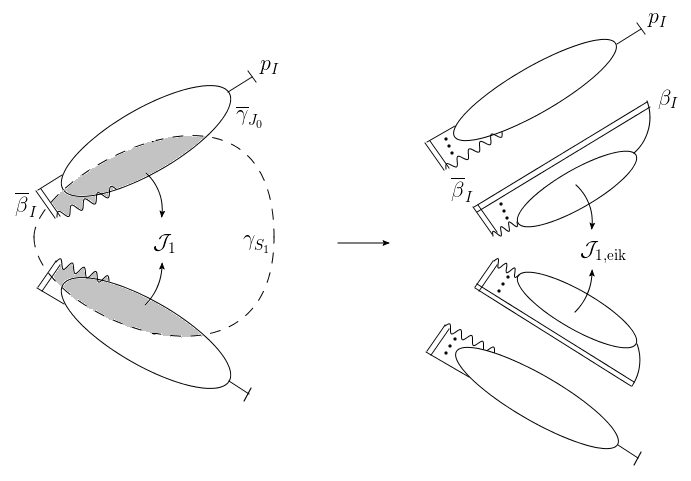}
\caption{The pictorial representation of eq.\ (\ref{hard&jet_forest_formula2}). The LHS describes the subgraphs $\overline{\gamma}_{J_0}^{ }$, $\gamma_{S_1}^{ }$ (enclosed by dashed curves) and $\mathcal{J}_1$ (shaded area). Note that $\gamma_{S_1}^{ }$ may have a nonzero intersection with $\overline{\gamma}_{H_0}^{ }$ as well, which is not drawn in the figure. The RHS describes the factorized expression. Due to the soft-collinear approximations $\text{sc}^{(\sigma_1^*)}$, $\mathcal{J}_1$ can be rewritten as $\mathcal{J}_{1,\text{eik}}$, which is decoupled from the rest of $\overline{\gamma}_{J_0}^{ }$.}
\label{sc_factorize_jets}
\end{figure}

To do this, we use the same idea to obtain eq.\ (\ref{graph_forest_formula3}): given a forest $F_{HJ}$ in (\ref{sum_graphs_factorization_hard&jet}), one identifies all the pinch surfaces of $\gamma_{H_0J_0}^{ }$ that have the largest reduced hard subgraph, and then selects the smallest pinch surface out of the set, which has the largest soft subgraph. We denote this pinch surface as $\sigma_1^*$, and the corresponding hard (jet, soft) subgraph as $\gamma_{H_1}^{ }$ ($\gamma_{J_1}^{ }$, $\gamma_{S_1}^{ }$). The approximation operator $t_{\sigma_1^*}$, by construction, acts on $\overline{\gamma}_{J_0}^{ }$ by soft-collinear approximations only. Again using the Ward identity, we factorize the subgraph $\mathcal{J}_1 \equiv \gamma_{S_1}^{ }\bigcap \overline{\gamma}_{J_0}^{ }$ from the rest of $\gamma_{H_0J_0}^{ }$, which can then be rewritten as:
\begin{eqnarray} \label{hard&jet_forest_formula2}
\Gamma_{H_0J_0}^{(n)} =&& \sum_{\gamma_{H_0J_0}^{(n)}} \sum_{\sigma_1^* [\gamma_{H_0J_0}^{(n)}]} \sum_{F_>\in \mathcal{F}_>[\sigma_1^*]} \prod_{\sigma_>\in F_>} \left( -t_{\sigma_>} \right) \left( -t_{\sigma_1^*} \right) \sum_{F_<\in \mathcal{F}_<[\sigma_1^*]} \prod_{\sigma_<\in F_<} \left( -t_{\sigma_<} \right) \gamma_{H_0J_0}^{(n)} \nonumber\\
=&& \sum_{i=0}^n \sum_{\mathcal{J}_{1,\text{eik}}^{(i)}} \sum_{F_>\in \mathcal{F}_> \left[ \mathcal{J}_{1,\text{eik}}^{(i)} \right]} \prod_{\sigma_>\in F_>} (-t_{\sigma_>}) \mathcal{J}_{1,\text{eik}}^{\ (i)} \cdot\Gamma_{H_0J_1}^{(n-i)}.
\end{eqnarray}
This relation resembles (\ref{amplitude_forest_formula1}), and from construction the jet factor $\mathcal{J}_{1,\text{eik}}^{ } = \mathcal{J}_{0,\text{eik}}^{ }$ in (\ref{script_J_eik_definition}). Here $\sigma_>$ denotes the pinch surfaces that have the same hard subgraph as $\sigma_1^*$ but smaller soft subgraphs, and $\mathcal{J}_{1,\text{eik}}\equiv \mathcal{J}_{1,\text{eik}}\left( \left\{ \beta_A,\overline{\beta}_A \right\}_{A=1}^N \right)$ denotes the graph of $\mathcal{J}_1$ with eikonal lines in the directions of $\beta_A^\mu$'s and $\overline{\beta}_A^\mu$'s as its framework, which is shown in figure\ \ref{sc_factorize_jets}.

Finally, the new ``hard-and-jet'' part $\Gamma_{H_0J_1}$ reads
\begin{eqnarray} \label{new_hard&jet_forest_formula1}
\Gamma_{H_0J_1}^{(n-i)}\equiv \sum_{\gamma_{H_0J_1}^{(n-i)}} \sum_{\sigma_{HJ}^*} \left( -t_{\sigma_{HJ}^*} \right) \sum_{F_{HJ}\in \mathcal{F}[\gamma_{H_0J_1}^{(n-i)}]} \prod_{\sigma_{HJ}\in F_{HJ}} \left( -t_{\sigma_{HJ}} \right) \gamma_{H_0J_1}^{(n-i)},
\end{eqnarray}
where the graph of $\gamma_{H_0J_1}^{ }$ is a combination of $\gamma_{H_0}^{ }$ and $\gamma_{J_1}^{ }$. Note that part of $\gamma_{S_1}^{ }$ may be also in $\gamma_{H_0J_1}^{ }$, which is implicit in figure\ \ref{sc_factorize_jets}. After summing over all the orders of $\Gamma_{H_0J_0}^{(n)}$, we obtain
\begin{eqnarray} \label{hard&jet_forest_formula3}
\Gamma_{H_0J_0}\equiv \sum_{n=0}^\infty \Gamma_{H_0J_0}^{(n)} =\left( \sum_{i=0}^\infty \sum_{\mathcal{J}_{1,\text{eik}}^{(i)}} \sum_{F_>\in \mathcal{F}_> \left[ \mathcal{J}_{1,\text{eik}}^{(i)} \right]} \prod_{\sigma_>\in F_>} (-t_{\sigma_>}) \mathcal{J}_{1,\text{eik}}^{\ (i)} \right) \cdot\left( \sum_{j=0}^\infty \Gamma_{H_0J_1}^{(j)} \right).
\end{eqnarray}
In this relation, all the approximations acting inside $\mathcal{J}_{1,\text{eik}}$ are from $t_{\sigma_>}$, which are soft-collinear. Then we can carry out the same steps from eq.\ (\ref{sum_graphs_factorization_soft&jet}) to (\ref{soft_forest_formula2}), and obtain an analog of (\ref{soft_forest_formula2}), where the eikonal jet factor $\mathcal{J}_{1,\text{eik}}^{ }$ is in place of $\gamma_{S_0,\text{eik}}^{ }$, giving
\begin{eqnarray} \label{soft_forest_formula3}
\sum_{i=0}^\infty \sum_{\mathcal{J}_{1,\text{eik}}^{(i)}} \sum_{F_>\in \mathcal{F}_> \left[ \mathcal{J}_{1,\text{eik}}^{(i)} \right]} \prod_{\sigma_>\in F_>} (-t_{\sigma_>}) \mathcal{J}_{1,\text{eik}}^{\ (i)}= \frac{\mathcal{J}_{1,\text{eik}} }{\mathcal{J}_{1,\text{eik}}} =1.
\end{eqnarray}
Then in eq.\ (\ref{new_hard&jet_forest_formula1}), this identity implies that $\Gamma_{H_0J_0}= \Gamma_{H_0J_1}$.

We now rephrase the calculations from eq.\ (\ref{hard&jet_forest_formula2}) to (\ref{soft_forest_formula3}) in an iterative way. That is, in the subgraph $\gamma_{H_0J_i}^{ }$ ($i=0,1,...$) we focus on a subset of all its pinch surfaces that have the largest hard subgraph, select a specific one with the largest soft subgraph, and denote this pinch surface as $\sigma_{i+1}^*$ and the corresponding hard (jet, soft) subgraph as $\gamma_{H_{i+1}}^{ }$ ($\gamma_{J_{i+1}}^{ }$, $\gamma_{S_{i+1}}^{ }$). According to the soft-collinear approximations from $t_{\sigma_{i+1}^*}$, the subgraph $\mathcal{J}_{i+1}\equiv \gamma_{S_{i+1}}^{ }\bigcap \overline{\gamma}_{J_0}^{ }$ is factorized from the remaining part $\gamma_{H_0J_{i+1}}^{ }\equiv \gamma_{H_0}^{ }\bigcup \gamma_{J_{i+1}}^{ }$ to form $\mathcal{J}_{i+1,\text{eik}}^{ }$. By construction, $\mathcal{J}_{i+1,\text{eik}}^{ }= \mathcal{J}_{i,\text{eik}}^{ }= ...= \mathcal{J}_{0,\text{eik}}^{ }$, and due to the soft-collinear approximations inside $\mathcal{J}_{i+1,\text{eik}}$, this subgraph contributes $1$ (as a multiplicative factor) after we sum over all the possible graphs and subtraction terms. In other words, we have $\Gamma_{H_0J_i}= \Gamma_{H_0J_{i+1}}$, so
\begin{eqnarray} \label{hard&jet_recursion}
\Gamma_{H_0J_0}= \Gamma_{H_0J_1}=...= \Gamma_{H_0J_{i+1}}.
\end{eqnarray}
In eqs.\ (\ref{hard&jet_forest_formula2})--(\ref{soft_forest_formula3}) we have $i=0$, and the procedure described above works for $i=1,2,...$, as long as there are soft-collinear approximations inside $\gamma_{J_{i+1}}^{ }$. Suppose we have carried out this calculation iteratively, until at a special value of $i$ where no soft-collinear approximations exist inside $\gamma_{J_{i+1}}^{ }$.

Now we can carry out the same procedure to obtain eq.\ (\ref{sum_graphs_factorization_hard&jet}) by applying (\ref{factorization_formula_hard&jet_approximations}), and rewrite $\Gamma_{H_0J_{i+1}}$ into a factorized form:
\begin{eqnarray} \label{sum_graphs_factorization_hard&jet_recursion}
\Gamma_{H_0J_{i+1}}&&= \sum_{h=0}^\infty \sum_{\gamma_{H_0}^{(h)}} \sum_{j=0}^\infty \sum_{\gamma_{J_{i+1}}^{(j)}} \sum_{F_{HJ}\in \mathcal{F}[\gamma_{H_0J_{i+1}}^{(h+j)}]} \prod_{\sigma_{HJ}\in F_{HJ}} \left( -t_{\sigma_{HJ}} \right) \left( \overline{\gamma}_{J_{i+1}}^{(j)} \cdot \overline{\gamma}_{H_0}^{(h)} \right) \nonumber\\
&&= \left( \sum_{j=0}^\infty \sum_{\gamma_{J_{i+1}}^{(j)}} \overline{\gamma}_{J_{i+1}}^{(j)} \right)\cdot \left( \sum_{h=0}^\infty \sum_{\gamma_{H_0}^{(h)}} \sum_{F_H\in \mathcal{F}[\gamma_{H_0}^{(h)}]} \prod_{\sigma_H\in F_H} \left( -t_{\sigma_H} \right) \overline{\gamma}_{H_0}^{(h)} \right).
\end{eqnarray}
In the first line, $\overline{\gamma}_{J_{i+1}}$ is obtained by attaching the scalar-polarized gauge bosons of the subgraph $\left( \gamma_{J_{i+1}}^{ }\setminus \big( \gamma_{H_0}^{ }\cap \gamma_{J_{i+1}}^{ }\big) \right)$ to eikonal lines, which are in the directions of $\overline{\beta}_A^\mu$ $(A=1,...,N)$. We come to the second equality because the operators $t_{\sigma_{HJ}}$ here do not act as soft-collinear approximations on $\overline{\gamma}_{J_{i+1}}^{(j)}$ anymore; also there are no hard-collinear approximations in $\overline{\gamma}_{J_{i+1}}^{(j)}$. Since all the approximations now act on $\gamma_{H_0}^{(h)}$, it is equivalent to regard the forests as the sets of pinch surfaces of $\gamma_{H_0}^{(h)}$, and denote them as $F_H$. We extract the factor involving $\overline{\gamma}_{J_{i+1}}^{(j)}$, and define it as
\begin{eqnarray} \label{script_J_part_definition}
\mathcal{J}_{\text{part}} \left( \left\{ p_A, \overline{\beta}_A \right\}_{A=1}^N \right) \equiv \sum_{j=0}^\infty \sum_{\overline{\gamma}_{J_{i+1}}^{(j)}} \overline{\gamma}_{J_{i+1}}^{(j)}.
\end{eqnarray}
The subscript ``part'' is an abbreviation of ``partonic'', because each function $\mathcal{J}_{\text{part}}$ is indeed a sum over partonic correlation functions. 

Finally, we insert eqs.\ (\ref{soft_forest_formula2}), (\ref{hard&jet_recursion}) and (\ref{sum_graphs_factorization_hard&jet_recursion}) into (\ref{amplitude_forest_formula2}), drop the subscript $0$ and obtain the gauge-invariant factorized form \cite{Sen1983, Cls11book, FgeSwtz14}:
\begin{eqnarray} \label{full_amplitude_factorization_result}
\mathcal{M}= &&\frac{\gamma_{S,\text{eik}}^{ }\left( \beta_1,...,\beta_N \right) }{\mathcal{J}_{\text{eik}} \left( \left\{ \beta_A,\overline{\beta}_A \right\}_{A=1}^N \right)}\cdot \mathcal{J}_{\text{part}} \left( \left\{ p_A,\overline{\beta}_A \right\}_{A=1}^N \right)\nonumber\\
&&\times \sum_{h=0}^\infty \sum_{\gamma_{H}^{(h)}} \sum_{F_{H}\in \mathcal{F}[\gamma_{H}^{(h)}]} \prod_{\sigma_{H}\in F_{H}} \left( -t_{\sigma_{H}} \right) \gamma_{H}^{(h)} \left( p_1,...,p_N \right).
\end{eqnarray}
From eqs.\ (\ref{sum_graphs_factorization_soft&jet})--(\ref{full_amplitude_factorization_result}), we see that all the IR divergences that appear in tables\ \ref{single_approximated_PS_summary} and \ref{repetitive_approximated_PS_summary} organize themselves to become the divergences along eikonal lines, which are in the directions of $\beta_A^\mu$ and $\overline{\beta}_A^\mu$. The divergences that do not exist in any original amplitude $\mathcal{A}^{(n)}$, are also cancelled between the numerator and the denominator. To see this, we further rewrite (\ref{full_amplitude_factorization_result}) as follows:
\begin{eqnarray} \label{full_amplitude_factorization_result_rewrite1}
\mathcal{M}= &&\frac{\gamma_{S,\text{eik}}^{ }\left( \beta_1,...,\beta_N \right) }{\mathcal{J}_{\text{eik}}^{1/2} \left( \left\{ \beta_A,\overline{\beta}_A \right\}_{A=1}^N \right)}\cdot \frac{\mathcal{J}_{\text{part}} \left( \left\{ p_A,\overline{\beta}_A \right\}_{A=1}^N \right)}{\mathcal{J}_{\text{eik}}^{1/2} \left( \left\{ \beta_A,\overline{\beta}_A \right\}_{A=1}^N \right)}\nonumber\\
&&\times \sum_{h=0}^\infty \sum_{\gamma_{H}^{(h)}} \sum_{F_{H}\in \mathcal{F}[\gamma_{H}^{(h)}]} \prod_{\sigma_{H}\in F_{H}} \left( -t_{\sigma_{H}} \right) \gamma_{H}^{(h)} \left( p_1,...,p_N \right).
\end{eqnarray}
In this form, both the $\beta_A$-direction collinear divergences in the first term and the $\overline{\beta}_A$-divergences in the second term are cancelled by the two factors $\mathcal{J}_{\text{eik}}^{1/2}$ in the denominator. This cancellation follows from the exponentiation of IR divergences for the cusp and jet functions \cite{Gtr83, FrkTlr84, DxnMgnStm08, Wht16, FcnGrdMly19}. Therefore, the first factor has only soft divergences, the second factor has only collinear divergences, and the third factor is IR finite. The renormalization of each of these functions has also been studied widely \cite{EdgStm15, BrNrSt81, KdnkOddStm98}.

We comment on another form of eq.\ (\ref{full_amplitude_factorization_result_rewrite1}). Collins in eqs.\ (10.118) and (10.119) of his book \cite{Cls11book}, derived a special factorization formula of the Sudakov form factor, in which the jet functions are normalized to absorb the soft contributions. Now we rewrite (\ref{full_amplitude_factorization_result_rewrite1}) by applying the same normalization as Collins suggests, and extend his result to arbitrary number of external lines. That is,
\begin{eqnarray} \label{full_amplitude_factorization_result_rewrite2}
\mathcal{M}= &&\frac{\gamma_{S,\text{eik}}^{ }\left( \beta_1,...,\beta_N \right) }{\left [\frac{\mathcal{J}_{\text{eik}} \left( \left\{ \beta_A,\overline{\beta}_A \right\}_{A=1}^N \right)\mathcal{J}_{\text{eik}} \left( \left\{ n_A,\beta_A \right\}_{A=1}^N \right)}{\mathcal{J}_{\text{eik}} \left( \left\{ n_A,\overline{\beta}_A \right\}_{A=1}^N \right)} \right ]^{1/2}} \cdot \mathcal{J}' \left( \left\{ \beta_A,\overline{\beta}_A \right\}_{A=1}^N \right) \nonumber\\
&&\times \sum_{h=0}^\infty \sum_{\gamma_{H}^{(h)}} \sum_{F_{H}\in \mathcal{F}[\gamma_{H}^{(h)}]} \prod_{\sigma_{H}\in F_{H}} \left( -t_{\sigma_{H}} \right) \gamma_{H}^{(h)} \left( p_1,...,p_N \right),
\end{eqnarray}
where $n_A$ is a spacelike vector for each $A(=1,...,N)$, and $\mathcal{J}'$ is the jet functions with Collins' normalization factor, which reads
\begin{eqnarray} \label{jet_function_Collins_normalization}
\mathcal{J}'= \frac{\mathcal{J}_{\text{part}} \left( \left\{ p_A,\overline{\beta}_A \right\}_{A=1}^N \right)}{\left[ \frac{ \mathcal{J}_{\text{eik}} \left( \left\{ \beta_A,\overline{\beta}_A \right\}_{A=1}^N \right)\mathcal{J}_{\text{eik}} \left( \left\{ n_A,\overline{\beta}_A \right\}_{A=1}^N \right)}{\mathcal{J}_{\text{eik}} \left( \left\{ n_A,\beta_A \right\}_{A=1}^N \right)} \right ]^{1/2}}.
\end{eqnarray}
For the Sudakov form factor where $N=2$, the soft eikonal function $\gamma_{S,\text{eik}}^{ }$ is equal to the square root of each of the $\mathcal{J}_{\text{eik}}$'s in the denominator. So the first factor in eq.\ (\ref{full_amplitude_factorization_result_rewrite2}) is unity, and eqs.\ (\ref{full_amplitude_factorization_result_rewrite2}) and (\ref{jet_function_Collins_normalization}) reduces to the result for the form factor automatically.

\section{Next-to-next-to-leading-order examples}
\label{NNLO_examples}

\begin{figure}[t]
\centering
\includegraphics[width=15cm]{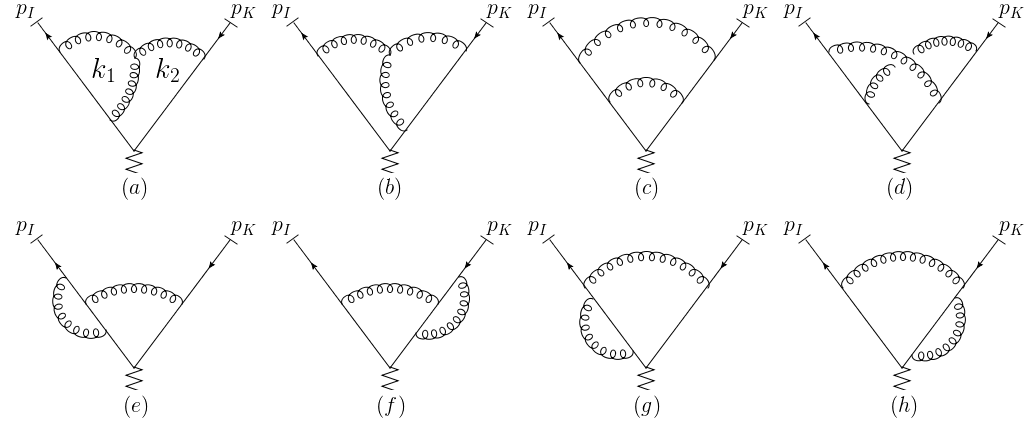}
\caption{The two-loop QCD corrections to $\gamma^*,W^{\pm},Z \rightarrow q\bar{q}$.}
\label{all_2_loop_graphs}
\end{figure}

With the rationales explained in sections\ \ref{pinch_surface_amplitudes_approximations}--\ref{factorization_subtraction_terms}, we shall visualize how they work through in next-to-next-to-leading-order (NNLO) calculations. Namely, we consider the $\gamma^*,W^{\pm},Z \rightarrow q\bar{q}$ processes in QCD at two loops. There are eight Feynman graphs in total, as are shown in figure\ \ref{all_2_loop_graphs}.

This section is arranged as follows. In section\ \ref{2_loop_example_forest_formula} we start by studying figure\ \ref{all_2_loop_graphs}$(a)$, showing its associated pinch surfaces and forests. After evaluating four selected subtraction terms, we analyze all the IR regions appearing in the forest formula, and exhibit how they form pairwise cancellations. In section\ \ref{2_loop_example_factorization}, we evaluate a term that contributes to the final result in section\ \ref{partonic_amplitudes_factorized_forms}, i.e. eq.\ (\ref{full_amplitude_factorization_result}), to see how such a factorized expression is formed by summing over gauge-invariant set of graphs.

\subsection{Infrared regions, forests and IR cancellations}
\label{2_loop_example_forest_formula}

With the final-state momenta being $p_I^\mu$ and $p_K^\mu$, which are not necessarily back-to-back, the expression for figure\ \ref{all_2_loop_graphs}$(a)$ is:
\begin{eqnarray} \label{figure_(a)_expression}
\mathcal{A}^\mu=\int \left [ dk \right ]&&\overline{u}\left( p_I \right) \gamma^\alpha \frac{\slashed{p}_I+\slashed{k}_1}{\left( p_I+k_1 \right)^2} \gamma^\beta \frac{\slashed{p}_I+\slashed{k}_2}{\left( p_I+k_2 \right)^2} \gamma^\mu \frac{\slashed{p}_K-\slashed{k}_2}{\left( p_K-k_2 \right)^2} \gamma^\gamma v\left ( p_K \right ) \nonumber\\
&&\cdot \frac{1}{k_1^2}\frac{1}{k_2^2}\frac{1}{\left( k_2-k_1 \right)^2} V_{\alpha\beta\gamma} \left( k_1, k_2-k_1, -k_2 \right).
\end{eqnarray}
The superscript $\mu$ of $\mathcal{A}$ is from the vector index of the gauge boson ($\gamma^*,W^{\pm},Z$), which we will omit in the upcoming text. On the RHS of eq.\ (\ref{figure_(a)_expression}), $k_1^\mu$ ($k_2^\mu$) is the clockwise momentum of the left (right) loop, as is shown in the figure, $V_{ \alpha \beta \gamma}\left( p, q, k \right)$ is the kinetic factor of the 3-gluon vertex, and $\int \left [ dk \right ]$ is the integration measure with the other constants absorbed. Throughout this subsection, we use:
\begin{equation} \label{measure&vertex_def}
\begin{split}
V_{\alpha\beta\gamma} \left(p,q,r\right) &\equiv g_{\alpha\beta} \left( p-q \right)_\gamma +g_{\beta\gamma} \left( q-r \right)_\alpha +g_{\gamma\alpha} \left( r-p \right)_\beta,\\
\int \left [ dk \right ] &\equiv \frac{ig_\text{W}^{ } V_{\text{CKM}}^{ } \alpha_S^2}{64\pi^6} (N_\text{c}^2-1) \int d^4k_1 d^4k_2,
\end{split}
\end{equation}
where $g_\text{W}$ is the electroweak coupling, $V_{\text{CKM}}^{ }$ is the CKM matrix, and $N_\text{c}$ is the number of colors. Note that the $+i\epsilon$ terms in the denominators are suppressed from now on.

We denote the leading pinch surfaces of $\mathcal{A}$, from eq.\ (\ref{figure_(a)_expression}), as follows:
\begin{eqnarray}
&\sigma_1\text{ (SS),}\ &\text{if }k_1^\mu\text{ and }k_2^\mu\text{ are both soft; }\nonumber\\
&\sigma_2\text{ (C}_1\text{S),}\ &\text{if }k_1^\mu\text{ is collinear to }\beta_I^\mu \text{ and }k_2^\mu\text{ is soft; }\nonumber\\
&\sigma_3\text{ (S}\text{C}_2),\ &\text{if }k_1^\mu\text{ is soft and }k_2^\mu\text{ is collinear to }\beta_K^\mu;\ \nonumber\\
&\sigma_4\text{ (C}_1\text{C}_1),\ &\text{if }k_1^\mu\text{ and }k_2^\mu\text{ are both collinear to }\beta_I^\mu;\ \nonumber\\
&\sigma_5\text{ (C}_2\text{C}_2),\ &\text{if }k_1^\mu\text{ and }k_2^\mu\text{ are both collinear to }\beta_K^\mu;\ \nonumber\\
&\sigma_6\text{ (C}_1\text{C}_2),\ &\text{if }k_1^\mu\text{ is collinear to }\beta_I^\mu \text{ and }k_2^\mu\text{ is collinear to }\beta_K^\mu;\ \nonumber\\
&\sigma_7\text{ (C}_1\text{H),}\ &\text{if }k_1^\mu\text{ is collinear to }\beta_I^\mu \text{ and }k_2^\mu\text{ is hard; }\nonumber\\
&\sigma_8\text{ (H}\text{C}_2),\ &\text{if }k_1^\mu\text{ is hard and }k_2^\mu\text{ is collinear to }\beta_K^\mu.\nonumber
\end{eqnarray}
Taking into account the orderings allowed by the nesting requirements, the set of forests for figure\ \ref{all_2_loop_graphs} is:
\begingroup
\allowdisplaybreaks
\begin{eqnarray} \label{2_loop_forest}
\mathcal{N}=&&\Big\{ \varnothing,\ \left\{ \sigma_i \right\}_{i=1,...,8},\ \left\{ \sigma_1,\sigma_i \right\}_{i=2,..,8},\ \left\{ \sigma_2,\sigma_4 \right\},\ \left\{ \sigma_2,\sigma_6\right\},\ \left\{ \sigma_2,\sigma_7 \right\},\ \left\{ \sigma_2,\sigma_8 \right\},\nonumber\\
&&\left\{ \sigma_3,\sigma_5 \right\},\ \left\{ \sigma_3,\sigma_6 \right\},\ \left\{ \sigma_3,\sigma_7 \right\},\ \left\{ \sigma_3,\sigma_8 \right\},\ \left\{ \sigma_4,\sigma_7 \right\},\ \left\{ \sigma_5,\sigma_8 \right\},\ \left\{ \sigma_6,\sigma_7 \right\},\ \left\{ \sigma_6,\sigma_8 \right\},\nonumber\\
&&\left\{ \sigma_1,\sigma_2,\sigma_4 \right\},\ \left\{ \sigma_1,\sigma_2,\sigma_6 \right\},\ \left\{ \sigma_1,\sigma_2,\sigma_7 \right\},\ \left\{ \sigma_1,\sigma_2,\sigma_8 \right\},\ \left\{ \sigma_1,\sigma_3,\sigma_5 \right\},\ \left\{ \sigma_1,\sigma_3,\sigma_6 \right\},\nonumber\\
&&\left\{ \sigma_1,\sigma_3,\sigma_7 \right\},\ \left\{ \sigma_1,\sigma_3,\sigma_8 \right\},\ \left\{ \sigma_1,\sigma_4,\sigma_7 \right\},\ \left\{ \sigma_1,\sigma_5,\sigma_8 \right\},\ \left\{ \sigma_1,\sigma_6,\sigma_8 \right\},\ \left\{ \sigma_1,\sigma_7,\sigma_8 \right\},\nonumber\\
&&\left\{ \sigma_2,\sigma_4,\sigma_7 \right\},\ \left\{ \sigma_2,\sigma_6,\sigma_7 \right\},\ \left\{ \sigma_2,\sigma_6,\sigma_8 \right\},\ \left\{ \sigma_3,\sigma_5,\sigma_7 \right\},\ \left\{ \sigma_3,\sigma_5,\sigma_8 \right\},\ \left\{ \sigma_3,\sigma_6,\sigma_7 \right\},\nonumber\\
&&\left\{ \sigma_3,\sigma_6,\sigma_8 \right\},\ \left\{ \sigma_1,\sigma_2,\sigma_4,\sigma_7 \right\},\ \left\{ \sigma_1,\sigma_2,\sigma_6,\sigma_7 \right\},\ \left\{ \sigma_1,\sigma_2,\sigma_6,\sigma_8 \right\},\ \left\{ \sigma_1,\sigma_3,\sigma_5,\sigma_8 \right\},\nonumber\\
&&\left\{ \sigma_1,\sigma_3,\sigma_6,\sigma_7 \right\},\ \left\{ \sigma_1,\sigma_3,\sigma_6,\sigma_8 \right\} \Big\}.
\end{eqnarray}
\endgroup

Each forest corresponds to a subtraction term. Now we evaluate some representatives among them: $t_{\sigma_1}\mathcal{A}$, $t_{\sigma_3}\mathcal{A}$, $t_{\sigma_3}t_{\sigma_1}\mathcal{A}$, $t_{\sigma_6}t_{\sigma_3}t_{\sigma_1}\mathcal{A}$ and $t_{\sigma_7}t_{\sigma_6}t_{\sigma_3}t_{\sigma_1}\mathcal{A}$. To begin, we analyze the IR regions of these subtraction terms, and see how they are cancelled pairwise.

\begin{figure}[t]
\centering
\includegraphics[width=15cm]{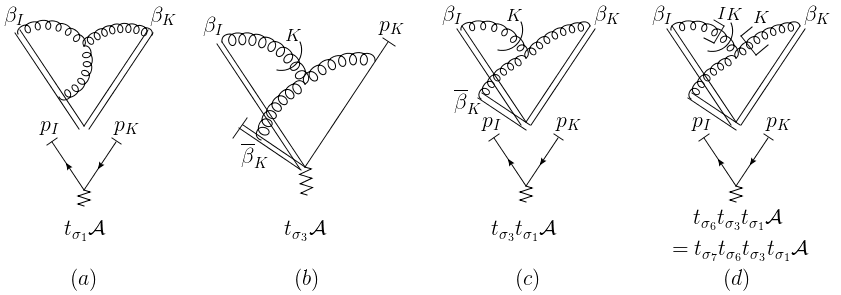}
\caption{Pictorial representations of the selected subtraction terms.}
\label{selected_subtraction_terms}
\end{figure}

We start from $t_{\sigma_1}\mathcal{A}$:
\begin{eqnarray} \label{2_loop_sigma1_amplitude_expression}
t_{\sigma_1}\mathcal{A}=\int \left [ dk \right ]&&\overline{u}\left( p_I \right) \overline{\slashed{\beta}}_I \beta_I^\alpha \frac{\slashed{p}_I+\left( k_1\cdot\beta_I \right) \overline{\slashed{\beta}}_I}{\left( p_I+\left( k_1\cdot\beta_I \right)\overline{\beta}_I \right)^2} \overline{\slashed{\beta}}_I \beta_I^\beta \frac{\slashed{p}_I+\left( k_2\cdot\beta_I \right) \overline{\slashed{\beta}}_I}{\left( p_I+\left( k_2\cdot\beta_I \right)\overline{\beta}_I \right)^2}\nonumber\\
&&\cdot \gamma^\mu \frac{\slashed{p}_K-\left( k_2\cdot\beta_K \right) \overline{\slashed{\beta}}_K}{\left( p_K+\left( k_2\cdot\beta_K \right)\overline{\beta}_K \right)^2} \overline{\slashed{\beta}}_K \beta_K^\gamma v\left ( p_K \right )\nonumber\\
&&\cdot \frac{1}{k_1^2}\frac{1}{k_2^2}\frac{1}{\left( k_2-k_1 \right)^2} V_{\alpha\beta\gamma} \left( k_1, k_2-k_1, -k_2 \right) \nonumber\\
=\int \left [ dk \right ] &&\Big(\overline{u}\left( p_I \right) \gamma^\mu v\left( p_K \right) \Big) \frac{\beta_I^\alpha}{k_1\cdot\beta_I}\frac{\beta_I^\beta}{k_2\cdot\beta_I}\frac{\beta_K^\gamma}{-k_2\cdot\beta_K} \frac{1}{k_1^2} \frac{1}{k_2^2} \frac{1}{\left( k_2-k_1 \right)^2}\nonumber\\
&&\cdot V_{\alpha\beta\gamma} \left( k_1, k_2-k_1, -k_2 \right).
\end{eqnarray}
The factor $\frac{\beta_I^\alpha}{k_1\cdot\beta_I}\frac{\beta_I^\beta}{k_2\cdot\beta_I}\frac{\beta_K^\gamma}{-k_2\cdot\beta_K}$ can be seen as the contribution from eikonal lines. Graphically, eq.\ (\ref{2_loop_sigma1_amplitude_expression}) is equivalent to figure\ \ref{selected_subtraction_terms}$(a)$. From the figure, it is direct to obtain the pinch surfaces of $t_{\sigma_1} \mathcal{A}$. The fermion propagators are replaced here by the eikonal lines, which are separately in the same directions ($\beta_I^\mu$ and $\beta_K^\mu$) as the final states of $\mathcal{A}$. With this the only change, we know that the pinch surfaces of $t_{\sigma_1} \mathcal{A}$ can be labelled as those of $\mathcal{A}$. Namely,
\begin{eqnarray} \label{2_loop_sigma1_pinch_surface}
\text{PS of }t_{\sigma_1}\mathcal{A}:\ \text{SS},\ \text{C}_1\text{S},\ \text{SC}_2,\ \text{C}_1\text{C}_1,\ \text{C}_2\text{C}_2,\ \text{C}_1\text{C}_2,\ \text{C}_1\text{H},\ \text{and }\text{HC}_2.
\end{eqnarray}
The only difference from the pinch surfaces of $\mathcal{A}$ lies in the intrinsic coordinates of $\text{C}_1\text{S}$ and $\text{SC}_2$ above. That is, the $\beta_I$-component of $k_1^\mu$ in $\text{C}_1\text{S}$ and the $\beta_K$-component of $k_2^\mu$ in $\text{SC}_2$ are unbounded due to the soft-collinear approximations in $t_{\sigma_1}$. This follows from our previous discussion in section\ \ref{pinch_surfaces_from_single_approximation} (more precisely, the paragraph below eq.\ (\ref{cii_result})). Figure\ \ref{selected_subtraction_terms} offers another direct way to see this result.

As another example, the expression for $t_{\sigma_3} \mathcal{A}$ is
\begin{eqnarray} \label{2_loop_sigma3_amplitude_expression}
t_{\sigma_3}\mathcal{A}=\int \left [ dk \right ] &&\overline{u}\left( p_I \right) \frac{\beta_I^\alpha}{k_1\cdot\beta_I} \frac{\overline{\beta}_K^\beta}{k_2\cdot\overline{\beta}_K} \gamma^\mu \frac{\slashed{p}_K-\slashed{k}_2}{\left( p_K-k_2 \right)^2} \gamma^\gamma v\left( p_K \right) \frac{1}{k_1^2} \frac{1}{k_2^2} \frac{1}{\left( k_2-\left(k_1\cdot\beta_K\right) \overline{\beta}_K \right)^2}\nonumber\\
&&\cdot V_{\alpha\beta\gamma} \left( \left(k_1\cdot\beta_K\right) \overline{\beta}_K, k_2-\left(k_1\cdot\beta_K\right) \overline{\beta}_K, -k_2 \right),
\end{eqnarray}
which is depicted in figure\ \ref{selected_subtraction_terms}$(b)$. Notice that here we have two eikonal lines (in the directions of $\beta_I^\mu$ and $\overline{\beta}_K^\mu$) and a partonic line as the framework. The pinch surfaces of $t_{\sigma_3} \mathcal{A}$ are different from those of $\mathcal{A}$, because we have an eikonal line in a new direction $\overline{\beta}_K^\mu$, and there is a soft approximation on the loop momentum $k_1^\mu$. Therefore, the region where $k_1^\mu$ is collinear to $\beta_I^\mu$ and $k_2^\mu$ is collinear to $\overline{\beta}_K^\mu$, which we denote as $\text{C}_1\overline{\text{C}}_2$, is a new pinch surface of $t_{\sigma_3} \mathcal{A}$, while $\text{C}_1\text{C}_1$ is no longer a pinch surface. In other words,
\begin{eqnarray} \label{2_loop_sigma3_pinch_surface}
\text{PS of }t_{\sigma_3}\mathcal{A}:\ \text{SS},\ \text{C}_1\text{S},\ \text{SC}_2,\ \text{C}_1\overline{\text{C}}_2,\ \text{C}_2\text{C}_2,\ \text{C}_1\text{C}_2,\ \text{C}_1\text{H},\ \text{and }\text{HC}_2.
\end{eqnarray}

From the rules of repetitive approximation, eqs.\ (\ref{denominator_change}) and (\ref{fermion_numerator_change}), we find that
\begin{eqnarray} \label{2_loop_sigma3sigma1_amplitude_expression}
t_{\sigma_3}t_{\sigma_1}\mathcal{A}=\int \left [ dk \right ] &&\Big(\overline{u}\left( p_I \right) \gamma^\mu v\left( p_K \right) \Big) \frac{\beta_I^\alpha}{k_1\cdot\beta_I}\frac{\overline{\beta}_K^\beta}{k_2\cdot\overline{\beta}_K}\frac{\beta_K^\gamma}{-k_2\cdot\beta_K} \frac{1}{k_1^2} \frac{1}{k_2^2} \frac{1}{\left( k_2-\left(k_1\cdot\beta_K\right) \overline{\beta}_K \right)^2}\nonumber\\
&&\cdot V_{\alpha\beta\gamma} \left( \left(k_1\cdot\beta_K\right) \overline{\beta}_K, k_2-\left(k_1\cdot\beta_K\right) \overline{\beta}_K, -k_2 \right),
\end{eqnarray}
\begin{eqnarray} \label{2_loop_sigma6sigma3sigma1_amplitude_expression}
t_{\sigma_6}t_{\sigma_3}t_{\sigma_1}\mathcal{A}=&&t_{\sigma_7}t_{\sigma_6}t_{\sigma_3}t_{\sigma_1}\mathcal{A} \nonumber\\
=&&\int \left [ dk \right ]\Big(\overline{u}\left( p_I \right) \gamma^\mu v\left( p_K \right) \Big) \frac{\beta_I^\alpha}{k_1\cdot\beta_I}\frac{\overline{\beta}_K^\beta}{k_2\cdot\overline{\beta}_K}\frac{\beta_K^\gamma}{-k_2\cdot\beta_K} \frac{1}{k_1^2} \frac{1}{k_2^2} \frac{1}{-2 \left( k_1\cdot\overline{\beta}_I \right) \left( k_2\cdot\overline{\beta}_K \right)}\nonumber\\
&&\cdot V_{ \alpha' \beta \gamma'} \Big( \left( k_1\cdot\overline{\beta}_I \right)\left( \beta_I\cdot\beta_K \right) \overline{\beta}_K, \left( k_2\cdot \overline{\beta}_K \right)\beta_K-\left( k_1\cdot\overline{\beta}_I \right) \overline{\beta}_K,\nonumber\\
&&\ \ \ \ \ \ \ \ \ \ \ \left( k_2\cdot \overline{\beta}_K \right)\beta_K \Big ) \overline{\beta}_K^{\alpha'} \left( \beta_K\cdot\beta_I \right) \overline{\beta}_{I\alpha} \beta_K^{\gamma'} \overline{\beta}_{K\gamma},
\end{eqnarray}
as are shown in figures\ \ref{selected_subtraction_terms}$(c)$ and $(d)$. The pinch surfaces of $t_{\sigma_3}t_{\sigma_1}\mathcal{A}$ are identical to those of $t_{\sigma_3}\mathcal{A}$, because the only difference between them is to replace the partonic line in $(b)$ by an eikonal line in the same direction in $(c)$. Meanwhile there is a new pinch surface appearing in $t_{\sigma_6}t_{\sigma_3}t_{\sigma_1}\mathcal{A}$, i.e. $\overline{\text{C}}_1\text{C}_2$, while $\text{C}_2\text{C}_2$ is no longer pinched. In summary, the pinch surfaces of these subtraction terms are
\begin{eqnarray} \label{2_loop_sigma3sigma1_pinch_surface}
\text{PS of }t_{\sigma_3}t_{\sigma_1}\mathcal{A}:\ \text{SS},\ \text{C}_1\text{S},\ \text{SC}_2,\ \text{C}_1\overline{\text{C}}_2,\ \text{C}_2\text{C}_2,\ \text{C}_1\text{C}_2,\ \text{C}_1\text{H},\ \text{and }\text{HC}_2;
\end{eqnarray}
\begin{eqnarray} \label{2_loop_sigma6sigma3sigma1_pinch_surface}
\text{PS of }t_{\sigma_6}t_{\sigma_3}t_{\sigma_1}\mathcal{A}:\ \text{SS},\ \text{C}_1\text{S},\ \text{SC}_2,\ \text{C}_1\overline{\text{C}}_2,\ \overline{\text{C}}_1\text{C}_2,\ \text{C}_1\text{C}_2,\ \text{C}_1\text{H},\ \text{and }\text{HC}_2.
\end{eqnarray}

In this notation, we list all the IR regions of the approximated amplitudes generated from figure\ \ref{all_2_loop_graphs}$(a)$ in the following long table, and distinguish them by enclosing them with numbered rectangles. Any two IR regions (from two subtraction terms) with the same lower index outside the rectangle cancel. By carrying out a laborious but direct check, it can be seen that all these listed terms cancel in a pairwise way, as expected.

\begin{center}
    
\begin{longtable}{| c || c |} 
\caption[An optional table caption ...]{Approximated amplitudes and their associated IR regions}\\
 \hline
 Approximated & \multirow{2}{*}{Regions of IR divergences} \\
 Amplitudes & \\
 \hline
 \hline
 \endfirsthead
 \multicolumn{2}{l}{\textit{Continued from previous page}} \\
 \hline
 Approximated & \multirow{2}{*}{Regions of IR divergences} \\
 Amplitudes & \\
 \hline
 \hline
 \endhead
 \hline \multicolumn{2}{r}{\textit{Continued on next page}} \\
 \endfoot
 \hline
 \endlastfoot
 $\mathcal{A}$ & $\fbox{SS}_1$ $\fbox{$\text{C}_1$S}_2$ $\fbox{S$\text{C}_2$}_3$ $\fbox{$\text{C}_1\text{C}_1$}_4$ $\fbox{$\text{C}_2\text{C}_2$}_5$ $\fbox{$\text{C}_1\text{C}_2$}_6$ $\fbox{$\text{C}_1$H}_7$ $\fbox{H$\text{C}_2$}_8$ \\
 \hline
 $t_{\sigma_1}\mathcal{A}$ & $\fbox{SS}_1$ $\fbox{$\text{C}_1$S}_9$ $\fbox{S$\text{C}_2$}_{10}$ $\fbox{$\text{C}_1\text{C}_1$}_{11}$ $\fbox{$\text{C}_2\text{C}_2$}_{12}$ $\fbox{$\text{C}_1\text{C}_2$}_{13}$ $\fbox{$\text{C}_1$H}_{14}$ $\fbox{H$\text{C}_2$}_{15}$ \\
 \hline
 $t_{\sigma_2}\mathcal{A}$ & $\fbox{SS}_{16}$ $\fbox{$\text{C}_1$S}_2$ $\fbox{S$\text{C}_2$}_{17}$ $\fbox{$\text{C}_1\text{C}_1$}_{18}$ $\fbox{$\overline{\text{C}}_1\text{C}_2$}_{19}$ $\fbox{$\text{C}_1\text{C}_2$}_{20}$ $\fbox{$\text{C}_1$H}_{21}$ $\fbox{H$\text{C}_2$}_{22}$ \\
 \hline
 $t_{\sigma_3}\mathcal{A}$ & $\fbox{SS}_{23}$ $\fbox{$\text{C}_1$S}_{24}$ $\fbox{S$\text{C}_2$}_3$ $\fbox{$\text{C}_1\overline{\text{C}}_2$}_{25}$ $\fbox{$\text{C}_2\text{C}_2$}_{26}$ $\fbox{$\text{C}_1\text{C}_2$}_{27}$ $\fbox{$\text{C}_1$H}_{28}$ $\fbox{H$\text{C}_2$}_{29}$ \\
 \hline
 $t_{\sigma_4}\mathcal{A}$ & $\fbox{SS}_{30}$ $\fbox{$\text{C}_1$S}_{31}$ $\fbox{S$\overline{\text{C}}_1$}_{32}$ $\fbox{$\text{C}_1\text{C}_1$}_4$ $\fbox{$\overline{\text{C}}_1\overline{\text{C}}_1$}_{33}$ $\fbox{$\text{C}_1\overline{\text{C}}_1$}_{34}$ $\fbox{$\text{C}_1$H}_{35}$ $\fbox{H$\overline{\text{C}}_1$}_{36}$ \\
 \hline
 $t_{\sigma_5}\mathcal{A}$ & $\fbox{SS}_{37}$ $\fbox{$\overline{\text{C}}_2$S}_{38}$ $\fbox{S$\text{C}_2$}_{39}$ $\fbox{$\overline{\text{C}}_2\overline{\text{C}}_2$}_{40}$ $\fbox{$\text{C}_2\text{C}_2$}_5$ $\fbox{$\overline{\text{C}}_2\text{C}_2$}_{41}$ $\fbox{$\overline{\text{C}}_2$H}_{42}$ $\fbox{H$\text{C}_2$}_{43}$ \\ \hline
 $t_{\sigma_6}\mathcal{A}$ & $\fbox{SS}_{44}$ $\fbox{$\text{C}_1$S}_{45}$ $\fbox{S$\text{C}_2$}_{46}$ $\fbox{$\text{C}_1\overline{\text{C}}_2$}_{47}$ $\fbox{$\overline{\text{C}}_1\text{C}_2$}_{48}$ $\fbox{$\text{C}_1\text{C}_2$}_6$ $\fbox{$\text{C}_1$H}_{49}$ $\fbox{H$\text{C}_2$}_{50}$ \\
 \hline
 $t_{\sigma_7}\mathcal{A}$ & $\fbox{SS}_{51}$ $\fbox{$\text{C}_1$S}_{52}$ $\fbox{S$\text{C}_2$}_{53}$ $\fbox{$\text{C}_1\text{C}_1$}_{54}$ $\fbox{$\overline{\text{C}}_1\text{C}_2$}_{55}$ $\fbox{$\text{C}_1\text{C}_2$}_{56}$ $\fbox{$\text{C}_1$H}_7$ $\fbox{H$\text{C}_2$}_{57}$ \\
 \hline
 $t_{\sigma_8}\mathcal{A}$ & $\fbox{SS}_{58}$ $\fbox{$\text{C}_1$S}_{59}$ $\fbox{S$\text{C}_2$}_{60}$ $\fbox{$\text{C}_1\overline{\text{C}}_2$}_{61}$ $\fbox{$\text{C}_2\text{C}_2$}_{62}$ $\fbox{$\text{C}_1\text{C}_2$}_{63}$ $\fbox{$\text{C}_1$H}_{64}$ $\fbox{H$\text{C}_2$}_8$ \\
 \hline
 $t_{\sigma_2}t_{\sigma_1}\mathcal{A}$ & $\fbox{SS}_{16}$ $\fbox{$\text{C}_1$S}_9$ $\fbox{S$\text{C}_2$}_{17}$ $\fbox{$\text{C}_1\text{C}_1$}_{65}$ $\fbox{$\overline{\text{C}}_1\text{C}_2$}_{19}$ $\fbox{$\text{C}_1\text{C}_2$}_{66}$ $\fbox{$\text{C}_1$H}_{67}$ $\fbox{H$\text{C}_2$}_{68}$ \\
 \hline
 $t_{\sigma_3}t_{\sigma_1}\mathcal{A}$ & $\fbox{SS}_{23}$ $\fbox{$\text{C}_1$S}_{24}$ $\fbox{S$\text{C}_2$}_{10}$ $\fbox{$\text{C}_1\overline{\text{C}}_2$}_{25}$ $\fbox{$\text{C}_2\text{C}_2$}_{69}$ $\fbox{$\text{C}_1\text{C}_2$}_{70}$ $\fbox{$\text{C}_1$H}_{71}$ $\fbox{H$\text{C}_2$}_{72}$ \\
 \hline
 $t_{\sigma_4}t_{\sigma_1}\mathcal{A}$ & $\fbox{SS}_{30}$ $\fbox{$\text{C}_1$S}_{73}$ $\fbox{S$\overline{\text{C}}_1$}_{32}$ $\fbox{$\text{C}_1\text{C}_1$}_{11}$ $\fbox{$\overline{\text{C}}_1\overline{\text{C}}_1$}_{33}$ $\fbox{$\text{C}_1\overline{\text{C}}_1$}_{74}$ $\fbox{$\text{C}_1$H}_{75}$ $\fbox{H$\overline{\text{C}}_1$}_{76}$ \\
 \hline
 $t_{\sigma_5}t_{\sigma_1}\mathcal{A}$ & $\fbox{SS}_{37}$ $\fbox{$\overline{\text{C}}_2$S}_{38}$ $\fbox{S$\text{C}_2$}_{77}$ $\fbox{$\overline{\text{C}}_2\overline{\text{C}}_2$}_{40}$ $\fbox{$\text{C}_2\text{C}_2$}_{12}$ $\fbox{$\overline{\text{C}}_2\text{C}_2$}_{78}$ $\fbox{$\overline{\text{C}}_2$H}_{79}$ $\fbox{H$\text{C}_2$}_{80}$ \\
 \hline
 $t_{\sigma_6}t_{\sigma_1}\mathcal{A}$ & $\fbox{SS}_{44}$ $\fbox{$\text{C}_1$S}_{81}$ $\fbox{S$\text{C}_2$}_{82}$ $\fbox{$\text{C}_1\overline{\text{C}}_2$}_{83}$ $\fbox{$\overline{\text{C}}_1\text{C}_2$}_{84}$ $\fbox{$\text{C}_1\text{C}_2$}_{13}$ $\fbox{$\text{C}_1$H}_{85}$ $\fbox{H$\text{C}_2$}_{86}$ \\
 \hline
 $t_{\sigma_7}t_{\sigma_1}\mathcal{A}$ & $\fbox{SS}_{51}$ $\fbox{$\text{C}_1$S}_{87}$ $\fbox{S$\text{C}_2$}_{88}$ $\fbox{$\text{C}_1\text{C}_1$}_{89}$ $\fbox{$\overline{\text{C}}_1\text{C}_2$}_{90}$ $\fbox{$\text{C}_1\text{C}_2$}_{91}$ $\fbox{$\text{C}_1$H}_{14}$ $\fbox{H$\text{C}_2$}_{92}$ \\
 \hline
 $t_{\sigma_8}t_{\sigma_1}\mathcal{A}$ & $\fbox{SS}_{58}$ $\fbox{$\text{C}_1$S}_{93}$ $\fbox{S$\text{C}_2$}_{94}$ $\fbox{$\text{C}_1\overline{\text{C}}_2$}_{95}$ $\fbox{$\text{C}_2\text{C}_2$}_{96}$ $\fbox{$\text{C}_1\text{C}_2$}_{97}$ $\fbox{$\text{C}_1$H}_{98}$ $\fbox{H$\text{C}_2$}_{15}$ \\
 \hline
 $t_{\sigma_4}t_{\sigma_2}\mathcal{A}$ & $\fbox{SS}_{99}$ $\fbox{$\text{C}_1$S}_{31}$ $\fbox{S$\overline{\text{C}}_1$}_{100}$$\fbox{$\text{C}_1\text{C}_1$}_{18}$ $\fbox{$\overline{\text{C}}_1\overline{\text{C}}_1$}_{101}$$\fbox{$\text{C}_1\overline{\text{C}}_1$}_{34}$ $\fbox{$\text{C}_1$H}_{102}$$\fbox{H$\overline{\text{C}}_1$}_{36}$ \\
 \hline
 $t_{\sigma_6}t_{\sigma_2}\mathcal{A}$ & $\fbox{SS}_{103}$$\fbox{$\text{C}_1$S}_{45}$ $\fbox{S$\text{C}_2$}_{104}$$\fbox{$\text{C}_1\overline{\text{C}}_2$}_{47}$ $\fbox{$\overline{\text{C}}_1\text{C}_2$}_{105}$$\fbox{$\text{C}_1\text{C}_2$}_{20}$ $\fbox{$\text{C}_1$H}_{106}$$\fbox{H$\text{C}_2$}_{107}$ \\
 \hline
 $t_{\sigma_7}t_{\sigma_2}\mathcal{A}$ & $\fbox{SS}_{108}$$\fbox{$\text{C}_1$S}_{52}$ $\fbox{S$\text{C}_2$}_{109}$$\fbox{$\text{C}_1\text{C}_1$}_{110}$$\fbox{$\overline{\text{C}}_1\text{C}_2$}_{111}$$\fbox{$\text{C}_1\text{C}_2$}_{112}$$\fbox{$\text{C}_1$H}_{21}$ $\fbox{H$\text{C}_2$}_{113}$ \\
 \hline
 $t_{\sigma_8}t_{\sigma_2}\mathcal{A}$ & $\fbox{SS}_{114}$$\fbox{$\text{C}_1$S}_{59}$ $\fbox{S$\text{C}_2$}_{115}$$\fbox{$\text{C}_1\overline{\text{C}}_2$}_{61}$ $\fbox{$\overline{\text{C}}_1\text{C}_2$}_{116}$$\fbox{$\text{C}_1\text{C}_2$}_{117}$$\fbox{$\text{C}_1$H}_{118}$$\fbox{H$\text{C}_2$}_{22}$ \\
 \hline
 $t_{\sigma_5}t_{\sigma_3}\mathcal{A}$ & $\fbox{SS}_{119}$$\fbox{$\overline{\text{C}}_2$S}_{120}$$\fbox{S$\text{C}_2$}_{39}$ $\fbox{$\overline{\text{C}}_2\overline{\text{C}}_2$}_{121}$$\fbox{$\text{C}_2\text{C}_2$}_{26}$ $\fbox{$\overline{\text{C}}_2\text{C}_2$}_{41}$ $\fbox{$\overline{\text{C}}_2$H}_{42}$ $\fbox{H$\text{C}_2$}_{122}$ \\
 \hline
 $t_{\sigma_6}t_{\sigma_3}\mathcal{A}$ & $\fbox{SS}_{123}$$\fbox{$\text{C}_1$S}_{124}$$\fbox{S$\text{C}_2$}_{46}$ $\fbox{$\text{C}_1\overline{\text{C}}_2$}_{125}$$\fbox{$\overline{\text{C}}_1\text{C}_2$}_{48}$ $\fbox{$\text{C}_1\text{C}_2$}_{27}$ $\fbox{$\text{C}_1$H}_{126}$$\fbox{H$\text{C}_2$}_{127}$ \\
 \hline
 $t_{\sigma_7}t_{\sigma_3}\mathcal{A}$ & $\fbox{SS}_{128}$$\fbox{$\text{C}_1$S}_{129}$$\fbox{S$\text{C}_2$}_{53}$ $\fbox{$\text{C}_1\overline{\text{C}}_2$}_{130}$$\fbox{$\overline{\text{C}}_1\text{C}_2$}_{55}$ $\fbox{$\text{C}_1\text{C}_2$}_{131}$$\fbox{$\text{C}_1$H}_{28}$ $\fbox{H$\text{C}_2$}_{132}$ \\
 \hline
 $t_{\sigma_8}t_{\sigma_3}\mathcal{A}$ & $\fbox{SS}_{133}$$\fbox{$\text{C}_1$S}_{134}$$\fbox{S$\text{C}_2$}_{60}$ $\fbox{$\text{C}_1\overline{\text{C}}_2$}_{135}$$\fbox{$\text{C}_2\text{C}_2$}_{136}$$\fbox{$\text{C}_1\text{C}_2$}_{137}$$\fbox{$\text{C}_1$H}_{138}$$\fbox{H$\text{C}_2$}_{29}$ \\
 \hline
 $t_{\sigma_7}t_{\sigma_4}\mathcal{A}$ & $\fbox{SS}_{139}$$\fbox{$\text{C}_1$S}_{140}$$\fbox{S$\overline{\text{C}}_1$}_{141}$$\fbox{$\text{C}_1\text{C}_1$}_{54}$ $\fbox{$\overline{\text{C}}_1\overline{\text{C}}_1$}_{142}$$\fbox{$\text{C}_1\overline{\text{C}}_1$}_{143}$$\fbox{$\text{C}_1$H}_{35}$ $\fbox{H$\overline{\text{C}}_1$}_{144}$ \\
 \hline
 $t_{\sigma_8}t_{\sigma_5}\mathcal{A}$ & $\fbox{SS}_{145}$$\fbox{$\overline{\text{C}}_2$S}_{146}$$\fbox{S$\text{C}_2$}_{147}$$\fbox{$\overline{\text{C}}_2\overline{\text{C}}_2$}_{148}$$\fbox{$\text{C}_2\text{C}_2$}_{62}$ $\fbox{$\overline{\text{C}}_2\text{C}_2$}_{149}$$\fbox{$\overline{\text{C}}_2$H}_{150}$$\fbox{H$\text{C}_2$}_{43}$ \\
 \hline
 $t_{\sigma_7}t_{\sigma_6}\mathcal{A}$ & $\fbox{SS}_{151}$$\fbox{$\text{C}_1$S}_{152}$$\fbox{S$\text{C}_2$}_{153}$$\fbox{$\text{C}_1\overline{\text{C}}_2$}_{154}$$\fbox{$\overline{\text{C}}_1\text{C}_2$}_{155}$$\fbox{$\text{C}_1\text{C}_2$}_{56}$ $\fbox{$\text{C}_1$H}_{49}$ $\fbox{H$\text{C}_2$}_{57}$ \\
 \hline
 $t_{\sigma_8}t_{\sigma_6}\mathcal{A}$ & $\fbox{SS}_{156}$$\fbox{$\text{C}_1$S}_{157}$$\fbox{S$\text{C}_2$}_{158}$$\fbox{$\text{C}_1\overline{\text{C}}_2$}_{159}$$\fbox{$\overline{\text{C}}_1\text{C}_2$}_{160}$$\fbox{$\text{C}_1\text{C}_2$}_{63}$ $\fbox{$\text{C}_1$H}_{64}$ $\fbox{H$\text{C}_2$}_{50}$ \\
 \hline
 $t_{\sigma_4}t_{\sigma_2}t_{\sigma_1}\mathcal{A}$ & $\fbox{SS}_{99}$ $\fbox{$\text{C}_1$S}_{73}$ $\fbox{S$\overline{\text{C}}_1$}_{100}$$\fbox{$\text{C}_1\text{C}_1$}_{65}$ $\fbox{$\overline{\text{C}}_1\overline{\text{C}}_1$}_{101}$$\fbox{$\text{C}_1\overline{\text{C}}_1$}_{74}$ $\fbox{$\text{C}_1$H}_{161}$$\fbox{H$\text{C}_2$}_{76}$ \\
 \hline
 $t_{\sigma_6}t_{\sigma_2}t_{\sigma_1}\mathcal{A}$ & $\fbox{SS}_{103}$$\fbox{$\text{C}_1$S}_{81}$ $\fbox{S$\text{C}_2$}_{104}$$\fbox{$\text{C}_1\overline{\text{C}}_2$}_{83}$ $\fbox{$\overline{\text{C}}_1\text{C}_2$}_{105}$$\fbox{$\text{C}_1\text{C}_2$}_{66}$ $\fbox{$\text{C}_1$H}_{162}$$\fbox{H$\text{C}_2$}_{163}$ \\
 \hline
 $t_{\sigma_7}t_{\sigma_2}t_{\sigma_1}\mathcal{A}$ & $\fbox{SS}_{108}$$\fbox{$\text{C}_1$S}_{87}$ $\fbox{S$\text{C}_2$}_{109}$$\fbox{$\text{C}_1\text{C}_1$}_{164}$$\fbox{$\overline{\text{C}}_1\text{C}_2$}_{111}$$\fbox{$\text{C}_1\text{C}_2$}_{165}$$\fbox{$\text{C}_1$H}_{67}$ $\fbox{H$\text{C}_2$}_{166}$ \\
 \hline
 $t_{\sigma_8}t_{\sigma_2}t_{\sigma_1}\mathcal{A}$ & $\fbox{SS}_{114}$$\fbox{$\text{C}_1$S}_{93}$ $\fbox{S$\text{C}_2$}_{115}$$\fbox{$\text{C}_1\overline{\text{C}}_2$}_{95}$ $\fbox{$\overline{\text{C}}_1\text{C}_2$}_{116}$$\fbox{$\text{C}_1\text{C}_2$}_{167}$$\fbox{$\text{C}_1$H}_{168}$$\fbox{H$\text{C}_2$}_{68}$ \\
 \hline
 $t_{\sigma_5}t_{\sigma_3}t_{\sigma_1}\mathcal{A}$ & $\fbox{SS}_{119}$$\fbox{$\overline{\text{C}}_2$S}_{120}$$\fbox{S$\text{C}_2$}_{77}$ $\fbox{$\overline{\text{C}}_2\overline{\text{C}}_2$}_{121}$$\fbox{$\text{C}_2\text{C}_2$}_{69}$ $\fbox{$\overline{\text{C}}_2\text{C}_2$}_{78}$ $\fbox{$\overline{\text{C}}_2$H}_{79}$ $\fbox{H$\text{C}_2$}_{169}$ \\
 \hline
 $t_{\sigma_6}t_{\sigma_3}t_{\sigma_1}\mathcal{A}$ & $\fbox{SS}_{123}$$\fbox{$\text{C}_1$S}_{124}$$\fbox{S$\text{C}_2$}_{82}$ $\fbox{$\text{C}_1\overline{\text{C}}_2$}_{125}$$\fbox{$\overline{\text{C}}_1\text{C}_2$}_{84}$ $\fbox{$\text{C}_1\text{C}_2$}_{70}$ $\fbox{$\text{C}_1$H}_{170}$$\fbox{H$\text{C}_2$}_{171}$ \\
 \hline
 $t_{\sigma_7}t_{\sigma_3}t_{\sigma_1}\mathcal{A}$ & $\fbox{SS}_{128}$$\fbox{$\text{C}_1$S}_{129}$$\fbox{S$\text{C}_2$}_{88}$ $\fbox{$\text{C}_1\overline{\text{C}}_2$}_{130}$$\fbox{$\overline{\text{C}}_1\text{C}_2$}_{90}$ $\fbox{$\text{C}_1\text{C}_2$}_{172}$$\fbox{$\text{C}_1$H}_{71}$ $\fbox{H$\text{C}_2$}_{173}$ \\
 \hline
 $t_{\sigma_8}t_{\sigma_3}t_{\sigma_1}\mathcal{A}$ & $\fbox{SS}_{133}$$\fbox{$\text{C}_1$S}_{134}$$\fbox{S$\text{C}_2$}_{94}$ $\fbox{$\text{C}_1\overline{\text{C}}_2$}_{135}$$\fbox{$\text{C}_2\text{C}_2$}_{174}$$\fbox{$\text{C}_1\text{C}_2$}_{175}$$\fbox{$\text{C}_1$H}_{176}$$\fbox{H$\text{C}_2$}_{72}$ \\
 \hline
 $t_{\sigma_7}t_{\sigma_4}t_{\sigma_1}\mathcal{A}$ & $\fbox{SS}_{139}$$\fbox{$\text{C}_1$S}_{177}$$\fbox{S$\overline{\text{C}}_1$}_{141}$$\fbox{$\text{C}_1\text{C}_1$}_{89}$ $\fbox{$\overline{\text{C}}_1\overline{\text{C}}_1$}_{142}$$\fbox{$\text{C}_1\overline{\text{C}}_1$}_{178}$$\fbox{$\text{C}_1$H}_{75}$ $\fbox{H$\overline{\text{C}}_1$}_{179}$ \\
 \hline
 $t_{\sigma_8}t_{\sigma_5}t_{\sigma_1}\mathcal{A}$ & $\fbox{SS}_{145}$$\fbox{$\overline{\text{C}}_2$S}_{146}$$\fbox{S$\text{C}_2$}_{180}$$\fbox{$\overline{\text{C}}_2\overline{\text{C}}_2$}_{148}$$\fbox{$\text{C}_2\text{C}_2$}_{96}$ $\fbox{$\overline{\text{C}}_2\text{C}_2$}_{181}$$\fbox{$\overline{\text{C}}_2$H}_{182}$$\fbox{H$\text{C}_2$}_{80}$ \\
 \hline
 $t_{\sigma_7}t_{\sigma_6}t_{\sigma_1}\mathcal{A}$ & $\fbox{SS}_{151}$$\fbox{$\text{C}_1$S}_{183}$$\fbox{S$\text{C}_2$}_{184}$$\fbox{$\text{C}_1\overline{\text{C}}_2$}_{185}$$\fbox{$\overline{\text{C}}_1\text{C}_2$}_{186}$$\fbox{$\text{C}_1\text{C}_2$}_{91}$ $\fbox{$\text{C}_1$H}_{85}$ $\fbox{H$\text{C}_2$}_{92}$ \\
 \hline
 $t_{\sigma_8}t_{\sigma_6}t_{\sigma_1}\mathcal{A}$ & $\fbox{SS}_{156}$$\fbox{$\text{C}_1$S}_{187}$$\fbox{S$\text{C}_2$}_{188}$$\fbox{$\text{C}_1\overline{\text{C}}_2$}_{189}$$\fbox{$\overline{\text{C}}_1\text{C}_2$}_{190}$$\fbox{$\text{C}_1\text{C}_2$}_{97}$ $\fbox{$\text{C}_1$H}_{98}$ $\fbox{H$\text{C}_2$}_{86}$ \\
 \hline
 $t_{\sigma_7}t_{\sigma_4}t_{\sigma_2}\mathcal{A}$ & $\fbox{SS}_{191}$$\fbox{$\text{C}_1$S}_{140}$$\fbox{S$\overline{\text{C}}_1$}_{192}$$\fbox{$\text{C}_1\text{C}_1$}_{110}$$\fbox{$\overline{\text{C}}_1\overline{\text{C}}_1$}_{193}$$\fbox{$\text{C}_1\overline{\text{C}}_1$}_{143}$$\fbox{$\text{C}_1$H}_{102}$$\fbox{H$\overline{\text{C}}_1$}_{144}$ \\
 \hline
 $t_{\sigma_7}t_{\sigma_6}t_{\sigma_2}\mathcal{A}$ & $\fbox{SS}_{194}$$\fbox{$\text{C}_1$S}_{152}$$\fbox{S$\text{C}_2$}_{195}$$\fbox{$\text{C}_1\overline{\text{C}}_2$}_{154}$$\fbox{$\overline{\text{C}}_1\text{C}_2$}_{196}$$\fbox{$\text{C}_1\text{C}_2$}_{112}$$\fbox{$\text{C}_1$H}_{106}$$\fbox{H$\text{C}_2$}_{113}$ \\
 \hline
 $t_{\sigma_8}t_{\sigma_6}t_{\sigma_2}\mathcal{A}$ & $\fbox{SS}_{197}$$\fbox{$\text{C}_1$S}_{157}$$\fbox{S$\text{C}_2$}_{198}$$\fbox{$\text{C}_1\overline{\text{C}}_2$}_{159}$$\fbox{$\overline{\text{C}}_1\text{C}_2$}_{199}$$\fbox{$\text{C}_1\text{C}_2$}_{117}$$\fbox{$\text{C}_1$H}_{118}$$\fbox{H$\text{C}_2$}_{107}$ \\
 \hline
 $t_{\sigma_8}t_{\sigma_5}t_{\sigma_3}\mathcal{A}$ & $\fbox{SS}_{200}$$\fbox{$\overline{\text{C}}_2$S}_{201}$$\fbox{S$\text{C}_2$}_{147}$$\fbox{$\overline{\text{C}}_2\overline{\text{C}}_2$}_{202}$$\fbox{$\text{C}_2\text{C}_2$}_{136}$$\fbox{$\overline{\text{C}}_2\text{C}_2$}_{149}$$\fbox{$\overline{\text{C}}_2$H}_{150}$$\fbox{H$\text{C}_2$}_{122}$ \\
 \hline
 $t_{\sigma_7}t_{\sigma_6}t_{\sigma_3}\mathcal{A}$ & $\fbox{SS}_{203}$$\fbox{$\text{C}_1$S}_{204}$$\fbox{S$\text{C}_2$}_{153}$$\fbox{$\text{C}_1\overline{\text{C}}_2$}_{205}$$\fbox{$\overline{\text{C}}_1\text{C}_2$}_{155}$$\fbox{$\text{C}_1\text{C}_2$}_{131}$$\fbox{$\text{C}_1$H}_{126}$$\fbox{H$\text{C}_2$}_{132}$ \\
 \hline
 $t_{\sigma_8}t_{\sigma_6}t_{\sigma_3}\mathcal{A}$ & $\fbox{SS}_{206}$$\fbox{$\text{C}_1$S}_{207}$$\fbox{S$\text{C}_2$}_{158}$$\fbox{$\text{C}_1\overline{\text{C}}_2$}_{208}$$\fbox{$\overline{\text{C}}_1\text{C}_2$}_{160}$$\fbox{$\text{C}_1\text{C}_2$}_{137}$$\fbox{$\text{C}_1$H}_{138}$$\fbox{H$\text{C}_2$}_{127}$ \\
 \hline
 $t_{\sigma_7}t_{\sigma_4}t_{\sigma_2}t_{\sigma_1}\mathcal{A}$ & $\fbox{SS}_{191}$$\fbox{$\text{C}_1$S}_{177}$$\fbox{S$\overline{\text{C}}_1$}_{192}$$\fbox{$\text{C}_1\text{C}_1$}_{164}$$\fbox{$\overline{\text{C}}_1\overline{\text{C}}_1$}_{193}$$\fbox{$\text{C}_1\overline{\text{C}}_1$}_{178}$$\fbox{$\text{C}_1$H}_{161}$$\fbox{H$\overline{\text{C}}_1$}_{179}$ \\
 \hline
 $t_{\sigma_7}t_{\sigma_6}t_{\sigma_2}t_{\sigma_1}\mathcal{A}$ & $\fbox{SS}_{194}$$\fbox{$\text{C}_1$S}_{183}$$\fbox{S$\text{C}_2$}_{195}$$\fbox{$\text{C}_1\overline{\text{C}}_2$}_{185}$$\fbox{$\overline{\text{C}}_1\text{C}_2$}_{196}$$\fbox{$\text{C}_1\text{C}_2$}_{165}$$\fbox{$\text{C}_1$H}_{162}$$\fbox{H$\text{C}_2$}_{166}$ \\
 \hline
 $t_{\sigma_8}t_{\sigma_6}t_{\sigma_2}t_{\sigma_1}\mathcal{A}$ & $\fbox{SS}_{197}$$\fbox{$\text{C}_1$S}_{187}$$\fbox{S$\text{C}_2$}_{198}$$\fbox{$\text{C}_1\overline{\text{C}}_2$}_{189}$$\fbox{$\overline{\text{C}}_1\text{C}_2$}_{199}$$\fbox{$\text{C}_1\text{C}_2$}_{167}$$\fbox{$\text{C}_1$H}_{168}$$\fbox{H$\text{C}_2$}_{163}$ \\
 \hline
 $t_{\sigma_8}t_{\sigma_5}t_{\sigma_3}t_{\sigma_1}\mathcal{A}$ & $\fbox{SS}_{200}$$\fbox{$\overline{\text{C}}_2$S}_{201}$$\fbox{S$\text{C}_2$}_{180}$$\fbox{$\overline{\text{C}}_2\overline{\text{C}}_2$}_{202}$$\fbox{$\text{C}_2\text{C}_2$}_{174}$$\fbox{$\overline{\text{C}}_2\text{C}_2$}_{181}$$\fbox{$\overline{\text{C}}_2$H}_{182}$$\fbox{H$\text{C}_2$}_{169}$ \\
 \hline
 $t_{\sigma_7}t_{\sigma_6}t_{\sigma_3}t_{\sigma_1}\mathcal{A}$ & $\fbox{SS}_{203}$$\fbox{$\text{C}_1$S}_{204}$$\fbox{S$\text{C}_2$}_{184}$$\fbox{$\text{C}_1\overline{\text{C}}_2$}_{205}$$\fbox{$\overline{\text{C}}_1\text{C}_2$}_{186}$$\fbox{$\text{C}_1\text{C}_2$}_{172}$$\fbox{$\text{C}_1$H}_{170}$$\fbox{H$\text{C}_2$}_{173}$ \\
 \hline
 $t_{\sigma_8}t_{\sigma_6}t_{\sigma_3}t_{\sigma_1}\mathcal{A}$ & $\fbox{SS}_{206}$$\fbox{$\text{C}_1$S}_{207}$$\fbox{S$\text{C}_2$}_{188}$$\fbox{$\text{C}_1\overline{\text{C}}_2$}_{208}$$\fbox{$\overline{\text{C}}_1\text{C}_2$}_{190}$$\fbox{$\text{C}_1\text{C}_2$}_{175}$$\fbox{$\text{C}_1$H}_{176}$$\fbox{H$\text{C}_2$}_{171}$ \\
 \hline
\end{longtable}

\end{center}

We comment that one will not encounter exotic pinch surfaces in this two-loop calculations. The simplest example for a soft-exotic configuration to emerge is shown in figure\ \ref{soft-exotic_example}, which is three-loop; the simplest example for a hard-exotic configuration to emerge is shown in figure\ \ref{hard-exotic_PS_example}, which corresponds to the ladder graph, figure\ \ref{all_2_loop_graphs}$(c)$. But this does not lead to more subtleties than our calculations shown above; such cancellations can be directly seen from the discussion in section\ \ref{pairwise_cancellation_exotic}.

\subsection{Factorization of the subtraction terms}
\label{2_loop_example_factorization}

In this subsection we verify that at NNLO, the factorized expression of the full amplitude, eq.\ (\ref{full_amplitude_factorization_result}), can be directly obtained from the forest formula. From another point of view, we also verify that if we write out all the forest formulas of the $\mathcal{O}(\alpha_S^2)$ subgraphs, some of their terms cancel and the remaining terms can be reorganized into (\ref{full_amplitude_factorization_result}).

First, we expand eq.\ (\ref{full_amplitude_factorization_result}) and only focus on its $\mathcal{O}(\alpha_S^2)$ terms, i.e.
\begin{eqnarray} \label{expansion_full_amplitude_factorization_result}
\mathcal{M}^{(2)}&&= \bigg[ \gamma_{S,\text{eik}}^{(2)} \cdot \overline{\gamma}_H^{(0)} + \mathcal{J}_{\text{part}}^{(2)} \cdot \overline{\gamma}_H^{(0)} -\mathcal{J}_{\text{eik}}^{(2)} \cdot \overline{\gamma}_H^{(0)} + \gamma_{S,\text{eik}}^{(1)} \cdot \mathcal{J}_{\text{part}}^{(1)} \cdot \overline{\gamma}_H^{(0)} \nonumber\\
&&- \mathcal{J}_{\text{part}}^{(1)} \cdot \mathcal{J}_{\text{eik}}^{(1)} \cdot \overline{\gamma}_H^{(0)} + \mathcal{J}_{\text{eik}}^{(1)} \cdot \mathcal{J}_{\text{eik}}^{(1)} \cdot \overline{\gamma}_H^{(0)} + \gamma_{S,\text{eik}}^{(1)} \cdot \mathcal{J}_{\text{eik}}^{(1)} \cdot \overline{\gamma}_H^{(0)} \nonumber\\
&&+ \gamma_{S,\text{eik}}^{(1)} \cdot \sum_{F_H} \prod_{\sigma_H} \left(-t_{\sigma_H}\right) \overline{\gamma}_H^{(1)} + \mathcal{J}_{\text{part}}^{(1)} \cdot \sum_{F_H} \prod_{\sigma_H} \left(-t_{\sigma_H}\right) \overline{\gamma}_H^{(1)} \nonumber\\
&&- \mathcal{J}_{\text{eik}}^{(1)} \cdot \sum_{F_H} \prod_{\sigma_H} \left(-t_{\sigma_H}\right) \overline{\gamma}_H^{(1)} \bigg] + \sum_{F_H} \prod_{\sigma_H} \left(-t_{\sigma_H}\right) \overline{\gamma}_H^{(2)}.
\end{eqnarray}

As is explained in section\ \ref{partonic_amplitudes_factorized_forms}, the last term in eq.\ (\ref{expansion_full_amplitude_factorization_result}) corresponds to the case where $\sigma_0^* =\eta$ in (\ref{graph_forest_formula3}), which comes from the remaining term $R\left[ \mathcal{A}^{(n)} \right]$ in (\ref{graph_forest_formula1}). Each term in the square bracket is associated with a set of subtraction terms of the forest formula, which we check directly below:
{
\allowdisplaybreaks
\begin{eqnarray}
\gamma_{S,\text{eik}}^{(2)} \cdot \overline{\gamma}_H^{(0)} &&= -\sum_\mathcal{A} \left( -t_{\text{SS}^{ }} \right) \mathcal{A};\label{one_term_from_expansion_first} \\
\mathcal{J}_{\text{part}}^{(2)} \cdot \overline{\gamma}_H^{(0)} &&= -\sum_\mathcal{A} \left( -t_{\text{C}_1\text{C}_1}^{ } -t_{\text{C}_2\text{C}_2}^{ } -t_{\text{C}_1\text{C}_2}^{ } \right) \mathcal{A}; \\
-\mathcal{J}_{\text{eik}}^{(2)} \cdot \overline{\gamma}_H^{(0)} &&= -\sum_\mathcal{A} \left( -t_{\text{C}_1\text{C}_1}^{ } -t_{\text{C}_2\text{C}_2}^{ } -t_{\text{C}_1\text{C}_2}^{ } \right) \left( -t_{\text{SS}^{ }} \right) \mathcal{A}; \\
\gamma_{S,\text{eik}}^{(1)} \cdot \mathcal{J}_{\text{part}}^{(1)} \cdot \overline{\gamma}_H^{(0)} &&= -\sum_\mathcal{A} \left( -t_{\text{C}_1\text{S}}^{ } -t_{\text{S}\text{C}_2}^{ } \right) \mathcal{A}; \\
-\mathcal{J}_{\text{part}}^{(1)} \cdot \mathcal{J}_{\text{eik}}^{(1)} \cdot \overline{\gamma}_H^{(0)} &&= -\sum_\mathcal{A} \Big[ \left( -t_{\text{C}_1\text{C}_1}^{ } -t_{\text{C}_1\text{C}_2}^{ } \right) \left( -t_{\text{C}_1\text{S}}^{ } \right)\nonumber\\
&&\hspace{1cm} + \left( -t_{\text{C}_1\text{C}_1}^{ } -t_{\text{C}_2\text{C}_2}^{ } \right) \left( -t_{\text{S}\text{C}_2}^{ } \right) \Big] \mathcal{A}; \\
-\mathcal{J}_{\text{eik}}^{(1)} \cdot \mathcal{J}_{\text{eik}}^{(1)} \cdot \overline{\gamma}_H^{(0)} &&= -\sum_\mathcal{A} \Big[ \left( -t_{\text{C}_1\text{C}_1}^{ } -t_{\text{C}_1\text{C}_2}^{ } \right) \left( -t_{\text{C}_1\text{S}}^{ } \right)\nonumber\\
&&\hspace{1cm} + \left( -t_{\text{C}_1\text{C}_1}^{ } -t_{\text{C}_2\text{C}_2}^{ } \right) \left( -t_{\text{S}\text{C}_2}^{ } \right) \Big] \left( -t_{\text{SS}^{ }} \right) \mathcal{A}; \\
\gamma_{S,\text{eik}}^{(1)} \cdot \mathcal{J}_{\text{eik}}^{(1)} \cdot \overline{\gamma}_H^{(0)} &&= -\sum_\mathcal{A} \left( -t_{\text{C}_1\text{S}}^{ } -t_{\text{S}\text{C}_2}^{ } \right) \left( -t_{\text{SS}^{ }} \right) \mathcal{A}; \\
\gamma_{S,\text{eik}}^{(1)} \cdot \sum_{F_H} \prod_{\sigma_H} \left(-t_{\sigma_H}\right) \overline{\gamma}_H^{(1)} &&= -\sum_\mathcal{A} \left( -t_{\text{SH}}^{ } \right) \sum_{F_<} \prod_{\sigma_<} \left( -t_{\sigma_<}^{ } \right) \mathcal{A}; \label{one_term_from_expansion_last_but_three} \\
\mathcal{J}_{\text{part}}^{(1)} \cdot \sum_{F_H} \prod_{\sigma_H} \left(-t_{\sigma_H}\right) \overline{\gamma}_H^{(1)} &&= -\sum_\mathcal{A} \left( -t_{\text{C}_1\text{H}}^{ } -t_{\text{H}\text{C}_2}^{ } \right) \sum_{F_<} \prod_{\sigma_<} \left( -t_{\sigma_<}^{ } \right) \mathcal{A};\label{one_term_from_expansion} \\
\mathcal{J}_{\text{eik}}^{(1)} \cdot \sum_{F_H} \prod_{\sigma_H} \left(-t_{\sigma_H}\right) \overline{\gamma}_H^{(1)} &&= -\sum_\mathcal{A} \left( -t_{\text{C}_1\text{H}}^{ } -t_{\text{H}\text{C}_2}^{ } \right) \left( -t_{\text{SH}}^{ } \right) \sum_{F_<} \prod_{\sigma_<} \left( -t_{\sigma_<}^{ } \right) \mathcal{A} \label{one_term_from_expansion_last}.
\end{eqnarray}
}

In eqs.\ (\ref{one_term_from_expansion_last_but_three})--(\ref{one_term_from_expansion_last}), $\sigma_<$ are the pinch surfaces contained in those with hard loops: for example, $\sigma_< \subset \text{SH}$ in (\ref{one_term_from_expansion_last_but_three}). Note that in eq.\ (\ref{one_term_from_expansion}), there is one requirement on the pinch surfaces denoted by $\sigma_<$: their soft subgraphs must not overlap with $\mathcal{J}_{\text{part}}^{(1)}$. Otherwise, a $\mathcal{J}_{\text{eik}}^{(1)}$ will be factorized from $\mathcal{J}_{\text{part}}^{(1)}$ in the sum over different graphs, and we will not get the LHS. Therefore, the forest formula provides us with all the elements to obtain eq.\ (\ref{full_amplitude_factorization_result}).

Among all the subtraction terms of the forest formulas corresponding to the eight graphs in figure\ \ref{all_2_loop_graphs}, one may find that some terms are missing throughout eqs.\ (\ref{one_term_from_expansion_first})--(\ref{one_term_from_expansion_last}). The term $t_{\text{C}_1\text{H}}^{ } t_{\text{SS}}^{ } \mathcal{A}_{(a)}$ is an example. (As is explained above, it does not appear in (\ref{one_term_from_expansion}); obviously, it does not appear in the other equations either.) But in the sum over different graphs, all such terms will cancel. For example, we will see
\begin{eqnarray} \label{an_example_on_factorization_cancellation}
\sum_\mathcal{A} t_{\text{C}_1\text{H}}^{ } t_{\text{SS}}^{ } \mathcal{A}- \sum_\mathcal{A} t_{\text{C}_1\text{H}}^{ } t_{\text{C}_1\text{S}}^{ } t_{\text{SS}}^{ } \mathcal{A} =0,
\end{eqnarray}
where we sum over all the graphs that are compatible with the pinch surfaces appearing in the approximation operators. Such cancellations are related to the pattern in eq.\ (\ref{soft_forest_formula3}).

\begin{figure}[t]
\centering
\includegraphics[width=15cm]{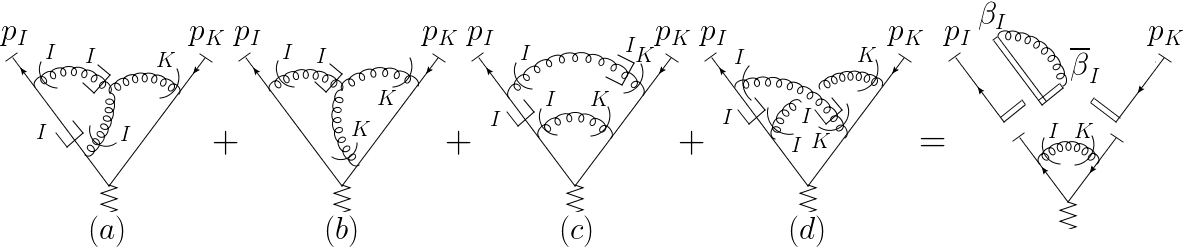}
\caption{The graphic contributions to $\sum_\mathcal{A} t_{\text{C}_1\text{H}}^{ } t_{\text{SS}}^{ } \mathcal{A}$ is equal to a factorized form.}
\label{an_example_on_factorization_cancellation_1st}
\end{figure}

In order to show eq.\ (\ref{an_example_on_factorization_cancellation}), we notice that the first term is contributed by figures\ \ref{all_2_loop_graphs}$(a)$-$(d)$. From our analysis in section\ \ref{factorization_presence_approximations}, the sum over these graphs are in a factorized form. We express this factorization in figure\ \ref{an_example_on_factorization_cancellation_1st}.

Explicit evaluation of the graphs on the LHS agrees with this conclusion, and we do not present it here for simplicity. In the same way, the second term in eq.\ (\ref{an_example_on_factorization_cancellation}) can be reorganized in the same factorized form as well. Here the only contributions are from figures\ \ref{all_2_loop_graphs}$(a)$, $(b)$ and $(d)$, and we have figure\ \ref{an_example_on_factorization_cancellation_2nd}:
\begin{figure}[t]
\centering
\includegraphics[width=14cm]{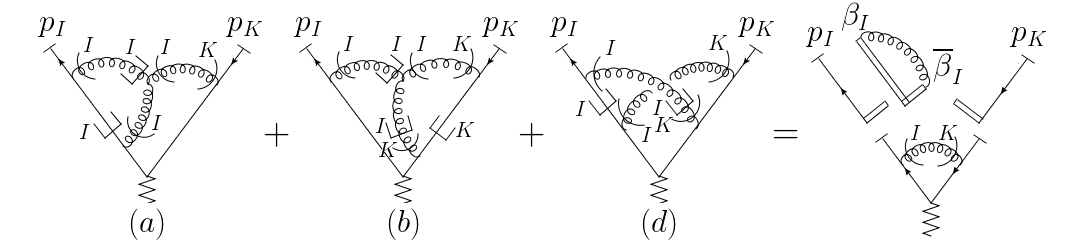}
\caption{The graphic contributions to $\sum_\mathcal{A} t_{\text{C}_1\text{H}}^{ } t_{\text{C}_1\text{S}}^{ } t_{\text{SS}}^{ } \mathcal{A}$ is equal to a factorized form.}
\label{an_example_on_factorization_cancellation_2nd}
\end{figure}

As a result, eq.\ (\ref{an_example_on_factorization_cancellation}) is verified, and other cancellations can be shown in the same way. With the help of these cancellations, we complete our verification of (\ref{full_amplitude_factorization_result}) at $\mathcal{O} \left( \alpha_S^2 \right)$.

\section{Summary and outlook}\label{summary_outlook}

In this paper, we have developed a new forest formula for wide-angle scattering in momentum space, eq.\ (\ref{forest_formula_amplitude}), extending the previous work in coordinate space \cite{EdgStm15} and for the Sudakov form factor \cite{Cls11book}. We first studied the pinch surfaces of the approximated amplitudes, which are generated by hard-collinear and soft-collinear approximations. There are many differences between these pinch surfaces and those of the original amplitude, which can be generalized to the case with repetitive approximations. Despite the differences, we have shown that the divergences of these new pinch surfaces are at worst logarithmic, through explicit power counting for each case.

In order to clarify the pairwise cancellations of the divergences near overlapping pinch surfaces, we studied the maximal region enclosed by $\sigma_m$ and $\rho^{\left\{ \sigma_n...\sigma_1 \right\}}$, where $\sigma_m$ is the smallest one of all the pinch surfaces in forest $\left\{\sigma_1,...,\sigma_n\right\}$ that are not contained in $\rho^ {\left\{\sigma_n...\sigma_1\right\}}$. We proved that this region, which we call an ``enclosed pinch surface'' $\tau\equiv \text{enc} \left[ \sigma, \rho^{\left\{ \sigma \right\}} \right]$ \cite{EdgStm15}, is indeed a leading pinch surface of the original amplitude, for both decay and scattering processes. The analysis involves studying a new algebra of normal spaces of pinch surfaces, which we developed for this purpose.

We then showed the pairwise cancellation within the forest formula. That is, for any IR divergent region of any subtraction term in the forest formula, we can uniquely find another subtraction term that cancels the divergence near that region. The proof includes the coincidence of pinch surfaces of the two subtraction terms in each pair, and the exactness of $t_\tau$ at that pinch surface. We verified these two aspects case by case.

Finally we made use of the forest formula and the gauge theory Ward identity, to rewrite any full amplitude of the decay or scattering processes into a gauge-invariant factorized expression with soft, jet and hard functions \cite{Cls11book, Sen1983, FgeSwtz14, ClsSprStm04, StmTjd-Yms03}, where all the IR divergences are organized along eikonal lines in the directions of $\beta^\mu$ and $\overline{\beta}^\mu$. In the obtained eq.\ (\ref{full_amplitude_factorization_result_rewrite1}), the first factor has only soft divergences, the second factor has only collinear divergences, and the last term has no IR divergences. All the other divergences in the numerators are cancelled by the denominators.

Our findings on the pinch surfaces of the approximated amplitudes may help with the study of Soft-Collinear Effective Theory (SCET) \cite{BurFlmLk00, BurPjlSwt02-1, BurPjlSwt02-2, BchBrgFrl15book}, because the approximated amplitudes are equivalent to the expanded integrals obtained by means of the method of regions \cite{BnkSmn97, Jtz11}, which are in one-to-one correspondence to the SCET Feynman graphs.

As we have mentioned in section\ \ref{discussion_forest_formula}, some other methods to construct subtraction terms also involve the forest-like structure. In principle, we should be able to also use equivalent forests \cite{ClsSpr81} instead of forests to subtract IR divergences identically. These ``freedoms'' suggest a common and general mathematical structure (like the Hopf algebra \cite{Krm97, BrkKrm15}) in these different approaches.

All we have studied in this project centers on wide-angle scattering, in which one can show by contour deformation that the Glauber regions are not pinched \cite{Sen1983, Cls11book, ClsStm81, BdwBskLpg81, Zeng15}, so the soft momenta are scaled as eq.\ (\ref{momenta_scaling}). For other kinematics like forward scatterings in the Regge limit, we also need to take the Glauber regions into account \cite{Sen83, DnghEMnfOvsy14, Flm14, BurLgeOvsy11, RstStw16, DDcFcnMgn15, CrHGrdRch18, CrH15, DDcDuhrGrd11, Vldmr17, FryeHndtHfdPaulSwtz19}.

There is another direction to extend this work, which so far is based on QCD amplitudes, in the future. It should be possible to generalize it to processes with jets in the final states, which can then be implemented to subtract IR divergences of cross sections. Beyond this, it may also be extended to weighted cross sections (angularity, $N$-jettiness, etc.) \cite{LkkMltNchm17}. These topics are left for future research.

\acknowledgments

I am truly indebted to Professor George Sterman. By no means will this project be finished without his inculcation, encouragement, and suggestions on the logic and language here. I also want to express my sincere appreciation to my colleagues Kyle Lee and Vivek Saxena for their assistance, and the people at QCD Masterclass 2019, especially Professors Bowen Xiao, St\'ephane Peign\'e, Alfred Mueller, Stefan Weinzierl, Predrag Cvitanovi\'c and Simone Marzani, who had offered me invaluable advice. This project is also supported by the National Science Foundation under Grants PHY-1620628 and PHY-1915093.

\appendix

\section{Power counting with repetitive approximations}
\label{power_counting_repetitive_approximation}

In this appendix we provide some details of the power counting in section\ \ref{divergences_are_logarithmic}, i,e, the ``regular \& overlapping'' and ``soft-exotic'' cases for the pinch surfaces of the approximated amplitude $t_{\sigma_n}...t_{\sigma_1} \mathcal{A}$, with $n\geqslant 2$. To shorten the notations, we will drop the superscript and use $\rho$ to denote the pinch surfaces $\rho^{ \left\{ \sigma_n ... \sigma_1 \right\} }$.

\subsection{Overlapping \& regular}
\label{overlapping_regular_repetitive}

We recall that the name ``overlapping \& regular'' refers to the case where $\rho^{ \left\{ \sigma_n ... \sigma_1 \right\} }$ overlaps with some $\sigma_m\ (1\leqslant m\leqslant n)$. In section\ \ref{divergences_are_logarithmic} we have already discussed the case of $n=1$, with figure\ \ref{structure_for_overlapping_pc} representing for the general configuration of $J_I^{(\sigma)} \bigcap J_K^{(\rho)}$; here we shall consider $n>1$, and discuss the general configuration of $\gamma \equiv J_I^{(\sigma_m)} \bigcap J_K^{(\rho)}$. The power counting procedure, will be found similar to that from eq.\ (\ref{overlapping_regular_pc1}) to (\ref{overlapping_regular_pc6}), implying that the divergences of $\rho$ are still at worst logarithmic.

The external and internal propagators of $\gamma$ can still be marked using the same notations in section\ \ref{divergences_are_logarithmic}. We denote by $m_J^{ }$ the number of the propagators in the subgraph $S^{(\sigma_m)} \bigcap J_K^{(\rho)}$, denote by $m_S^{ }$ the number of the propagators in $J_I^{(\sigma_m)} \bigcap S^{(\rho)}$, denote by $n_J^{ }$ the number of the propagators that are attached to $H^{(\sigma_m)}$ but not to $H^{(\rho)}$, denote by $n_H^{ }$ the number of the propagators attached to $H^{(\rho)}$ but not to $H^{(\sigma_m)}$, and denote by $m_H^{ }$ the number of the propagators, each one of which has an endpoint attached to both $H^{(\rho)}$ and $H^{(\sigma_m)}$.

Compared with figure\ \ref{structure_for_overlapping_pc}, the differences in the case we study here are caused by the approximations from other pinch surfaces nested with $\sigma_m$, i.e. $t_{\sigma_i}$ ($i\neq m$). For example, some of the $m_J^{ }$ propagators described above may be lightlike in the direction of $\beta_J^\mu$ ($\neq \beta_I^\mu$) and attach to the hard subgraph $H^{(\sigma_r)}$ in $\sigma_r\ (\supset \sigma_m)$. In this case there is an $\text{hc}_J^{(\sigma_r)}$ following an $\text{sc}_I^{(\sigma_m)}$ acting on each of them, and we denote the number of all these propagators by $m_{J,2}^{ }$. In comparison, we denote by $m_{J,1}^{ }$ the number of propagators with a single approximation $\text{sc}_I^{(\sigma_m)}$ on each of them. By definition, $m_J^{ }= m_{J,1}^{ }+ m_{J,2}^{ }$. For both types of these propagators, the projected momenta are in the direction of $\overline{\beta}_I^\mu$, so every one of these $m_J^{ }$ propagators provides a $\overline{\beta}_I^\mu$ vector to form product with the subgraph $\gamma$.

In the same way, we study the approximations that are from other pinch surfaces and acting on the $m_S^{ }$, $n_J^{ }$, $n_H^{ }$ and $m_H^{ }$ propagators. The notations denoting the relevant propagators, as well as the vectors they provide to $\gamma$, are summarized in table\ \ref{structure_for_overlapping_pc_repetitive}. Note that the content (approximations and orderings) of the table is based on nesting requirements. Also, there are possibly hard-collinear and/or soft-collinear approximations inside $\gamma$, but they do not affect the power counting of $\gamma$ as a whole.
\begin{table}[t]
\captionsetup{justification=centering,margin=0.2cm}
\caption{The notations we use to mark the propagators with different approximations. In the last column, we denote the produced vectors that form products with $\gamma$.}
\begin{center}
\begin{tabular}{ |c||c|c|c|c|c| } 
\hline
 \multirow{2}{*}{} & Approx. & Approx. from & \multirow{2}{*}{Notation} & \multirow{2}{*}{Vector}\\ 
 & from $\sigma_m$ & other pinch surfaces & & \\
 \hline
 \hline
 \multirow{2}{*}{$m_J^{ }$} & \multirow{2}{*}{$\text{sc}_I^{(\sigma_m)}$} & None & $m_{J,1}^{ }$ & \multirow{2}{*}{$\overline{\beta}_I^\mu$} \\ \cline{3-4}
  & & $\text{hc}_J^{(\sigma_p)}\ (\sigma_p \supset \sigma_m)$ & $m_{J,2}^{ }$ & \\
 \hline
 \multirow{2}{*}{$m_S^{ }$} & \multirow{2}{*}{None} & None & $m_{S,1}^{ }$ & \multirow{2}{*}{$\beta_I^\mu$} \\ \cline{3-4}
 & & $\text{hc}_I^{(\sigma_q)}\ (\sigma_q \supset \sigma_m)$ & $m_{S,2}^{ }$ & \\  \hline
 \multirow{2}{*}{$n_J^{ }$} & \multirow{2}{*}{$\text{hc}_I^{(\sigma_m)}$} & None & $n_{J,1}^{ }$ & \multirow{2}{*}{$\overline{\beta}_I^\mu$} \\ \cline{3-4}
 & & $\text{sc}_K^{(\sigma_r)}\ (\sigma_r \subset \sigma_m)$ & $n_{J,2}^{ }$ & \\ \hline
 \multirow{2}{*}{$m_H^{ }$} & \multirow{2}{*}{$\text{hc}_I^{(\sigma_m)}$} & None & $m_{H,1}^{ }$ & \multirow{2}{*}{$\overline{\beta}_I^\mu$} \\ \cline{3-4}
 & & $\text{sc}_L^{(\sigma_s)}\ (\sigma_s \subset \sigma_m)$ & $m_{H,2}^{ }$ & \\ \hline
 \multirow{3}{*}{$n_H^{ }$} & \multirow{3}{*}{None} & None & $n_{H,1}^{ }$ & \multirow{2}{*}{$\beta_I^\mu$} \\ \cline{3-4}
 & & $\text{sc}_I^{(\sigma_t)}\ (\sigma_t \subset \sigma_m)$ & $n_{H,2s}^{ }$ & \\ \cline{3-5}
 & & $\text{hc}_I^{(\sigma_u)}\ (\sigma_u \supset \sigma_m)$ & $n_{H,2h}^{ }$ & $\overline{\beta}_I^\mu$ \\ \hline
\end{tabular}
\end{center}
\label{structure_for_overlapping_pc_repetitive}
\end{table}

From table\ \ref{structure_for_overlapping_pc_repetitive}, it is obvious that we can evaluate the degree of divergence of $\rho$ by carrying out almost the same calculations from eq.\ (\ref{overlapping_regular_pc1}) to (\ref{overlapping_regular_pc6}), with the only difference being in the $n_{H,{2h}}^{ }$ term. The final result is:
\begin{eqnarray} \label{overlapping_regular_result_repetitive}
p^{(\rho)} \left(\gamma\right)=m_H^{ }+n_J^{ }+n_{H,{2h}}^{ }-m_S^{ }.
\end{eqnarray}
Apparently, we still have $p^{(\rho)} \left(\gamma\right) \geqslant 0$, which means that the divergence is at worst logarithmic with repetitive approximations included. In order to achieve the equality, one more requirement is needed compared with the two below eq.\ (\ref{overlapping_regular_pc6}) in section\ \ref{divergences_are_logarithmic}: $n_{H,{2h}}^{ }=0$.

\subsection{Soft-exotic}
\label{soft-exotic_repetitive}

Now we consider the case where there is a soft-exotic configuration in $\rho$. By definition, this occurs when a lightlike momentum in $\rho$ is projected soft by one of the approximations among $t_{\sigma_n}, ..., t_{\sigma_1}$. We suppose that specific projection is provided by $t_{\sigma_m}$, and discuss the two possibilities of this approximation: hard-collinear and soft-collinear. Apparently, the figures\ \ref{structure_for_soft-exotic_pc}$(a)$ and $(b)$ in section\ \ref{divergences_are_logarithmic} still describe these two possibilities, after we replace the $\sigma$ there by $\sigma_m$. (If we only consider the approximations from $t_{\sigma_m}$.)

We also need to take into account the approximations from other pinch surfaces $\sigma_i\ (i\neq m)$, as we have done in appendix\ \ref{overlapping_regular_repetitive}. Taking figure\ \ref{structure_for_soft-exotic_pc}$(a)$ as the example, we find that among all the $m_J^{ }$ propagators with a soft-collinear approximation $\text{sc}_I^{(\sigma_m)}$, some of them may be projected by a hard-collinear approximation $\text{hc}_K^{(\sigma_p)}\ (\sigma_p \supset \sigma_m)$. We denote the number of such propagators by $m_{J,2}^{ }$, and the number of the remaining propagators with only $\text{sc}_I^{(\sigma_m)}$ by $m_{J,1}^{ }$. For both these types of propagators, only the $\overline{\beta}_I$-components of their momenta will be left after the approximations, because the projected momenta are of the form:
\begin{eqnarray}
\left( k\cdot\beta_I \right)\overline{\beta}_I^\mu\text{, or }\left( k\cdot\overline{\beta}_K \right) \left( \beta_K\cdot\beta_I \right)\overline{\beta}_I^\mu.
\end{eqnarray}
In other words, every one of the $m_J^{ }$ propagators provides a $\overline{\beta}_I^\mu$ to the red subgraph in figure\ \ref{structure_for_soft-exotic_pc}$(a)$. Similarly, we can prove that every one of the $m_S^{ }$ propagators provides a $\beta_I^\mu$, and every one of the $m$ propagators provides a $\overline{\beta}_I^\mu$ to the red subgraph. Therefore, the power counting procedure in the case we study here is totally identical with that in section\ \ref{divergences_are_logarithmic}.

In the same way, we can analyze the other case, where $t_{\sigma_m}$ exerts a soft-collinear approximation. For simplicity, we shall not review it here since the power counting procedure is exactly the same with that for figure\ \ref{structure_for_soft-exotic_pc}$(b)$. In conclusion, a pinch surface $\rho^{\left\{ \sigma_n...\sigma_1 \right\}}$ with soft-exotic configurations, is at worst logarithmically divergent.

\section{Some details in section \ref{subgraphs_enclosed_pinch_surfaces}}
\label{enclosed_PS_details}

In this appendix, we provide the reader with some details that are omitted in section\ \ref{subgraphs_enclosed_pinch_surfaces}. First, we emphasize that the approximations in the definition of enclosed pinch surfaces, eq.\ (\ref{enc_definition}), are necessary. After that we provide a detailed proof to Theorem 4, which can be seen as the generalization of Theorem 3 to repetitive approximations.

\subsection{Necessity of the approximations}
\label{enclosed_PS_details_approximations_importance}

Our method to demonstrate the necessity of the approximations is by contradiction. Namely, we suppose that the $\rho^{\left\{ \sigma \right\}}$ in eq.\ (\ref{enc_definition_single}) is replaced by $\rho$, as is shown in (\ref{enc_definition_trial}). Then we come to a counterexample, which suggests that we cannot relate the subgraphs of $\sigma$, $\rho$ and $\tau$ as simply as eq.\ (\ref{enc_subgraph_relations}).

\begin{figure}[t]
\centering
\includegraphics[width=14cm]{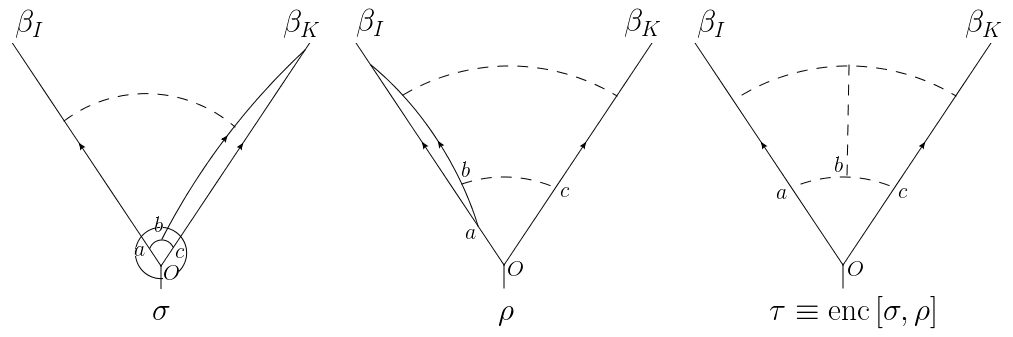}
\caption{An example of $\sigma$, $\rho$, and $\tau\equiv \text{enc}\left[ \sigma,\rho \right]$.}
\label{enclosed_PS_without_approx}
\end{figure}
Our example is given in figure\ \ref{enclosed_PS_without_approx}. If eq.\ (\ref{enc_definition_trial}) is the definition of enclosed pinch surfaces, all the loop momenta should be soft in $\tau$ by definition. However, eq.\ (\ref{enc_subgraph_relations}) does not work here. On the one hand, both the propagators $Oa$ and $ab$ belong to $H^{(\sigma)}\bigcap J_I^{(\rho)}$, and therefore to $J_I^{(\tau)}$ if (\ref{enc_subgraph_relations}) is assumed; on the other hand, $Oa$ is collinear to $\beta_I^\mu$ in $\tau$ while $ab$ is soft in $\tau$ from the definition (\ref{enc_definition_trial}). Then we see that (\ref{enc_subgraph_relations}), which relates the subgraphs of $\sigma$, $\rho$ and $\tau$, does not work here.

This problem is automatically cured in our original definition, eq.\ (\ref{enc_definition_single}), because every momentum entering $b$ is soft in $\rho^{ \left\{ \sigma \right\} }$ and $ab$ is a soft propagator. From our previous knowledge in case (Ciii) of section\ \ref{pinch_surfaces_from_single_approximation}, this corresponds to the soft-exotic configuration, as we have introduced. We can then easily verify that (\ref{enc_subgraph_relations}) holds.

\subsection{Proof of Theorem 4}
\label{enclosed_PS_details_theorem_4_proof}

Now we prove Theorem 4 (eq.\ (\ref{enc_to_prove_generalize})), which generalizes Theorem 3 in section\ \ref{subgraphs_enclosed_pinch_surfaces} to repetitive approximations. Recalling the definition of the $\star$-operation in table\ \ref{star_definition}, we start by rewriting (\ref{enc_to_prove_generalize}) as
\begin{eqnarray} \label{enc_to_prove_before'_generalize}
\mathcal{N}_\tau \left(l_i\right) \star \mathcal{N}_\tau \left(l_j\right) = &&\left (\mathcal{N}_{\sigma_m} \left( l_i \right) \star \mathcal{N}_{\sigma_m} \left( l_j \right) \right) \nonumber\\
&&\oplus \left (\mathcal{N}_{\rho^{\left\{ \sigma_n...\sigma_1 \right\}}}\left( t_{\sigma_n}...t_{\sigma_1} l_i \right) \star \mathcal{N}_{\rho^{\left\{ \sigma_n...\sigma_1 \right\}}}\left( t_{\sigma_n}...t_{\sigma_1} l_j \right) \right).
\end{eqnarray}
Then we insert the defining property of $\mathcal{N}_\tau$, eq.\ (\ref{enc_definition}) into the LHS of (\ref{enc_to_prove_before'_generalize}), and obtain an equivalent relation:
\begin{eqnarray} \label{enc_to_prove'_generalize_appendx}
&&\left (\mathcal{N}_{\sigma_m} \left(l_i\right) \oplus \mathcal{N}_{\rho^{\left\{\sigma_n...\sigma_1\right\}}} \left(l_i\right) \right) \star \left (\mathcal{N}_{\sigma_m} \left(l_j\right) \oplus \mathcal{N}_{\rho^{\left\{\sigma_n...\sigma_1\right\}}} \left(l_j\right) \right) \nonumber\\
&&\hspace{0.8cm}=\left (\mathcal{N}_{\sigma_m} \left( l_i \right) \star \mathcal{N}_{\sigma_m} \left( l_j \right) \right) \oplus \left (\mathcal{N}_{\rho^{\left\{ \sigma_n...\sigma_1 \right\}}}\left( t_{\sigma_n}...t_{\sigma_1} l_i \right) \star \mathcal{N}_{\rho^{\left\{ \sigma_n...\sigma_1 \right\}}}\left( t_{\sigma_n}...t_{\sigma_1} l_j \right) \right).
\end{eqnarray}
This is exactly eq.\ (\ref{enc_to_prove'_generalize}), which we prove below.

From the expression of eq.\ (\ref{enc_to_prove'_generalize_appendx}), it is clear that as long as one knows the explicit projections that $t_{\sigma_n}...t_{\sigma_1}$ exerts on $l_i^\mu$ and $l_j^\mu$, one can check its correctness. Once (\ref{enc_to_prove'_generalize_appendx}) is verified for all the possible projections, the proof of Theorem 4 is finished. We will discuss the possibilities by first classifying them into the following two cases: 1, $t_{\sigma_m}=\text{id}$ for $l_i^\mu$ and $l_j^\mu$; 2, $t_{\sigma_m}\neq \text{id}$ for $l_i^\mu$ or $l_j^\mu$.

We highlight the requirement of Theorem 4, that $\sigma_m$ is the smallest pinch surface not contained in $\rho^{\left\{ \sigma_n...\sigma_1 \right\}}$. This implies that all the pinch surfaces $\sigma_1,...,\sigma_{m-1}$ are contained in $\rho^{\left\{ \sigma_n...\sigma_1 \right\}}$, which is a crucial property that we will repeatedly refer to throughout the proof. 
\bigbreak
\textbf{1, $t_{\sigma_m}=\text{id}$ in eq.\ (\ref{enc_to_prove'_generalize_appendx}). }Because $l_i^\mu$ and $l_j^\mu$ flow into the same vertex, $t_{\sigma_m}=\text{id}$ implies that the propagators carrying $l_i^\mu$ and $l_j^\mu$ must be simultaneously in a hard, soft or jet subgraph in $\sigma_m$.

\begin{itemize}
\item[$\textit{(1)}$ ]{}For the subcase of $l_i^\mu$ and $l_j^\mu$ soft in $\sigma_m$, both sides of eq.\ (\ref{enc_to_prove'_generalize_appendx}) become the entire 4-dimensional space, and (\ref{enc_to_prove'_generalize_appendx}) holds automatically.
\end{itemize}

\begin{itemize}
\item[$\textit{(2)}$ ]{}For the subcase of $l_i^\mu$ and $l_j^\mu$ hard in $\sigma_m$, $\mathcal{N}_{\sigma_m} \left( l_i \right) \star \mathcal{N}_{\sigma_m} \left( l_j \right)=\varnothing$, and eq.\ (\ref{enc_to_prove'_generalize_appendx}) becomes the following relation, which we aim to prove:
\begin{eqnarray} \label{identical_t_sigma_hard}
\mathcal{N}_{\rho^{\left\{\sigma_n...\sigma_1\right\}}} \left(l_i\right) \star \mathcal{N}_{\rho^{\left\{\sigma_n...\sigma_1\right\}}} \left(l_j\right) = \mathcal{N}_{\rho^{\left\{ \sigma_n...\sigma_1 \right\}}}\left( t_{\sigma_n}...t_{\sigma_1} l_i \right) \star \mathcal{N}_{\rho^{\left\{ \sigma_n...\sigma_1 \right\}}}\left( t_{\sigma_n}...t_{\sigma_1} l_j \right)&& \nonumber\\
= \mathcal{N}_{\rho^{\left\{ \sigma_{n}...\sigma_1 \right\}}}\left( t_{\sigma_{m-1}}...t_{\sigma_1} l_i \right) \star \mathcal{N}_{\rho^{\left\{ \sigma_{n}...\sigma_1 \right\}}}\left( t_{\sigma_{m-1}}...t_{\sigma_1} l_j \right).\hspace{0.5cm}&&
\end{eqnarray}
The second equality is due to the fact that $l_i^\mu$ and $l_j^\mu$ are both hard in $\sigma_m$, and consequently, all the projections on them should correspond to the pinch surfaces that are contained in $\sigma_m$. Since any projected momentum always has equal or more normal coordinates compared with the original one, apparently LHS $\subseteq$ RHS. Next we argue that if $l_i^\mu$ and $l_j^\mu$ are hard in $\sigma_m$, LHS $\not\subset$ RHS. We do this by examining the possible actions of $t_{\sigma_{m-1}} ... t_{\sigma_1}$ on $l_i^\mu$ and $l_j^\mu$.

First, suppose $t_{\sigma_{m-1}} ... t_{\sigma_1}$ acts as hard-collinear approximations on both $l_i^\mu$ and $l_j^\mu$. Then at the specific pinch surface(s) where $t_{\sigma_{m-1}} ... t_{\sigma_1}$ exerts these approximations, at least one of the following three configurations will emerge: (1) $l_i^\mu$ is lightlike and $l_j^\mu$ is hard; (2) $l_j^\mu$ is lightlike and $l_i^\mu$ is hard; (3) $l_i^\mu$ and $l_j^\mu$ are both lightlike, in different directions. No matter which configuration we encounter, according to our observation that all the pinch surfaces $\sigma_1,...,\sigma_{m-1}$ are contained in $\rho^{\left\{ \sigma_n...\sigma_1 \right\}}$, the momenta $t_{\sigma_{m-1}} ... t_{\sigma_1} l_i^\mu$ and $t_{\sigma_{m-1}} ... t_{\sigma_1} l_j^\mu$ in $\rho^{\left\{ \sigma_n...\sigma_1 \right\}}$ are either hard or lightlike, and if they are both lightlike, they cannot be in the same direction. As a result, the RHS of eq.\ (\ref{identical_t_sigma_hard}) is empty, and LHS $\subset$ RHS cannot occur.

For the same reason, we observe that $t_{\sigma_{m-1}} ... t_{\sigma_1}$ cannot act as soft-collinear approximations on both $l_i^\mu$ and $l_j^\mu$, otherwise we get the following implications: (1) $l_i^\mu$ is lightlike and $l_j^\mu$ is soft at some $\sigma_{m'}\ (1\leqslant m'\leqslant m-1)$; (2) $l_i^\mu$ is soft and $l_j^\mu$ is lightlike at some $\sigma_{m''}\ (1\leqslant m''\leqslant m-1)$. Apparently, $\sigma_{m'}$ and $\sigma_{m''}$ are not nested, thus cannot be the pinch surfaces appearing in $t_{\sigma_{m-1}} ... t_{\sigma_1}$.

Finally the only possibility left is that the approximation on $l_i^\mu$ is hard-collinear provided by $t_{\sigma_{m_1}}$, while that on $l_j^\mu$ is soft-collinear provided by $t_{\sigma_{m_2}}$ ($1\leqslant m_2 < m_1 \leqslant m-1$). Then we can rewrite eq.\ (\ref{identical_t_sigma_hard}) as
\begin{eqnarray} \label{identical_t_sigma_hard_hc_sc}
\mathcal{N}_{\rho^{\left\{\sigma_n...\sigma_1\right\}}} \left(l_i\right) \star \mathcal{N}_{\rho^{\left\{\sigma_n...\sigma_1\right\}}} \left(l_j\right) = \mathcal{N}_{\rho^{\left\{ \sigma_{n}...\sigma_1 \right\}}}\left( ^I\widehat{l}_i \right) \star \mathcal{N}_{\rho^{\left\{ \sigma_{n}...\sigma_1 \right\}}}\left( ^I\widetilde{l}_j \right),
\end{eqnarray}
where the hat and tilde are defined in (\ref{hat_tilde_definition}). We also know that in $\sigma_{m_1}$, $l_i^\mu$ is collinear to $\beta_I^\mu$ while $l_j^\mu$ is hard. As we have explained, $\sigma_{m_1} \subseteq \rho^{\left\{\sigma_n...\sigma_1\right\}}$, so $l_i^\mu$ is either collinear to $\beta_I^\mu$ or hard, while $l_j^\mu$ must be hard in $\rho^{\left\{\sigma_n...\sigma_1\right\}}$. The RHS of (\ref{identical_t_sigma_hard_hc_sc}) is then empty, and again LHS $\not\subset$ RHS. Therefore this possibility is also eliminated, and we have finished verifying (\ref{identical_t_sigma_hard}).
\end{itemize}

\begin{itemize}
\item[$\textit{(3)}$ ]{}For the case of $l_i^\mu$ and $l_j^\mu$ being lightlike in a certain direction in $\sigma_m$, say $\beta_I^\mu$, eq.\ (\ref{enc_to_prove'_generalize_appendx}) becomes the following relation, which we aim to prove:
\begin{eqnarray} \label{identical_t_sigma_collinear}
&&\left (\mathcal{N}^{(I)} \oplus \mathcal{N}_{\rho^{\left\{\sigma_n...\sigma_1\right\}}} \left(l_i\right) \right) \star \left (\mathcal{N}^{(I)} \oplus \mathcal{N}_{\rho^{\left\{\sigma_n...\sigma_1\right\}}} \left(l_j\right) \right) \nonumber\\
&&\hspace{1cm}=\mathcal{N}^{(I)} \oplus \left (\mathcal{N}_{\rho^{\left\{ \sigma_n...\sigma_1 \right\}}}\left( t_{\sigma_n}...t_{\sigma_1} l_i \right) \star \mathcal{N}_{\rho^{\left\{ \sigma_n...\sigma_1 \right\}}}\left( t_{\sigma_n}...t_{\sigma_1} l_j \right) \right).
\end{eqnarray}
To show this, we begin by observing that if there does not exist a lightlike vector $v^\mu(\neq \overline{\beta}_I^\mu)$ in the space $\mathcal{N}_{\rho^{\left\{ \sigma_n...\sigma_1 \right\}}}\left( t_{\sigma_n}...t_{\sigma_1} l_i \right) \star \mathcal{N}_{\rho^{\left\{ \sigma_n...\sigma_1 \right\}}}\left( t_{\sigma_n}...t_{\sigma_1} l_j \right)$, the RHS of (\ref{identical_t_sigma_collinear}) is equal to $\mathcal{N}^{(I)}$. (If $v^\mu(\neq \overline{\beta}_I^\mu)$ is not part of $\mathcal{N}_{\rho^{\left\{ \sigma_n...\sigma_1 \right\}}}\left( t_{\sigma_n}...t_{\sigma_1} l_i \right) \star \mathcal{N}_{\rho^{\left\{ \sigma_n...\sigma_1 \right\}}}\left( t_{\sigma_n}...t_{\sigma_1} l_j \right)$, one of $\mathcal{N}_{\rho^{\left\{ \sigma_n...\sigma_1 \right\}}}\left( t_{\sigma_n}...t_{\sigma_1} l_i \right)$ or $\mathcal{N}_{\rho^{\left\{ \sigma_n...\sigma_1 \right\}}}\left( t_{\sigma_n}...t_{\sigma_1} l_j \right)$ may still contain $v^\mu$ but not both.) Then because
\begin{eqnarray} \label{theorem4_proof_case3_a_lemma}
\mathcal{N}_{\rho^{\left\{ \sigma_n...\sigma_1 \right\}}} \left( t_{\sigma_n}...t_{\sigma_1} l\right) \supseteq \mathcal{N}_{\rho^{\left\{ \sigma_n...\sigma_1 \right\}}} \left(l\right)
\end{eqnarray}
for any $l^\mu$, either $\mathcal{N}_{\rho^{\left\{ \sigma_n...\sigma_1 \right\}}} \left(l_i\right)$ or $\mathcal{N}_{\rho^{\left\{ \sigma_n...\sigma_1 \right\}}} \left(l_j\right) $ does not contain $v^\mu$ as a result. Then the LHS will also be equal to $\mathcal{N}^{(I)}$, and (\ref{identical_t_sigma_collinear}), and hence (\ref{enc_to_prove'_generalize_appendx}), is valid.

We will now argue that eq.\ (\ref{identical_t_sigma_collinear}) holds in the alternative case,
\begin{eqnarray} \label{theorem4_proof_case3_assumption1}
\exists v^\mu\neq\overline{\beta}_I^\mu,\text{ }v^2=0,\ v^\mu \in \mathcal{N}_{\rho^{\left\{ \sigma_n...\sigma_1 \right\}}}\left( t_{\sigma_n}...t_{\sigma_1} l_i \right) \star \mathcal{N}_{\rho^{\left\{ \sigma_n...\sigma_1 \right\}}}\left( t_{\sigma_n}...t_{\sigma_1} l_j \right).
\end{eqnarray}

With eq.\ (\ref{theorem4_proof_case3_assumption1}) satisfied, the RHS of (\ref{identical_t_sigma_collinear}) is equal to the entire 4-dim space. Again, because of (\ref{theorem4_proof_case3_a_lemma}), the LHS is also the full space, unless
\begin{eqnarray} \label{theorem4_proof_case3_assumption2}
v^\mu \notin \mathcal{N}_{\rho^{\left\{ \sigma_n...\sigma_1 \right\}}} \left(l_i\right)\text{ and/or }\mathcal{N}_{\rho^{\left\{ \sigma_n...\sigma_1 \right\}}} \left(l_j\right).
\end{eqnarray}
We now show that eq.\ (\ref{theorem4_proof_case3_assumption2}) is inconsistent with (\ref{theorem4_proof_case3_assumption1}), and hence that (\ref{identical_t_sigma_collinear}) holds in this case as well. To proceed, we suppose that $v^\mu \notin \mathcal{N}_{\rho^{\left\{ \sigma_n...\sigma_1 \right\}}} \left(l_i\right)$. Taken together, (\ref{theorem4_proof_case3_assumption1}) and (\ref{theorem4_proof_case3_assumption2}) show that $t_{\sigma_n}...t_{\sigma_1}$ cannot be an identity operator on $l_i^\mu$. Since $l_i^\mu$ and $l_j^\mu$ are collinear to $\beta_I^\mu$ in $\sigma_m$, $t_{\sigma_n}...t_{\sigma_1}$ either acts as an $\text{hc}_I$ or $\text{sc}_I$ on $l_i^\mu$. However, the approximation cannot be $\text{hc}_I$, otherwise it would produce a $\overline{\beta}_I^\mu$ in $\mathcal{N}_{\rho^{\left\{ \sigma_n...\sigma_1 \right\}}}\left( t_{\sigma_n}...t_{\sigma_1} l_i \right)$. By comparing that $v^\mu \in \mathcal{N}_{\rho^{\left\{ \sigma_n...\sigma_1 \right\}}}\left( t_{\sigma_n}...t_{\sigma_1} l_i \right)$ and $v^\mu \notin \mathcal{N}_{\rho^{\left\{ \sigma_n...\sigma_1 \right\}}} \left(l_i\right)$, we would have $v^\mu= \overline{\beta}_I^\mu$, which is ruled out by (\ref{theorem4_proof_case3_assumption1}). In other words, we must have
\begin{eqnarray}
t_{\sigma_n}...t_{\sigma_1} l_i^\mu = \left( l_i\cdot\beta_I \right) \overline{\beta}_I^\mu,\text{ and }v^\mu = \beta_I^\mu,
\end{eqnarray}
which means that $t_{\sigma_n}...t_{\sigma_1}$ acts as $\text{sc}_I$ from some $t_{\sigma_{m'}}$, where $\sigma_{m'}\in \left\{ \sigma_n, ..., \sigma_1 \right\}$, and $\sigma_{m'} \subset \sigma_m$. Since $\sigma_m$ is the smallest pinch surface not contained in $\rho^{\left\{ \sigma_n...\sigma_1 \right\}}$, we also know that $\sigma_{m'} \subseteq \rho^{\left\{ \sigma_n...\sigma_1 \right\}}$.

In summary, we know that $l_i^\mu$ is soft while $l_j^\mu$ is collinear to $\beta_I^\mu$ in $\sigma_{m'}$. So $l_j^\mu$ is either collinear to $\beta_I^\mu$ or hard in $\rho^{\left\{ \sigma_n...\sigma_1 \right\}}$. Meanwhile, from eq.\ (\ref{theorem4_proof_case3_assumption1}), $v^\mu= \beta_I^\mu \in \mathcal{N}_{\rho^{\left\{ \sigma_n...\sigma_1 \right\}}}\left( t_{\sigma_n}...t_{\sigma_1} l_j \right)$, so $t_{\sigma_n}...t_{\sigma_1} l_j^\mu$ must be either collinear to $\overline{\beta}_I^\mu$, or soft in $\rho^{\left\{ \sigma_n...\sigma_1 \right\}}$.

The action of $t_{\sigma_n}...t_{\sigma_1}$ on $l_j^\mu$ hence also cannot be the identity operator, and is either $\text{hc}_I$ or $\text{sc}_I$. By comparing the possibilities of $l_j^\mu$ and $t_{\sigma_n}...t_{\sigma_1} l_j^\mu$ in $\rho^{\left\{ \sigma_n...\sigma_1 \right\}}$, namely, $l_j^\mu$ is either collinear to $\beta_I^\mu$ or hard, while $t_{\sigma_n}...t_{\sigma_1} l_j^\mu$ is either collinear to $\overline{\beta}_I^\mu$ or soft, we conclude that $t_{\sigma_n}...t_{\sigma_1}$ also acts as $\text{sc}_I$ on $l_j^\mu$. In other words,
\begin{eqnarray}
t_{\sigma_n}...t_{\sigma_1} l_j^\mu = \left( l_j\cdot\beta_I \right) \overline{\beta}_I^\mu.
\end{eqnarray}

Now we come to a contradiction: eq.\ (\ref{theorem4_proof_case3_assumption2}) is not consistent with (\ref{theorem4_proof_case3_assumption1}). The reason is that $t_{\sigma_n}...t_{\sigma_1}$ cannot simultaneously provide two $\text{sc}_I$'s on both $l_i^\mu$ and $l_j^\mu$, otherwise the pinch surfaces in the forest $\left\{ \sigma_n, ..., \sigma_1 \right\}$ will not be nested. This has been explained in \textit{(2)} above. Thus eq.\ (\ref{identical_t_sigma_collinear}), and hence (\ref{enc_to_prove'_generalize_appendx}), hold in this case as well.

\end{itemize}

In conclusion, we have shown that Theorem 4 holds, whenever $t_{\sigma_m}$ is an identity operator on the loop momenta whose linear combination gives the momentum of a propagator.
\bigbreak
\textbf{2, $t_{\sigma_m} \neq \text{id}$ in eq.\ (\ref{enc_to_prove'_generalize_appendx}). }When $t_{\sigma_m}$ provides approximations on $l_i^\mu$ or $l_j^\mu$, the possible configurations of $l_i^\mu$ and $l_j^\mu$ have been shown in figure\ \ref{momenta_confluence}, which we draw here again.
\begin{figure}[t]
\centering
\includegraphics[width=13cm]{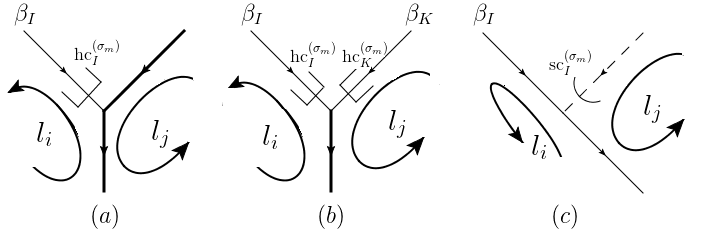}
\caption{The three cases where the loop momenta $l_i^\mu$ or $l_j^\mu$ (or both) are projected by the approximation operator $t_{\sigma_m}$.}
\label{momenta_confluence_appendx}
\end{figure}

The treatment on these configurations is similar to that in eqs.\ (\ref{enclosed_normal_coordinates_requirement1})--(\ref{enclosed_normal_coordinates_requirement2}), That is, for figure\ \ref{momenta_confluence_appendx}$(a)$ we combine the definition of the $\star$-symbol in table\ \ref{star_definition} with the rules of $\text{hc}_I^{(\sigma)}$, i.e. eq.\ (\ref{hard-collinear_approximation}), to rewrite (\ref{enc_to_prove'_generalize_appendx}) as
\begin{eqnarray} \label{enclosed_normal_coordinates_requirement1_appendx}
&&\left (\mathcal{N}^{(I)} \oplus \mathcal{N}_{\rho^{\left\{ \sigma_n...\sigma_1 \right\}}} \left(l_i\right) \right) \star \mathcal{N}_{\rho^{\left\{ \sigma_n...\sigma_1 \right\}}} \left(l_j\right) \nonumber\\
&&\hspace{1cm} =\mathcal{N}_{\rho^{ \left\{ \sigma_n...\sigma_1 \right\} }}\left( t_{\sigma_n}...t_{\sigma_1} l_i \right) \star \mathcal{N}_{\rho^{ \left\{ \sigma_n...\sigma_1 \right\} }} \left (t_{\sigma_n}...t_{\sigma_1} l_j \right).
\end{eqnarray}
Similarly for figure\ \ref{momenta_confluence_appendx}$(b)$, eq.\ (\ref{enc_to_prove'_generalize_appendx}) becomes
\begin{eqnarray} \label{enclosed_normal_coordinates_requirement1'_appendx}
&&\left (\mathcal{N}^{(I)} \oplus \mathcal{N}_{\rho^{\left\{ \sigma_n...\sigma_1 \right\}}} \left(l_i\right) \right) \star \left (\mathcal{N}^{(K)} \oplus \mathcal{N}_{\rho^{\left\{ \sigma_n...\sigma_1 \right\}}} \left(l_j\right) \right) \nonumber\\
&&\hspace{1cm} =\mathcal{N}_{\rho^{ \left\{ \sigma_n...\sigma_1 \right\} }}\left( t_{\sigma_n}...t_{\sigma_1} l_i \right) \star \mathcal{N}_{\rho^{ \left\{ \sigma_n...\sigma_1 \right\} }} \left( t_{\sigma_n}...t_{\sigma_1} l_j \right).
\end{eqnarray}
For figure\ \ref{momenta_confluence_appendx}$(c)$, eq.\ (\ref{enc_to_prove'_generalize_appendx}) becomes
\begin{eqnarray} \label{enclosed_normal_coordinates_requirement2_appendx}
\mathcal{N}^{(I)} \oplus \mathcal{N}_{\rho^{\left\{ \sigma_n...\sigma_1 \right\}}} \left(l_i\right) = \mathcal{N}^{(I)} \oplus \left( \mathcal{N}_{\rho^{ \left\{ \sigma_n...\sigma_1 \right\} }}\left( t_{\sigma_n}...t_{\sigma_1} l_i \right) \star \mathcal{N}_{\rho^{ \left\{ \sigma_n...\sigma_1 \right\} }} \left( t_{\sigma_n}...t_{\sigma_1} l_j \right) \right).
\end{eqnarray}

In the following, we will verify eqs.\ (\ref{enclosed_normal_coordinates_requirement1_appendx})--(\ref{enclosed_normal_coordinates_requirement2_appendx}) for an arbitrary forest. Compared with our analysis in section\ \ref{subgraphs_enclosed_pinch_surfaces}, now we need to consider the approximations from other pinch surfaces $\sigma_i\ (i\neq m)$. This can be done by brute force: for each configuration in figure\ \ref{momenta_confluence_appendx}, we list all the possibilities of the net effect of $t_{\sigma_n}...t_{\sigma_1}$, and for each possibility we can rewrite (\ref{enc_to_prove'_generalize_appendx}) into an explicit form, to check whether it is correct. For example, with the presence of an $\text{hc}_I$, there are six possibilities corresponding to figure\ \ref{momenta_confluence_appendx}$(a)$, which are shown in figure\ \ref{case_a_net_approximations}. Eq.\ (\ref{enc_to_prove'_generalize_appendx}) can then be written into (\ref{enc_to_prove'_generalize_appendx_a1})--(\ref{enc_to_prove'_generalize_appendx_a6}), using (\ref{hat_plus_tilde_definition}) for $^{KI}\widetilde{\widehat{l}}_i$:
\begin{figure}[t]
\centering
\includegraphics[width=15.1cm]{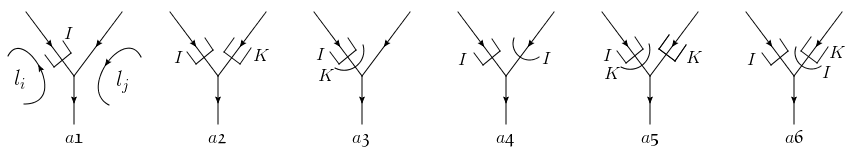}
\caption{The six possibilities of net approximations, whose associated pinch surfaces are nested with figure\ \ref{momenta_confluence_appendx}$(a)$.}
\label{case_a_net_approximations}
\end{figure}
{
\allowdisplaybreaks
\begin{eqnarray}
\label{enc_to_prove'_generalize_appendx_a1}
a\mathfrak{1}: &&\left (\mathcal{N}^{(I)} \oplus \mathcal{N}_{\rho^{(a\mathfrak{1})}} \left(l_i\right) \right) \star \mathcal{N}_{\rho^{(a\mathfrak{1})}} \left(l_j\right) = \mathcal{N}_{\rho^{(a\mathfrak{1})}} \left( ^{I}\widehat{l}_i \right) \star \mathcal{N}_{\rho^{(a\mathfrak{1})}} \left( l_j \right), \\
\label{enc_to_prove'_generalize_appendx_a2}
a\mathfrak{2}: &&\left (\mathcal{N}^{(I)} \oplus \mathcal{N}_{\rho^{(a\mathfrak{2})}} \left(l_i\right) \right) \star \mathcal{N}_{\rho^{(a\mathfrak{2})}} \left(l_j\right) = \mathcal{N}_{\rho^{(a\mathfrak{2})}} \left( ^{I}\widehat{l}_i \right) \star \mathcal{N}_{\rho^{(a\mathfrak{2})}} \left( ^{K}\widehat{l}_j \right), \\
\label{enc_to_prove'_generalize_appendx_a3}
a\mathfrak{3}: &&\left (\mathcal{N}^{(I)} \oplus \mathcal{N}_{\rho^{(a\mathfrak{3})}} \left(l_i\right) \right) \star \mathcal{N}_{\rho^{(a\mathfrak{3})}} \left(l_j\right) = \mathcal{N}_{\rho^{(a\mathfrak{3})}} \left( ^{KI}\widetilde{\widehat{l}}_i \right) \star \mathcal{N}_{\rho^{(a\mathfrak{3})}} \left( l_j \right), \\
\label{enc_to_prove'_generalize_appendx_a4}
a\mathfrak{4}: &&\left (\mathcal{N}^{(I)} \oplus \mathcal{N}_{\rho^{(a\mathfrak{4})}} \left(l_i\right) \right) \star \mathcal{N}_{\rho^{(a\mathfrak{4})}} \left(l_j\right) = \mathcal{N}_{\rho^{(a\mathfrak{4})}} \left( ^{I}\widehat{l}_i \right) \star \mathcal{N}_{\rho^{(a\mathfrak{4})}} \left( ^{I}\widetilde{l}_j \right), \\
\label{enc_to_prove'_generalize_appendx_a5}
a\mathfrak{5}: &&\left (\mathcal{N}^{(I)} \oplus \mathcal{N}_{\rho^{(a\mathfrak{5})}} \left(l_i\right) \right) \star \mathcal{N}_{\rho^{(a\mathfrak{5})}} \left(l_j\right) = \mathcal{N}_{\rho^{(a\mathfrak{5})}} \left( ^{KI}\widetilde{\widehat{l}}_i \right) \star \mathcal{N}_{\rho^{(a\mathfrak{5})}} \left( ^{K}\widehat{l}_j \right), \\
\label{enc_to_prove'_generalize_appendx_a6}
a\mathfrak{6}: &&\left (\mathcal{N}^{(I)} \oplus \mathcal{N}_{\rho^{(a\mathfrak{6})}} \left(l_i\right) \right) \star \mathcal{N}_{\rho^{(a\mathfrak{6})}} \left(l_j\right) = \mathcal{N}_{\rho^{(a\mathfrak{6})}} \left( ^{I}\widehat{l}_i \right) \star \mathcal{N}_{\rho^{(a\mathfrak{6})}} \left( ^{IK}\widetilde{\widehat{l}}_j \right).
\end{eqnarray}
}
In these relations, which we shall verify below, we have used the notations of the figures as the upper indices of $\rho$ to denote pinch surfaces of certain approximated amplitudes. For example, $\rho^{(a\mathfrak{1})}$ is a pinch surface of the approximated amplitude whose approximation on $l_i^\mu$ is a single $\text{hc}_I$, as is shown in figure\ \ref{case_a_net_approximations}$(a\mathfrak{1})$.

\begin{figure}[t]
\centering
\includegraphics[width=8cm]{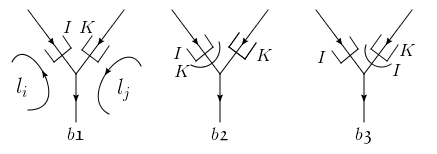}
\caption{The three possibilities of net approximations, whose associated pinch surfaces are nested with figure\ \ref{momenta_confluence_appendx}$(b)$.}
\label{case_b_net_approximations}
\end{figure}

Considering the pinch surfaces nested with each other including figure\ \ref{momenta_confluence_appendx}$(b)$, there are three possibilities of the net approximation, as are shown in figure\ \ref{case_b_net_approximations}.
From these possibilities, eq.\ (\ref{enc_to_prove'_generalize_appendx}) can be written as (\ref{enc_to_prove'_generalize_appendx_b1})--(\ref{enc_to_prove'_generalize_appendx_b3}), using $\mathcal{N}^{(I)} \star \mathcal{N}^{(K)} =\varnothing\ (I\neq K)$:
\begin{eqnarray}
\label{enc_to_prove'_generalize_appendx_b1}
b\mathfrak{1}: &&\left (\mathcal{N}^{(I)} \oplus \mathcal{N}_{\rho^{(b\mathfrak{1})}} \left(l_i\right) \right) \star \left (\mathcal{N}^{(K)} \oplus \mathcal{N}_{\rho^{(b\mathfrak{1})}} \left(l_j\right) \right) = \mathcal{N}_{\rho^{(b\mathfrak{1})}}\left( ^{I}\widehat{l}_i \right) \star \mathcal{N}_{\rho^{(b\mathfrak{1})}} \left( ^{K}\widehat{l}_j \right), \\
\label{enc_to_prove'_generalize_appendx_b2}
b\mathfrak{2}: &&\left (\mathcal{N}^{(I)} \oplus \mathcal{N}_{\rho^{(b\mathfrak{2})}} \left(l_i\right) \right) \star \left (\mathcal{N}^{(K)} \oplus \mathcal{N}_{\rho^{(b\mathfrak{2})}} \left(l_j\right) \right) = \mathcal{N}_{\rho^{(b\mathfrak{2})}} \left( ^{KI}\widetilde{\widehat{l}}_i \right) \star \mathcal{N}_{\rho^{(b\mathfrak{2})}} \left( ^{K}\widehat{l}_j \right), \\
\label{enc_to_prove'_generalize_appendx_b3}
b\mathfrak{3}: &&\left (\mathcal{N}^{(I)} \oplus \mathcal{N}_{\rho^{(b\mathfrak{3})}} \left(l_i\right) \right) \star \left (\mathcal{N}^{(K)} \oplus \mathcal{N}_{\rho^{(b\mathfrak{3})}} \left(l_j\right) \right) = \mathcal{N}_{\rho^{(b\mathfrak{3})}} \left( ^{I}\widehat{l}_i \right) \star \mathcal{N}_{\rho^{(b\mathfrak{3})}} \left( ^{IK}\widetilde{\widehat{l}}_j \right).
\end{eqnarray}

Finally, for figure\ \ref{momenta_confluence_appendx}$(c)$, there are four possibilities for the net approximation, as are shown in figure\ \ref{case_c_net_approximations}. Similarly, eq.\ (\ref{enc_to_prove'_generalize_appendx}) can be written as (\ref{enc_to_prove'_generalize_appendx_c1})--(\ref{enc_to_prove'_generalize_appendx_c4}), where we use that $\mathcal{N}_{\sigma_m} \left( l_i \right)= \mathcal{N}^{(I)}$ and $\mathcal{N}_{\sigma_m} \left( l_j \right)= \mathcal{N}^{(\text{soft})}$:
\begin{figure}[t]
\centering
\includegraphics[width=10.5cm]{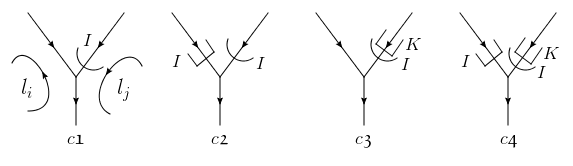}
\caption{The four possibilities of net approximations, whose associated pinch surfaces are nested with figure\ \ref{momenta_confluence_appendx}$(c)$, using $\mathcal{N}_{\sigma_m} \left( l_j \right)= \mathcal{N}^{(\text{soft})}$ here.}
\label{case_c_net_approximations}
\end{figure}
\begin{eqnarray}
\label{enc_to_prove'_generalize_appendx_c1}
c\mathfrak{1}: &&\mathcal{N}^{(I)} \oplus \mathcal{N}_{\rho^{(c\mathfrak{1})}} \left(l_i\right) = \mathcal{N}^{(I)} \oplus \left( \mathcal{N}_{\rho^{(c\mathfrak{1})}} \left( l_i \right) \star \mathcal{N}_{\rho^{(c\mathfrak{1})}} \left( ^{I}\widetilde{l}_j \right) \right), \\
\label{enc_to_prove'_generalize_appendx_c2}
c\mathfrak{2}: &&\mathcal{N}^{(I)} \oplus \mathcal{N}_{\rho^{(c\mathfrak{2})}} \left(l_i\right) = \mathcal{N}^{(I)} \oplus \left( \mathcal{N}_{\rho^{(c\mathfrak{2})}} \left( ^I\widehat{l}_i \right) \star \mathcal{N}_{\rho^{(c\mathfrak{2})}} \left( ^I\widetilde{l}_j \right) \right), \\
\label{enc_to_prove'_generalize_appendx_c3}
c\mathfrak{3}: &&\mathcal{N}^{(I)} \oplus \mathcal{N}_{\rho^{(c\mathfrak{3})}} \left(l_i\right) = \mathcal{N}^{(I)} \oplus \left( \mathcal{N}_{\rho^{(c\mathfrak{3})}} \left( l_i \right) \star \mathcal{N}_{\rho^{(c\mathfrak{3})}} \left( ^{IK}\widetilde{\widehat{l}}_j \right) \right), \\
\label{enc_to_prove'_generalize_appendx_c4}
c\mathfrak{4}: &&\mathcal{N}^{(I)} \oplus \mathcal{N}_{\rho^{(c\mathfrak{4})}} \left(l_i\right) = \mathcal{N}^{(I)} \oplus \left( \mathcal{N}_{\rho^{(c\mathfrak{4})}} \left( ^{I}\widehat{l}_i \right) \star \mathcal{N}_{\rho^{(c\mathfrak{4})}} \left( ^{IK}\widetilde{\widehat{l}}_j \right) \right).
\end{eqnarray}

In order to simplify the verification of eqs.\ (\ref{enc_to_prove'_generalize_appendx_a1})--(\ref{enc_to_prove'_generalize_appendx_c4}), we introduce the following two lemmas.

\textbf{Lemma 1: For a pinch surface $\rho^{ \left\{ ...\sigma_m ...\right\} }$ of an approximated amplitude, and a momentum $l^\mu$ which is collinear to $\beta_I^\mu$ in $\sigma_m$, we have}
\begin{eqnarray} \label{lemma_1_appendx}
\mathcal{N}^{(I)} \oplus \mathcal{N}_{\rho^{ \left\{ ...\sigma_m ...\right\} }} \left(l\right) = \mathcal{N}_{\rho^{ \left\{ ...\sigma_m ...\right\} }} \left(^I\widehat{l} \right).
\end{eqnarray}
\textit{Proof: }Since $l^\mu$ is collinear to $\beta_I^\mu$ in $\sigma_m$, from Theorem 2 in section\ \ref{subtraction_terms} we know that $l^\mu$ can only be hard, soft, collinear to $\beta_I^\mu$ or $\overline{\beta}_I^\mu$ in $\rho^{ \left\{ ...\sigma_m ...\right\} }$. If it is hard or collinear to $\beta_I^\mu$, then both sides are equal to $\mathcal{N}^{(I)}$. If it is soft or collinear to $\overline{\beta}_I^\mu$, then both sides are equal to the entire 4-dim space. So this relation always holds.

\textbf{Lemma 2: Given momenta $l_1^\mu$ and $l_2^\mu$, and $\rho^{(...)}$ being a pinch surface of either an amplitude or an approximated amplitude, if $l_1^\mu$ is either hard or collinear to $\beta_I^\mu$ in $\rho^{(...)}$, then for any $l_2^\mu$,}
\begin{eqnarray} \label{lemma_2_appendx}
\mathcal{N}_{\rho^{(...)}} \left( l_1 \right) \star \mathcal{N}_{\rho^{(...)}} \left( l_2 \right) &&= \mathcal{N}_{\rho^{(...)}} \left( l_1 \right) \star \mathcal{N}_{\rho^{(...)}} \left( ^{I}\widetilde{l}_2 \right)\\
&&=\left\{\begin{matrix}
\ \ \varnothing\ ,\ \ \text{for }^I\widetilde{l}_2 \neq 0, \nonumber\\ 
\mathcal{N}^{(I)},\ \ \text{for }^I\widetilde{l}_2 = 0, \nonumber
\end{matrix}\right.
\end{eqnarray}
where $^I\widetilde{l}_2 = 0$ implies either $l_2^\mu =0$ or $l_2^\mu \propto \beta_I^\mu$.

\textit{Proof: }The relation is trivial for $l_1^\mu$ hard, where both sides of eq.\ (\ref{lemma_2_appendx}) are $\varnothing$, so we consider $l_1^\mu$ collinear to $\beta_I^\mu$. We enumerate the types of $l_2^\mu$ in $\rho^{(...)}$. If it is soft, then $\mathcal{N}_{\rho^{(...)}} \left(l_2\right)= \mathcal{N}_{\rho^{(...)}} \left(^I\widetilde{l}_2\right)$, and the two sides automatically coincide. If it is hard or collinear to any direction other than $\beta_I^\mu$, then both sides are equal to $\varnothing$. Finally, if it is collinear to $\beta_I^\mu$, then both sides are equal to $\mathcal{N}^{(I)}$.

With the help of these lemmas, we are prepared to verify eqs.\ (\ref{enc_to_prove'_generalize_appendx_a1})--(\ref{enc_to_prove'_generalize_appendx_c4}) directly. We recall again that since $\sigma_m$ is the smallest pinch surface that is not contained in $\rho^{\left\{ \sigma_n...\sigma_1 \right\}}$, for every $\rho^{(...)}$ that we will encounter below, it contains $\sigma_k$ as long as $k<m$.

\begin{itemize}
  \item $a\mathfrak{1}:$ Eq.\ (\ref{enc_to_prove'_generalize_appendx_a1}) is the same with (\ref{enclosed_normal_coordinates_requirement1}) which we have already shown in section\ \ref{subgraphs_enclosed_pinch_surfaces}.
\end{itemize}
\begin{itemize}
  \item $a\mathfrak{2}:$ This net approximation in the figure is obtained by combining $t_{\sigma_m}$ and another operator $t_{\sigma_k}$ ($1\leqslant k< m$), with $\sigma_k$ being a pinch surface where $l_i^\mu$ is collinear to $\beta_I^\mu$ and $l_j^\mu$ is collinear to $\beta_K^\mu$. Since $\sigma_k \subseteq \rho^{(a\mathfrak{2})}$, $l_i^\mu$ is either collinear to $\beta_I^\mu$ or hard in $\rho^{(a\mathfrak{2})}$, while $l_j^\mu$ is either collinear to $\beta_K^\mu$ or hard in $\rho^{(a\mathfrak{2})}$.
  
  Then We apply Lemma 1 to eq.\ (\ref{enc_to_prove'_generalize_appendx_a2}) to rewrite the term in the bracket on the LHS as $\mathcal{N}_{\rho^{(a\mathfrak{2})}} \left( ^{I}\widehat{l}_i \right)$. As can be inferred from the previous paragraph, both sides of (\ref{enc_to_prove'_generalize_appendx_a2}) are equal to $\varnothing$, either from $\mathcal{N}^{(I)} \star \mathcal{N}^{(K)} =\varnothing$ or $\varnothing \star \mathcal{N}^{(K)} =\varnothing$.
\end{itemize}
\begin{itemize}
  \item $a\mathfrak{3}:$ To obtain this net approximation, we need a pinch surface $\sigma_k$ ($1\leqslant k< m$) with $l_i^\mu$ soft and $l_j^\mu$ collinear to $\beta_K^\mu$ there. So in $\rho^{(a\mathfrak{2})}$, $l_j^\mu$ is either hard or collinear to $\beta_K^\mu$. Then we apply Lemma 2 to the RHS of eq.\ (\ref{enc_to_prove'_generalize_appendx_a3}), using (\ref{hat_plus_tilde_definition}) for $^{KI}\widetilde{\widehat{l}}_i$, after which we get our previous result $a\mathfrak{1}$, eq.\ (\ref{enc_to_prove'_generalize_appendx_a1}).
\end{itemize}
\begin{itemize}
  \item $a\mathfrak{4}:$ Here we need a pinch surface $\sigma_k$ ($1\leqslant k< m$) with $l_i^\mu$ collinear to $\beta_I^\mu$ and $l_j^\mu$ soft there. So $l_i^\mu$ is either hard or collinear to $\beta_I^\mu$ in $\rho^{(a\mathfrak{4})}$. Again we apply Lemma 2 to the RHS of eq.\ (\ref{enc_to_prove'_generalize_appendx_a4}), and it returns to case $a\mathfrak{1}$.
\end{itemize}
\begin{itemize}
  \item $a\mathfrak{5}:$ We compare this case with $a\mathfrak{2}$. Here everything is the same except that in eq.\ (\ref{enc_to_prove'_generalize_appendx_a5}) we have $^{KI}\widetilde{\widehat{l}}_i$ rather than $^I\widehat{l}_i$ in (\ref{enc_to_prove'_generalize_appendx_a2}). But this difference is eliminated by applying Lemma 2 to the RHS, with $l_1= ^K\widehat{l}_j$ in (\ref{lemma_2_appendx}).
\end{itemize}
\begin{itemize}
  \item $a\mathfrak{6}:$ Again we compare this case with $a\mathfrak{2}$. Everything is the same except that in eq.\ (\ref{enc_to_prove'_generalize_appendx_a6}) we have $^{IK}\widetilde{\widehat{l}}_j$ rather than $^K\widehat{l}_j$ in (\ref{enc_to_prove'_generalize_appendx_a2}). But this difference is eliminated by applying Lemma 2 to the RHS, as well, with $l_1= ^I\widehat{l}_i$ in (\ref{lemma_2_appendx}).
\end{itemize}
\begin{itemize}
  \item $b\mathfrak{1}:$ Eq.\ (\ref{enc_to_prove'_generalize_appendx_b1}) is the same with (\ref{enclosed_normal_coordinates_requirement1'}) which we have already shown in section\ \ref{subgraphs_enclosed_pinch_surfaces}.
\end{itemize}
\begin{itemize}
  \item $b\mathfrak{2}:$ To obtain the net approximation in the figure, we need a pinch surface $\sigma_k$ ($1\leqslant k< m$) with $l_i^\mu$ being soft and $l_j^\mu$ being collinear to $\beta_K^\mu$. So $l_j^\mu$ is either hard or collinear to $\beta_K^\mu$ in $\rho^{(b\mathfrak{2})}$. By applying Lemma 2 to the RHS of eq.\ (\ref{enc_to_prove'_generalize_appendx_b2}), this relation returns to (\ref{enc_to_prove'_generalize_appendx_b1}) above.
\end{itemize}
\begin{itemize}
  \item $b\mathfrak{3}:$ This case can be treated in the same way as $b\mathfrak{2}$ above, by exchanging $\beta_I^\mu$ and $\beta_K^\mu$.
\end{itemize}
\begin{itemize}
  \item $c\mathfrak{1}:$ Eq.\ (\ref{enc_to_prove'_generalize_appendx_c1}) is the same with (\ref{enclosed_normal_coordinates_requirement2}) which we have already shown in section\ \ref{subgraphs_enclosed_pinch_surfaces}.
\end{itemize}
\begin{itemize}
  \item $c\mathfrak{2}:$ We focus on the brackets of the RHSs of eqs.\ (\ref{enc_to_prove'_generalize_appendx_c2}) and (\ref{enc_to_prove'_generalize_appendx_c1}). Obviously the only difference lies in the $\text{hc}_I$ on $l_i^\mu$. Therefore, if the two brackets are different due to this approximation, the difference must be within $\mathcal{N}^{(I)}$, because $\mathcal{N}_\rho \left( ^I\widehat{l}_i \right) \subseteq \mathcal{N}_\rho \left( l_i \right)$. Meanwhile, since they are both in the direct sums with $\mathcal{N}^{(I)}$, the difference is eliminated. In other words, the two relations (\ref{enc_to_prove'_generalize_appendx_c2}) and (\ref{enc_to_prove'_generalize_appendx_c1}) are identical.
\end{itemize}
\begin{itemize}
  \item $c\mathfrak{3}:$ From the configuration of the net approximation, we know there is a pinch surface $\sigma_k$ ($m<k \leqslant n$) where $l_j^\mu$ is collinear to $\beta_K^\mu$. From the result of Theorem 2 in section\ \ref{subtraction_terms}, $l_j^\mu$ can only be hard, soft, collinear to $\beta_K^\mu$ or $\overline{\beta}_K^\mu$ in $\rho^{(c\mathfrak{3})}$. Now we compare the brackets on the RHSs of eqs.\ (\ref{enc_to_prove'_generalize_appendx_c1}) and (\ref{enc_to_prove'_generalize_appendx_c3}). They can be different only when $l_j^\mu$ is collinear to $\overline{\beta}_K^\mu$, with the difference contained in $\mathcal{N}^{(I)}$. It is again eliminated since both the brackets are in the direct sums with $\mathcal{N}^{(I)}$. In other words, we return to our previous result (\ref{enc_to_prove'_generalize_appendx_c1}).
\end{itemize}
\begin{itemize}
  \item $c\mathfrak{4}:$ The comparison between eqs.\ (\ref{enc_to_prove'_generalize_appendx_c4}) and (\ref{enc_to_prove'_generalize_appendx_c3}) is exactly the same as that between (\ref{enc_to_prove'_generalize_appendx_c2}) and (\ref{enc_to_prove'_generalize_appendx_c1}). So we can directly use the analysis in $c\mathfrak{2}$ to prove it.
\end{itemize}

We have finished the discussion on all the possibilities of $t_{\sigma_m}$ in eq.\ (\ref{enc_to_prove'_generalize_appendx}). The analysis throughout this appendix is also sufficient for non-planar graphs, or other loop assignments in a planar graph. The reason has already been explained in (\ref{enc_to_prove'_nonplanar})--(\ref{enc_to_prove'_nonplanar_end}), so we do not repeat it here. In conclusion, we have verified Theorem 4, which implies the relations between subgraphs in eq.\ (\ref{enc_subgraph_relations_generalize}).

\section{Interpretations in position space}\label{interpretation_position_space}

This appendix aims to provide the position-space aspects of sections\ \ref{pinch_surface_amplitudes_approximations}--\ref{the_proof_of_cancellation} \cite{EdgStm15}. In position space, IR divergences are long-distance, which come from the integration measure over the four-coordinates. Meanwhile, given any integral representing a hard QCD process with massless partons, we can multiply every momentum component by a scale factor without changing the form of the integrand. Due to this scale invariance, the pictures of IR divergences for massless partons should be related to those of UV divergences. To this extent, in order to identify the IR divergences in position space, it suffices to study where the integrand is pinched, which by definition corresponds to the UV divergences. However, we emphasize that the momentum-space analysis is not simply a Fourier transformation from that in position-space, because it shows how the pinch surfaces not in the original amplitude $\mathcal{A}$ can emerge, and how loop momenta behave nearby, which is still necessary to understand factorization.

\begin{figure}[t]
\centering
\includegraphics[width=10cm]{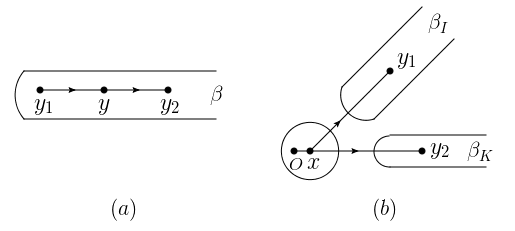}
\caption{Examples of the formation of pinches in position space.}
\label{pinches_position_space}
\end{figure}

First we study the pinches formed between parallel propagators. Referring to figure\ \ref{pinches_position_space}$(a)$, suppose $0$, $y_1^\mu$ and $y_2^\mu$ are all vertices of jet $J$ in the direction of $\beta^\mu$, i.e. $y_1^\mu \sim \left( y_1\cdot\overline{\beta} \right)\beta^\mu$ and $y_2^\mu \sim \left( y_2\cdot\overline{\beta} \right)\beta^\mu$, then for any intermediate vertex $y^\mu$ joining $y_1^\mu$ and $y_2^\mu$ through propagators, there is a factor in the denominator given by:
\begin{eqnarray}
&&\left [ -2\left( \left( y_1-y \right)\cdot\overline{\beta} \right)\left( \left( y_1-y \right)\cdot\beta \right)+\left( \mathbf{y}_{1\perp}-\mathbf{y}_\perp \right)^2 +i\epsilon \right] \nonumber\\
&&\hspace{1cm} \cdot\left [ -2\left( \left( y-y_2 \right)\cdot\overline{\beta} \right)\left( \left( y-y_2 \right)\cdot\beta \right)+\left( \mathbf{y}_\perp-\mathbf{y}_{2\perp} \right)^2 +i\epsilon \right ].
\end{eqnarray}
As long as $y_1\cdot\overline{\beta}< y\cdot\overline{\beta}< y_2\cdot\overline{\beta}$ (or reversed), the coordinates $(y\cdot\beta)$ and $\mathbf{y}_\perp$ are pinched at zero. At this pinch surface both $y_1y$ and $yy_2$ are jet propagators of $J$ (see figure\ \ref{pinches_position_space}$(a)$).

Then we study the pinches formed by the intersection of two jets. Suppose $y_1^\mu$ and $y_2^\mu$ are vertices in different jets $J_I$ and $J_K$, which are in the directions of $\beta_I^\mu$ and $\beta_K^\mu$. Meanwhile, $x^\mu$ is another vertex which is connected to $y_1$, $y_2$ and $0$ through propagators. Then in the denominator we have the following factor:
\begin{eqnarray}
\left( -x^2+i\epsilon \right) \left[ -\left(x-y_1\right)^2+i\epsilon \right] \left[ -\left(x-y_2\right)^2+i\epsilon \right].
\end{eqnarray}
A direct check shows that all the four components of $y^\mu$ can be pinched at zero. At this pinch surface $xy_1$ and $xy_2$ are separately jet propagators of $J_I$ and $J_K$, while $x^\mu$ is in the hard subgraph (see figure\ \ref{pinches_position_space}$(b)$).

In terms of the language of pinch surfaces, given any propagator $xy$ in an amplitude, all the four components of $\left( x-y \right)^\mu$ are normal coordinates if $xy$ is hard; its components $\left( x-y \right)\cdot \beta$ and $\mathbf{x}_\perp- \mathbf{y}_\perp$ are normal coordinates if it lightlike in the direction of $\beta^\mu$; none of its components are pinched if it is soft. We can also scale these normal coordinates as we have in eq.\ (\ref{momenta_scaling}), assuming the length scale $L\sim 1/Q$:
\begin{eqnarray} \label{positions_scaling}
\text{hard:}&& \left( x-y \right)^\mu \sim \left( \lambda, \lambda, \lambda, \lambda \right)L; \nonumber\\
\text{collinear:}&& \left( x-y \right)^\mu= \Big( (x-y)\cdot\overline{\beta},(x-y)\cdot\beta, (\mathbf{x}-\mathbf{y})_\perp \Big) \sim \left( 1, \lambda, \lambda^{\frac{1}{2}} \right)L; \\
\text{soft:}&& \left( x-y \right)^\mu \sim \left(1,1,1,1\right)L. \nonumber
\end{eqnarray}
The power counting according to (\ref{positions_scaling}), as is carried out in detail in \cite{Edg14}, suggests that such UV divergences are at worst logarithmic. This is in agreement with the conclusion in momentum space, as a direct result of scale invariance.

With the knowledge of UV singularities in position space, we shall study how to identify each pinch surface through approximations. In appendix\ \ref{approximations_position_space} we derive the position-space version of hard-collinear and soft-collinear approximations; then in appendix\ \ref{pinch_surfaces_approximated_amplitudes_position_space} we discuss how these approximations can change the pinch surfaces. The enclosed pinch surfaces introduced in section\ \ref{enclosed_pinch_surface}, and the pairwise cancellation of UV divergences in section\ \ref{the_proof_of_cancellation}, then follow similarly. This appendix is not supposed to be a full-fledged explanation on every detail; rather, it will only focus on some representative examples to illustrate the main ideas.

\subsection{Approximations}
\label{approximations_position_space}

Given a Feynman integral in position space, the numerators and denominators of the integrand are polynomials in normal coordinates. As the normal coordinates are scaled as eq.\ (\ref{positions_scaling}), each such polynomial can be approximated by keeping only the leading term, corresponding to the hard-collinear and soft-collinear approximations in position space. In the text below we shall derive the forms of these approximations, and show their equivalence with those we have studied in momentum space.

Given a leading pinch surface $\sigma$, suppose $x^\mu$ are the hard vertices, $y_A^{\mu}$ are the jet vertices in the direction of $\beta_A^\mu$, and $z^\mu$ are the soft vertices. Then for any jet subgraph $J_A^{(\sigma)}$, since the distances of the external propagators are of the form $\left( y-x \right)^\mu$, and the $\beta_A$-components of $y^\mu$ are much larger than its other components, only the $\overline{\beta}_A$-components of $x^\mu$ will show up in $J_A^{(\sigma)}$ in the leading term. Similarly, for any soft subgraph $S^{(\sigma)}$, the distances of the external propagators are of the form $\left( z-y \right)^\mu$, and since all the components of its internal vertices $z^\mu$ are of the same order, only the (largest) $\beta_A$-components of $y^\mu$ will show up in $S^{(\sigma)}$. That is to say, the hard-collinear and soft-collinear approximations provided by $t_\sigma$ should read
\begin{eqnarray}
\label{hard-collinear_approximation_position}
J_A^{(\sigma)}\left( \left\{ (y-x_i)^{\alpha_i} \right\} \right)_\eta^{\left\{ \mu_i \right\}} \xrightarrow[ ]{\text{hc}_A} && J_A^{(\sigma)}\left( \left\{ \left( y-(x_i\cdot\beta_A) \overline{\beta}_A \right)^{\alpha_i} \right\} \right)_{\left\{ \nu_i \right\},\eta}\cdot \prod_{j}^{ }\overline{\beta}_A^{\nu_j} \beta_A^{\mu_j}\nonumber \\
\cdot&& \left\{\begin{matrix}
\frac{1}{2}\left( \gamma\cdot\beta_A \right) \left( \gamma\cdot\overline{\beta}_A \right)\ \ \ \ \ \ \ \text{fermion line,}\\ 
\ \ \ \ \ \ \ \ \ \ \ \ 1\text{\ \ \ \ \ \ \ \ \ \ \ \ \ \ \ otherwise, etc.}
\end{matrix}\right.\\
S^{(\sigma)}\left( \left\{ (z-y_i)^{\alpha_i} \right\} \right)^{\left \{ \mu_i \right \}} \xrightarrow[ ]{\text{sc}_A} && S^{(\sigma)} \left( \left\{ \left( z-(y_i\cdot\overline{\beta}_A) \beta_A \right)^{\alpha_i} \right\} \right)_{\left \{ \nu_i \right \}}\cdot \prod_{j}^{ }\beta_A^{\nu_j} \overline{\beta}_A^{\mu_j}.
\label{soft-collinear_approximation_position}
\end{eqnarray}
Here, as in eqs.\ (\ref{hard-collinear_approximation}) and (\ref{soft-collinear_approximation}), $\eta$ is the polarization index of the physical parton, and $\mu_i$ are the polarization indices of the scalar-polarized gauge bosons. Unlike the approximations in momentum space which act on the hard and jet subgraphs, the approximations we have in position space act on the jet and soft subgraphs.

We now show that eqs.\ (\ref{hard-collinear_approximation_position}) and (\ref{soft-collinear_approximation_position}) imply the momentum-space approximation rules (\ref{hard-collinear_approximation}) and (\ref{soft-collinear_approximation}). In other words, the approximated amplitudes $t_\sigma\mathcal{A}$ in position space are related to $t_\sigma\mathcal{A}$ in momentum space through a standard Fourier transformation. To start with, we approximate the external vertices of the soft subgraph as $\left( y_i^A\cdot\overline{\beta}_A \right)\beta_A^\mu$, where $A$ labels the jets. Then the soft subgraph in position space can be rewritten as:
\begin{eqnarray} \label{sc_position_to_momentum}
&&S\left( \left \{ \left( y_i^A\cdot\overline{\beta}_A \right)\beta_A \right \}_{\substack{i=1,...,n_A \\ A=1,...,N}} \right)\nonumber\\
&&\hspace{1cm}= \int \left( \prod_{A=1}^{N} \prod_{i=1}^{n_A} \frac{d^4p_{Ai}}{\left( 2\pi \right)^4}\right )e^{-i\sum_i^{ } \left( y_i^A\cdot\overline{\beta}_A \right) \left(\beta_A\cdot p_{Ai}\right)} S\left( \left \{ p_{Ai} \right \}_{\substack{i=1,...,n_A \\ A=1,...,N}}\right ).
\end{eqnarray}
Here we have ignored the vector indices of $S$ for simplicity, and $n_A$ is the number of soft propagators attached to $J_A$. In the last step, we have Fourier-transformed $S$ into momentum space, so $p_{Ai}$ represents the external momenta of $J_A$ that are soft in $\sigma$. The momentum-conservation delta functions have been absorbed into the definition of $S$. Eq.\ (\ref{sc_position_to_momentum}) implies that the approximated subgraph $S$ in position space is equal to the Fourier transformation of the original $S$ in momentum space, with its external vertices projected onto the attached jets. As the next step, we combine the phases of (\ref{sc_position_to_momentum}) with the approximated jet subgraph $J_A$ in position space, integrate over the vertices $y_i^A$'s, and express the remaining position dependence on $x_j^A$'s as a momentum transform, i.e.
\begin{eqnarray} \label{hc_position_to_momentum}
&&\int \left( \prod_{i=1}^{n_A} d^4y_i^A \right)e^{-i\sum_i^{ }\left( y_i^A\cdot\overline{\beta}_A \right)\left (\beta_A\cdot p_{Ai} \right)} J_A\left( \left \{ y_i^A \right \}_{i=1,...,n_A},\left \{ \left( x_j^A\cdot\beta_A\right) \overline{\beta}_A \right \}_{j=1,...,m_A} \right) \nonumber\\
\equiv&& J_A\left( \left \{ \left (p_{Ai}\cdot\beta_A \right)\overline{\beta}_A \right \}_{i=1,...,n_A},\left \{ \left( x_j^A\cdot\beta_A\right) \overline{\beta}_A \right \}_{j=1,...,m_A} \right)\nonumber\\
=&& \int \left( \prod_{i=1}^{n_A} \frac{d^4q_{Aj}}{\left( 2\pi \right)^4} \right)e^{-i\sum_j^{ }\left( x_j^A\cdot\beta_A \right)\left (\overline{\beta}_A\cdot q_{Aj} \right)} J_A\left( \left \{ \left (p_{Ai}\cdot\beta_A \right)\overline{\beta}_A \right \}_{i=1,...,n_A},\left \{ q_{Aj} \right \}_{j=1,...,m_A} \right). \nonumber\\
&&
\end{eqnarray}
Here $m_A$ is the number of scalar-polarized gauge bosons of $J_A$ that are attached to the hard subgraph. Eq.\ (\ref{hc_position_to_momentum}) implies that after we integrate over the jet-soft vertices, the approximated jet subgraph in position space becomes that in momentum space, with the incident soft momenta projected onto $\overline{\beta}_A^\mu$.

Finally we integrate over the four components of the jet-hard vertices $x_j^A$. The remaining factors in $t_\sigma\mathcal{A}$ is then equal to:
\begin{eqnarray} \label{hard_position_to_momentum}
&&\int \left( \prod_{A=1}^{N} \prod_{j=1}^{m_A} d^4x_j^A \right) e^{-i\sum_j^{ } \left( x_j^A\cdot\beta_A \right) \left(\overline{\beta}_A\cdot q_{Aj}\right)} H\left( \left \{ x_j^A \right \}_{\substack{j=1...m_A \\ A=1...N}}\right )\nonumber\\
&&\hspace{6cm}\equiv H\left( \left \{ \left( q_{Aj}\cdot\overline{\beta}_A \right)\beta_A \right \}_{\substack{j=1...m_A \\ A=1...N}}\right ),
\end{eqnarray}
which is exactly the hard subgraph of $t_\sigma\mathcal{A}$ in momentum space. Again, the momentum-conservation delta functions are absorbed into $H$. Combining eqs.\ (\ref{sc_position_to_momentum})--(\ref{hard_position_to_momentum}), we see that (\ref{hard-collinear_approximation_position}) and (\ref{soft-collinear_approximation_position}) are equivalent to (\ref{hard-collinear_approximation}) and (\ref{soft-collinear_approximation}).

One can also derive the rules for repetitive approximations in position space directly from what we have in section\ \ref{subtraction_terms}. As is shown in figure\ \ref{requirement_nontrivial_structure}, with the presence of the operator $t_{\sigma_2} t_{\sigma_1}$, the line momentum $\left( p_I+q \right)^\mu$ in $\mathcal{A}$ becomes $\left( p_I+ (q\cdot\overline{\beta}_K) (\beta_K\cdot\beta_I) \overline{\beta}_I \right)^\mu$ in $t_{\sigma_2} t_{\sigma_1} \mathcal{A}$, where $t_{\sigma_1}$ gives a soft-collinear and $t_{\sigma_2}$ a hard-collinear approximation. Then if we perform Fourier transformations to rewrite the propagators in position space, we will encounter the following factor:
\begin{eqnarray}
e^{-i\left(p\cdot\overline{\beta}_K\right) \left(\beta_K\cdot\beta_I\right) \left(\overline{\beta}_I\cdot x\right)}.
\end{eqnarray}
We can interpret the exponent from a position-space point of view: the vertex $x^\mu$ is initially projected by a hard-collinear approximation $\text{hc}_K^{(\sigma_2)}$, and then projected by a soft-collinear approximation $\text{sc}_I^{(\sigma_1)}$. In other words, for figure\ \ref{requirement_nontrivial_structure},
\begin{eqnarray} \label{repetitive_approximation_rule_position}
t_{\sigma_2}t_{\sigma_1} \left( y^\mu-x^\mu \right) = y^\mu -(x\cdot\overline{\beta}_I) (\beta_I\cdot\beta_K) \overline{\beta}_K^\mu.
\end{eqnarray}
This rule of repetitive approximations is in agreement with the prescription given in \cite{EdgStm15}, and through a direct calculation, it satisfies the requirements, eqs.\ (\ref{requirement1}) and (\ref{requirement2}).

\subsection{Pinch surfaces of the approximated amplitudes}
\label{pinch_surfaces_approximated_amplitudes_position_space}

The analysis in the previous subsection ensures that one can study the pinch surfaces of the approximated amplitudes in a way similar to what we had in section\ \ref{pinch_surface_amplitudes_approximations}. Rather than giving a detailed discussion as we have in the text, we will only list several heuristic examples of $\rho^{\left\{ \sigma \right\}}$, and see how they are interpreted in position space. In order to make our analysis general, we will assume that $\beta_I^\mu$ and $\beta_K^\mu$ are not back-to-back, i.e. $\beta_I^\mu \neq \overline{\beta}_K^\mu$.

\begin{figure}[t]
\centering
\includegraphics[width=12cm]{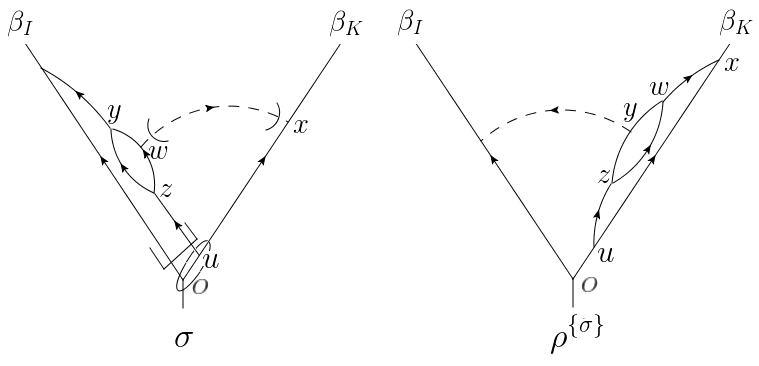}
\caption{An example of the pinch surfaces $\sigma$ (of $\mathcal{A}$) and $\rho^{ \left\{\sigma\right\} }$ (of $t_\sigma \mathcal{A}$), where the vertices $w^\mu$, $y^\mu$ and $z^\mu$ are in alignment to $\overline{\beta}_I^\mu$ rather than $\beta_K^\mu$ in $\rho^{ \left\{\sigma\right\} }$.}
\label{regular_pinch_surface1_position}
\end{figure}
The first example is figure\ \ref{regular_pinch_surface1_position} (the momentum-space version has been studied in the upper row of figure\ \ref{regular_pinch_surface1}), where $\sigma$ is the pinch surface of $\mathcal{A}$ and $\rho^{ \left\{\sigma\right\} }$ that of $t_\sigma \mathcal{A}$. Here we claim that in $\rho^{\left\{ \sigma \right\}}$ where $xw$, $wy$, $yz$ and $wz$ are all lightlike propagators, $x^\mu$ and $u^\mu$ are pinched in alignment to $\beta_K^\mu$, while $w^\mu$, $y^\mu$ and $z^\mu$ are pinched in alignment to $\overline{\beta}_I^\mu$, as noted in \cite{EdgStm15}. To see this, we focus on the relevant factors that appears in the denominator:
\begin{eqnarray}
&&\left[ -\left(\left (x\cdot\overline{\beta}_K \right)\beta_K-\left (w\cdot\overline{\beta}_I \right)\beta_I\right)^2 +i\epsilon \right] \left[ -\left(y-w\right)^2 +i\epsilon \right]\left[ -\left(z-w\right)^2 +i\epsilon \right]\nonumber\\
\cdot&&\left[ -\left(y-z\right)^2 +i\epsilon \right]\left[ -\left( z-(u\cdot\beta_I)\overline{\beta}_I \right)^2 +i\epsilon \right] \left[ -\left( x-(u\cdot\beta_K)\overline{\beta}_K \right)^2 +i\epsilon \right] \left[ -u^2+i\epsilon \right]. \nonumber\\
&&
\end{eqnarray}
Since these denominators are associated with jet propagators in $\rho^{ \left\{\sigma\right\} }$, each term in the bracket should vanish at the pinch surface. A direct analysis shows that given $x^\mu = \left( x\cdot\overline{\beta}_K \right) \beta_K^\mu$ in $\rho^{ \left\{\sigma\right\} }$ this happens for
\begin{align} \label{vertex_pinched_position}
\begin{split}
y^\mu &= \left( y\cdot\beta_I \right) \overline{\beta}_I^\mu;\ z^\mu = \left( z\cdot\beta_I \right) \overline{\beta}_I^\mu;\ w^\mu = \left( w\cdot\beta_I \right) \overline{\beta}_I^\mu;\ u^\mu = \left( u\cdot\overline{\beta}_K \right) \beta_K^\mu.
\end{split}
\end{align}
As a direct result, the propagators $wy,\ wz,\ yz\ \text{and}\ zu$ are all parallel to $\overline{\beta}_I^\mu$ in $\rho^{\left \{ \sigma \right \}}$. Note that although $u^\mu$ is in alignment with $\beta_K^\mu$, only its $\overline{\beta}_I$-component appears in the distance of the propagator $zu$, so $zu$ is also lightlike. This is in agreement with our crucial observation in momentum space, which was stated as Theorem 1 in section\ \ref{pinch_surfaces_from_single_approximation}.

\begin{figure}[t]
\centering
\includegraphics[width=12cm]{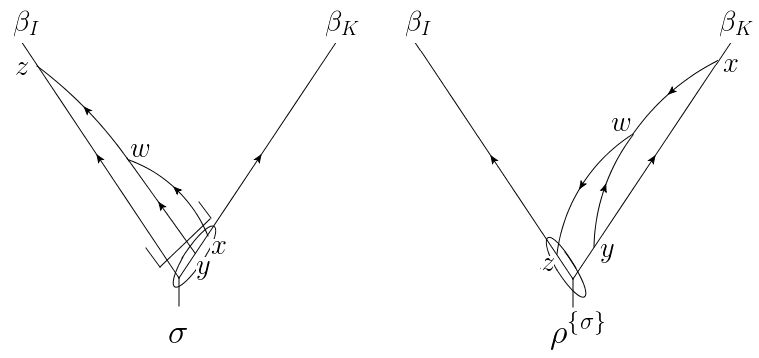}
\caption{Another example on $\rho^{\left\{\sigma\right\}}$ in position space.}
\label{regular_pinch_surface2_position}
\end{figure}
Another example is in figure\ \ref{regular_pinch_surface2_position}, where we claim that $x^\mu$, $y^\mu$ and $w^\mu$ are all in alignment to $\overline{\beta}_I^\mu$. To see this, we focus on the denominators of $t_\sigma \mathcal{A}$ that involve $w$ and $y$, which reads:
\begin{eqnarray}
\left[ -\left(z-w\right)^2 +i\epsilon \right] \left[ -\left(w-\left( x\cdot\beta_I \right)\overline{\beta}_I \right)^2 +i\epsilon \right] \left[ -\left(w-\left( y\cdot\beta_I \right)\overline{\beta}_I\right)^2 +i\epsilon \right] \left[ -y^2+i\epsilon \right].\nonumber\\
\end{eqnarray}
Again, since these denominators are associated with jet propagators in $\rho^{ \left\{\sigma\right\} }$, each term in the bracket should vanish. Given $x^\mu = \left( x\cdot\overline{\beta}_K \right) \beta_K^\mu$ in $\rho^{ \left\{\sigma\right\} }$, we see that this occurs for
\begin{align} \label{regular_pinch_surface2_position_conclusion}
\begin{split}
y^\mu &= \left( y\cdot\overline{\beta}_K \right) \beta_K^\mu;\ \ w^\mu = \left( w\cdot\beta_I \right) \overline{\beta}_I^\mu;\ \ z^\mu = 0.
\end{split}
\end{align}
As a direct result, the propagators $wx$, $wy$ and $wz$ are all parallel to $\overline{\beta}_I^\mu$ in $\rho^{ \left\{\sigma\right\} }$. Note that only the $\overline{\beta}_I$-components of $x^\mu$ and $y^\mu$ enters the expression of $wx$ and $wy$, so we have the conclusion above.

Figures\ \ref{regular_pinch_surface1_position} and \ref{regular_pinch_surface2_position} are the position-space examples of the regular configurations of $\rho^{\left\{ \sigma \right\}}$, and we can also study the exotic configurations. An example of the soft-exotic configuration is shown in figure\ \ref{soft-exotic_example_position}, whose momentum-space version was studied in the upper row of figure\ \ref{soft-exotic_example}. After the action of soft- and hard-collinear approximations, the denominators of $t_\sigma \mathcal{A}$ which involve $w^\mu$ read:
\begin{figure}[t]
\centering
\includegraphics[width=12cm]{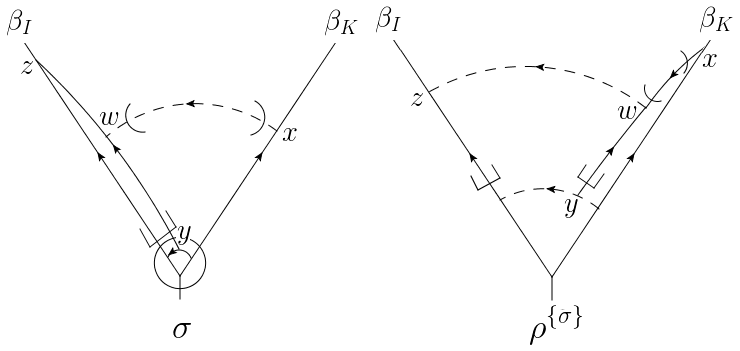}
\caption{The position-space version of soft-exotic configurations.}
\label{soft-exotic_example_position}
\end{figure}

\begin{eqnarray} \label{soft-exotic_denominator_position}
\left [ -\left( \left( x\cdot\overline{\beta}_K \right)\beta_K-\left( w\cdot\overline{\beta}_I \right)\beta_I \right)^2+i\epsilon \right ]\left[ -\left( w-z \right)^2+i\epsilon \right] \left[ -\left( w-\left( y\cdot\beta_I \right)\overline{\beta}_I \right)^2+i\epsilon \right].\nonumber\\
\end{eqnarray}
In order to assure that the approximated denominators in eq.\ (\ref{soft-exotic_denominator_position}) corresponding to the lines $wx$ and $wy$ both vanish at $\rho^{ \left\{\sigma\right\} }$, we only need $w^\mu= \left( w\cdot\beta_I \right) \overline{\beta}_I^\mu$, even if $w\cdot\beta_I \neq 0$. Meanwhile, the position of $y^\mu$ is not constrained in $\rho^{ \left\{\sigma\right\} }$. In other words, $y^\mu$ is a soft vertex although it is an endpoint of the jet line $wy$. This is the position-space interpretation of the soft-exotic configuration.

\begin{figure}[t]
\centering
\includegraphics[width=11.5cm]{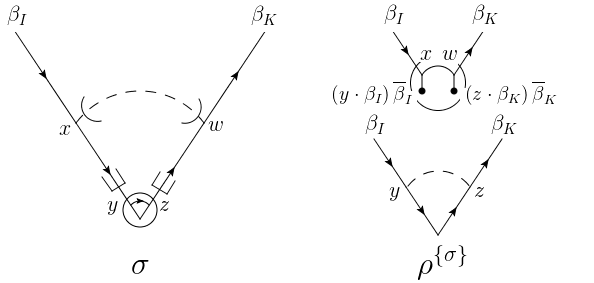}
\caption{The position-space version of hard-exotic configurations.}
\label{hard-exotic_PS_example_position}
\end{figure}

Figure\ \ref{hard-exotic_PS_example_position} provides an example of the hard-exotic configuration. In $\sigma$, the propagator $xw$ is soft, $xy$ is lightlike in the direction of $\beta_I^\mu$ and $wz$ is lightlike in the direction of $\beta_K^\mu$. Meanwhile, they all become hard in $\rho^{ \left\{\sigma\right\} }$. According to the hard-collinear approximations in $t_\sigma$, $y^\mu$ ($z^\mu$) is projected onto $\overline{\beta}_I^\mu$ ($\overline{\beta}_K^\mu$) in $xy$ ($wz$), but unchanged in the other propagators. So we can express $\rho^{ \left\{\sigma\right\} }$ as the RHS of figure\ \ref{hard-exotic_PS_example_position}, which yields the following factor in the denominator of $t_\sigma \mathcal{A}$:
\begin{eqnarray}
&&\left [ -\left( x-\left( y\cdot\beta_I \right)\overline{\beta}_I \right)^2+i\epsilon \right ]\left( -y^2+i\epsilon \right)\nonumber\\
&&\hspace{0.8cm} =\left[ 2\left( x\cdot\overline{\beta}_I \right)\left( \left( y-x \right)\cdot\beta_I \right) +\mathbf{x}_{I\perp}^2+i\epsilon \right ]\left [ -2\left( y\cdot\overline{\beta}_I \right)\left( y\cdot\beta_I \right) +\mathbf{y}_{I\perp}^2+i\epsilon \right ].
\end{eqnarray}

Since $xy$ is hard in $\rho^{ \left\{\sigma\right\} }$, we have $\left( x-\left( y\cdot\beta_I \right)\overline{\beta}_I \right)^\mu \sim \mathcal{O}(\lambda)$, so $x\cdot\beta_I \sim \mathcal{O}(\lambda)$. Then we examine the poles in the $y\cdot\beta_I$-plane. There are two poles:
\begin{eqnarray}
x\cdot\beta_I- \frac{\mathbf{x}_{I\perp}^2 +i\epsilon}{2x\cdot \overline{\beta}_I}\text{, and }\frac{\mathbf{y}_{I\perp}^2 +i\epsilon} {2y\cdot \overline{\beta}_I}.
\end{eqnarray}
Clearly a pinch is formed when $y^\mu= \left( y\cdot\overline{\beta}_I \right)\beta_I^\mu$ is finite, and $y\cdot \overline{\beta}_I$ is of the same sign as $x\cdot \overline{\beta}_I$. Both poles are of order $\lambda$ from the origin if $\mathbf{y}_{I\perp}^2 \sim\lambda$, as appropriate for a collinear pinch surface for $y^\mu$. Similarly $z^\mu= \left( z\cdot\overline{\beta}_K \right)\beta_K^\mu$. This implies a pinch surface structure ``inside'' $H^{(\sigma)}$, and renders the position-space interpretation of the hard-exotic configuration. These interpretations in figures\ \ref{regular_pinch_surface1_position}--\ref{hard-exotic_PS_example_position} are compatible with the results in section\ \ref{pinch_surfaces_from_single_approximation}, and are fundamental in proving the pairwise cancellation of UV divergences in the position-space forest formula.

\bibliographystyle{JHEP}
\bibliography{refs}% refs is the name of your reference file (no need to write the .bib extension)

\end{document}